Imperial College London
Department of Computing

# Investigations into Elasticity in Cloud Computing

Rui Han

Supervised by Professor Yi-ke Guo

# Copyright Declaration

# Abstract

The pay-as-you-go model supported by existing cloud infrastructure providers is appealing to most application service providers to deliver their applications in the cloud. Within this context, elasticity of applications has become one of the most important features in cloud computing. This elasticity enables real-time acquisition/release of compute resources to meet application performance demands. In this thesis we investigate the problem of delivering cost-effective elasticity services for cloud applications.

Traditionally, the **application level elasticity** addresses the question of how to scale applications up and down to meet their performance requirements, but does not adequately address issues relating to minimising the costs of using the service. With this current limitation in mind, we propose a scaling approach that makes use of cost-aware criteria to detect the bottlenecks within multi-tier cloud applications, and scale these applications only at bottleneck tiers to reduce the costs incurred by consuming cloud infrastructure resources. Our approach is generic for a wide class of multi-tier applications, and we demonstrate its effectiveness by studying the behaviour of an example electronic commerce site application.

Furthermore, we consider the characteristics of the algorithm for implementing the business logic of cloud applications, and investigate the **elasticity** at the **algorithm level**: when dealing with large-scale data under resource and time constraints, the algorithm's output should be elastic with respect to the resource consumed. We propose a novel framework to guide the development of elastic algorithms that adapt to the available budget while guaranteeing the quality of output result, e.g. prediction accuracy for classification tasks, improves monotonically with the used budget. We demonstrate the application of the framework by developing two elastic data mining algorithms as examples. Experimental evaluations have been performed using prediction accuracy as the quality measure on real datasets. The results show that both algorithms indeed exhibit consistent increase in quality.

# Acknowledgements

Without the support and contributions from the following people, this thesis would not be possible:

My supervisor, Professor Yi-ke Guo, for his great patience, profound expertise, inspiration and perceptive comments.

Great thanks to all my lovely friends and colleagues in Imperial College London, Professor Moustafa M. Ghanem, Dr. Li Guo, Mr. Lei Nie, Dr. Roy Clements, Dr. Michelle Osmond, Mr. Shicai Wang, Ms. Shuri Pentu, Ms. Xian Yang, Ms. Shulin Yan, Mr. Xiangchuan Tian, Dr. Yajie Ma, Ms. Michelle Osmond, Ms. Diana O'Malley, Mr. Florian Guitton, Mr. Yang Li, Mr. Chun-Hsiang Lee, Mr. Weikun Wang, Mr. Fangjing Hu, Mr. Andreas Williams, Mr. Pavlos Mitsoulis-Ntompos, Dr. Chao Wu, Mr. Dilshan Silva, Mr. Orestis Tsinalis, Ms. Xinyu Liu, Mr. Xiaoping Fan, Dr. Anthony Rowe, Dr. Ioannis Pandis, Ms. Mihaela Mabes, and Ms. Tania Buckthorp. They have provided much help and brought much happiness to my life at Imperial.

至我最亲爱的父亲、母亲和妻子

To my dearest Baba, Mama, and Wife

"A person who never made a mistake never tried anything new."

"We cannot solve our problems with the same thinking we used when we created them."

"It's not that I'm so smart, it's just that I stay with problems longer."

"Anyone who doesn't take truth seriously in small matters cannot be trusted in large ones either."

"If you can't explain it simply, you don't understand it well enough."

"Learn from yesterday, live for today, hope for tomorrow. The important thing is not to stop questioning."

Albert Einstein (1879-1955)

# Contents

# List of Tables

# List of Figures

# Chapter 1

# Introduction

## 1.1   Background

Cloud computing has gained unquestionable commercial success in recent years and will continue its rapid development over the next decade. It refers to a new paradigm for delivering on-demand and elastic compute resources from hardware to system software via the internet in a cost-effective manner supported by virtualization and data center technologies [1]. As the internet becomes faster, more reliable and more ubiquitous, the pay-as-you-go model supported by existing Infrastructure-as-a-Service (IaaS) cloud providers is appealing to most application owners (application service providers that consume cloud infrastructure resources), because it removes the costs of buying, installing and maintaining a dedicated infrastructure for running their application. Moreover, most IaaS providers allow the application owners to scale up and down the resources used based on the performance demands of their applications, thus letting them pay only for the amount of resources they use. This model is appealing for deploying applications that provide services for third parties, e.g. traditional e-commerce sites, financial services applications, online healthcare applications, gaming applications, media servers and bioinformatics applications.

Within this context, elasticity (on-demand resource provision) of applications has become one of the most important features of a cloud platform. This elasticity enables application owners to acquire and release resources on demand so that they are billed only for the resources they use. In this thesis, we explore theoretical and empirical work on cloud elasticity management that addresses topics of interest to application owners in delivering cost-effective application services in the cloud.

## 1.2   Challenges in Elasticity Management for Cloud Applications

Addressing the issue of elasticity management effectively requires taking a closer look at the structure of most common services and applications deployed in IaaS clouds to provide services to other parties. Such applications are typically implemented as multi-tier applications running on distributed software platforms, i.e. server components of the applications. For example, in an e-commerce site, there are at least three tiers to be scaled: a frontend web server for handling HTTP requests; a middle-tier application server for implementing business logic; and a backend database with data store and processing. In the IaaS providers' cloud, each of these tiers can be implemented as a number of Virtual Machine (VM) instances used to host the applications' server components. These IaaS providers charge the application owners for using these VMs per unit time (e.g. hour or minute).

### 1.2.1   Application level elasticity

Many applications hosted in a cloud environment may have varying resource demands as they are expected to serve a wide range of incoming requests. Those involve different *volumes* and *types* of workloads. First, the volume of workload fluctuates periodically (e.g. the request arrival rate is high at 1:00 pm and low at 1:00 am) or due to external events (e.g. the incoming requests increase due to the release of some promotional activities). Secondly, the *types* of workload differ depending on the behaviour of end users (customers of application services). For example, in an e-commerce site, if webpage browsing is the end users' main action over a period of time, web servers might become stressed and their resources saturated, but the application and database servers continue to perform well. By contrast, if a large number of users make orders because of some discount or special offer, the application and database servers might become stressed and their number needs to be increased as a result. Thus, depending on the volume and type of business involved, servers at each tier of the application can be stressed by heavy workloads, or can become idle due to light workloads.

In this context, **application level elasticity** denotes the scaling of an application up and down in order to meet its Quality of Service (QoS) requirements such as response time. Specifically, if the workload of a service increases (e.g. more end users start submitting requests at the same time), the application owner ideally can scale up the resources used to maintain the QoS. On the other hand, when the workload eases down, they can then scale the resources used down to save deployment costs.

Although some existing scaling techniques [2-15] address the question of how to maintain an applications' QoS, they rarely consider the equally important aspect of cloud applications — the cost of using the resources. Applications deployed in a cloud environment require both good performance and cost-efficient resource usage. Hence, the key challenge in cost-effective elasticity at the **application level**, is to consider workload changes in both volume and type, as well as resource costs in scaling, thus managing the bottlenecks in the application to be scaled. The target of such cost-efficient elasticity is to minimise application owners' operational costs while still meeting the desired application performance.

## 1.2.2   Algorithm level elasticity

The problem of elasticity management is further complicated when the characteristics of the algorithm for implementing the business logic of cloud applications are considered. Traditional *application level elasticity* assumes that there is a one-off answer in the algorithm's output — either it produces a result or it fails to do so. This assumption needs to be revisited when dealing with large-scale datasets under resource and time constraints. For example, large e-commerce sites such as Amazon.com (www.amazon.com), CDNOW (www.cdnow.com), eBay (www.ebay.com), and Moviefinder.com (www.moviefinderonline.com/) offer millions of products to end users. Recommender systems have been widely deployed in these e-commerce sites to guide users to items they are interested in and thereby increase sales. Today's recommender systems usually need to handle hundreds of thousands of requests per second, and each request is required to be given a satisfactory recommendation within a specified short real-time period. However, existing recommendation algorithms such as collaborative filtering (CF) need to scan millions of potential items in the database and cannot give an exact answer within the given time budget.

To deal with such problems, end users are usually willing to accept approximate results produced using their available time budget. Typically, such approximate results can be produced either by restricting the size of the input data fed to exact algorithms, or by using approximating algorithms over full datasets. Elasticity management at the **algorithm level**, coupled with the pay-as-you-go cloud business models, give rise to various new challenges about how we design programs and algorithms in order to use such approximations efficiently and successfully. The first is to investigate trade-offs between the quality of output result, e.g. the prediction accuracy in recommendations, and available resource and time budgets. The second, and more important, challenge is to organise the computation of the algorithm to produce a result whose quality, based on some metric, improves monotonically with the consumed time budget. Finally, when more budget is available, the algorithm should allow users to obtain a refined result by starting from a previously obtained result. Less computation should be needed when staring from a result of better quality.

## 1.3 Contributions Offered by This Work

This thesis seeks to make a number of useful contributions, to the need for elasticity of cloud computing at both application and algorithm levels. These are:

**Application level elasticity**. We propose a scaling approach that is both cost-aware and workload-adaptive, thus allowing application owners to perform more efficient application elasticity management in the cloud. This approach features three key elements:

- *Cost-aware criteria*: a flexible analytical model is developed to capture the behavior of multi-tier applications. Cost-aware criteria are introduced to measure the effect of cost of resources on every unit of response time.

- *Workload-adaptive scaling*: using the above criteria, a Cost-Aware Scaling (CAS) algorithm is designed to handle changing workloads of multi-tier applications by adaptively scaling up and down bottlenecked tiers within applications.

- *Experimental evaluation*: the proposed cost-aware scaling approach is implemented and tested. The test results show: (1) the CAS algorithm responds to changing workloads effectively by scaling applications up and down appropriately to meet their QoS requirements; (2) deployment costs are reduced compared to other scaling techniques.

**Algorithm level elasticity**. We investigate the concept of algorithmic elasticity in the context of data mining, which has been applied in a large number of commercial fields, including insurance, health care, banking, and marketing. A generic approach for developing elastic data mining algorithms that enables quality monotonicity with respect to allocated time budgets is proposed. The concrete contributions of the presented approach are:

- *Formal definition of elastic algorithms*: we define a class of elastic algorithms that generate a range of approximate results whose result quality, based on the same measure, is proportional to their resource consumption. We formally define the properties of elastic algorithm and discuss the meaning of elasticity in the context of algorithmic elasticity.

- *A generic framework for designing quality-monotonic algorithms*: we provide a generic framework for designing quality-monotonic elastic algorithms. Such algorithms are suitable for a wide class of data mining problems. Our framework for designing such algorithms comprises two components: a *coding component* and a *mining component*. The coding component applies compression techniques to proactively map a dataset into a set of codes with smaller lengths; that is, those require shorter processing time. By processing a code with suitable length, the mining component can then produce a useful approximate result within a specified time budget. Moreover, we use an information-theory approach to define a **resolution** measure that permits an examination of how allocated budget affects the resolution of a code. We then define the key property that the coding component must meet to support a quality-monotonic data mining algorithm. This property indicates a code needing a longer processing time but which has a higher resolution, thus producing a result with better quality.

- *Case studies of developing two elastic mining algorithms*: we demonstrate the validity and practicality of our approach by designing two elastic mining algorithms. The core of the algorithms uses standard naïve k-Nearest-Neighbour (kNN) classification [16] or neighbourhood-based CF [17] over an R-tree coding component [18]. The codes produced by the R-tree are the nodes at different depths that successively approximate the training set at different levels of granularity. We present extensive experimental evaluation of real datasets to demonstrate that both algorithms indeed exhibit consistent increase in quality, i.e. prediction accuracy. Furthermore, we also compare the performance of each algorithm with existing time-adaptive mining algorithms. The comparison results show that the elastic algorithms not only demonstrate a steady improvement of quality but also produce better overall qualities for the majority of results when using the same or slightly smaller computational costs.

## 1.4   Thesis Organisation

Chapter 2 presents basic concepts in cloud computing useful for the analysis of challenges in application level elasticity. It also introduces the elasticity management of cloud applications and provides a detailed overview of the properties of typical multi-tier applications in the cloud. This chapter then describes the design and implementation of the imperial Smart Scaling engine (iSSe) implemented to support the dynamic scaling of such applications. Finally, it illustrates the challenges to be addressed in application level elasticity by describing some current examples and scaling techniques.

Chapter 3 explains the proposed CAS algorithm and its details. By applying queueing systems to model such applications, cost-aware criteria are designed to analyse the effect of cost on every unit of response time. Based on these cost-aware criteria, the CAS algorithm is developed to lower cost by detecting the bottlenecks in multi-tier applications and scaling up or down only at these tiers. By testing the scaling behaviours of an example e-commerce site, we evaluate the effectiveness of the proposed scaling approach and compare its performance with traditional scaling approaches.

Chapter 4 first elaborates on the reason for investigating algorithmic elasticity. We also review some existing methodologies for designing time-adaptive algorithms that are, traditionally used in real-time systems and, that can provide insights on how to design elastic algorithms for cloud computing. Finally, we present a formal definition for a class of elastic algorithms and their key properties.

Chapter 5 proposes a generic framework for developing elastic mining algorithms by defining their two components and associated properties. An elastic kNN algorithm consisting of an R-tree coding component and a naïve kNN classification component is developed according to this framework. We give formal proofs that the R-tree coding component satisfies the property that supports the quality-monotonicity of the elastic kNN algorithm. We also present experimental evaluation of a list of real datasets to demonstrate that the algorithm indeed guarantees the steady increase of quality (prediction accuracy) to resource consumption.

Chapter 6 presents a typical application of an elastic data mining algorithm in delivering elastic recommendation services in e-commerce sites. An elastic CF algorithm is developed by using a basic neighbourhood-based CF algorithm over an R-tree coding method that hierarchically aggregates user rating information. The effectiveness of the elastic CF algorithm is evaluated by extensive comparative experiments on large datasets with millions of ratings in a variety of settings.

Finally, Chapter 7 summarises this work and present a discussion of directions for future research on cloud elasticity management.

## 1.5    Statement of Originality

I Rui Han declare that this thesis is my own work. All use of the previously published and unpublished work of others has been acknowledged in the text and references are given in the bibliography.

## 1.6    Publications

The following publications arose during the course of this PhD study. I mark the publications where I am the first author using *. I also indicate how these papers fit into the contents of this thesis, however the research on workflow technology is not described in the thesis.

**Application level elasticity**.

**A Deployment Platform for Dynamically Scaling Applications in the Cloud**\*. In: *The IEEE Third International Conference on Cloud Computing Technology and Science (CloudCom 2011), Athens,*

*Greece, 2011.* Han et al. [19].

This paper presents a deployment platform to simply the process of deploying and scaling applications in cloud environments. Using an e-commerce site as an example, we discuss how the automatic deployment process is supported using different service components in the platform. This work is presented in Chapter 2.

**Enabling cost-aware and adaptive elasticity of multi-tier cloud applications**\*. In: *Future Generation Computer Systems (Impact factor: 1.864), Elsevier, North-Holland, 2012*. Han et al. [20].

This paper proposes an elasticity management approach that makes use of cost-aware criteria to detect and manage the bottlenecks within multi-tier cloud-based applications. Using queueing system as an analytical model, we present a workload-adaptive scaling algorithm that reduces the costs incurred by users of cloud infrastructure resources, allowing them to scale their applications only at bottleneck tiers. This work is presented in Chapter 3.

**Lightweight Resource Scaling for Cloud Applications**\*. In: *The 12th IEEE/ACM International Symposium on Cluster, Cloud and Grid Computing (CCGrid 2012), Ottawa, Canada, 2012*. Han et al. [21].

Motivated by the problems that traditional VM-level scaling typically incurs both considerable resource overheads and extra management costs, especially for applications with rapidly fluctuating demands. This paper proposes a lightweight approach to enable the fine-grained scaling of cloud applications at the hardware resource level itself (CPUs, memory, I/O, etc) in addition to VM-level scaling. This work is discussed at the end of Chapter 3.

**Elastic−TOSCA: Supporting Elasticity of Cloud Application in TOSCA**\*. In: *The Fourth International Conference on Cloud Computing, GRIDs, and Virtualization (CLOUD COMPUTING 2013). Valencia, Spain, 2013*. Han et al. [22].

This paper enriches Topology and Orchestration Specification for Cloud Applications (TOSCA), which is an emerging framework aims to enhance the portability of cloud applications, and defines Elastic-TOSCA to support the dynamic scaling of cloud applications. In the Elastic-TOSCA framework, we also provide a detailed example to describe how Elastic-TOSCA can be used to support easily a dynamic scaling approach based on a queueing system model. This work is presented in Chapter 2.

**Programming Directives for Elastic Computing**. In: *IEEE Internet Computing (Impact factor: 2.00), IEEE Computer Society, 2012*. Dustdar et al. [23].

This paper describes how the TOSCA can be integrated with Simple-Yet-Beautiful Language (SYBL), which defines possible directives and runtime functions used for dynamic scaling of cloud applications, to enhance the elasticity control of cloud applications.

**Algorithm level elasticity**.

**Does the Cloud need new algorithms? An introduction to elastic algorithms**. In: *The IEEE 4th International Conference on Cloud Computing Technology and Science (CloudCom 2012). Taiwan, 2012*. Guo et al. [24].

The extended journal version of this paper invited by the conference organiser: **Towards Elastic Algorithms as a New Model of Computation for the Cloud**. In: *International Journal of Next-generation Computing, Taiwan, 2013* [25].

In this paper, Professor Yike Guo, Professor Moustafa M. Ghanem, and I for the first time introduce the elastic algorithm in which the computation itself is organised in a "pay-as-you-go" fashion. In contrast to conventional algorithms, in which computation is a deterministic process that only produces an "all-or-nothing" result, an elastic algorithm can generate a range of approximate results corresponding to its resource consumption. As more resources are consumed, better results will be derived. In the paper, we formalise the properties of elasticity and also formalise the desirable properties for elastic algorithms themselves. This work is presented in Chapter 4.

**Elastic Algorithms for Guaranteeing Quality Monotonicity in Big Data Mining**\*. In: *the IEEE International Conference on Big Data 2013 (IEEE BigData 2013), Santa Clara, CA, USA, 2013*. Han et al. [26].

This work proposes a framework for developing elastic data mining algorithms. Based on Shannon's entropy, an information-theory approach is introduced to investigate how result quality is affected by the allocated budget. This is then used to guide the development of algorithms that adapt to the available time budgets while guaranteeing better quality results if the budget is increased. This work is presented in Chapter 5.

**Developing Anytime SVM Training Algorithms for Large-Scale Data Classification**\*. In: *2014*

*International Conference on Artificial Intelligence and Software Engineering (AISE2014), Phuket, Thailand, 2014*. Han et al. [27].

This paper presents an Anytime Programming Library (APL) to simplify the development of anytime support vector machines (SVM) training algorithms. The effectiveness of APL is demonstrated by developing three different anytime SVM training algorithms and experiments have been conducted to evaluate the effectiveness of these algorithms. This work is discussed at the end of Chapter 5.

**Workflow technology**.

**Formal Modelling and Performance Analysis of Clinical Pathway***. In: *NETTAB 2011 workshop focused on Clinical Bioinformatics. Pavia, Italy, 2011*. Han et al. [28].

The extended journal version of this paper invited by the conference organiser: **Modelling and performance analysis of clinical pathways using the stochastic process algebra PEPA**. In: *BMC bioinformatics (Impact factor: 3.02), Jo Appleford-Cook, 2012*. [29].

This paper introduces a clinical pathway management approach, whose core element is the stochastic model Performance Evaluation Process Algebra (PEPA). PEPA can unambiguously describe a variety of elements in a clinical pathway. Using PEPA, the clinical pathway can be quantitatively analysed and this analysis can provide useful information to facilitate clinical pathway management. A real-world stroke clinical pathway, obtained from Charing Cross hospital of Imperial College London, is employed to demonstrate the effectiveness of the approach.

**Applying genetic algorithm to optimise personal worklist management in workflow systems**. In: *International Journal of Production Research (Impact factor: 1.460). Taylor & Francis, 2013*. Ren et al. [30].

This paper proposes an approach that applies genetic algorithm (GA) to manage activity instances in personal worklists in workflow management systems. The approach is applied according to activity instances' probabilities of satisfying deadlines and costs of violating deadlines. The approach can schedule activity instances among multiple executors' personal worklists and recommend to each executor a list of activity instances that can be successfully executed, while minimising the overall deadline violation cost. The effectiveness of our approach is demonstrated using real-world data collected from three manufacturing enterprises.

# Part I.

# Application Level Elasticity

# Chapter 2

# Cloud Computing and Application Level Elasticity

## 2.1 Introduction

In this chapter, we first introduce some basic concepts of cloud computing in Section 2.2 and formally define multi-tier applications in Section 2.3. We then explain the three roles in elasticity management of cloud applications and discuss the basic target of application level elasticity in Section 2.4. Next, Section 2.5 introduces a platform developed to support the dynamic scaling of cloud applications, and Section 2.6 discusses existing scaling techniques for cloud applications. Finally, Section 2.7 illustrates the need for developing a new approach to address two key challenges in cost-effective elasticity at the application level.

## 2.2 Basic Concepts of Cloud Computing

Cloud computing is an internet-based access model that applies virtualization and data center technologies to provide on-demand and highly scalable compute resources ranging from hardware (e.g. processing powers, storage, and network bandwidths) to system software (e.g. operation systems and middleware) for cloud consumers [31]. A cloud is an elastic execution environment that supports multiple application owners to host and deliver application services by consuming metered virtualized resources with minimal management effort [32, 33]. In the following, we explain several basic concepts of cloud computing, including virtualization technology, service model, deployment model, and resource pricing scheme.

**Virtualization technology**

Virtualisation, or hardware virtualisation, is a key enabling technology in cloud computing. This technology usually takes place in an actual *physical machine* (PM) (i.e. host machine), in which multiple VMs (i.e. guest machines) can be created and the software executed on these VMs are independent of the underlying hardware environment. Each VM executes like a real computer on an operating system such as Linux Ubuntu or Microsoft Windows. Typically, virtualization technology can be divided into three types. First, *full virtualization* provides the complete simulation of hardware environment to support the running of an isolated and unmodified operating system. Second, *partial virtualization* simulates part of the actual hardware environment and a modified operating system can run in this simulated environment. Third, *paravirtualization* does not simulate hardware but isolates address space; that is, each VM is allocated a separated address to run.

In mainstream cloud platforms, full virtualization is the most widely applied technique and partial virtualization is usually used to improve the performance of virtualisation. For example, Amazon Web Service (AWS) [2] employs Xen [34], a hypervisor or virtual machine monitor (VMM), to support the simultaneous creation and execution of multiple VMs at small overheads. Using the full virtualization technology, Xen enables complete isolation between different VMs and allows them to share virtualized hardware resources safely.

With virtualization technology, the pools of PMs within a cloud data center can support a large number of VMs and provide centralised administration of all these VMs. In the data center, a VM can be controlled more flexibly than a PM because the VM's configurations including CPU number, memory size and I/O capacity can be easily adjusted; the VM can be added/removed as needed; the VM can be migrated from one PM to another one with small overheads; and all the VMs are allowed to share the storage device in the data center. Thus, the provision of VMs in cloud environments can achieve high resource utilisation, high availability, and low cost of compute resources.

**Service models**

At present, services in cloud computing can be offered according to three models: IaaS, platform as a service (PaaS), and software as a service (SaaS).

*IaaS* is the most basic and popular model of provisioning low-level compute resources such as VMs, storage, and network resources (bandwidth and IP address). For example, AWS [2] provides a library of VM disk images in Amazon Elastic Compute Cloud (Amazon EC2) and online storage in Amazon Simple Storage Service (Amazon S3). Cloud users usually consume IaaS services in a pay-as-you-go fashion: the cost of service is decided by the amount of allocated resources (e.g. the size of a VM or the storage space) and the consumed time units (e.g. 10 hours).

In the *PaaS* model, cloud providers such as Google App Engine [35] and Windows Azure Cloud Services [36] build a platform to integrate both hardware resource and system software. PaaS services are mainly designed to assist programmers to develop, run, and scale their software applications conveniently and relieve them from the complexity of managing the underlying hardware and software resources.

In the *SaaS* model, cloud providers directly provide services of application software to end users and dynamically scale up and down the application at run-time to meet users' changing demand. For example, Google Apps [37] offers a wide range of web-based applications such as Gmail and Google Docs.

**Deployment models**

Typically, a cloud can be deployed using four models:

A *public cloud* makes all its resources and applications available for open use by the general public via the internet.

A *private cloud* is built to share resources among multiple consumers in a single organisation. Either the organisation or a third part can manage and operate this cloud.

A *community cloud* is constructed for several organisations with common demands to share their infrastructures.

A *hybrid cloud* is a combination of two or three clouds (public, private, or community clouds) in order to take the advantage of multiple deployment models.

**Resource pricing scheme in IaaS cloud**

In the cloud, there are currently two dominant ways for IaaS providers to provide VMs under a *fixed* pricing scheme. The first *on-demand* VM covers the case of occasional and short-term needs for resources. Users are charged based on the amount of time they consume the resources, e.g. pay by hour. In addition, the second *reserved* VM is design for the long-term use of resources. Users need to pay an upfront fee to make a long-term subscription such as one or three years, thereby getting some discount in the hourly usage fee of the reserved VM.

From the perspective of cloud IaaS providers, after having met the resource requirements of *on-demand* and *reserved* VMs, they may still have some unused capacity. These providers therefore offer consumers a *third* spot pricing scheme so as to make profit from their excess compute capacity. Consumers can use a spot VM once their bid equals or exceeds the dynamically changing spot price and they only need to pay their bidding price for using this VM. From the perspective of cloud consumers, the main benefit of choosing the spot pricing scheme is that they can use a spot VM at a significantly lower price than that of the fixed pricing scheme. For example, in Amazon EC2 [2], the spot price of a standard VM is approximately 10% of its on-demand price and 10% to 30% of its reserved price. However, the main disadvantage of this spot pricing scheme is that when the offered spot price rises above the consumers' bid, the spot VM is instantly terminated without notice. This means there is no warranty for the application data and configuration in case the machine is terminated.

In this work, we focus on studying *IaaS* cloud and we summarise the three key characteristics of such cloud service model as follows [31].

- *Resource pooling and broad network access*. In a cloud data center, infrastructure resources such as VM, storage, and networking are pooled and shared among multiple cloud consumers. They can access these resources at any location over the network using different devices including PCs, smart phones, Macs, and laptops.

- *On-demand and elastic resource provision*. In a cloud platform, physical infrastructures are transformed into elastic virtual infrastructures, thus allowing the consumption of compute resources on-demand like a utility such as water and electricity. Cloud consumers can get any quantity of resources at any time according to their demand, and the acquisition/release of these resources can be completely rapidly with minimal management effort.

- *Metered resources*. Mainstream cloud IaaS providers such as AWS usually provide metered

compute resources in terms of VM and charge consumers according to the capacity (hardware configuration) of the offered VM and the time unit (e.g. hour) of using this VM, either using the fixed or the spot pricing scheme.

The approaches proposed in this thesis have been implemented and evaluated on the Imperial College (IC) Cloud workstation [38], which is an IaaS cloud platform developed based on Xen hypervisor. The IC Cloud workstation consists of four components:

- *User frontend*: this component acts as the interface of cloud consumers and allows them to interact to other components of the system. It provides the graphical user interface (GUI) for consumers to acquire, control, release, and pay the VM, storage, and network resources.

- *VM pool*: this is the key component of the IC Cloud workstation that comprises four parts: (1) *VM life-cycle manager* is responsible for creating, booting, pausing, resuming, shutting down, and destroying/removing VMs; (2) *network manager* allocates IP addresses to VMs by maintaining a mapping table between available IP addresses and the virtual media access control (MAC) addresses of VMs in each VM pool; (3) *storage manager* is responsible for all storage related management of VMs; (4) *monitor* has three functionalities: backing up the running VMs; checking these VMs' healthy status periodically; and keeping records of VM resource usage including CPU, memory, and I/O utilisations.

- *VM pool controller*: this component manages multiple VM pools in the IC Cloud workstation using three functional units: (1) *security manager* applies different security mechanisms to guarantee secure access of VMs; (2) *scheduler* controls the VM allocation among different VM pools; (3) *service-level agreement (SLA)&Billing manager* charges the consumed resources to maintain applications' QoS requirement.

- *Storage Controller*: it provide similar services as Amazon's Simple Storage Service (S3) [2]; that is, allowing consumers upload and store their data.

## 2.3 Definition of Multi-tier Cloud Applications

At present, irresistible trends promoted by mainstream IaaS cloud enterprises such as AWS [2], GoGrid

[39], and IBM [40] encourage application service providers to migrate their on-premise application to the cloud. International Data Corporation (IDC) forecasts the cloud market can have a 25.3% annual growth rate and by 2014, 65% of new applications are expected to be delivered via the cloud [41]. Also, small and medium enterprises (SMEs) are expected to settle 10% of their applications in the cloud by 2014 and this ratio is expected to be 15% by 2015.

A cloud application can either be an infrastructure application or an end-user application [42]. Examples of infrastructure applications are a Domain Name System (DNS) server, an email server or a database. Applications of this sort often have simple structures such as one or two tiers. By contrast, the structure of an *end-user* application is more complex. Examples of end-user applications are e-commerce sites, financial services applications, online healthcare applications, and gaming applications. For example, consider the typical infrastructure for a *multi-tier* e-commerce application as shown in Figure 2.1. This application is composed of five tiers of *server components* (or servers) that work together to handle requests from end users. Specifically, the HAProxy LB server receives requests and forwards them to the Apache HTTP Server for handling static HTTP requests such as browsing texts and images, or to the Tomcat application server for processing dynamic requests such as recommending and ranking products. The requests processed in the Tomcat application server also represent the business logic of the e-commerce service. The requests in Apache and Tomcat servers are then further relayed for database queries. Amoeba LB server distributes databases requests for MySQL databases such as inserting and updating database tables. The MySQL database stores all persistent data that represent the e-commerce site service's state such as customer and order records. In this work, we study the *end-user* application because it can also incorporate the simpler *infrastructure* application scenario.

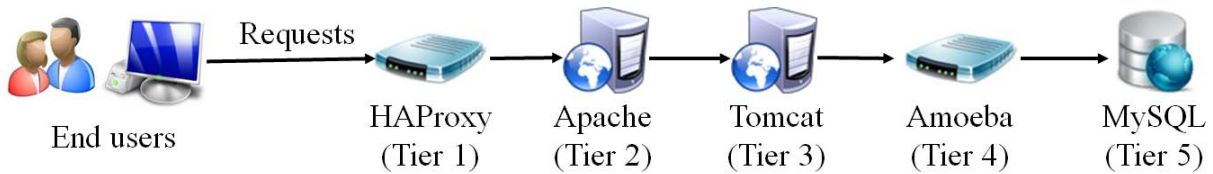

Figure 2.1: Multi-tier infrastructure of an e-commerce site.

In a multi-tier application, servers are categorized into different tiers according to their functionalities, as listed in Figure 2.2: servers at the two LB tiers such as HAProxy and Amoeba distribute requests to servers at the service or storage tiers; servers at the service tier such as Apache and Tomcat are responsible for handling HTTP requests and implementing business logic; and servers at the storage tier such as MySQL database are used for managing application data. In addition, servers at the service or storage tier can be further divided into different sub tiers.

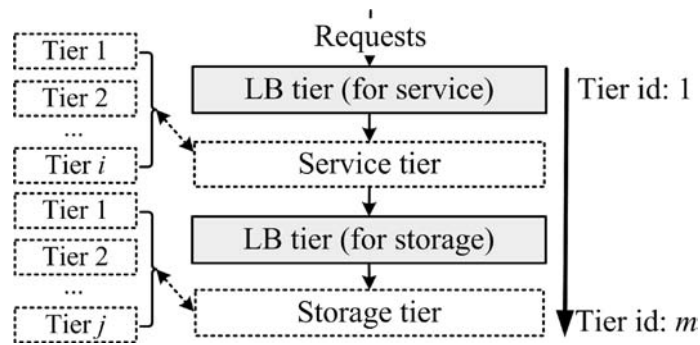

Figure 2.2: The multi-tier architecture of cloud applications.

Typically, each application has a set of demands and constraints specified by the application service providers in the form of a SLA. A QoS demand (requirement) is defined by the required response time for a request. This required *response time* denotes the maximum acceptable response delay when a request traverses the application; that is, the time interval between the arrival and departure moments of the request. This required time can either be an average response time of all requests (used in this work) or a high percentile of response time distribution (e.g. 90% of the requests' response time should be less than 2 seconds). A *cost constraint* is the budget of the total application deployment. In addition, each tier has a *resource constraint* that restricts the maximum number of servers in this tier. For instance, the maximum number of Tomcat servers is 10.

**Definition 2.1 (A multi-tier application).** *A multi-tier application consists of two parts: (1) the server set S including all servers of the application and (2) the demand set D capturing the requiring specified in the SLA.*

The multi-tier architecture guarantees the modularity of cloud applications and facilities the control of their tiers. An application's server set can be divided into multiple subsets and each subset consists of servers belonging to the *same* tier. Each server is marked by a unique tier id. For instance, in Figure 2.1's an e-commerce site, the tier ids of HAProxy, Apache, Tomcat, Amoeba and MySQL are 1 to 5, respectively. Starting from HAProxy that acts as end users' communication interface, servers at each tier first receive in-processing requests from previous tiers, process these requests locally and then transmit them to the next tier.

**Definition 2.2 (Server's tier id).** *In a m-tier application, each server s has a unique tier ID, denoted by* $id(s)$. *Servers' tier ids are numbered consecutively from* 1 *to m according to these servers' tier types: the LB tier (for service), the service tier, the LB tier (for storage), and the storage tier.*

**Definition 2.3 (Server subsets).** *In a m-tier application, the server set S can be divided into m server subsets:* $S_1 \cup S_2 \cup \ldots \cup S_m$, *which are sorted in strictly ascending order according to the tier id of their servers. That is: each subset $S_i$ consists of servers belonging to tier i (i=1,…,m) and any pair of servers s and s', we have $id(s) < id(s')$ if $s \in S_i$ and $s' \in S_{i+1}$.*

## 2.4   Elasticity Management of Cloud Applications

Elasticity management, also known as on-demand scaling or dynamic provisioning, of applications is one of the most important features in cloud computing. We define the **application level elasticity** as the ability to adaptively *scale resources up and down* in order to meet the application performance demand such as the required response time. Typically, there are three roles in elasticity management of cloud applications, as shown in Figure 2.3.

- **Cloud IaaS providers**

  These providers are infrastructure owners or resellers such as AWS [2] and GoGrid [39] who supply infrastructure resources including compute resources such as VMs; storage resources such as online storage; and network resources such as IP addresses and bandwidths.

- **Application owners**

  Application owners, or application service providers, are *consumers* of the IaaS clouds who use the purchased resources to deploy their applications and provide these application services for third parties via the internet. In other words, application owners are the *consumers* of the IaaS clouds and, at the same time, the *providers* of their own application services.

- **End users**

  Application end users, or customers of applications, directly consume application services. For example, end users can browse web pages, enquire and rank items, and make orders in e-commerce sites.

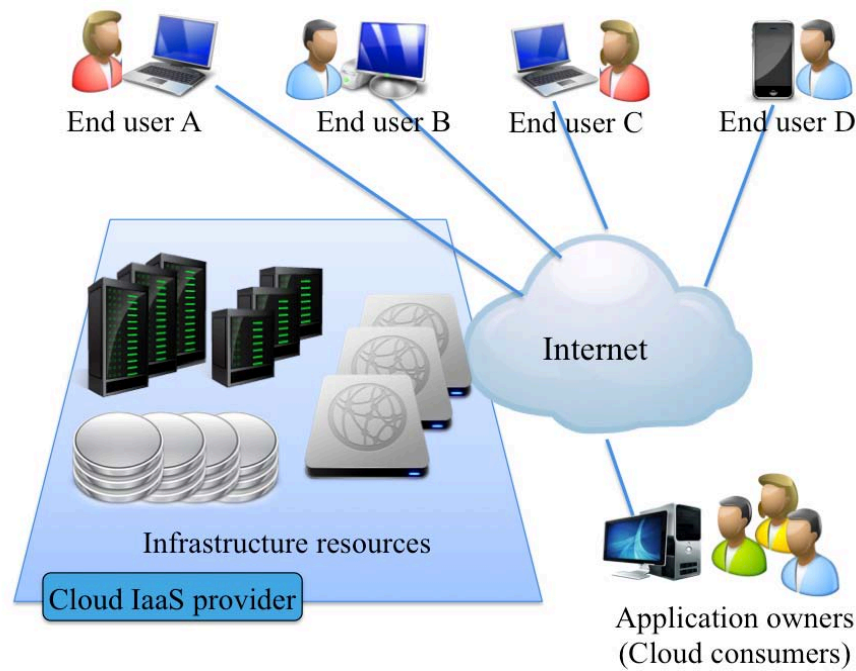

Figure 2.3: The three roles in cloud elasticity management.

The basic target of **application level elasticity** is to assist application owners to *scale* their own applications up and down to achieve the *smallest possible deployment cost* (the cost incurred by consuming infrastructure resources) in the cloud, while still maintaining *QoS* (e.g. response time) they provide to their end users. The elasticity management at the application level has triple meanings.

- First, dynamic scaling is needed to respond to end users' changing requirements and automatically scales applications to restore acceptable performance without having to shut down the delivered services.

- Secondly, the scaling should be conducted to maintain an application's QoS under varying resource demands of servers at different tiers of the application. An interesting point here is that the amount of resources consumed by these servers depends on the behaviour of the application end users themselves. For example, if webpage browsing and product searching are these end users' main action over a period of time, Apache web servers and Tomcat application servers might become stressed and their resources saturated. By contrast, the database component would continue to perform well. When scaling up and down an application, it is therefore crucial to discover the real bottlenecks that may be caused at any, or all, of the servers.

- Finally, in addition to QoS requirements, the **cost** of using the infrastructure resources themselves is the equally important aspect of cloud applications. Hence, elasticity management should guarantee both good performance and cost-efficient resource usage when scaling cloud applications.

In Section 2.5, we tackle the first issue in elasticity management by proposing a platform that enables the automatic scaling of cloud applications.

## 2.5 iSSe: A Platform to Support Elastic Management of Cloud Applications

Simplifying the process of managing (deploying and scaling) applications is almost essential in the cloud. However, existing techniques can automate applications' *initial* deployment but have not yet adequately addressed their *dynamic scaling* problem. For such an issue, we propose a platform [19] to support the automatic scaling of cloud applications. The context and motivation are first introduced in Section 2.5.1. We then extend the TOSCA, an emerging framework for describing cloud application and enhancing their portability across infrastructure services, to define the Elastic-TOSCA framework to manage the dynamic scaling of cloud applications (Section 2.5.2). Based on Elastic-TOSCA, an intelligent platform, called iSSe, is implemented to automate the scaling process of cloud applications in the IC Cloud workstation (Section 2.5.3).

### 2.5.1 Context and Motivation

Since mainstream cloud providers such as AWS [2] usually provide application owners with standalone VM images, application owners have to manually conduct a series of deployment tasks before they can deliver services in the cloud. For example, Figure 2.4 shows the 15 manual steps needed to deploy an e-commerce site with five servers. To deploy each server, application owners need to purchase a VM; install the software; and configure it (e.g. after installing the Tomcat software, its login name and password need to be set). Such manual deployment process incurs three problems. First, the manual deployment is a time-consuming process, in which a lot of time is wasted in tedious tasks such as installing, configuring, and integrating applications. Second, the complexity nature of these tasks makes them error-prone. Finally, professional knowledge is required and external consultant hours for domain experts and solution architects are often expensive.

**Task 1**: Purchase a suitable VM
**Task 2**: Require a public IP address and attach it to the VM
**Task 3**: Install and configure the HAProxy server
**Task 4**: Enable its access for all IP addresses
**Task 5**: Purchase a suitable VM
**Task 6**: Install and configure the Apache server
**Task 7**: Register the Apache in HAProxy for load balancing
**Task 8**: Purchase a suitable VM
**Task 9**: Install and configure the Tomcat server
**Task 10**: Register the Tomcat in HAProxy for load balancing
**Task 11**: Purchase a suitable VM
**Task 12**: Install and configure the Amoeba server
**Task 13**: Purchase a suitable VM
**Task 14**: Install and configure the MySQL database server
**Task 15**: Register the MySQL in Amoeba for load balancing

Figure 2.4: A manual process of deploy an example e-commerce site.

Recent work, whether used in practice or described in the literature, has tried to simplify the deployment process by providing pre-defined packages capable of being automatically deployed. In cloud computing, a representative technique in simplifying deployment is proposed by RightScale [10]. This technique integrates applications with VM images to generate server templates that can be automatically deployed. In addition, some other enterprises such as 3Tera [43] provide visual user portals to facilitate the design of application deployment plans.

In addition, researchers have proposed a number of approaches to simply the deployment process of cloud applications. Konstantinou et al. [44] introduce a model-based architecture using virtual solution model (VSM) to provide abstract and platform-independent deployment plans. When a VSM is bound to a cloud platform, it can be transformed into an executable deployment plan. Chieu et al. [45] present a cloud provisioning system that preloads applications in VMs to generate basic application images. This system allows application owners to specify complex deployment scenarios by combining these application images. In addition, Xabriel et al. use a meta model based approach to automate applications' initial deployment and their approach supports static deployment modifications at the design time [46]. Hughes et al. propose a framework to support individual applications' self-management, including setup, configuration, recovery, and scaling up and down [47].

Although existing techniques can serve well for automating applications' initial deployment, they still leave deployed application' dynamic scaling at run-time for human intervention. This manual redeployment requires services to be put offline and this is sometimes unacceptable for the end users. The deployment platform proposed in this section, therefore, attempts to solve this problem by supporting both applications' initial deployment and dynamic scaling.

## 2.5.2   Elastic-TOSCA for Supporting Elasticity of Cloud Application

In this section, we investigate the enrichment of TOSCA to support the dynamic scaling of cloud applications. We first introduce the basic TOSCA framework briefly, and then describe how it is extended to define Elastic-TOSCA. Elastic-TOSCA is extension of TOSCA that supports the specification of dynamic scaling plans, which enable guiding scale-up/down of cloud applications at run-time.

**Basic Introduction of TOSCA**

TOSCA is an emerging framework for describing components' dependencies and deployment plans of cloud applications. Proposed by the Organisation for the Advancement of Structured Information Standards (OASIS) [48, 49], TOSCA is designed to simplify the life cycle management of cloud applications in a vendor-neutral manner so as to enhance their portability. Such portability is enabled through specifying the operational behaviours of cloud applications, e.g. how servers are deployed or removed and how they are connected, in a uniform way independent of the cloud platform used. This uniform description provides application owners with flexibility when deploying and migrating their applications and associated components across different IaaS providers, thus finding an IaaS provider capable of offering better performance or cheaper resources.

TOSCA server templates are described in XML and can be used for describing cloud application, including server components and their linking relationships [48, 49]. Figure 2.5 shows the high-level structure of a TOSCA server template describing a five-tier e-commerce service using four sections: "Topology template", "Node types", "Relationship type" and "Plans". The "Topology template" section defines the whole cloud application, including its nodes (server components in the application) and the dependency between these servers. The "Node types" section defines the properties of one server, e.g. its owner and the configuration of its hosted VM (CPU numbers, memory size, disk capacity and operating system). The "Relationship type" section specifies the relationship between two servers. In the shown example, a HAProxy server and an Apache server are connected, where the HAProxy is the source node and the Apache is the target node. Finally, the "Plans" section defines the process model for initially deploying a new application and also for removing a running application. For example, Figure 2.6 shows a "Deploying new applications" plan, which is used for initially deploying a five-tier application of e-commerce site. This plan comprises five deployment actions that each action deploys one server to one tier of the application.

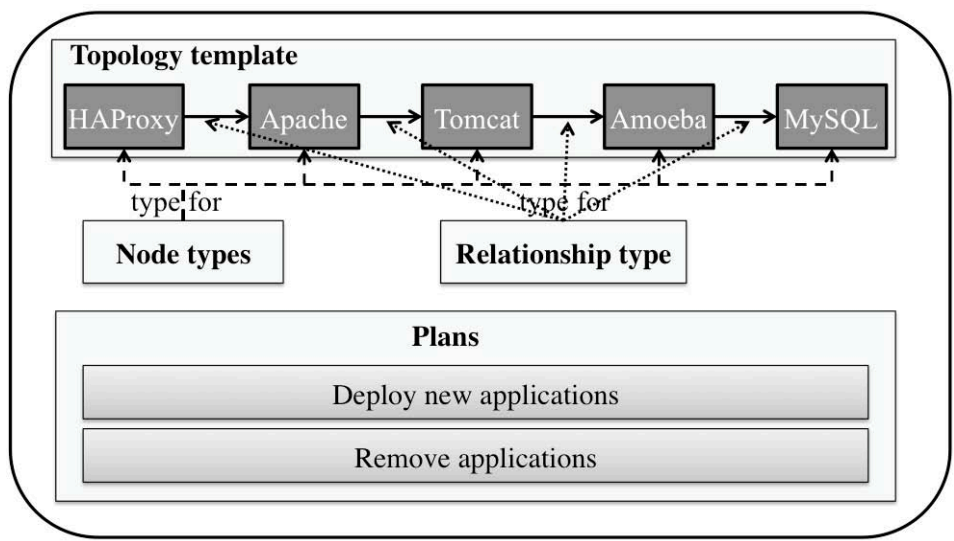

Figure 2.5: An example server template in basic TOSCA.

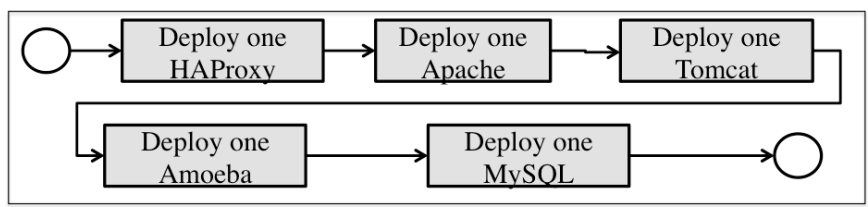

Figure 2.6: A plan for deploying the five-tier e-commerce site.

Currently, TOSCA supports the specification of key activities required for the *initial deployment* of cloud applications and also the activities required to *shut down* the application. However, it does not provide support for the equally important aspect of specifying how application level elasticity can be managed at run-time, e.g. by enabling the specification of how resources can be added or removed at run-time based on workload variation. For such an issue, we enrich and extend the existing TOSCA framework to support such elasticity management activities in a vendor-neutral manner.

**Elastic-TOSCA: Extensions of TOSCA to Support Elasticity**

We extend the basic TOSCA framework and enrich it with the information required for guiding dynamic scaling of cloud applications, allowing application owners to specify different scaling strategies. For example, an owner could define a scaling up/down strategy based on performance requirements, budgets and QoS requirements specified in the SLA.

Using the Elastic-TOSCA framework, we generate a new Elastic-TOSCA-based XML document that includes monitoring information structures and new plans for scaling up/down. Figure 2.7 shows an example server template in Elastic-TOSCA, including two new sections ("Monitoring Information" and "SLA&Constraints") as well as extensions to the "Plans" section, corresponding to three components needed for guiding dynamic scaling of an e-commerce site. Note that the specification and extension of these sections follows TOSCA extensibility mechanism [48, 49], which guarantees that the extended sections are independent of cloud IaaS providers.

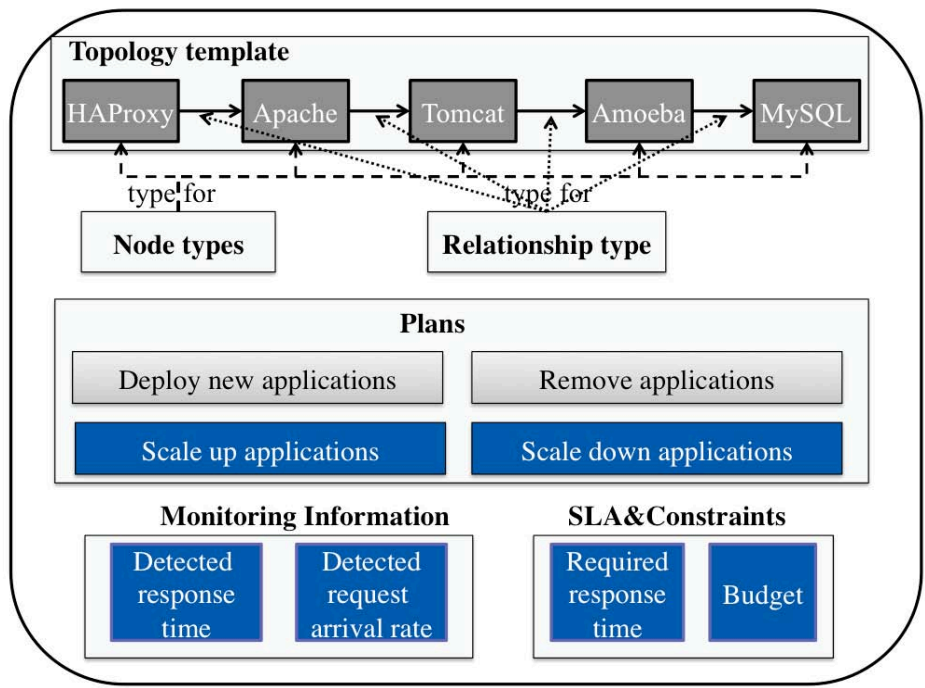

Figure 2.7: An example server template in Elastic-TOSCA.

The "Monitoring Information" section mainly specifies a running application's current status and underlying infrastructures. In the example fragment in Figure 2.8, this section records the detected response time and the request arrival rate.

```
<MonitoringInformationTemplate id="ApplicationMonitor"
                               name="Application Monitor"
                               InformationType="Monitor">
        <DetectedResponseTime>1.2Seconds</DetectedResponseTime>
        <RequestArrivalRate>15RequestsPerSec</RequestArrivalRate>
</MonitoringInformationTemplate>
```

Figure 2.8: An example "Monitoring Information" section in Elastic-TOSCA.

The "SLA&Constraints" section describes QoS requirements and any constraint on quality, budget, and other aspects of the application. In the example shown in Figure 2.9, this section specifies the end users' required QoS: the required response time and the application owner's constraints: the total deployment budget (the maximal cost to support the running of the application).

```
<SLAAndConstraintsTemplate id="ApplicationSLA"
                           name="Application SLA"
                           InformationType="SLA">
        <RequiredResponseTime>1.2Seconds</RequiredResponseTime>
        <Budget>10Dollars</Budget>
</SLAAndConstraintsTemplate >
```

Figure 2.9: An example "SLA&Constraints" section in Elastic-TOSCA.

Finally, we extend the "Plan" section in basic TOSCA to define more types of plans that handle the application's dynamic scaling cases. Figure 2.7 shows the Elastic-TOSCA definition of two types of scaling plans — "Scale up applications" and "Scale down applications". Each type can have multiple scaling plans. Each plan describes a specific scaling scenario that consists of a list of scaling tasks, and each task corresponds to a deployment action. For example, Figure 2.10 shows a fragment of a plan for scaling up an e-commerce site. This plan is used for adding one Apache server and two Tomcat servers to the application. Note that for each scaling case, a scaling plan and its scaling tasks are generated dynamically. The information needed to generate documents describing the scaling tasks (e.g. a server's user name, password and VM configuration) is obtained from the "Node types" section of Elastic-TOSCA.

```
<Plan id="ScalingUpAnE_CommerceApplication"
            name="A Scaling Up Plan of E-commerce Website"
            planType="http://docs.oasis-open.org/tosca/ns/2013/01/...."
            languageUsed="http://www.omg.org/spec/BPMN/2.0/">
      <PlanModel>
            <process name="DeployNewApplication" id="p1">
                  <task id="t1" name="DeployAnApacheServer"/>
                  <task id="t2" name="DeployATomcatServer"/>
                  <task id="t3" name="DeployATomcatServer"/>
                  <sequenceFlow id="s1" targetRef="t2" sourceRef="t1"/>
                  <sequenceFlow id="s2" targetRef="t3" sourceRef="t2"/>
            </process>
      </PlanModel>
</Plan>
```

Figure 2.10: An example scaling up plan in Elastic-TOSCA.

### 2.5.3   iSSe to Support Automatic Scaling of Cloud Applications

Based on the Elastic-TOSCA framework, iSSe [19, 20] is implemented to enable the automatic scaling of cloud applications. As shown in Figure 2.11, iSSe acts as middleware between cloud IaaS provider (the IC Cloud workstation) and application owners. At the client side, the *Application Deployment Portal* assists application owners to define deployment specifications and executes deployment on their behalf. In addition to this portal, the other five service components in iSSe work together to support automatic scaling of cloud applications.

The *Application Owner Portal*, as shown in Figure 2.12, is designed to assist application owners to configure their services; select servers from the iSSe *Repository of servers*; define VM configurations; and design their topology. In Figure 2.12's example, an e-commerce site with five tiers of servers is designed. Using the portal, application owners can configure the Tomcat sever by specifying its VM configuration (CPU number and memory capacity); number of servers; IaaS provider to be deployed; and software configuration (login user and password). This portal also allows them to specify the QoS requirement such as the required response time and other constraints such as the deployment budget in the SLA. To enable interaction with Elastic-TOSCA, the information is stored in the "Topology template", "Node types", and "Relationship types" sections in Elastic-TOSCA server templates.

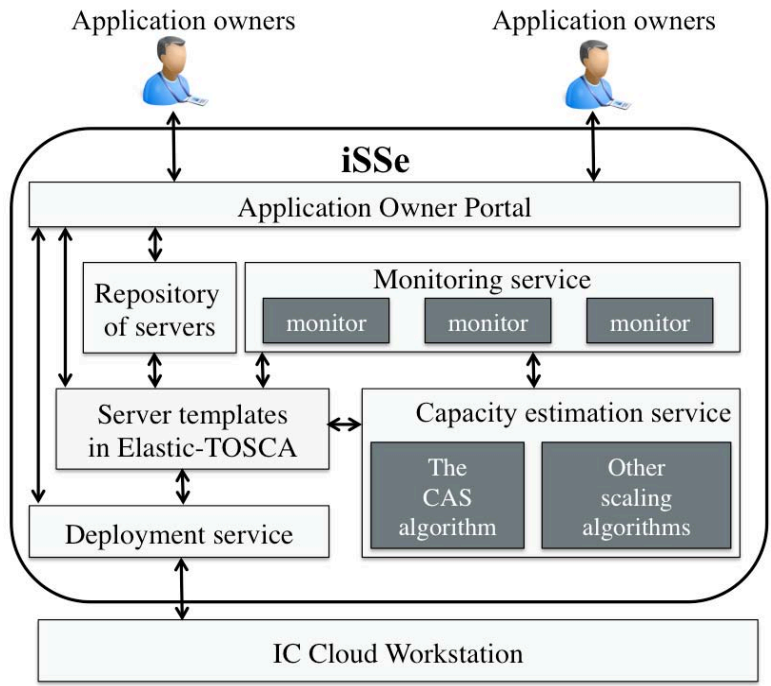

Figure 2.11: iSSe for supporting dynamic scaling using Elastic-TOSCA.

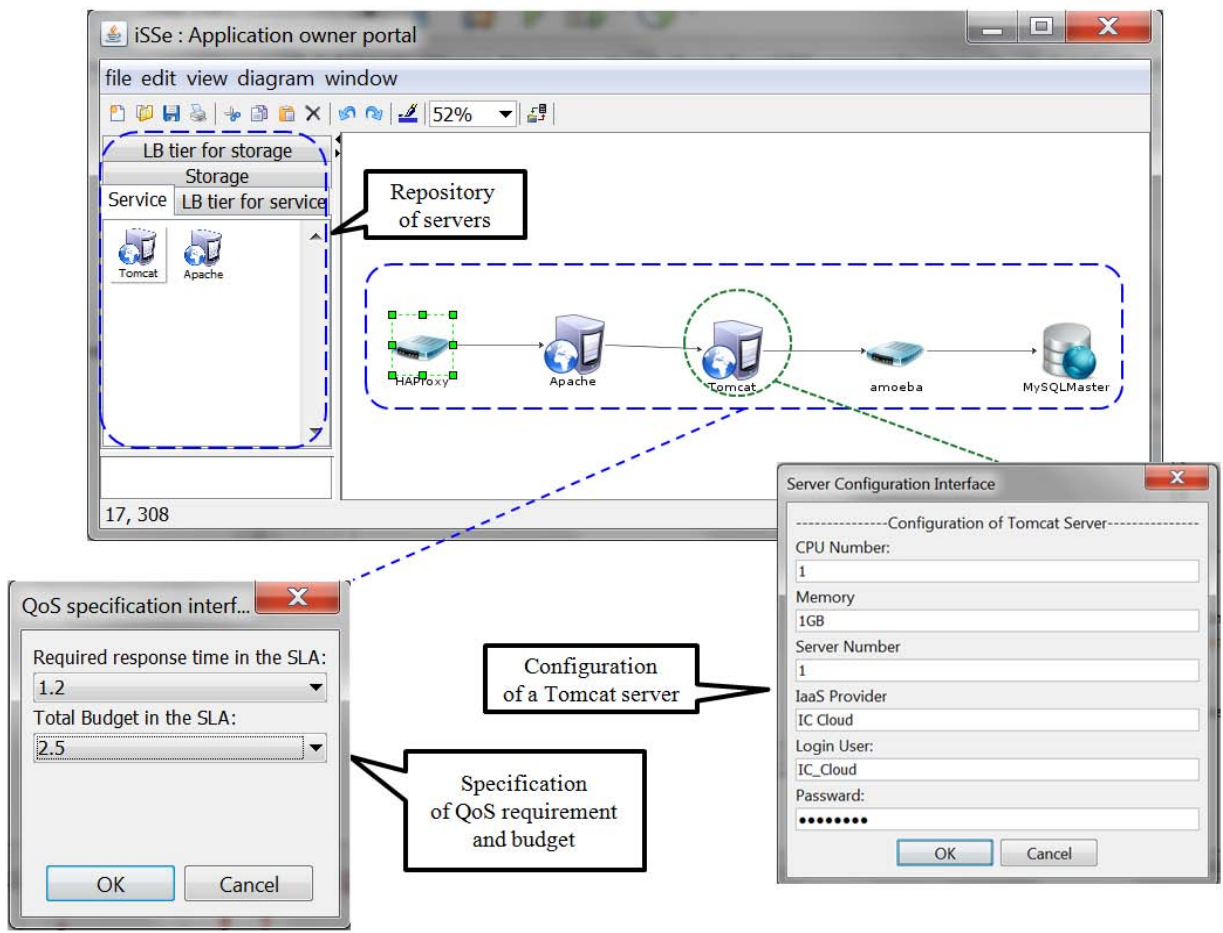

Figure 2.12: Application owner portal in iSSe.

The *Repository of servers* shown in Figure 2.12 contains a library of servers. These servers can be pre-designed by domain experts and solution architects based on best practice. This repository also provides a flexible registration mechanism. For example, if application owners want to use JBoss web server but cannot find it in the repository. They can register it in the platform, after which they can even package the JBoss web server into a server template. This newly registered JBoss server can then be dynamically scaled in the same way as other pre-defined servers.

The *Monitoring Service* monitors each running application using a monitor (see Figure 2.13). This monitor examines the incoming requests and servers at different tiers over a finite interval (e.g. 60 seconds) and records information including the request arrival rate and the response time of the application. The collected information is used to update the "Monitoring information" section in Elastic-TOSCA server templates. This information is then used to decide whether a scaling up (or down) is needed. For instance, a scaling up is triggered if the observed response time exceeds the required response time in the SLA. When a scaling is triggered, the *Capacity Estimation Service* conducts a scaling algorithm (e.g. the CAS algorithm introduced in Chapter 3). This algorithm estimates the number of servers to be scaled using the information in the server templates to generate a scaling up/down plan. Specifically, the approach estimates the type and number of servers to be scaled, and then updates the "Plans" section by adding the generated scaling plan.

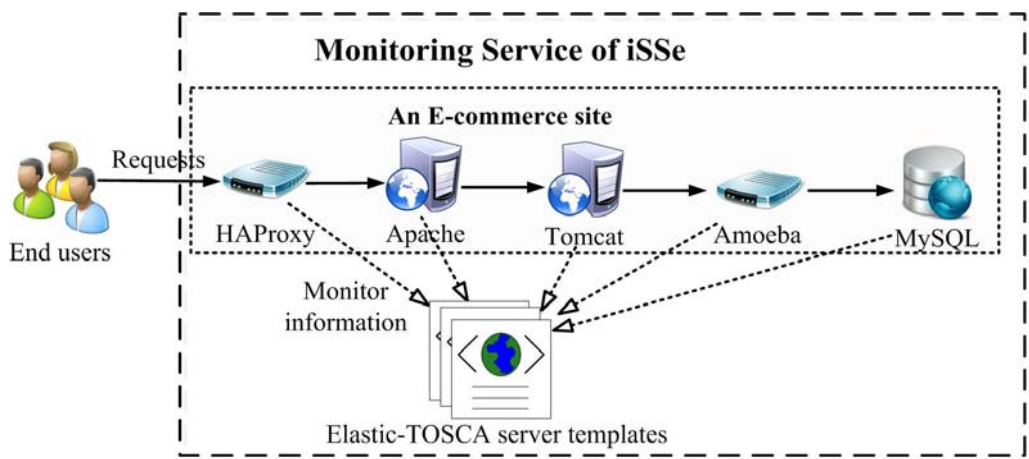

Figure 2.13: Monitoring service in iSSe.

Finally, the iSSe *Deployment Service* automates the scaling actions of *addition* and *removal* of servers (VMs) by calling auto scaling APIs of the IC Cloud workstation [38]. In iSSe, each server is installed in a standalone VM. This VM consists of installed software such as Tomcat software, as well as pre-loaded auto-running scripts used in automating the scaling process. Figure 2.14 shows an example specification of the Tomcat server in Elastic-TOSCA. This specification has three sections. The first section lists the server's basic information, such as its tier id, its deployment platform and price. The second section specifies the server's VM configuration and the third section defines software user settings including login name and password.

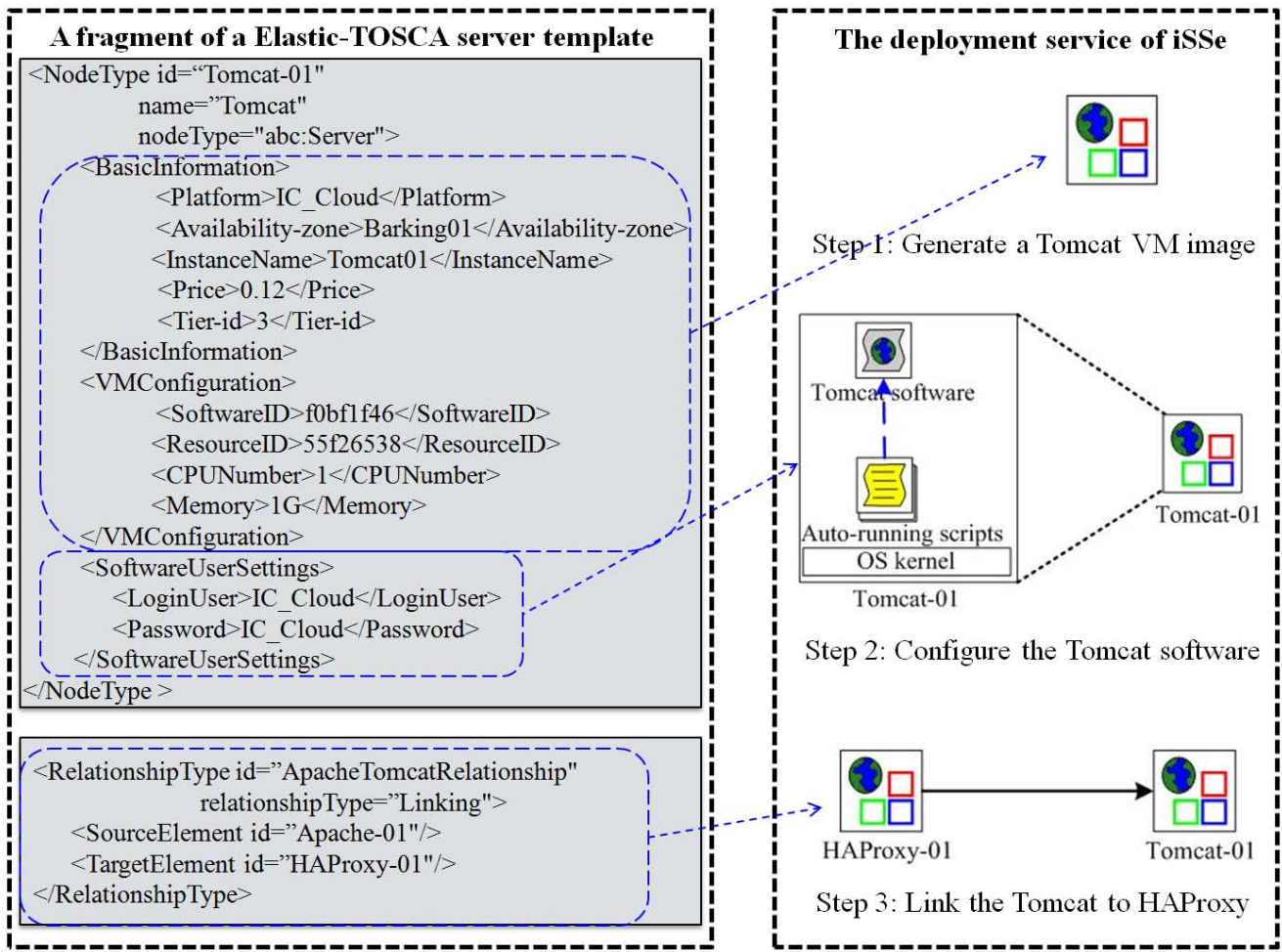

Figure 2.14: Deployment service in iSSe.

By interpreting the servers' specifications, the automation of the *addition* activity involves three steps, as shown in Figure 2.14. At step 1, the *Deployment Service* first generates a Tomcat VM image (a VM with pre-installed Tomcat), as specified in the sections of "BasicInformation" and "VM Configuration". At step 2, the VM is first started, after which the auto-running script preloaded in this VM uses the parameters in the "SoftwareUserSettings" section to configure the Tomcat software: the login name and password are set. At step 3, the Tomcat is linked to its input load-balancing server HAProxy-01, as specified in the "RelationshipType" section of the server template.

Typically, the *addition* operations take a few minutes to complete. Specifically, at step 1, the generation of a VM image (step 1) can be finished within a few seconds. This is because after a server is packaged, most existing cloud platforms such as AWS usually keep a certain number of images ready for use. In addition, step 2 and 3 can be completed within 1 or 2 minutes. The time is mainly consumed in starting up the VM and executing the scripts to configure the server software. However, there are two situations when the *addition* activity may need slightly longer time to complete. The first situation is when a server at the storage tier is added. In this case, the server may need some time to update/replicate data. An example is when a newly added MySQL slave needs to synchronize with the MySQL master. The second situation arises due to the auto-running script of a LB server (e.g. a HAProxy), which executes once every few minutes. Hence when a new service/storage server (e.g. Tomcat) is added, it needs to wait until the LB server's script runs again to detect and register this new server.

Automating the *removal* activity is achieved by conducting the reverse operations of the *addition* activity: the running Tomcat server is disconnected from its input server HAProxy and removed from the application. Note that this server's running VM is only shut down when its existing billing period ends. For example, most mainstream providers such AWS today bill their users by the hour. However, some scaling algorithm can charge VMs in a finer granularity than hour in order to achieve the cost-efficient scaling. For example, a server $s_1$ is added to an application at $t$=0 (time unit is minute), it is removed at $t$=5 and added at $t$=15; it is removed again at $t$=30 and added back to the application at $t$=45, then it keeps running until $t$=60. Assumed that, server $s_1$ is charged by minute, its cost is less than the server $s_2$ that keeps running during an hour: $c(s_1)=(35/60)c(s_2)$ because server $s_1$ only runs, or it is charged, 35 minutes. This fine-grained pricing strategy is supported by the IC Cloud workstation (e.g. a VM can be billed by minute).

## 2.6   Existing Scaling Techniques for Cloud Applications

In this section, we present an overview of the existing scaling techniques used to support elasticity management of cloud applications. We first discuss some pre-cloud techniques for scaling multi-tier applications, and then introduce current research on scaling cloud applications.

### 2.6.1  Traditional Scaling Techniques Before Clouds

Scaling of applications has been studied extensively before clouds. Early work considers single-tier application and focuses on transforming performance target into underlying hardware resources such as CPU and memory [3-5]. Further investigations classify an application into multiple tiers [6-9]. They then break down the total response time by each tier and conduct the worst-case capacity estimation to ensure applications meeting the peak workload. Overall, the single-tier model can be viewed as a special case of multi-tier model and the latter model can guide the scaling in a more accurate way.

However, although scaling of traditional applications, which are often hosted on physical servers, shares many commonalities with that of cloud applications, these two types of scaling techniques have different emphases. Conventional techniques mainly concentrate on how to schedule compute nodes to meet QoS requirement of applications by predicting their *long-term* demand changes. In contrast, clouds focus on providing metered resources on-demand and on quickly scaling applications up and down whenever application demand changes, thus achieving cost-efficient resource provisioning. Further investigations, therefore, are needed to address challenges brought by this requirement for high elasticity. For example, a CloudSim toolkit is proposed to simulate an application in order to accurately estimate its required resources before the actual deployment in clouds [50]. Another example, Merino et al. introduce a hybrid grid-cloud architecture and propose a market-based economic mechanism to assist grid users to perform scaling with cloud resources [51].

### 2.6.2  General Research on Supporting Scaling of Cloud Applications

At present, numerous efforts are contributed to scale cloud applications. Some resource provisioning system [52], frameworks [53, 54], tool suite for estimating deployment cost [55],  lifecycle management toolkit [56] are proposed to manage cloud resources based on the idea of autonomic control [57]. Rather than developing concrete scaling methods, these studies generally discuss higher-level concerns in building an effective provisioning system. Examples are performance metrics used in resource allocation,

SLA analyser, performance monitor and VM allocator.

Current cloud IaaS providers usually maintain an auto-scaling queue of pre-booted and pre-configured VMs that can be allocated to applications quickly, thus guaranteeing a *short time* in scaling up these applications. In [58], Doughertya et al. study how to determine an optimal queue size that can ensure the application scaling up can be severed quickly, while minimising the power consumption of maintaining pre-booted VMs. They propose a model-driven engineering (MDE) approach that transforms the problem of VM configurations into constraint satisfaction problems (CSPs). The approach then decides the VM configurations in the auto-scaling queueing by analysing scaling requirements of applications, VM power consumptions and operating costs. In [59], Sharma et al. develop a prototype of provision system in clouds that helps application owners minimise their deployment costs when deploying and reconfiguring applications. Specifically, the system considers the price differences in different VMs in terms of per CPU core cost and determines the server configurations by solving the integer linear program problem.

## 2.6.3  Policy-based Scaling Techniques

Most of the cloud providers (e.g. AWS [2]) and vendors (e.g. RightScale [10]) employ pre-defined policies (or rules) to guide application scaling. Take RightScale [10] for example, application owners need to manually define an application's rules of triggering scaling after its deployment. These rules specify the minimum and maximum numbers of servers in the application, the condition to scale these servers, the number of servers in each scaling and even the scaling speed. Server monitor triggers these rules to perform scaling. Inheriting from the RightScale, UniCloud extends the policy-based scaling by considering more issues such as work priority and CPU speed [11]. In addition, Nathania et al. introduce four types of policies to manage the allocation of VMs for different types of applications [12].

Policy-based scaling allocates additional servers in scaling up whenever some performance metrics exceed a threshold and removes redundant servers in scaling down whenever these metrics are less than a threshold. This scaling mechanism assumes that application owners have particular knowledge of the application being executed to define proper policies and this assumption is sometimes not applicable to application owners. In addition, this kind of scaling is designed to meet applications' QoS requirement and the cost-effective application scaling is not achievable in many cases.

### 2.6.4   Scaling Techniques Using Analytical Models

Many researchers apply the analytical modelling technique to help application owners make scaling decisions by informing them the performance analysis results of applications. Xiong et al. model an application by a network of queuing models and conduct the performance analysis to show relationships among workloads, server number and QoS level [60]. In [61], Ghosh et al. divide an application into three types of sub analytic models: resource provisioning decision model, VM provisioning model and run-time model. By iteratively solving each individual sub-model, their analysis obtains two results: response time and service availability. In addition, Ghosh et al. utilises a stochastic reward net to model an application and gives two analysis results: job rejection rate and response delay [62]. In [63], Pal et al. propose a pricing framework with economic models designed for multiple cloud providers in the marketplace, in which each cloud provider is modelled as a queueing system. Using this queueing system, the framework aims at informing application owners the price and its related QoS level. In [64], Huu et al. introduce several network provisioning strategies based on a cost estimation model. These strategies are used by application owners to predict the amount of resources and their deploying cost for an application.

Moreover, a variety of application scaling approaches are proposed using the analysis results of queueing models. In [13], Bacigalupo et al. model an application by a queueing model with three tiers, namely application, database and database disk tiers. Each tier is then solved to analyse the expected response time, throughput and utilisation of a server. Using these results, a resource management algorithm is proposed to scale applications in dynamic-urgent clouds.

Similar to [6-9], Bi et al. break down an application's total response time to each tier [14]. They then calculate the number of servers allocated to the application that subject to constraints of average response time and arrival rate.

Other stochastic models are also applied in guiding scaling of applications. In [65], Iqbala et al. present a methodology supporting both the scaling up and down of a two-tier web application. Scaling up is based on a reactive model that actively profiles CPU utilisations of VMs. Scaling down uses a regression-model-based predictive mechanism. In [66], Islam et al. present a prediction-model based scaling approach. The prediction model is built using existing machine learning algorithms including neural network and linear regression. In addition, Li et al. use a network flow model to analyse applications and introduce an approach to assist application owners in making a trade-off between cost and QoS [67]. Zhen et al. model the scaling of cloud applications as a Class Constrained Bin Packing (CCBP) problem, in which a server is modelled as a bin and an application is modelled as a class [68].

**Towards cost-effective application level elasticity**. To the best of our knowledge, previous scaling techniques either provide generally information to application owners and rely on them to make proper scaling decisions [60-64], or propose scaling approaches using the analysis results of analytical models [6-9, 13-15]. This sort of investigations has solved the scaling of applications to meet their QoS satisfactorily. However, existing scaling approaches rarely consider the equally important issue of cloud computing − the cost. Applications deployed in highly scalable cloud environment require not only just good performance but also cost-efficient resource provision. An elastic scaling approach for more cost-efficient cloud elasticity management, therefore, could be a desirable advance.

## 2.7    Challenges in Cost-Effective Application Level Elasticity

This section illustrates the two key challenges to achieve cost-effective application level elasticity for multi-tier applications deployed in IaaS clouds. Without loss of generality, we use an example based on an e-commerce site to capture the typical behaviour of such applications. Also for simplicity, we focus only on applications that are deployed in a single IaaS cloud provider and a single server component of an application is assumed to be installed on one dedicated VM. An e-commerce site is designed to support end users to search and browse; register and login as members; add interested goods to cart and shop online. The workload for such applications depends on the number of end users submitting requests at the same time. The workload may be composed of different types of requests that need to be handled by different parts of the application. For example, some end users may be browsing the web site itself, while others may be querying the product catalogue or making a payment transaction.

**Challenge 1: Cost-aware scaling.**

In a highly scalable cloud environment in which compute resources are consumed as a utility such as water and electricity [69], application owners would expect to spend the smallest cost for the desired application performance. To this aim, the elastic scaling must take cost-aware criteria into consideration and use them to guide scaling of applications. Take Figure 2.1's e-commerce site application for example, these criteria should be aware of both the cost of adding a server (e.g. an Apache or a MySQL) and the performance effect brought by this scaling up (e.g. reducing response time).

**Challenge 2: Workload-adaptive scaling.**

Once the e-commerce site is deployed, the five tiers of servers of this application are hosted across different VMs to support a small number of customers. When the demand increases, the application should be scaled up. An interesting point here is that this scaling process is greatly influenced by the behaviour (i.e. the type of workload) of the application itself. We examine three typical types of workloads, in which each workload places varying demands on different tiers of the application. In Figure 2.15(a), the typical "Shopping" workload simultaneously stresses the service and storage tiers and so the numbers of servers in all three tiers (Apache, Tomcat, and MySQL) are increased. By contrast, in the primarily "Browsing" workload (Figure 2.15(b)), end users mainly browse web pages and preview products. This workload mainly stresses the two service tiers including the Apache and Tomcat servers, so their resources are saturated and the number of these servers needs to be increased. Finally, the primarily "Ordering" workload mainly stresses the storage tier including the MySQL database and so the number of these database servers needs increasing (Figure 2.15(c)).

Based on the above analysis, two types of uncertainties exist in the application workload: (1) the *volume* of workload, which is denoted in terms of the arrival rate of incoming requests, namely the number of incoming requests per time unit; (2) the *type* of workload, such as Figure 2.15's three types of workload. In this context, the elastic scaling must be adaptive to the changing workload, and such adaptive scaling has twofold meanings. First of all, bottleneck tiers of applications should be automatically identified for both scaling up and down. Secondly, scaling should be performed as an iterative process because fixing a bottleneck tier may create another bottleneck at a different tier of the application. For instance, in Figure 2.15(a)'s workload, the bottleneck is shifted to the storage tier of MySQL databases if multiple Apache and Tomcat servers are added to the two service tiers.

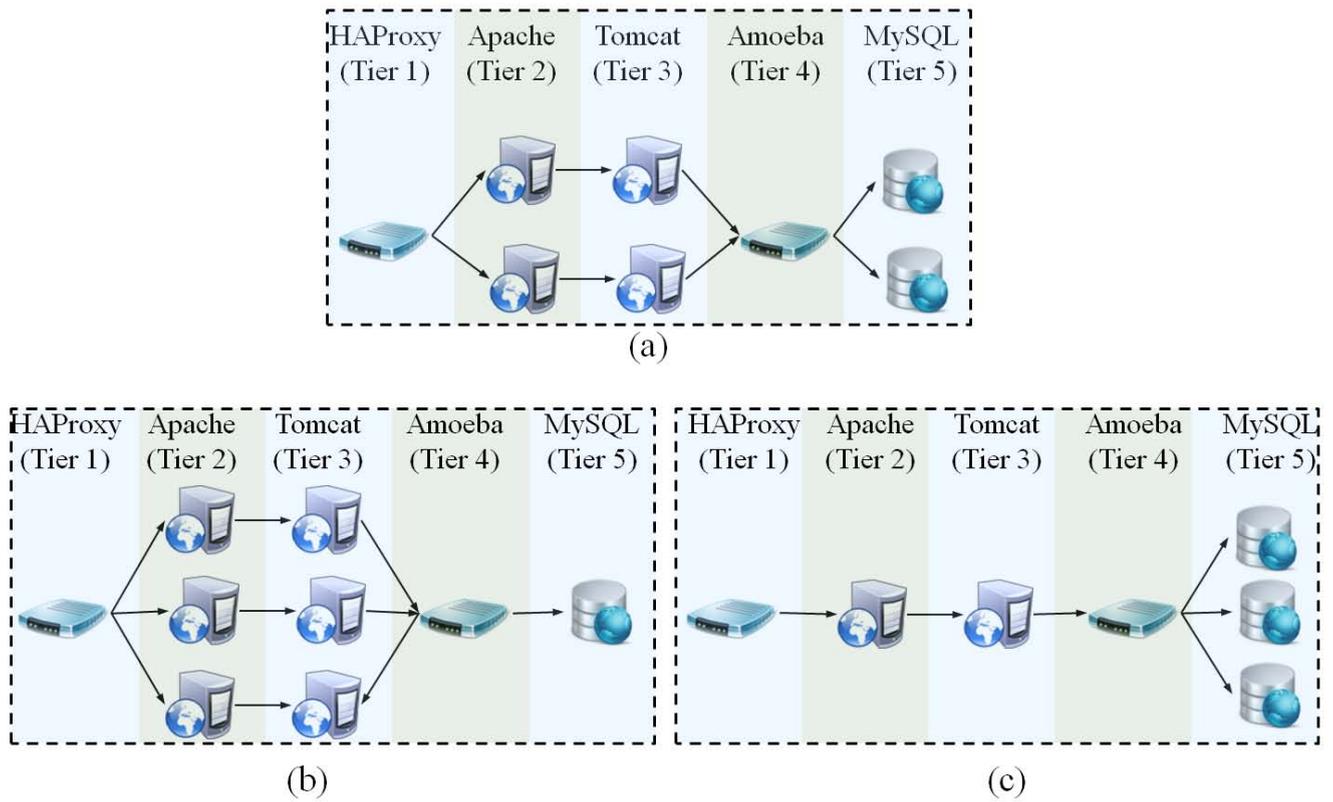

Figure 2.15: Three scenarios of scaling up for three types of workloads. (a) Scenario 1: the typical "Shopping" workload. (b) Scenario 2: the primarily "Browsing" workload. (c) Scenario 3: the primarily "Ordering" workload.

# Chapter 3

# A Cost-aware Scaling Algorithm for Multi-tier Cloud Applications

## 3.1 Introduction

In this chapter, we propose a CAS algorithm that addresses both challenges in cost-effective application level elasticity. We first present an overview of the CAS algorithm for scaling multi-tier cloud-based applications in Section 3.2. In Section 3.3, queueing system is employed as the analytical model to capture the behaviour of such multi-tier applications. Cost-aware criteria are then introduced to measure the effect of cost of resources on every unit of response time. Based on the above criteria, the Cost-Aware-Capacity-Estimation (CACE)-For-Scaling-Up (Section 3.4) and CACE-For-Scaling-Down (Section 3.5) algorithms are designed to handle changing workloads of multi-tier applications by adaptively scaling up and down bottlenecked tiers within these applications. Using a case study of managing the elasticity of a five-tier e-commerce site application, we evaluate the effectiveness of the CAS algorithm in Section 3.6. Finally, we present some discussion of the CAS algorithm in Section 3.7.

## 3.2 The CAS Algorithm Overview

For clarity, we first lists the main parameters of a $m$-tier application used throughout the CAS algorithm in Table 3.1. The values of these parameters are obtained in a variety of ways: (1) single server's deployment cost (price) is decided by cloud IaaS providers (vendors); (2) the required response time and deployment budget are specified by the application owner in a SLA; (3) the response time of the application and arrival rates of end users' requests are collected online by the monitors; 4) each server's service rate and the requests' route probabilities between different tiers are analysed offline by profiling of user behaviours. The data for the offline profiling can be obtained through simulation of servers (e.g. simulating by CloudSim [70]), analysing servers' execution logs and consulting related deployment documents.

The CAS algorithm, detailed below, is started after an application is initially deployed and keeps running until the application is terminated (line 2 to 11). Whenever a violation in required response time is detected (line 6), the algorithm triggers a capacity estimation for scaling up (line 7) to obtain the updated server set $S$. Similarly, whenever a decrease in the arrival rate of incoming requests is detected (line 9), algorithm triggers a capacity estimation for scaling down (line 10) to update the server set $S$. Using the result of the capacity estimation, the algorithm adds new servers (line 8) or removes redundant servers (line 11) in parallel to quickly scale the application within a few minutes. The time complexity of each scaling is decided by the capacity estimation algorithms introduced in the following sections.

Table 3.1. Main parameters of the CAS algorithm

| Parameters | Descriptions | Data source |
|:---:|:---:|:---:|
| $c(s)$ | A single server $s$'s deployment cost | 1. Cloud IaaS providers |
| $r^{sla}$ | The required response time | 2. SLA |
| $c^b$ | The budget of the total deployment | |
| $\lambda_e$ | Arrival rate of end users' requests | 3. Online measurement |
| $r_e$ | The monitored response time | |
| $\mu$ | A server $s$'s service rate | 4. Offline profiling |
| $p_{ij}$ | The routing probability that a request leaves tier $i$ and proceeds to tier $j$ | |
| $p_{0i}$ | The probability that a request enters tier $i$ from end-users | |
| $p_{i0}$ | The probability that a request at tier $i$ leaves the application | |

**The CAS Algorithm**

**Input:** $S$, $D$ and $r^{sla}$.

1. **Begin**

2.     **while** (the application is not completed)

3.         Monitor $\lambda_e$ and $r_e$ once every few seconds;

4.         Let $\lambda_e{}'$ be the last monitored request arrival rate;

5.         Let $S'=S_1' \cup S_2' \cup, \dots , \cup S_m'$ be the server set before scaling;

6.         **if** $r_e > r^{sla}$, **then** // the required response time is violated

7.             $S$ = CACE-For-Scaling-Up $(S', \lambda_e, r^{sla})$;

                // $S = S_1 \cup S_2 \cup, \dots , \cup S_m$ is the updated server set with $m$ subsets

8.             Simultaneously add each server $s$ where $s \in S_i$ and $s \notin S_i'$; //scaling up

9.         **else if** $\lambda_e < \lambda_e{}'$, **then** // the arrival rate decreases

10.            $S$ = CACE-For-Scaling-Down $(S', \lambda_e, r^{sla})$;

11.            Simultaneously remove each server $s$ where $s \notin S_i$ and $s \in S_i'$.  //scaling down

12. **End**

The CAS algorithm applies an automatic reactive scaling mechanism similar to mechanisms applied by Amazon WS [2] and RightScale [10], but the scaling mechanism in our work needs no pre-defined rules to trigger scaling. In contrast, the CAS algorithm replies on online monitors to detect the changes in workload (request arrival rate) and performs corresponding scaling. In addition, traditional *predictive* scaling methods are motivated by the long-term workload variations that can be predicted using application profiling [6-9]. These methods usually allocate servers to an application well ahead of the expected workload increase because compute resources are difficult to obtain on-demand in traditional infrastructures such as grids. In contrast, cloud infrastructures provide metered resources on-demand. Within this context, the *reactive* mechanism is applied in this work to quickly scale applications up and down whenever the user demand changes. This makes the CAS algorithm suitable for scaling applications with both long-term and predictable workload variations and short-term and unpredictable variations.

## 3.3   Criteria for Capacity Estimation

In the scaling up or down of an application, addition or removal of a server influences both the *response time* and *deployment cost* of the application. The CAS algorithm aims at spending as small cost as possible to meet the required response time. To this aim, the cost-aware criteria are designed to analyse the effect of cost on every unit of response time. To achieve this, we develop a performance model of the multi-tier applications based on queueing system. Given the hardware and software configuration of an application and the required response time, the queueing system model can determine how much capacity is needed to service different volumes and types of workload. Furthermore, given resource price (servers' deployment cost) in cloud environment, the queueing system model can identify bottlenecks in scaling this multi-tier application, thus minimising its deployment cost incurred by consuming cloud infrastructure resources.

## 3.3.1 Modelling of Multi-tier Applications Based on Queueing System

**One server component in the application**

Typically, a queueing system can be described as an $A/X/n$, where $A$ represents the distribution of interarrival time of customers; $X$ denotes the distribution of service time; and $n$ is the number of servers [71]. In this work, an M/M/1 (*M* for Markov) queueing system is applied to model a *server* in a cloud application and the *customers* in the queueing system are represented by the *incoming requests* from application end users. In an M/M/1 queueing system, there is only one server $s$ and the arrival requests are determined by a Poisson process. During the Poisson process, requests arrive at intervals and the interarrival time denotes the period between two successive requests. In this queueing system, both the interarrival time and the service time of requests follow exponential distributions. For example, Figure 3.1's M/M/1 queueing system is used to model a Tomcat server in an e-commerce site application. In the queueing system, incoming requests either wait in the queue or are served by the Tomcat server.

An M/M/1 queueing system can be described by its request arrival rate $\lambda$ (the inverse of interarrival time, e.g. 7 requests/second) and its service rate $\mu$ (the inverse of request service time, e.g. 14 requests/second). The M/M/1 queueing system discussed in this work makes the following assumptions:

- The arrival rate $\lambda$ denotes the long-term average arrival rate of incoming requests.

- The service rate $\mu$ is independent of the arrival process and this rate is also state independent; that is, $\mu$ is a constant rate that is independent of the number of requests in the waiting queue.

- It's easy to see that if the service rate $\mu$ is smaller than the request arrival rate $\lambda$ (i.e. $\mu/\lambda < 1$) in the queueing system, the system is unstable: the queue will become longer and longer. Hence, we assume the M/M/1 queueing system is at *stable* state (i.e. $\mu/\lambda > 1$ and the queue is at equilibrium). Based on this *steady-state* assumption, we have the departure rate $d$ of the queueing system's output requests equals the arrival rate $\lambda$ of this system's incoming requests.

- The queueing system also has three optional components: B/K/SD, where $B$ represents the queue capacity; $K$ represents the size of incoming requests; and SD stands for the service discipline of the requests in the waiting request. We assumed that B/K/SD=$\infty/\infty$/FIFO, where FIFO denotes the first-in first-out principle (or equivalently first-come first-served (FCFS)) that a request coming first is served first.

- All the incoming requests come from the same class. In the waiting queue, these requests are put in one queue and managed using the FIFO queueing discipline.

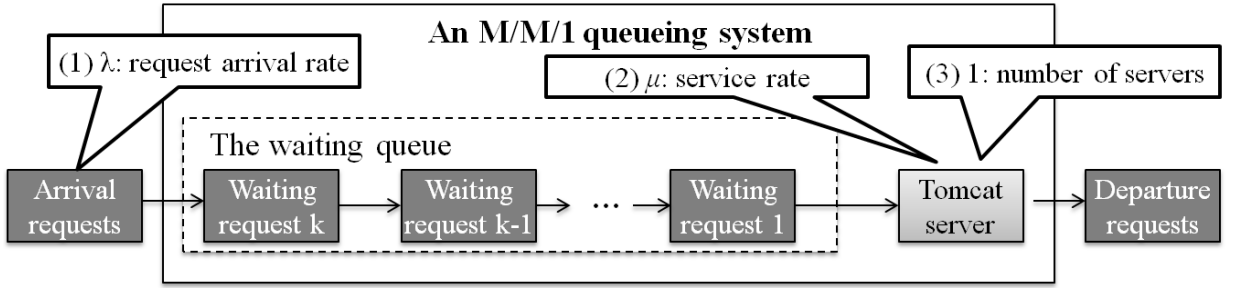

Figure 3.1: An example of an M/M/1 queueing system.

Typically, the *state* of an M/M/1 queueing system can be represented by a single value $k$, i.e. the number of requests in the system ($k \geq 0$). By mapping this M/M/1 queueing system into a continuous time Markov chain (CTMC), the system can be described using a birth–death process. This process has two features: (1) at any time point, there is at most one event happens: either a new request arrives or an existing request leaves the system; (2) the arrival rate $\lambda$ and the service rate $\mu$ are independent of the current state. Let $p_k$ be the steady state probability that the state is $k$. By solving the CTMC [72], we have:

$$p_0 = 1 - \frac{\lambda}{\mu} \qquad (3.1)$$

And

$$p_k = (\frac{\lambda}{\mu})^k \times p_0 \tag{3.2}$$

where $k > 0$ and $\mu > \lambda$. According to Equations (3.1) and (3.2), the expected number $\bar{k}$ of requests in the queueing system,

$$\bar{k} = \sum_{k=0}^{\infty}(k \times p_k) = p_0 \times \sum_{k=1}^{\infty}(k \times (\frac{\lambda}{\mu})^k) = \frac{\lambda}{\mu - \lambda} \tag{3.3}$$

where the summation $\sum_{k=1}^{\infty}(k \times (\frac{\lambda}{\mu})^k) = \frac{\frac{\lambda}{\mu}}{(1-\frac{\lambda}{\mu})^2}$.

By applying the Little's law, which states that in a steady state system, the expected number $\bar{k}$ of requests in the system is equal to the average request arrival rate $\lambda$ *multiplied* by these requests' expected response time $r(s)$, we have,

$$\bar{k} = \lambda \times r(s)$$

Thus, by substituting Equations (3.3), we can calculate a request's expected response time $r(s)$ in an M/M/1 queueing system with server $s$,

$$r(s) = \frac{\bar{k}}{\lambda} = \frac{1}{\mu - \lambda} \tag{3.4}$$

Note that this response time including both the request's waiting time in the queue and its time of being served. For example, in an M/M/1 queueing system of a Tomcat server, the service rate $\mu = 14$ requests/second, the arrival rate $\lambda = 7$ requests/second, a request's expect response time in the system is approximately 0.14 seconds.

**A tier of servers in the application**

Suppose a tier of $m$-tier application has $n$ parallel and independent servers, each server can be modelled as an M/M/1 queueing system and hence this tier can be described by $n$ M/M/1 queueing systems. For convenience, we assume that the servers at the same tier are *homogeneous*; that is, they have the same service rate (relaxation of the assumption is possible and the proposed estimation algorithm is still applicable). Based on this assumption, the requests of this tier are equally distributed to each server by the LB server, as illustrated in Figure 3.2. Similarly, the departure requests of this tier equal the summarised departure requests of all $n$ servers.

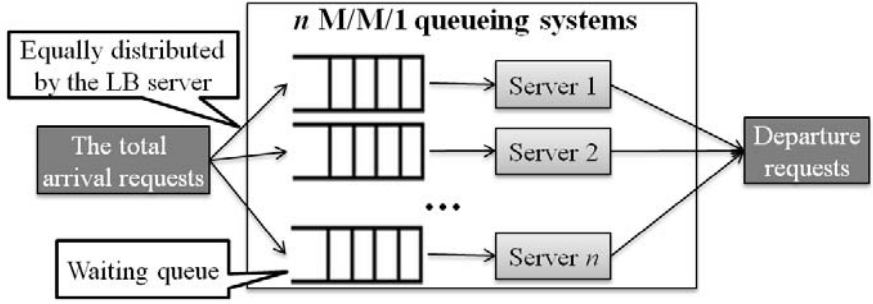

Figure 3.2: An example of $n$ M/M/1 queueing systems.

In a $m$-tier application, let $S_i$ be the set of servers at tier $i$ ($1 \leq i \leq m$) and $n=|S_i|$ be its number of servers. Let $n \times \lambda$ be the total request arrival rate of tier $i$, so $\lambda$ is the request arrival rate of a single server $s$ at this tier. Since a request at tier $i$ can only be assigned to one of its $n$ servers to be served and all the requests are equally distributed, these requests' expected response time at tier $i$ is equivalent to the expected response time at server $s$. We can estimate this response time using sever $s$'s M/M/1 queueing system. Note that the scaling algorithm uses this estimated response time to guide the scaling process only when new servers are added to or existing servers are removed from the application; otherwise the tier's response time equals the monitored response time.

**Definition 3.1 (A tier's response time).** *Consider a tier $i$ of servers in a multi-tier application, in which each server $s$ is modelled as an M/M/1 queueing system with arrival rate $\lambda$ and service rate $\mu$. A request's expected response time of this tier, denoted by $r(S_i)$, is equivalent to the expected response time at server $s$:*

$$r(S_i) = r(s)$$

*According to Equation* (3.4),

$$r(S_i) = \frac{1}{\mu - \lambda} \tag{3.5}$$

**The whole multi-tier application**

The whole $m$-tier application is modelled as a network of M/M/1 queueing systems. Specifically, this queueing network consists of a series of $m$ tiers (layers) and each tier consists of one or multiple M/M/1 queueing systems. For example, Figure 3.3's queueing network has five tiers and each tier has one or multiple queueing systems. In addition, each queueing system in the network contains a single-server queue and this system is used to analyse a server component in the application.

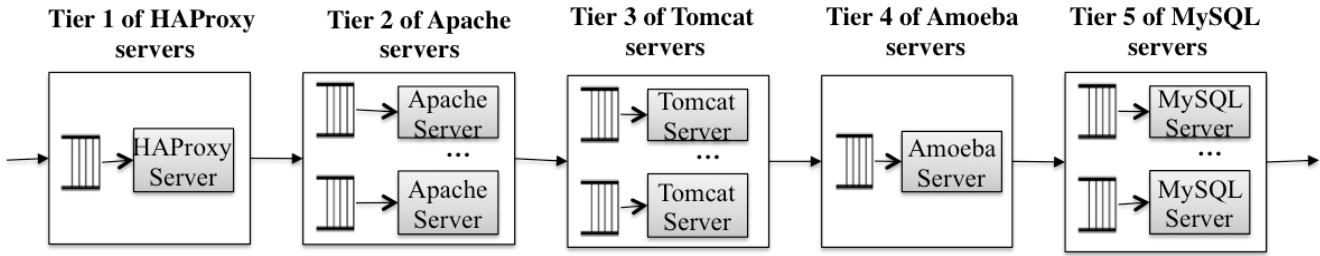

Figure 3.3: The queueing network of a five-tier e-ecommerce site.

Based on the description of multi-tier cloud applications and their servers in Definition 2.1, 2.2 and 2.3, the following two assumptions have been made in the queueing network used for the analysis of such application.

**Assumption 1. The open queueing network with no feedback.**

This assumption indicates that when a request enters the queueing network, it traverses the network in a sequel way; that is, this request either visits a subsequent tier of the network or leaves the network. Based on this assumption, in a steady-state queueing system with a Poisson arrival process with rate parameter $\lambda$, the departure process also follows a Poisson process with the same rate parameter $\lambda$.

Note that there are two situations to consider when modelling a multi-tier application using an open queueing network. First, a request may visit a previous tier after it is processed at one tier. Second, the cache mechanism in sever-side software causes a request at one tier to immediately leave the application without visiting servers at the following tiers. For example, a Tomcat application server can cache database queries, thus avoiding the visiting to the database sever at the subsequent tier. In the open queueing network, both situations mean completion of the request. Specifically, the situation of visiting previous tiers can be described as the current request leaves the application and a new request enters the network; the impact of cache can be reflected by appropriately setting the transition probability of a tier (cached requests leave the network from this tier) and service time (this time incorporates situations of both cache hits and misses).

**Assumption 2. Constraints on the arrival and departure requests at different tiers of the queueing network.**

According to Definition 2.2, there are four types of tiers in a multi-tier application with increasing tier ids: LB tiers (for service), service tier, LB tier (for storage), and storage tier. These tiers have different constraints on their arrival and departure requests in the queueing network:

- **Constraint on arrival requests**: requests from end users can only enter the network from the LB tiers.

- **Constraint on departure requests**: requests can leave the network at either the service tier or the storage tier.

- **Constraint on intermediate requests**: a request can walk through the network at a sequential path; visit several tiers; and only visit one server at one tier. Specifically, let $p_{ij}$ be the transition probability that a request departures from a queueing system at tier $i$ and proceeds to the next queueing system at tier $j$. This constraint requires that the tier id of $j$ is larger than the id of tier $i$. In addition, a request has several possibilities of choosing the next tier and we assume each possibility is assigned a fixed probability. For example, in an e-commerce site, a request leaving the tier of HAProxy (i.e. the LB server for service) has a 50% probability to choose the tier of Apache servers and another 50% probability to proceed to the tier of Tomcat servers.

Under the above constraints, the arrival and departure rates of requests at each tier can be analysed using the arrival rate $\lambda_e$ of end users' requests and these requests' transition probabilities between different tiers of the network. Take Figure 2.1's five-tier e-commerce site as an example, Figure 3.4(a) to Figure 3.4(e) show the arrival and departure requests in these five tiers of the queueing network respectively. In Figure 3.4(a), the tier 1 (HAProxy server) of the application directly receives requests from end users with arrival rate $\lambda_e$ and this tier's departure rate $d_1 = \lambda_e$. Tier 1's departure requests are distributed to two service tiers, namely tier 2 (Apache servers) and tier 3 (Tomcat servers), of the application with probabilities $p_{12}$ and $p_{13}$ where $p_{12} + p_{13}$=1. Similarly, in Figure 3.4(d), another LB tier of Amoeba server can receive requests from two service tiers and forward these requests to the storage tier of MySQL servers. In Figure 3.4 (b)'s tier 2, the request arrival rate is the product of tier 1's departure rate $d_1$ and the transition probability $p_{12}$ that a request leaves tier 1 and proceeds to tier 2. After processing these requests in the Apache servers, the departure requests of tier 2 can either leave the application (with probability $p_{20}$) or go to other tiers (either tier 3 with probability $p_{23}$ or tier 4 with probability $p_{24}$, where $p_{20}+p_{23}+p_{24}$=1. Tier 3 (Tomcat) and tier 5 (MySQL) have similar behaviours to tier 2.

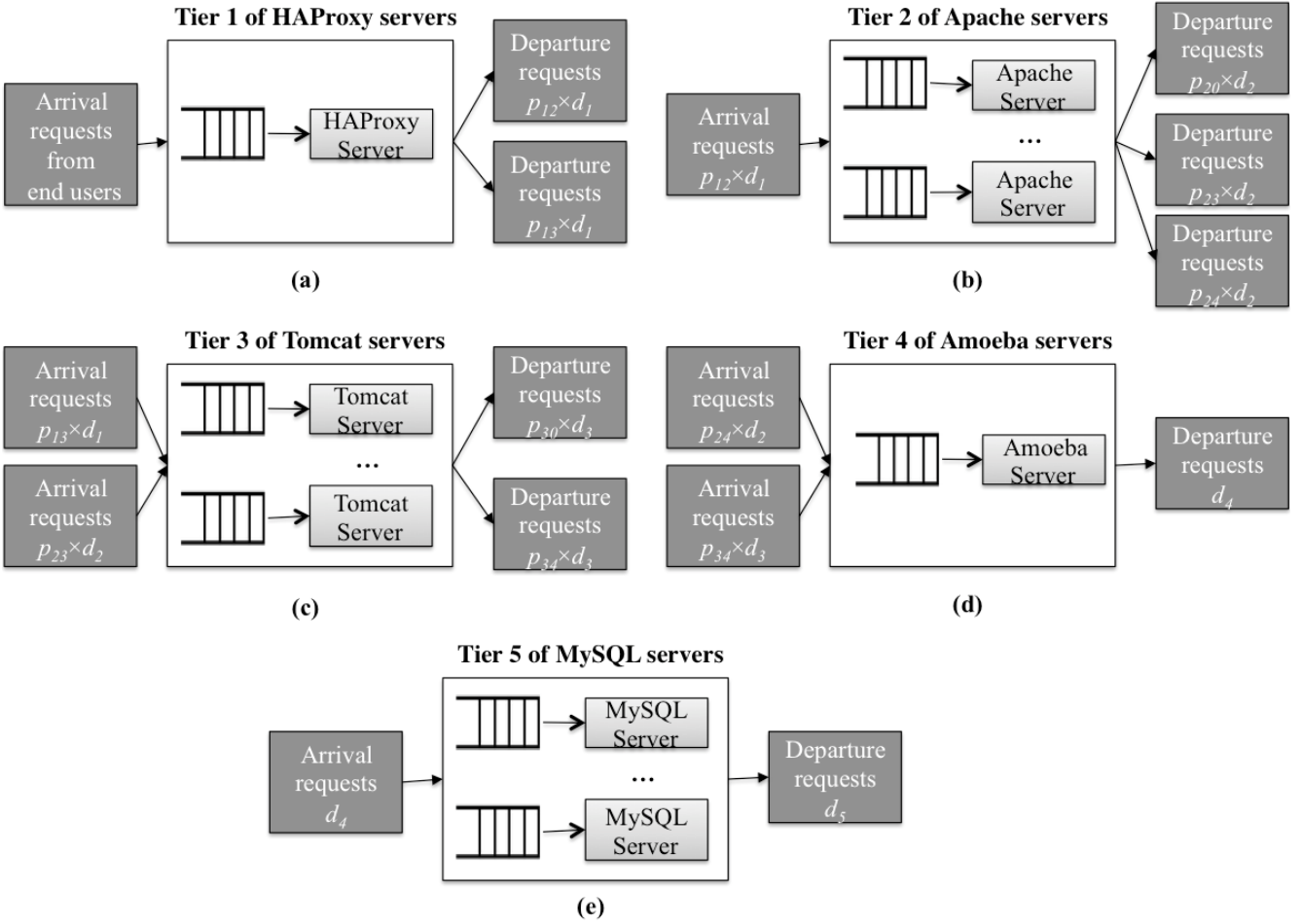

Figure 3.4: Different tiers of queueing systems in the queueing network. We can see (a) the tier 1 of HAProxy server; (b) the tier 2 of Apache servers; (c) the tier 3 of Tomcat servers; (d) the tier 4 of Amoeba server; (e) the tier 5 of MySQL servers.

In conclusion, using an open queueing network to model a $m$-tier application, end users' requests enter the network from the tier of HAProxy server (i.e. tier 1). When these requests are processed, they are immediately forwarded to other tiers, or leave the application after traversing one or multiple tiers. The application's response time, therefore, is the sum of each tier's response time.

**Definition 3.2 (Total response time of a multi-tier application).** *Given a $m$-tier application whose server set $S$ can be divided into m subsets of servers: $S_1$, $S_2$, ..., $S_m$. The total response time of the application, denoted by $r^t(S)$, is equal to the sum of response time at each tier:*

$$r^t(S) = \sum_{i=1}^{m} r(S_i) \qquad (3.6)$$

**Different types of workloads**

In the above analysis, we consider requests that come from the same type of workload; that is, these requests follow the same path (transition probabilities) through the network. However, in reality, there are various kinds of requests that form different types of workloads, which can be described by different transition probabilities of requests through the network. For example, there are three types of workloads ("Shopping", "Browsing", and "Ordering") in Figure 3.5's five-tier e-commerce site and different workloads are described by different request flows in the queueing networks. In Figure 3.5(a), the "Shopping" workload has the same amount of requests on both tier 2 and tier 3: 50% of departure requests at tier 1 (HAProxy) enter tier 2 (Apache) and the remaining 50% of requests go to tier 3 (Tomcat). After processing these requests, the departure requests at both tiers proceed to tier 4 (Amoeba). In contrast, in the "Browsing" workload (Figure 3.5(b)), end users mainly visit pages hosted in Apache and Tomcat servers and only 50% of these servers' departure requests will go to tier 4. Finally, in the "Ordering" workload (Figure 3.5(c)), requests from end users mainly stress the Tomcat servers at tier 2 and MySQL servers at tier 5. Hence, only 25% of end-user requests are distributed to tier 1 of Apache servers.

Note that we use slightly different terms to describe the queueing systems used to model multi-tier applications, which are summarised in Table 3.2.

Table 3.2. Terms in the queueing systems used to model a multi-tier application

| Classical queueing system | Queueing system used to model multi-tier applications |
| --- | --- |
| Customer | Request |
| Service center in a queue | Server (component) in a queue |
| A layer of queueing network | A tier of queueing network |

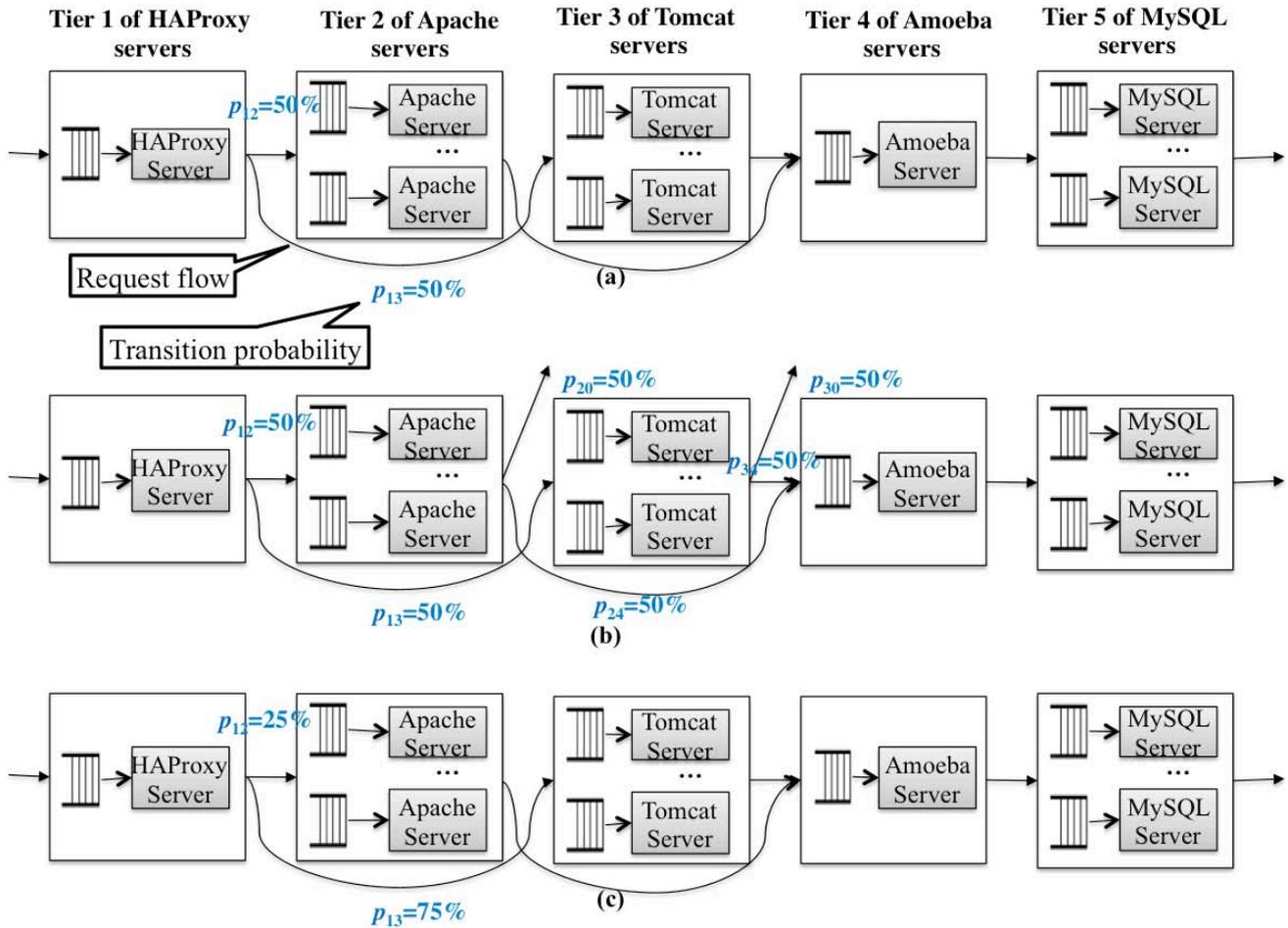

Figure 3.5: Three example queueing networks of the e-ecommerce site for three types of workloads. (a) The typical "Shopping" workload. (b) The primarily "Browsing" workload. (c) The primarily "Ordering" workload.

### 3.3.2 Definition of Cost-aware Criteria

In the queueing network of a multi-tier application, the parameters for analysing the application can be divided into three categories. (1) *Workload parameters*: the workload characterization can be described by two types of parameters: workload *type* and *volume*. The request flow (e.g. a departure request of tier $i$ has a probability of 20% to go to tier $j$ and another probability of 80% to go to tier $k$) denotes the *type* of workload, and the requests' arrival rate denotes the *volume* of workload. (2) *Resource parameters*: these parameters describe the features of resources that can influence cost and performance of the application, including the VM configurations (CPU, memory and network bandwidth) of servers and the service time depending on these configurations. (3) *Application parameters*: these parameters specify the service principles and constraints that affect the performance of the application. In queueing systems, there are a variety of the optional service principles of waiting requests. Examples are FIFO, last in first out (LIFO) (the request with the shortest waiting time is served first), priority (requests are divided into different priorities). The maximal size of the waiting queue and the incoming requests can also be specified using parameter $B$ and $K$ in an M/M/1 queueing system.

Using the three parameters of a queueing network, the resource consumption (or equivalently the deployment cost) and the total response time of the application can be analysed. Based on the analysis results of the queueing systems, we can define the cost aware criteria for the CAS algorithm. The criterion for the CACE-For-Scaling-Up algorithm is designed to measure the *cost spent* in adding a server divided by the *decreased response time* because of this addition. Hence, this criterion is called consumed cost/decreased response time (CC/DRT) ratio.

**Definition 3.3 (CC/DRT ratio).** *Tier $i$'s CC/DRT ratio, denoted by $c^c(S_i)$, is the cost spent per unit time in decreasing response time through adding a server $s$ to tier $i$:*

$$c^c(S_i) = c(s)/(r(S_i) - r(S_i \cup \{s\})) \qquad (3.7)$$

The criterion for the CACE-For-Scaling-Down algorithm, called Saved Cost/Increased Response Time (SC/IRT) ratio (definition 3.4), is the *cost saved* by removing a server divided by the *increased response time* due to this removal.

**Definition 3.4 (SC/IRT ratio).** *Tier $i$'s SC/IRT ratio, denoted by $c^s(S_i)$, is the cost saved per unit time in increasing response time through removing a server $s$ from tier $i$:*

$$c^s(S_i) = c(s)/(r(S_i \setminus \{s\}) - r(S_i)) \qquad (3.8)$$

In elastic cloud environment, compute resources are consumed on-demand similar to traditional utilities such as water and electricity [69]. In this context, an application's cost includes the expense of deploying all servers (Definition 3.5) and these servers are usually charged in pay-as-you-go pricing model in clouds. In a cloud, this **cost** is usually measured by the cost (e.g. 10 cents or dollars) spent per time unit (e.g. minute or hour).

**Definition 3.5 (Total deployment cost of a multi-tier application).** *In a $m$-tier application with server*

*set $S$, the total cost needed to deploy the application is denoted by $c^t(S)$, where $c^t(S) = \sum_{i=1}^{m} |S_i| c(s_i)$ ($s_i \in S_i$) and $c(s_i)$ is the cost of a server $s_i$ at tier $i$.*

## 3.4 Capacity Estimation for Scaling Up

The CACE-For-Scaling-Up algorithm aims at adding servers to an application so as to reduce its response time below a specified response time, while keeping the deployment cost as small as possible. Given this motivation, the algorithm judges the tier with the smallest CC/DRT ratio as the *bottleneck* tier where a server needs to be added. Compared to other tiers, addition of servers to this tier can decrease response time with the smallest cost per unit time.

A detailed algorithm is given below. The algorithm first builds a candidate server set $S^C$. The candidate set consists of all eligible tiers' server subsets (an eligible tier is the tier that can be added at least one server in scaling up). For example, in the e-commerce site, the two LB tiers of HAProxy and Amoeba are ineligible tiers because these tiers do not need to add servers in most of the cases. The initial candidate set takes each tier's server subset as its element (line 2). The algorithm iteratively executes under two conditions: (1) the candidate set $S^C$ is not empty (that is, new servers can be added); (2) the total response time $r^t(S)$ is greater than the required time $r^{sla}$ (line 4 to 15). In each loop, the algorithm *first* tries to find a set $S^*$ of the tiers where adding a server to can make $r^t(S) \leq r^{sla}$ and ends the scaling up (line 5). If one or multiple tiers are found (line 6), the tier whose single server is the cheapest is selected as the bottleneck tier (line 7); otherwise, the algorithm selects the bottleneck tier with the *smallest* CC/DRT ratio (line 9 and 10). Subsequently, the algorithm judges whether adding a server to the bottleneck tier violates any constraint (line 11). If the addition is feasible, a server is added (line 12); otherwise, the selected tier is viewed as ineligible to be added a server and its server subset is removed from the candidate list (line 14).

The constraints checked in line 11 include the constraints specified in the SLA and servers' own constraints. Examples of the former constraints are cost constraint (the application's deployment budget) and resource constraint (each tier's maximum number of servers). An example of the latter constraint is server's replication constraint. For instance, there is at most one MySQL master server in an application, so MySQL master's replication constraint is 1.

Note that if $r^t(S)$ is still larger than the required time $r^{sla}$ while adding a server to any tier of the application is infeasible (line 18), the scaling process is halted and an exception handling is triggered to inform the application owner (line 19). The application owner can either relax the violated constraints or modify the response time target. For example, if the cost constraint is violated (that is, adding any server to the application exceeds the deployment budget), the application owner can either increase the budget and resume the scaling process, or accept the larger response time $r^t(S)$ and stop the scaling up.

**The CACE-For-Scaling-Up Algorithm**

**Input:** $S$, $\lambda_e$ and $r^{sla}$.

**Output:** updated $S$.

1. **Begin**

2.     $S^C = \{S_1, S_2, \ldots, S_m\}$; // the original candidate server set

3.     Compute $r^t(S)$ using Equations 3.5 and 3.6;

4.     **while** ($S^C$ is not empty and $r^t(S) > r^{sla}$) **do**

5.         Find subset $S^*$ from $S^C$;

6.         **if** ($S^*$ is not empty), **then**

7.             Select $S_i$ from $S^*$ with the smallest $c(s_j)$ where $s_j \in S_i$;

8.         **else**

9.             Compute each $c^c(S_k)$ where $S_k \in S^C$ using Equation 3.7;

10.            Select $S_i$ from $S^C$ with the smallest $c^c(S_i)$;

11.        **if** (Add a server $s$ to tier $i$ is feasible), **then**

12.            $S_i = S_i \cup \{s\}$; // add $s$ to tier $i$

13.        **else**

14.            $S^C = S^C \backslash S_i$; // remove server subset $S_i$ from the candidate list

15.            Compute $r^t(S)$ using Equations 3.5 and 3.6;

16.    **if** ($r^t(S) < r^{sla}$), **then**

17.        Return $S$.

18.    **else** // $r^t(S) > r^{sla}$ and $S^C$ is empty

19.        Halt the scaling process and trigger an exceptional handling.

20. **End**

**Proposition 3.1.** *The time complexity of the CACE-For-Scaling-Up algorithm is $O(m^2)$, where $m$ is the number of tiers for the application. Note that $m$ is usually a small number that ranges from 1 to 8.*

**Proof.** In the CACE-For-Scaling-Up algorithm, each time we conduct the estimation loop (line 4 to 15), it takes $O(m)$ to complete the traversal of all $m$ tiers to find the bottleneck tier (line 7 and 10). Other operations in the loop can be done in constant time. In each loop, either a server is added (line 12) or a server subset is removed (line 14). Hence, the algorithm can be completed within $O(C + m)$ loops, where it takes $O(C)$ to add finite servers ($C$ is a constant that is usually less than 20) and $O(m)$ to remove all subsets from the candidate list. The total time complexity, therefore, is $O(m^2)$. ∎

## 3.5 Capacity Estimation for Scaling Down

The CACE-For-Scaling-Down algorithm aims at removing servers from an application to save as largest costs possible, while still meeting the required response time. To this aim, this algorithm judges the tier with the largest SC/IRT ratio as the bottleneck tier where a server needs to be removed, because this removal can save the maximum cost per unit of response time increased. A detailed CACE-For-Scaling-Down algorithm is given below. The algorithm iteratively executes until no redundant servers can be removed. In each loop, the algorithm first finds all ineligible tiers, where removing one server from any of these tiers would make the total response time exceed the required response time in the SLA. The algorithm then removes all ineligible tiers' server subsets from the candidate server set (line 5). Consequently, the remaining tiers can be removed by at least one server. From these tiers, the algorithm selects the bottleneck tier with the *largest* SC/IRT ratio and removes a server (line 6 and 7). Note that the algorithm also checks the constraints which may forbid removal (line 9). For example, the HAProxy acts as end users' communication interface, so the server set of HAProxy must have at least one server.

**Proposition 3.2.** *The time complexity of the CACE-For-Scaling-Down algorithm is $O(m^2)$, where $m$ is application A's tier number.*

**Proof.** Similar to the CACE-For-Scaling-Up algorithm, it is not difficult to prove that the CACE-For-Scaling-Down algorithm has $O(m + C)$ loops in maximum and each loop takes $O(m)$ to complete all operations. Hence, the total time complexity is $O(m^2)$. ∎

**CACE-For-Scaling-Down Algorithm**

**Input:** $S$, $\lambda_e$ and $r^{sla}$.

**Output:** Updated $S$.

1. **Begin**

2.     $S^C = \{S_1, S_2, \ldots, S_m\}$; // the original candidate server set

3.     Compute $r^t(S)$ using Equations 3.5 and 3.6;

4.     **while** ($S^C$ is not empty and $r^t(S) \leq r^{sla}$) **do**

5.        Find and remove ineligible server subset from $S^C$;

6.        Compute each $c^s(S_k)$ where $S_k \in S^C$ using Equation 3.8;

7.        Select $S_i$ from $S^C$ with the largest $c^s(S_i)$;

8.        **if** (Remove a server $s$ from tier $i$ is feasible), **then**

9.           $S_i = S_i \backslash \{s\}$; // remove $s$ from tier $i$

10.       **else**

11.           $S^C = S^C \backslash S_i$; // remove subset $S_i$ from the candidate server set

12.       Compute $r^t(S)$ using Equations 3.5 and 3.6;

13.       Return $S$.

14. **End**

## 3.6    Evaluating the Effectiveness of the Cost-aware Scaling Algorithm

In this section, we first introduce the evaluation setup (section 3.6.1), following the results of simulation and evaluation. The first evaluation is designed to illustrate the effectiveness of our CAS algorithm in adapting changing workload volumes and types by effectively scaling applications up and down (section 3.6.2). Furthermore, the CAS algorithm's salient feature in delivering cost-efficient services is demonstrated by comparison with existing scaling techniques (section 3.6.3).

### 3.6.1    Evaluation Setup

**Hardware environment**. Our experiments are conducted in a data centre running the IC Cloud workstation [38]. The configuration used has four physical machines (PMs), each with eight CPU cores and 32 GB memories. The version of each processor is Quad-Core AMD Opteron(tm) Processor 2380, with 2.50 GHz clock frequency and 512 KB cache size. All four PMs share a 4.1 Tb centralised storage and are connected through a switched gigabit Ethernet LAN with speed 1000 mbs.

**Software environment**. The e-commerce site in Figure 2.1 was implemented and the scaling up and down of this application was tested. For convenience, each server component of the application was installed on a single dedicated VM with Linux Ubuntu operating system. In deployment, different servers have different VM configurations, as listed in Table 3.3. Two versions of the MySQL database (master and slave) are implemented to support a data replication model. A MySQL master is initially deployed and, when the tier of MySQL is scaled up, extra MySQL slaves are added and configured with replication from the MySQL master. Given a fixed VM configuration, the deployment of the Tomcat and Apache servers can be completed in a constant time. In the evaluation, the database has a fixed amount of data to be replicated; that is, the data replication time of MySQL slave is fixed. Thus, the deployment time of a MySQL server is also a constant time.

Table 3.3. Six types of servers' deployment information

| Server name | CPU | RAM (GB) | Cost (dollars/hour) | Replication constraint |
|:---:|:---:|:---:|:---:|:---:|
| HAProxy | 4 | 4 | 0.24 | 1 |
| Apache | 4 | 4 | 0.24 | infinite |
| Tomcat | 2 | 2 | 0.12 | infinite |
| Amoeba | 2 | 2 | 0.12 | 1 |
| MySQL master | 1 | 1 | 0.06 | 1 |
| MySQL slave | 1 | 1 | 0.06 | infinite |

**Application logic implementation**. According to the replication constraints in Table 3.3, only the three types of servers, namely Tomcat, Apache and MySQL slave, can be scaled and the numbers of other types of server do not change. Thus, we implemented the requests of the e-commerce site for these three types of servers, in which each one differs from the others in terms of the server-side operations. Specifically, the requests of *Apache* servers mainly handle static HTML pages and images. The requests of *Tomcat* servers represent the business logic of the e-commerce services and we implemented the kNN method as an example. The kNN method is widely applied in ranking, classification, and recommendation services provided by e-commerce sites. The requests of *MySQL* servers stand for different database operations including inserting, reading, updating and removing data. Note that the two LB servers, namely *HAProxy* and *Amoeba*, can distribute requests instantly. Thus, the response time of these two severs are not considered.

**Three types of simulated workloads**. We distinguish three types of workloads to represent the typical behaviours of end users (customers of e-commerce site). (1) The *typical "Shopping" workload* represents the whole shopping process and it comprises requests that visit all static web pages in Apache servers, dynamic pages in Tomcat servers and MySQL databases. Thus, this workload stresses all three tiers of servers. (2) The *primarily "Browsing" workload* represents the browsing behaviour of end users such as classifying and ranking products as well as online recommendations. This workload mainly stresses servers at the service tier (e.g. Apache and Tomcat) and only make light and short database queries. Finally, (3) the *primarily "Ordering" workload* stands for the ordering actions such as logging in and making orders. This workload makes large requests to the application server (i.e. Tomcat) and stresses MySQL servers for accessing and updating database tables.

By mixing the requests of Apache, Tomcat and MySQL servers, we implemented three types of workloads, where each one differs from the others in terms of the path/route of requests. Table 3.4 shows the route (transition probabilities) of requests for each workload. For example, in the "Shopping" workload, 50% of end-user requests are distributed to the Apache servers and another 50% requests are allocated to the Tomcat server. After these requests are processed, 100% of them are forwarded to the MySQL servers.

Table 3.4. Routes of requests in three types of workloads

| Workload | Apache | Tomcat | MySQL |
|----------|--------|--------|-------|
| Shopping | 50% | 50% | 100% |
| Browsing | 50% | 50% | 50% |
| Ordering | 25% | 75% | 100% |

**The client simulator**. To simulate the above workloads, we implemented a client emulator based on siege benchmark (www.joedog.org/siege-home/). After setting the test period, this emulator can simulate a number of concurrent end users. Each end user continuously generates a sequence of requests to stress the server-side application. After a request is completed, the simulated end user waits for a random interval before initiating the next request to simulate actual end users' thinking time. In experiment, the "thinking time" between two requests randomly varies between 0 and 3 seconds. In the evaluation, the VM with 1 CPU core and 1 GB memory is employed to run the emulator, and this VM and the VMs of all server are in the same data center. All the VMs communicate using private IP addresses, thus guaranteeing a steady environment for testing.

### 3.6.2   Effectiveness of the CAS Algorithm

This section demonstrates the effectiveness of our CAS algorithm in scaling up and down applications to handle changing workload *volumes* and *types*. In the experiment, the e-commerce site is initially deployed in one HAProxy, Apache, Tomcat, Amoeba and MySQL master. In the SLA of this application, the required response time $r^{sla}$ is assumed to be no more than 1.2 seconds and total deployment budget $c^b$ is 2.5 dollars/hour.

**Scaling for changing workload volumes**. In the first evaluation, we test the typical "Shopping" workload using five sessions: the first three sessions stepwise increase the number of simulated end users to initiate scaling up and the remaining two sessions gradually decrease this number to trigger scaling down, as shown in Figure 3.6. This variance of end user numbers denotes the changing workload volume. More specifically, the first session is generated at time=0 second and it lasts 600 seconds. During this period, the application is monitored once every 60 seconds and the arrival rate and response time of the application are detected.

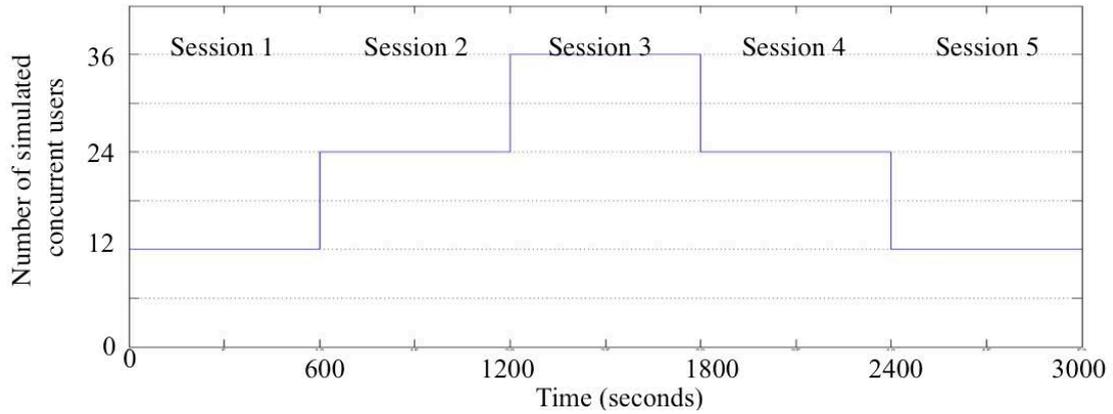

Figure 3.6: Five simulated sessions in the "Shopping" workload.

Figure 3.7 demonstrates the fluctuation in the total response time of the application observed in the evaluation of the "Shopping" workload. In the first three sessions, the response time is violated whenever the active session number is increased. For instance, when the concurrent user number is increased to 24 at time = 600 seconds and saturates all three tiers of Apache, Tomcat and MySQL servers. The scaling up is triggered and a new server is added to each tier. The violation typically lasts for 1 or 2 minutes because the addition of new servers consumes some time. By contrast, in the fourth and fifth sessions, the CAS algorithm scales down the application while meeting the required response time $r^{sla}$.

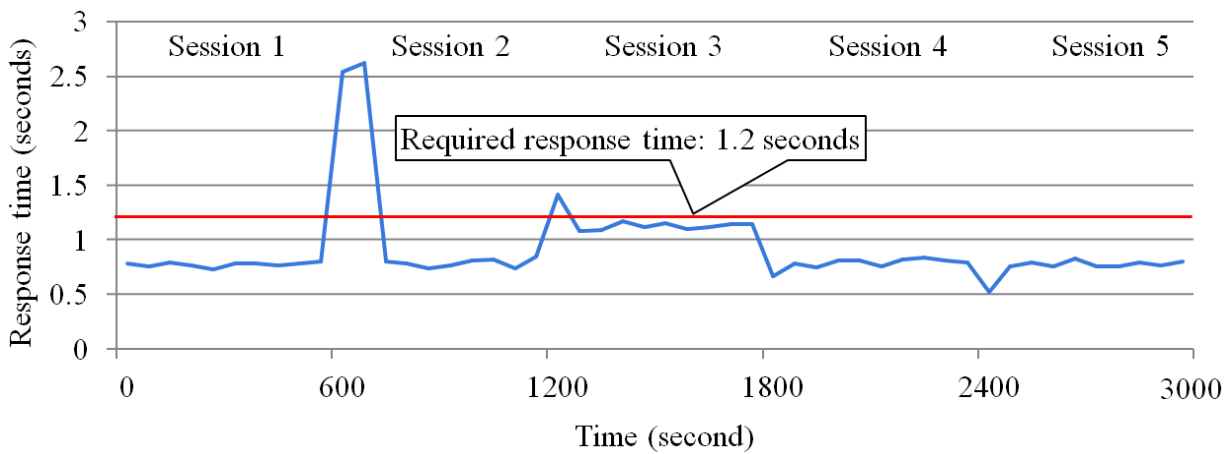

Figure 3.7: The total response time in the evaluation of the "Shopping" workload.

*Result: When the workload volume increases, the CAS algorithm can scale up the application to restore the required response time within 1 or 2 minutes. On the other hand, the algorithm can scale down the application when the workload volume decreases, while maintaining the response time target.*

**Scaling for changing workload types**. We repeat the first evaluation to test all three types of workloads.

Figure 3.8 shows the simulated "Browsing" and "Ordering" workloads and the number concurrent users at each session. Although these rates are similar to each other, workload of different types stresses different tiers of servers in the application.

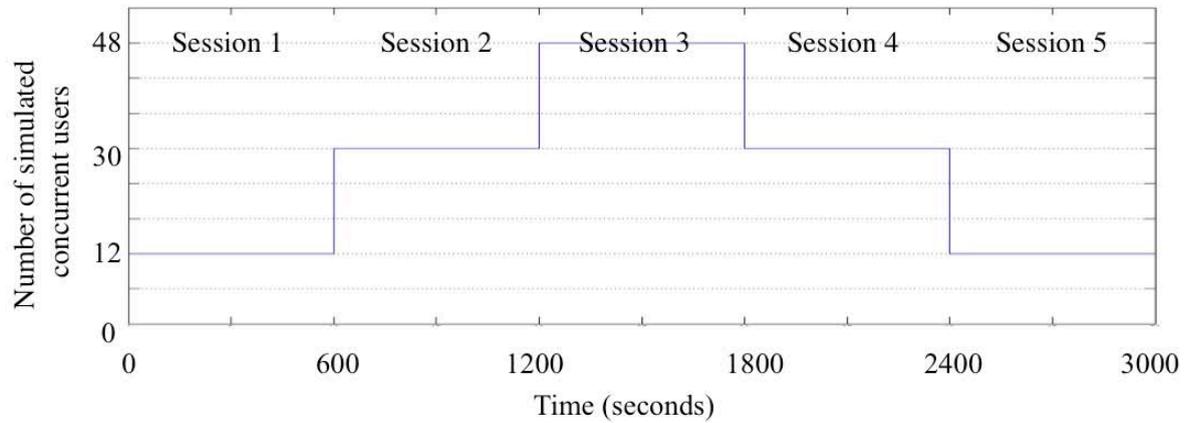

**(a) The "Browsing" workload.**

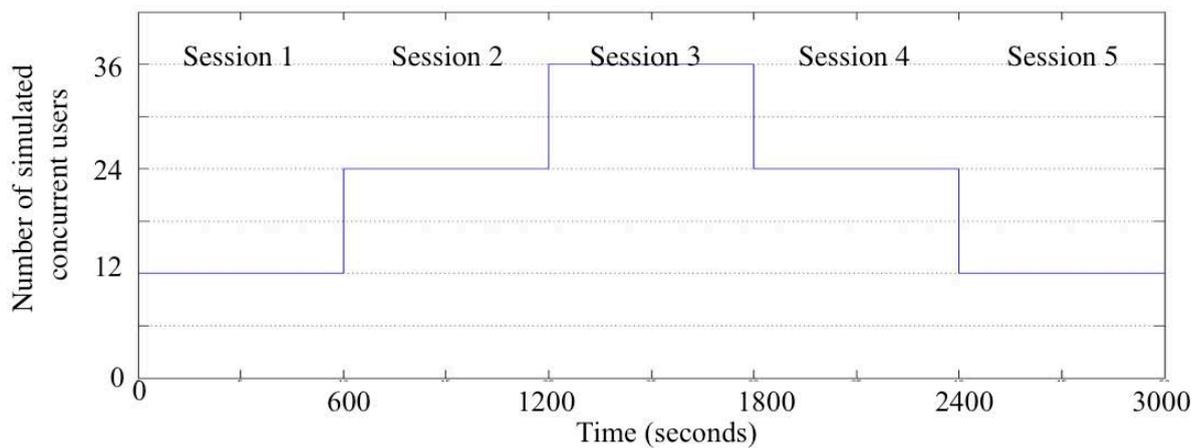

**(b) The "Ordering" workload.**

Figure 3.8: Five simulated sessions in the "Browsing" and "Ordering" workloads.

Given the above workloads, Figure 3.9 shows that the CAS algorithm can scale up and down the application to meet the required response time for the "Browsing" and "Ordering" workloads. Moreover, Figure 3.10 shows the number of servers at each tier of the application to support different types of workloads. This number adapted to the workload types. For example, in most cases of scaling up (down) for the "Browsing" workload, the tiers of Tomcat and Apache are saturated or idle and the number of these servers changes with the number of concurrent users (Figure 3.10(b)). By contrast, the numbers of Tomcat and MySQL are influenced significantly by the concurrent user number in the "Ordering" workload (Figure 3.10(c)). In addition, Figure 3.11 presents the total cost of deploying these servers and this cost is always kept within the budget (2.5 dollars/hour).

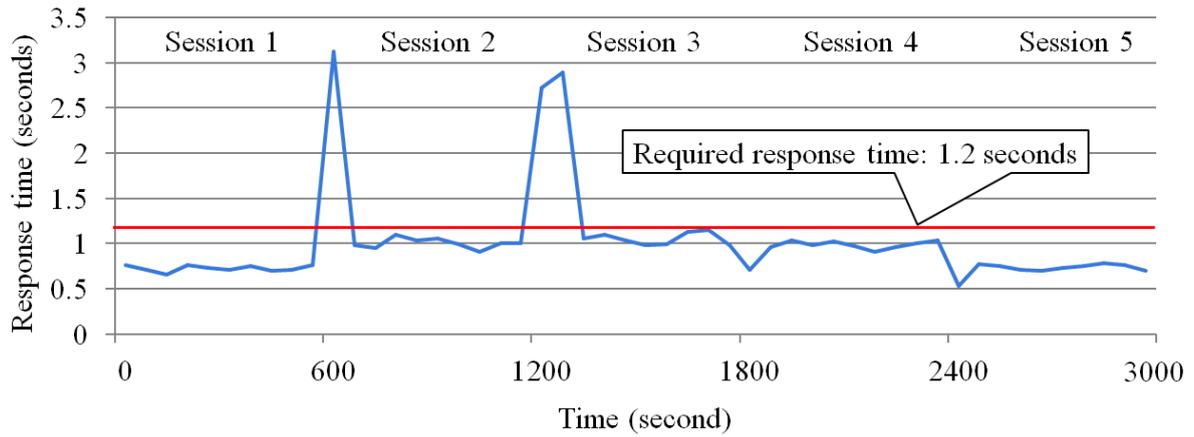

**(a) "Browsing" workload**

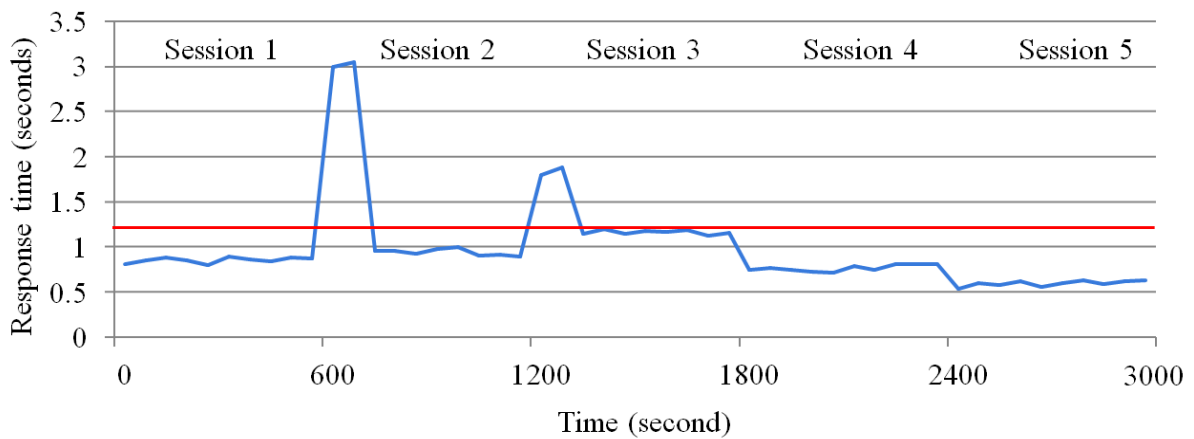

**(b) "Ordering" workload**

Figure 3.9: The total response time in the evaluation of the "Browsing" and "Ordering" workloads.

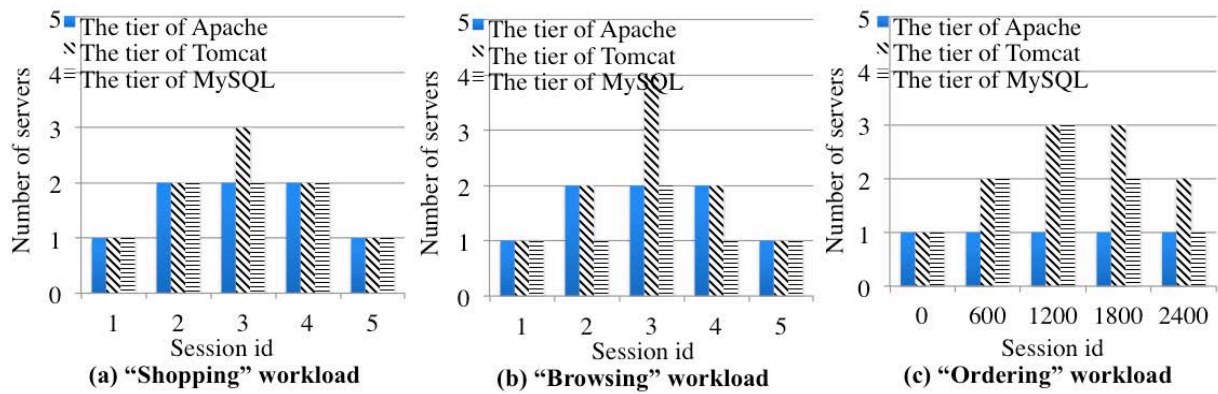

Figure 3.10: The number of servers in three types of workloads.

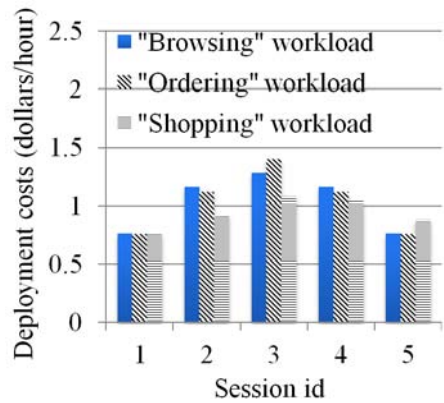

Figure 3.11: The total deployment cost in three types of workloads.

*Result: Our CAS algorithm is able to adapt with different volumes and types of workloads and identify the bottleneck tiers for each scaling. When the scaling is performed, only the bottleneck tiers are scaled up and down to maintain the response time target.*

### 3.6.3 Comparison with Existing Scaling Techniques

This section shows the effectiveness of our CAS algorithm to deliver cost-efficient services in scaling by comparing it with existing scaling algorithms. Typically, existing scaling techniques can be divided into two categories.

**Policy-based scaling (PBS) algorithms**

In the first category, applications are scaled using the pre-defined polices [2, 10-12]. These scaling algorithms can be termed PBS, which are adopted by mainstream cloud IaaS providers. The PBS algorithm first puts all tiers of Apache, Tomcat and MySQL servers into one server array and sets thresholds for scaling up and down for each tier of servers. If a majority of servers (i.e. larger than 50% of servers) vote for scaling up or down, a scaling up or down is performed to add/remove one servers to each tier of the application. In the evaluation, the scaling up thresholds (unit is second) of the Apache, Tomcat and MySQL servers are set to 0.5, 0.5 and 0.2, respectively; the scaling down thresholds are set to 0.25, 0.25 and 0.1, respectively.

**Tier-Dividing Scaling (TDS) algorithms**

In the second category, an application is modelled by a network of queueing systems including single or multiple tiers of servers. Each server's capacity is then analysed using the M/M/1 queueing system. Subsequently, each tier's required server number is calculated by breaking the application's total response time into per-tier response times. This type of scaling techniques [3-9, 14, 19], therefore, is termed TDS. The TDS algorithm applies *worst-case* capacity estimation to deploy sufficient servers capable of handling the peak workload. In the evaluation, the required response time (1.2 seconds) of the whole application is broken down into three per-tier required response times, which are 0.5 seconds, 0.5 seconds, 0.2 seconds for the tiers of Apache, Tomcat, MySQL master (slave), respectively. The worst-case capacity estimation is performed to deploy sufficient servers to meet the required response time in SLA even handling 130% of the detected request arrival rate.

In comparison, all three algorithms, namely CAS, PBS and TDS scaling algorithms, have the same initial application deployment: at each tier of the e-commerce site, one server is deployed.

**Effects of scaling on cost-efficient services**. We repeat section 3.6.2's experiment to test the CAS, PBS and TSD algorithms using three types of workloads. The effects of scaling on response time in the evaluations of the "Shopping", "Browsing" and "Ordering" workloads are displayed in Figure 3.12. It can be observed that all three algorithms can meet the required response time by adding more servers to handling larger workload volumes and removing servers when the request number decreases.

In addition, Figure 3.13 compares the total number of servers in the application deployed to handle different volumes and types of workloads using three scaling algorithms. Figure 3.14 presents the total cost of deploying these servers. The experiment results indicate that the CAS algorithm identifies the bottleneck tiers and only scales these tiers, thus this algorithm can meet the required response time with the smallest numbers of servers, i.e. the smallest deployment costs.

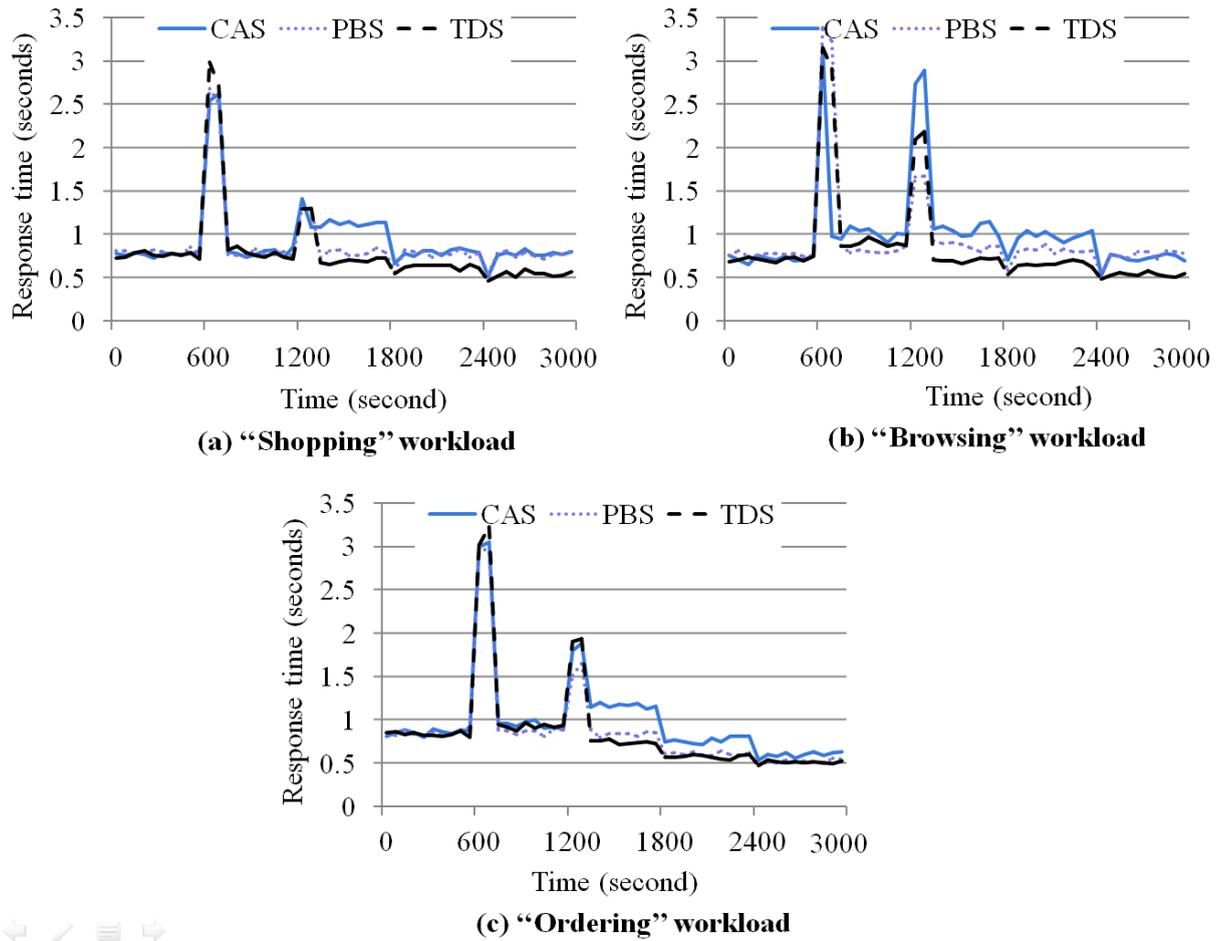

Figure 3.12: The total response time in the evaluation of three workloads using three algorithms.

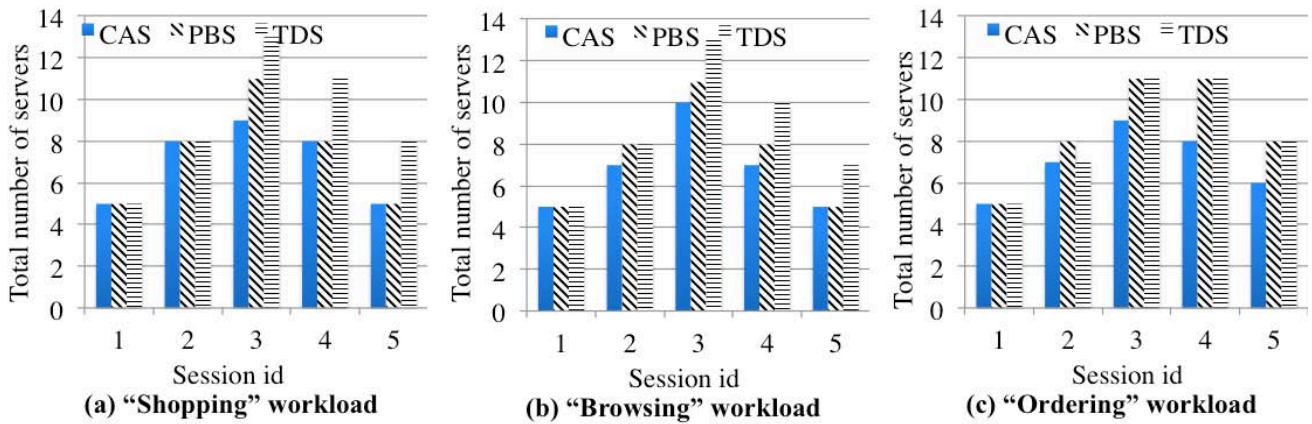

Figure 3.13: The total numbers of servers in three types of workloads using three algorithms.

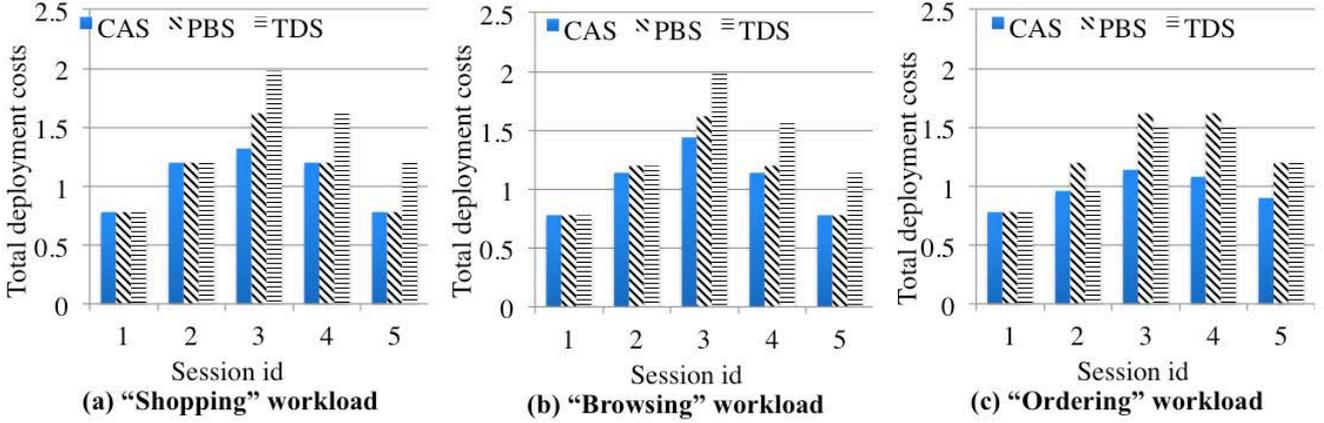

Figure 3.14: The total deployment costs in three types of workloads using three algorithms.

*Result: Using the cost-aware criteria, our CAS algorithm is able to provide cost-efficient resources in both scaling up and down by meeting the required response time using the smallest deployment cost.*

## 3.7 Discussion of the CAS Algorithm

### 3.7.1 Discussion of the Applicability of Other Queueing Systems

In this work, we apply the M/M/1 queueing system to analytically model the behaviour of multi-tier cloud applications and estimate their requests' response time. This choice of queueing system is based on the assumption that these applications' incoming requests are determined by a Poisson process. Based on the same assumption, we discuss another two queueing systems: M/G/1 and M/D/1, where G means *general* and indicates the service time can follow *arbitrary* distributions, and D denotes *deterministic* and represents *constant* service time.

Since the M/M/1 and M/D/1 queueing systems are two special cases of the M/G/1 queueing system, we first introduce how to use the M/G/1 queueing system to analyse a server $s$'s expected response time $r(s)$ at the steady state. Formally, let $\lambda$ be the request arrival rate and $\mu$ be the service rate. Let $x$ be the service time; $\bar{x} = \frac{1}{\mu}$ be the mean service time; and $var(x)$ be the variation of service time. This expected response time $r(s)$ can be calculated as [71]:

$$r(s) = \bar{x} + \frac{\lambda(1+C_x^2)}{2\mu^2(1-p)}. \qquad (3.9)$$

where $C_x^2 = \frac{var(x)}{\bar{x}^2}$ is the squared coefficient of variation of service time $x$ and $p = \frac{\lambda}{\mu}$ is the server utilisation.

Fixing the values of arrival rate $\lambda$ and mean service time $\bar{x}$, the expected response time $r(s)$ is determined by the squared coefficient of variation $C_x^2$; that is, the variation $var(x)$ of service time $x$. Hence, the larger the randomness of the service time $x$, i.e. the larger the variation $var(x)$, the longer the time $r(s)$. In the M/G/1 queueing system whose service time can follow arbitrary distributions, it is usually assume that $C_x^2 > 1$, which denotes high variance of service time. In contrast, in the M/D/1 queueing system, service time has no variations; that is, $var(x)$=0 and $C_x^2$=0. Hence, in the M/D/1 queueing system, the expected response time $r(s)$ is calculated as:

$$r(s) = \bar{x} + \frac{\lambda(1+C_x^2)}{2\mu^2(1-p)} = \bar{x} + \frac{\lambda}{2\mu^2(1-p)}. \qquad (3.10)$$

Finally, in the M/M/1 queueing system, $C_x^2$=1, because this queueing system's service time $x$ follows the exponential distribution, and the variance $var(x)$ equals the squared mean $\bar{x}^2$: $C_x^2 = \frac{var(x)}{\bar{x}^2} = 1$. Hence, the expected response time $r(s)$ in the M/M/1 queueing system is calculated as:

$$r(s) = \bar{x} + \frac{\lambda(1+1)}{2\mu^2(1-p)} = \frac{1}{\mu} + \frac{\lambda(1+1)}{2\mu^2\left(1-\frac{\lambda}{\mu}\right)} = \frac{1}{\mu} + \frac{2\lambda}{2\mu(\mu-\lambda)} = \frac{\mu-\lambda+\lambda}{\mu(\mu-\lambda)}$$

That is,

$$r(s) = \frac{1}{\mu-\lambda} \qquad (3.11)$$

Based on the above analysis, we use an example scenario of the "Shopping" workload in Figure 3.6 to illustrate the application of these three queueing systems in the CAS algorithm. According to Equations (3.9), (3.10), and (3.11), given the same arrival rate $\lambda$ and service rate $\mu$, the estimated response time in the M/D/1 queueing system is the smallest. For example, in a Tomcat server, suppose the arrival rate $\lambda$=3.27 requests/second and the service rate $\mu$=5.56 requests/second. Using the M/D/1, M/M/1, and M/G/1 queueing systems ($C_x^2$=2 in the M/G/1 queueing system), the estimated expected response times are 0.31 seconds, 0.44 seconds, and 0.57 seconds, respectively. However, the actual detected response time is 0.34 seconds. This means when applying the M/D/1 queueing system in the CAS algorithm, the estimated response time is smaller than the actual response time and the required response time in the SLA may be violated as a result. For example, in the "Shopping" workload, when the number of concurrent users increases to 24 at Session 2. Using the M/D/1 queueing system, the CAS algorithm estimates that the required response time (1.20 seconds) can be met by scaling up the application to deploy two Tomcat servers, one Apache server, and two MySQL servers. However, the actual response time is 1.37 seconds after this scaling up. Hence, the M/D/1 queueing system is not applicable to the CAS algorithm because it *underestimates* the response time.

Furthermore, we compare the M/M/1 and M/G/1 queueing systems using the five sessions in the "Shopping" workload. In the M/G/1 queueing system, the squared coefficient $C_x^2$ is set to 2. Figure 3.15 shows that applying these two queueing systems to guide the scaling of the e-commerce site application, both the estimated and actual response times are smaller than the required response time. Hence, both queueing systems are applicable to the CAS algorithm. We can also see that using the M/G/1 queueing system, the difference between the estimated and actual response times is larger. This is because the M/G/1 queueing system has larger squared coefficient $C_x^2$ of variation, thus incurring larger overestimation of the response time.

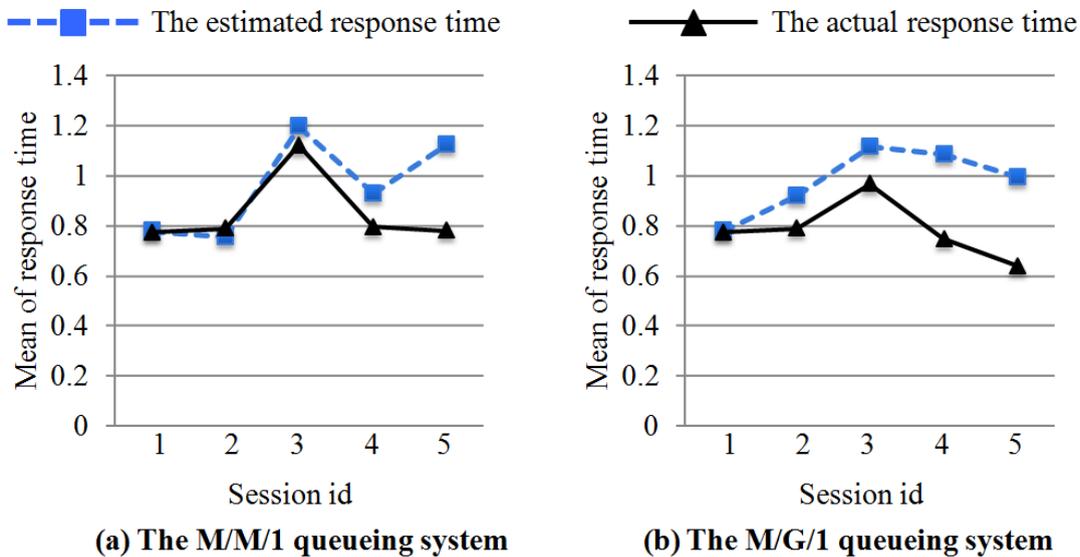

(a) The M/M/1 queueing system          (b) The M/G/1 queueing system

Figure 3.15: Comparison of the estimate and actual response times using the M/M/1 and M/G/1 queueing systems.

Figure 3.16 lists the number of servers at different tiers of the application after scaling up or down. The results indicate that when applying the M/M/1 queueing system in the CAS algorithm, application owners need to deploy a smaller or equal number of servers to meet the required response time. This is because although both queueing systems overestimate the response time, the estimation of the M/M/1 queueing system is more precise, thus guiding the scaling in a more accurate way. Since the CAS algorithm aims to minimise the deployment cost when scaling up and down the application, the M/M/1 queueing system is a better analytical model to guide this scaling in a more cost-effective manner. As illustrated in Figure 3.17, smaller or equal deployment costs are needed for the CAS algorithm when using the M/M/1 queueing system.

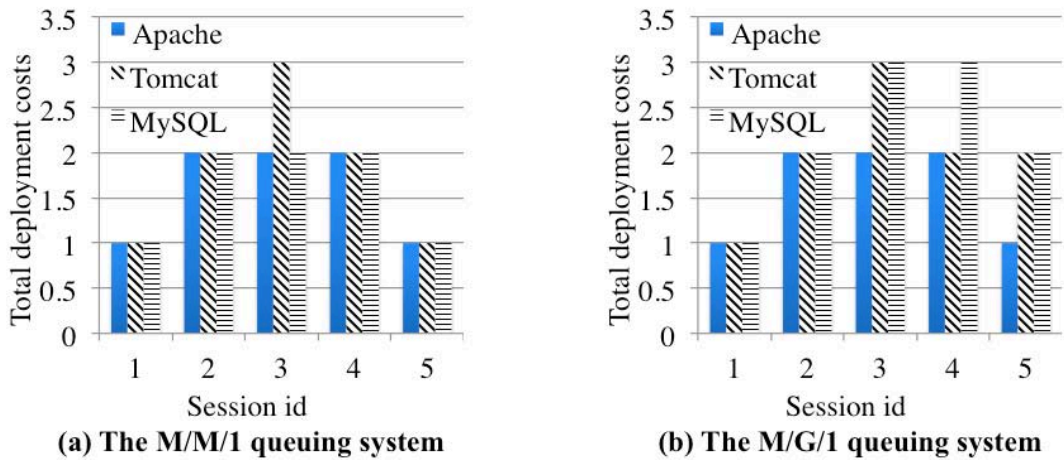

**(a) The M/M/1 queuing system**  **(b) The M/G/1 queuing system**

Figure 3.16: Comparison of the numbers of servers using the M/M/1 and M/G/1 queueing systems.

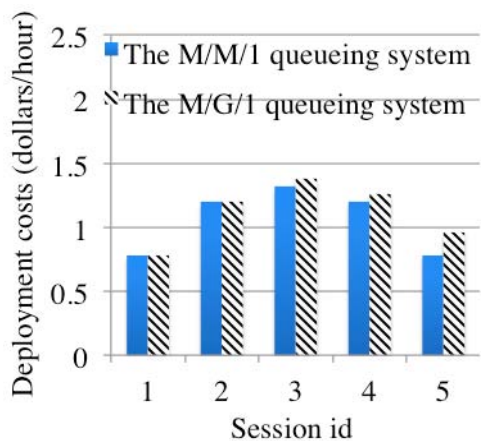

Figure 3.17: Comparison of the total deployment costs of the application using the M/M/1 and M/G/1 queueing systems.

## 3.7.2  Discussion of the CAS algorithm

The CAS algorithm presented in this chapter is based on reactive (immediate) scaling of multi-tier applications rather than using predictive mechanisms. The reactive scaling approaches are used by most providers such as Amazon EC2 [2] and RightScale [10] since they are simpler to support and require no prior knowledge of the workload characteristics. The CAS algorithm uses two methods to handle the possible errors in capacity estimation. First, it adds/removes only one server to/from the bottleneck tier in each estimation analysis and then iteratively conducts the next analysis. Even if a wrong tier is selected in an analysis, this would give prominence to the actual bottleneck tier and so this tier has a high probability to be selected in the next analysis. Secondly, the two capacity estimation algorithms are complementary to each other. For instance, if redundant servers are deployed in the scaling up, the CAS algorithm can trigger the scaling down to remedy this problem. The algorithm and approach fit multi-tier applications that are deployed to provide services for third parties, and where the variation in the workload changes based on the number of users using the application, and the type of requests they make. This is in contrast to other work that is based on scheduling workflow-based applications [73-77] across multiple resources and where the task durations for individual workflow steps can vary based on the input.

In this thesis we have investigated the use of the CAS algorithm in which the cost function was assumed to be the price, in monetary terms, that the user pays for the servers. The algorithm can be equally applied with other cost functions, for example one based on saving energy. Saving energy without sacrificing application owners' SLAs has a great economic incentive for cloud providers. For example, Amazon estimates that power-related costs occupy 42% of its total budget [78]. The CAS algorithm provides a good fit for power-aware scaling and can take data centers' power consumption into account using the cost function employed by cloud providers. For example, assume that a server $s$ is deployed at cloud provider $cp$ with $k$ physical machines (PMs) $\{pm_1, pm_2,\ldots, pm_k\}$ and its deployment cost is $c^{cp}(s)$. Let $pwr(u_j)=(p_j^{max}\text{-}p_j^{min})\, u_j+p_j^{min}$ be the power consumption of PM $pm_j$ ($1 \leq j \leq k$), where $p_j^{max}$ is $pm_j$'s power consumption at the peak load (e.g. 100% utilisation), $p_j^{min}$ is $pm_j$'s minimum power consumption with the least load (e.g. 1% utilisation) and $u_j$ is $pm_j$'s resource utilisation. Hence, the power-aware cost function $c^{pm}(cfg(s), pwr(u_j))$ measures server $s$'s deployment cost at $pm_j$, where $cfg(s)$ is $s$'s VM configuration. According to this cost function, the starting up or turning off of $s$'s VM and other VMs that are running in parallel on $pm_j$ influence this PM's resource utilisation, thus determining server $s$'s cost. Taking energy efficiency as a key indicator in scalings, the cloud provider $cp$ checks every PM and identifies the most cost-efficient PM for server $s$ : $c^{cp}(s) = min_{1 \le j \le k}\{c^{pm}(cfg(s), pwr(u_j))\}$.

### 3.7.3   Discussion of Lightweight Scaling Technique

The presented CAS algorithm is based on controlling (increasing or decreasing) the number of VMs that host an application's server components. The approach is appealing for a wide class of applications especially those that are based on multi-tier architectures, or server-side software platforms. For such applications, scaling up an application typically involves adding an extra software server, and hence an extra (server) VM in the cloud environment. We have also investigated another lightweight scaling approach [21] that forms a complement to the cost-aware scaling approach proposed in this chapter.

This lightweight approach conducts the fine-grained scaling of cloud applications at the resource level itself (CPU, memory, and I/O) in addition to the VM-level scaling [21]. The motivation of this approach is to avoid heavy-load operations such as creation/removal of VMs and improve resource utilisation of these VMs as application demands vary.

In **scaling up**, the approach supports three types of scaling up methods with different levels of priority. The two resource-level scaling up methods, called *self-healing* and *resource-level scaling*, have higher priority to be conducted. Both methods reduce the application's response time by increasing VMs' capacity using the available resources from these servers' hosting PMs. Note that these methods have some constraints. For example, a VM with a 32-bit operating system can only be allocated 4 GB memory in maximum. The scaling up process is completed if either of the two methods meets the response time target; otherwise, the *VM-level* scaling up method is triggered to add a VM from a new PM to the application. This new PM may have available resources and so the two resource-level scaling up methods can be executed again.

Specifically, the basic idea of the *self-healing* scaling up method is that if two VMs from the same application are hosted in the same PM, the idle resources of one VM can be used to release the overloaded resources in another VM. Hence, the application can be scaled up without incurring extra cost for its application owner. The *resource-level* scaling up method follows the observation that if the PMs that host the application's server components have available resources, these resources can be used to scale up while other applications hosted in these PMs are not affected. The first two scaling ups only require subtle changes (modifying VMs' capacity) to achieve the desired QoS of the application, thus only needing less running cost and, usually, less time in scaling up. In contrast, the *VM-level* scaling up method complements the two resource-level scaling methods and it has the lowest priority to be triggered. This scaling method detects the bottleneck tier from the application to add one server.

Similarly, in **scaling down**, the lightweight scaling approach first aims to remove as many VMs and resources from an application as possible, while still trying to hold its response time target. Specifically, the algorithm first performs the *VM-level* scaling down method to identify a bottleneck tier to remove a server. This removal VM is expected to save the maximum cost per unit of increased response time. The *VM-level* scaling down method keeps running until it is infeasible; that is, removing a VM would violate the response time target. This feasibility can be checked using application profiling and workload predication techniques. Next, the approach conducts the *resource-level* scaling down method. At each step, this scaling down method removes one unit of resource that has the largest available amount (i.e. the smallest utilisation) and the largest deployment cost.

# Part II.

# Algorithm Level Elasticity

# Chapter 4

# Algorithm Level Elasticity and Elastic Algorithm in Cloud Computing

## 4.1    Introduction

In this chapter, we first introduce motivations of algorithmic elasticity in Section 4.2 and discuss the related methodology in Section 4.3, thus introducing the need for developing a new class of elastic algorithms to enable such elasticity at the algorithm level. Section 4.4 then formally defines the properties of elastic algorithms and Section 4.5 discusses the meaning of elasticity in the context of algorithmic elasticity. Finally, we propose the key challenges to be addressed in the algorithm level elasticity in Section 4.6. Note that in Part II (Algorithm Level Elasticity), we still use *application owners* to denote consumers of cloud infrastructures and *end users* or *users* to represent customers that directly use application services deployed in the cloud.

## 4.2    Motivation of Algorithm Level Elasticity

Cloud computing has emerged as a cost-effective paradigm for delivering metered resources [79]. Within the cloud paradigm, hardware and/or software resources are provided as a utility that is shared between multiple application owners. Each application owner is provided a piece of solely owned virtual resource instances. Moreover, based on a pay-as-you-go business model, application owners are allowed to acquire and release resources on demand and are billed only for the resources they use. This feature, of being able to scale-up and down resources used on demand, is typically described as *elasticity* of resource provision. *Elasticity management* [80, 81], in this context, coupled with the new pay-as-you-go cloud business models, give rise to various new challenges that require revisiting our assumptions about how we design programs and algorithms.

## 4.2.1   Discussion of the Application Level Elasticity

For years, we had a simple view of algorithms that implements business logic in an application: the algorithm is a sequence of computational steps that should produce a *deterministic* result after consuming some resource. Since the algorithm's output result is prescribed, its computational properties are typically measured by its computational (time and space) complexity as the problem size grows. Similarly, the properties of its implementation are measured by performance metrics such as response time and throughput. This view forms the basis of the modern concept of QoS where software is provided as a service, and where application owners pay for the resources used to satisfy their QoS requirement.

To date, most research on supporting *elasticity management* in cloud environments has focused on either the provision of mechanisms that simplify the dynamic acquisition/release of resources based on the variation of the user's computational demand [19, 52-56], or on developing analytical-model-based capacity estimation and scheduling algorithms that help in minimising the execution costs of programs in a cloud environment [60].

A typical example of applying this traditional view in cloud environments can be seen in managing multi-tier web applications, such as e-commerce sites or other applications that server multiple end users. The QoS requirement for the implementation is typically expressed as response time, effectively measuring the performance of producing a result for each request. When the demand for application increases (measured by the number of requests submitted by end users), the application owner is traditionally willing to pay more so as to maintain the performance of the application as seen by its users. When the demand decreases, the application owner is not willing to pay for idle resources. Elasticity management at this level, called the *application level elasticity*, enables real-time acquisition/release of compute resources used, at each tier of the application, either up and down so as to meet the QoS requirements while minimising the monetary costs paid for the resources used [20, 21]. The resource usage, in this case, can be described as being elastic with respect to the user demand, and also with respect to the price of resources. We note that exploiting elasticity here does not change the *output* of the program as each user still receives the *exact prescribed result* of the computation. Rather, it is used mainly to change the performance characteristics of the application such as response time.

We note that elasticity management becomes slightly more complicated, albeit still manageable, if the price of resources used varies over time. Various cloud providers, such AWS, provide resource pricing schemes in which prices vary dynamically according to supply and demand conditions. For example, under a spot price scheme, application owners are allowed to bid for compute resources and gain access to them so long as the IaaS provider's offer price is lower than or equal to the owner's bid price. If the offer price becomes higher than the bid price, the provider reserves the right to terminate the owner's computation without notice. A key implication of such model is that application owners have to take into consideration such price fluctuations when making their resource provisioning decisions. They now not only have to minimise the cost of executing their computation but also have to ensure that the deadline for obtaining the computation's results is also met. Another practical implication is that the implementation of the algorithm now needs to incorporate check points where its execution can be suspended and then safely resumed. It even becomes desirable that some kind of *meaningful or useful approximate results* can be returned at such check points to ensure that the investments already made towards the computation are not lost if the application owner cannot resume the program later.

## 4.2.2 Towards the Algorithm Level Elasticity

Traditionally, elasticity within the cloud world is considered as an approach for on-demand scaling, or resource provisioning. It is exploited to enable real-time acquisition/release of compute resources to scale the performance of applications up and down so as to meet application owners' QoS requirements. In this traditional view, the "pay-as-you-go" is measured by how many times one algorithm is executed. No inherent elasticity within an algorithm itself is explored. Thus, given a cloud, we can address a question like "*How much it may cost to get my result by a particular time?*" However, there is no answer to the question such as "*Given a certain budget, what kind of result can I get from my algorithm?*" We may even feel that such a question is ill-formed since a traditional algorithm has a one-off answer only; either there is a result or we fail to get a result - there is no *elasticity* within such algorithms at all. Accordingly, cloud computing is mainly a business innovation rather than a new computation paradigm. Such a limitation has a severe impact on the exploration of the full power of cloud computing. For example, the spot price model is a powerful dynamic pricing mechanism for cloud. However, the practical use of the spot price model requires that an algorithm can be suspended and then safely resumed. It is also desirable that some kind of meaningful "approximate results" can be returned even using small amount of resources.

In this work, we depart from the application-oriented view of elasticity with a clear objective in mind; investigating the concept of elasticity at the *algorithm level*, rather than at the *application level*. We are motivated by the question of whether money can buy something else rather than just resources to improve performance? Given this motivation, we consider another form of *elasticity*: the elasticity of the algorithm's outputs with respect to the resource consumed. In this case, we may be willing to pay more (use more resources) to obtain better quality of results, not simply better performance. The challenge now becomes how to organise our computations to exploit such elasticity of result quality.

To address our objective we investigate algorithms that generate a sequence of improving *approximate results* whose result quality, based on same metric, is proportional to their resource consumption. As more resources are consumed, better results will be derived. We can illustrate this concept using an image rendering algorithm as an example. A conventional rendering algorithm is designed to generate the final result with the highest resolution. In a cloud environment the application owner has no option but to pay a high cost for using the resources required to produce this final result. However, on a limited budget, it may make more sense to adopt an incremental rendering method; that is, the algorithm could start by producing an approximate, but acceptable, result using a limited amount of resources (or budget) and return this result to the application owner. If the application owner has more budget, the algorithm can continue to refine the obtained image to improve its resolution by using more resources. In this case the quality of the result could be regarded as being elastic with respect to the resource usage, and we can easily call an algorithm with such behaviour an **elastic algorithm**. It is not difficult to see that similar elastic algorithms that trade off result quality with resource usage can be designed and used in a wide range of domains, including numerical, scientific and engineering computations; statistical estimation and prediction in data mining applications; heuristic search applications and database query processing applications in which generating approximate and cheaper answers may be acceptable to the application owner.

## 4.3  Related Methodologies

Our proposed concept of elastic algorithms builds on lessons learnt from previous methodologies used outside the cloud computing area, and especially those designed for developing time-adaptive algorithms in the context of real-time applications executing on environments with limited resources. We can summarise the key methodologies traditionally used in such applications into two camps as described below.

### 4.3.1  Time-Adaptive Algorithms

At present, many techniques have been developed to produce approximate results under resource and time constraints by restricting data size or computations. We call such algorithms "time-adaptive" algorithms.

**Incremental learning algorithm**

Incremental learning algorithm is a well-known methodology for machine learning under resource constraint. The basic idea of such algorithms is to incrementally learn and improve models from large and dynamic data, so as to produce a list of intermediate results to represent the acquired knowledge. An incremental learning algorithm can benefit from new raw data and can integrate it with previously acquired data in order to improve the quality of its learning result. We now review two major categories of existing incremental learning algorithms.

The first *category* of algorithms incrementally updates a model from new seen data. For example, on-line learning algorithms [82-84] learn a data instance at one time and assume that the true label of this instance is known soon after the learning. These algorithms then use the true label as feedback to update the learned model. At present, a variety of incremental learning algorithms belonging to the first category have been developed. In [85], Syed et al. propose a SVM training algorithm that divides the entire training set into several subsets and sequentially trains a list of SVM models using these subsets. At each training step, the support vectors (SVs) in the trained SVM model are maintained. These SVs are then added to the next training subset to represent previously trained data points. This incremental training algorithm is improved in [86] by considering weights in SVs. In [87], Wu et al. introduce another type of incremental SVM training algorithm by performing approximate matrix-factorization operations from coarse granularity to fine granularity. In [88], a decision tree is reconstructed whenever new points are added in order to support the incremental learning of this decision tree. In [89], the two steps of feature selection, namely the optimisation of free parameters and the selection of new features, are incrementally performed by one-step of gradient descent. In addition, some other incremental versions of popular data mining algorithms including PCA for subspace learning [90], k-means clustering [91], and the mining of sequential patterns [92] are proposed.

The second *category* of incremental learning algorithms is motivated by the boosting principle [93]. Such algorithms are based on learning multiple models along the accumulative data and assembling these models to generate an output. In [93], Schapire employs AdaBoost as a typical example and gives an overview of the boosting-based machine learning algorithms. In [94], Polikar et al. present an algorithm to train a neural network classifier. This algorithm maintains all the previously acquired classifiers and generates the output of predication using a weighted majority voting of these classifiers.

In conclusion, incremental learning algorithms have two motivations: (1) the entire data cannot be accessed (or new data are continuously coming); (2) there are insufficient resources such as memory to process the entire data at once. Thus, such algorithms are designed to sequentially process data and retain the acquired knowledge as a prior to support future learning process. However, most of existing incremental learning algorithms apply a simple data accumulation strategy (e.g. randomly sampling from the entire data) to passively accept new data. This means generating results whose qualities improve monotonically to the used resources is difficult.

**Anytime algorithm**

Anytime algorithms provide a generic approach to produce approximate results within time constraints. In [95, 96], Boddy and Dean first introduce the idea of anytime algorithm and apply it in the area of time-dependent planning. In their anytime framework, a planning problem consists a set of decision procedures, the quality of the planning problem is the expectation of the predicted system performance, and the cost is the computational time used in planning. They define an anytime planning as an expectation-driven iterative improvement process: the more time spent in planning, the better the expectation of future performance. In [97], Joshua Grass divides an anytime algorithm into three components: an iterative improvement function, a result evaluator, and a performance profile. Furthermore, Shlomo Zilberstein systematically summarises all desired features of anytime algorithm [98]: an anytime algorithm should produce a result with a *measurable* and *recognisable* quality. This quality should be a *nondecreasing* function of time and the quality improvement *diminishes* over time. The algorithm can be stopped to provide an intermediate result (*interruptibility*) and resumed with small overhead (*preemptability*). Finally, the output quality can be predicted precisely given input quality and computational time based on the profiling of the algorithm running history (*consistency*).

The basic idea of an anytime algorithm is that once an initial approximate result is produced, the algorithm can be interrupted and output a result at anytime. If the algorithm is not interrupted it can continue to update its result. The anytime principle is applicable to many areas, including the evaluation of probabilistic network [99], reasoning [100], relational database querying [101], sensing and planning [102, 103], and similarity measures in image alignment [104]. For example, in database query [101], when part of the data is inaccessible (e.g. caused by a network partition or a host failure) or there is insufficient time to process the entire data to give an exact answer, Vrbsky et al. present an anytime method that can provide an approximate answer of database query with limited available data and processing time. This method can further increase the accuracy of the answer when more processing time is given.

In recent years, the fast development of anytime algorithms has lead to a number of successful applications on data mining. A variety of anytime clustering and classification algorithms have been developed.

*Anytime clustering algorithms.* In [105], Vlachos et al. propose an anytime k-means clustering algorithm to cluster time-series data by applying Haar wavelet to transform high-dimensional data into low-dimensional data. This algorithm first finds the low-dimensional data's clustering centers, which approximately capture the shape of the clusters. It then uses these centers as the initial centers and conducts the next iteration of k-means clustering using higher-dimensional data to improve the quality of the clustering result. This quality is measured by the mean distance from all the data points to their nearest clustering centers. In [106], Kranen et al. develop an index structure for anytime clustering of data streams.

This index structure is designed to maintain a complete model without dropping any data stream item. Using this structure, an anytime k-means clustering algorithm is developed to automatically adapt to the stream speed and concept drift.

*Anytime classification algorithms.* In [107], Ueno et al. introduce an anytime kNN classification algorithm by ranking all training points according to their importance (i.e. their distances to the closest training point from the same class; that is, the smaller the distance, the more important of a training point). The algorithm first uses more important points to produce an initial result and gradually adds less important points to refine the result. In [108], Seidl et al. develop an anytime Bayes classifier based on R-tree index structure [18]. This algorithm employs Gaussian mixture model to describe the probability distribution of training points and calculate the likelihood used for Bayes classification. The algorithm also uses an R-tree to index different granularities of these Gaussian mixture models. Given a data point to be classified, the algorithm first descends the R-tree from the root nodes to calculate the likelihood in a coarse level of granularity. If more time is available, the algorithm then descends more R-tree nodes to estimate the likelihood in a finer level of granularity, thus giving more accurate classification. In [109], Yang et al. present another anytime Bayes classification algorithm. In this algorithm, a set of $d$-dimensional training points can produce $d$ super-parent-one-dependence estimators (SPODEs). Each SPODE takes one attribute as the super parent of the rest ($d$-1) attributes that are independent of each other. Given a test point to be classified, the algorithm initially gives a quick classification result by applying the naïve Bayes classifier; that is, all attributes are independent of each other in classification. Since an ensemble of SPODEs can guarantee high classification accuracy [110], the algorithm then adds one SPODE into the classification as one refinement step to produce an improved result. In [111], Kranen et al. try to use the above anytime classification algorithms in a data stream environment. They first define the confidence of a stream item and assume that there is a linear dependency between the confidence and the classification accuracy. Based on this confidence, they propose two scheduling approaches that either allocate more execution time to an item with a higher confidence (Batch approach), or let the item with the highest confidence to be classified first (first-in-first-out approach).

In general, anytime algorithms are interruptible algorithms designed to return a valid result whenever the algorithms are suspended. Such algorithms can continue refining the result when more time is allocated. They can hide quality decrease by storing and returning the best result obtained so far. Anytime algorithms, thus, do not guarantee effective use of extra resources.

**Flexible computation and imprecise computation**

Similar ideas are presented in flexible computation [112, 113] to discuss the trade-off between the quality of result and the computational time. This work calculates the expected value of computation (EVC) of partial results and tries to find the result with the maximum revenue (i.e. the benefit of the computation minus the cost of this computation) under the deadline constraint. In addition, imprecise computation presented in [114] first divides tasks into two categories: (1) mandatory tasks that must be completed before deadline to provide an acceptable/usable result, called imprecise result; (2) less important/optional tasks that can refine the imprecise result. Scheduling techniques are then proposed to complete all mandatory tasks before their hard deadline and leave optional tasks for further quality improvement when more time is available.

## 4.3.2   Resource-aware Algorithms

Resource-aware algorithms, e.g. [115, 116], focus on organising computation to produce the best possible results on devices with limited resources, e.g. mobile devices or sensor nodes. On such devices resources such as memory, processing cycles, communication bandwidth, and battery life may degrade, or vary, with time. The resource-aware methodology controls how an algorithm adapts its use of resource dynamically in response to such changes. In [115], three key control strategies are proposed, these are: (1) controlling the algorithm's input granularity, e.g. by changing the resolution or details of the input data structures; (2) controlling the algorithm's processing granularity, e.g. by performing less or more computation and (3) controlling the algorithm's output granularity, e.g. by controlling the resolution or detail of the output data structures.

Resource-aware algorithms typically require the implementation of a real-time resource monitor and a decision mechanism (e.g. a set of rules) for choosing between different implementations of individual steps in the algorithm's implementation. The reactive approach builds implicitly on knowledge of how the quality of result varies with resources but does not necessarily require the definition and use of an explicit quality function or metric.

## 4.4   Definition of Elastic Algorithms

Both the time-adaptive algorithm methodology and resource-aware algorithm methodology provide valuable insights on how we could develop elastic algorithms for enabling the **algorithm level elasticity** in cloud environment. The time-adaptive algorithm methodology provides a generic approach for designing algorithms that incrementally refine their output results along a computing process. In contrast, the resource-aware algorithms methodology provides the notion of trade-off between the quality of computation result and the available resource. Combining both schools, it is possible to investigate the development of the new paradigm of elastic algorithms for cloud environments in which the quality of the computation results improves as more resources are used. We also note that neither methodology explicitly formalises the notion of elasticity itself at the algorithm level. Moreover, neither provides a framework for developing and reasoning about the elasticity properties of the algorithm. Based on these observations, this section formalises our concept of elastic algorithms for cloud computing.

The basic idea of elastic algorithms is that application owners are still guaranteed useable results even with limited resource consumption. If more resources are put towards the computation, the algorithm guarantees results with better qualities by refining the previously obtained results. An elastic algorithm organises the computing process in an incremental manner that offers the application owner a selection of approximate results and allows them to obtain any of these results depending on their available budget. The quality of each result can be evaluated using a measurable function, e.g. prediction accuracy in classification or recommendation problems. The key property of an elastic algorithm is that it guarantees continuous improvement in the quality of the results produced as the application owner's resource consumption increases. This method of computation is particularly suited to the pay-as-you-go computing frameworks, such as cloud computing, where real-time scaling up and down of resources is supported and the resources used are measured in monetary terms [117].

The four properties of elastic algorithms can be formalised as follows:

**Definition 4.1 (Elastic algorithms)**. *An elastic algorithm offers a range of approximate results such that the following four properties hold:*

- ***Measurable quality***: *For any approximate result $ar$, there is a computable quality function $Q(ar)$.*

- ***Meaningful results***: *For any approximate result $ar$ , its quality measurement is non-negative: $Q(ar) \geq 0$.*

- ***Quality monotonicity***: *For any two approximate results $ar$ and $ar'$ obtained using two investments $I$ and $I'$ respectively, and with qualities of $Q(ar)$ and $Q(ar')$ respectively, we have:*

$$Q(ar) \geq Q(ar') \; if \; I > I'$$

- ***Accumulative computation***: *Starting from either of two approximate results $ar$ or $ar'$, suppose the algorithm needs an additional investment $\Delta I$ or $\Delta I'$ respectively to obtain a refined result $ar''$ with better quality. Then, we have:*

$$\Delta I \leq \Delta I' \; if \; Q(ar) > Q(ar')$$

The first property, *measurable quality*, means that an explicitly defined and measurable quality function can be computed for each approximate result. The second and third properties, *meaningful results* and *quality monotonicity,* mean that: each approximate result must be a complete, rather than partial, output from the computation so that it is useful to the application owner; there is a minimum acceptable quality threshold associated with the first produced result; and quality improves monotonically as more investments are made (i.e. more computational resources are consumed). The fourth property, *accumulative computation*, means that a particular result $ar''$ can be obtained by refining previous results $ar$ or $ar'$, and this refinement requires less investment if the starting result $ar$ has better quality.

We can illustrate the meaning of the above properties by considering an incremental image rendering algorithm [118].

*Measurable quality*: in incremental image rendering, the quality of the result can be quantitatively measured by the resolution (samples-per-pixel) of the generated image such as 100×100 pixels.

*Meaningful results*: the overall process of image rendering can produce a range of approximate results. Each approximate result $ar$ is a fully rendered image, where $Q(ar) \geq$1×1 pixels.

*Quality monotonicity*: in incremental image rendering, each new investment can guarantee to generate an image with a higher resolution/quality, e.g. from 100×100 pixels to 1000×1000 pixels.

*Accumulative computation*: in many image rendering algorithms, a resumable file can be used as the starting point to resume the image rendering process without calculating the samples that have already been processed. Suppose $ar$ is a previously obtained result and its resolution/quality $Q(ar) = m \times m$ pixels. Starting from the resumable file of result $ar$, the algorithm can get a refined result $ar''$ using an additional investment $\Delta I$. In contrast, starting from the resumable file of another result $ar'$ with resolution/quality $Q(ar') = m' \times m'$ pixels. If $Q(ar) > Q(ar')$, i.e. $m > m'$, the algorithm needs a larger investment $\Delta I'$ to get the same refined result $ar''$, where $\Delta I' > \Delta I$ and investment $(\Delta I' - \Delta I)$ represents the investment needed to render the image from $m' \times m'$ pixels to $m \times m$ pixels.

## 4.5 Elasticity in the Context of Algorithm Level Elasticity

The definition of elastic algorithms, in turn, requires us to formalise what we mean by elasticity and how it may impact the design of such algorithms. We note that elasticity has a precise meaning as an economic term: it is the measurement of how changing one economic variable affects others. The elasticity of y with respect to x, $E_x^y$, is defined as $E_x^y = \frac{\%\Delta y}{\%\Delta x} = \frac{\partial y}{\partial x} \cdot \frac{x}{y}$.

In the context of algorithm level elasticity, we consider the elasticity of quality with respect to investment: $E_I^Q = \frac{\%\Delta Q}{\%\Delta I}$. This elasticity characterises the key property of an elastic algorithm, where $\%\Delta Q$ is the percentage quality improvement and $\%\Delta I$ is the percentage investment increase.

**Definition 4.2 (Investment elasticity).** *Given an elastic algorithm, its investment elasticity, denoted by $E_I^Q$, is a function of quality $Q$ and investment $I$. Suppose that an approximate result $ar$ is produced using investment $I$, and is refined to another result $ar'$ using investment $\Delta I$. The qualities of results $ar$ and $ar'$ are $Q(ar)$ and $(ar')$ respectively, with $Q(ar') > Q(ar)$. The investment elasticity between these two results is calculated as the percentage quality improvement $\%\Delta Q$ divided by the percentage investment increase $\%\Delta I$:*

$$E_I^Q = \frac{\%\Delta Q}{\%\Delta I} = \frac{\partial Q}{\partial I} \cdot \frac{I}{Q} = \frac{(Q(ar') - Q(ar))/\,Q(ar)}{\Delta I/I}. \tag{4.1}$$

In a cloud environment, we can express this elasticity in different ways. Given an elastic algorithm, its quality $Q$ is a function of investment $I$ and its starting state $S$: $Q = q(I, S)$. This state can either be the initial state $S_0$ the algorithm or a state $S_{ar}$ representing a previously obtained result $ar$. For example, in an incremental image rendering algorithm, $S_{ar}$ is the resumable file of an rendered image $ar$. In addition, an investment is modelled by the consumed resource as well as by the price of this resource. Suppose that, the investment function $I = i(R, P)$ and the state function $S = s(R, P)$, where $R$ is the resource and $P$ is its price (cost for per unit of resource). We can further define resource elasticity and price elasticity.

**Resource elasticity**: the elasticity of quality with respect to the resources used:

$$E_R^Q = (\frac{\partial Q}{\partial I} \cdot \frac{\partial I}{\partial R} + \frac{\partial Q}{\partial S} \cdot \frac{\partial S}{\partial R}) \cdot \frac{R}{Q}. \tag{4.2}$$

**Price elasticity**: the elasticity of quality with respect to the price:

$$E_P^Q = (\frac{\partial Q}{\partial I} \cdot \frac{\partial I}{\partial P} + \frac{\partial Q}{\partial S} \cdot \frac{\partial S}{\partial P}) \cdot \frac{P}{Q}. \tag{4.3}$$

Assume that state $S$ is independent of resource $R$ and price $P$, that is, $\frac{\partial S}{\partial R} = 0$ and $\frac{\partial S}{\partial P} = 0$, we have $E_R^Q = \frac{\partial Q}{\partial I} \cdot \frac{\partial I}{\partial R} \cdot \frac{R}{Q}$ and $E_P^Q = \frac{\partial Q}{\partial I} \cdot \frac{\partial I}{\partial P} \cdot \frac{P}{Q}$ according to Equations (4.2) and (4.3). Given an example investment function $I = R \cdot P$, we have: $E_R^Q = \frac{\partial Q}{\partial I} \cdot \frac{\partial I}{\partial R} \cdot \frac{R}{Q} = \frac{\partial Q}{\partial I} \cdot P \cdot \frac{R}{Q} = \frac{\partial Q}{\partial I} \cdot \frac{P \cdot R}{Q} = \frac{\partial Q}{\partial I} \cdot \frac{I}{Q} = E_I^Q$ and $E_P^Q = \frac{\partial Q}{\partial I} \cdot R \cdot \frac{P}{Q} = \frac{\partial Q}{\partial I} \cdot \frac{I}{Q} = E_I^Q$; that is, three elasticities are equivalent: $E_I^Q = E_R^Q = E_P^Q$. We note that this equivalence holds only under the independence condition of state $S$ and the given investment function $I = R \cdot P$.

We use a simple theoretical example to understand the elasticity behaviour of an elastic algorithm. Table 4.1 shows an example of eight approximate results $ar_1$ to $ar_8$ produced using the algorithm. The qualities (the larger value, the better quality) and investments for these results are also listed in the table. In this example, we assume that a fixed investment $\Delta I = 1$ (dollar) is needed to produce an improved result $ar_{i+1}$ starting from the state of result $ar_i$ for any $0 \leq i < 7$. Figure 4.1(a) shows that the quality is a monotonic function of the investment $I$ in this algorithm. Figure 4.1(b) and (c) further show that the percentage investment increase $\%\Delta I = (I_{i+1} - I_i)/I_i$ and the percentage quality improvement $\%\Delta Q = (Q_{i+1} - Q_i)/Q_i$ are larger at the early stages of the computation (e.g. in results $ar_1$ to $ar_2$) and they diminish over time. In addition, Figure 4.1(d) illustrates the investment elasticities of seven pairs of results (e.g. "1 to 2" means the investment elasticity between results $ar_1$ and $ar_2$ and the starting state is $S_{ar_1}$ of result $ar_1$). These investment elasticities show that when starting from state $S_{ar_7}$ of result $ar_7$, the algorithm has the highest investment elasticity. This indicates application owners can obtain the largest percentage quality improvement when they make the same percentage investment increase if the algorithm starts from state $S_{ar_7}$.

Furthermore, we assume that the investment function $I = R \cdot P$, which means the investment is decided by the consumed resource and its price. Since $\Delta I$ is fixed, the resource consumption $\Delta R$ is also fixed; that is, the same amount of resource is needed to produce an improved result $ar_{i+1}$ starting from state $S_{ar_i}$ for $0 \leq i < 7$. Suppose that, $\Delta R$ equals the execution of the algorithm for an hour in a VM. Figure 4.2(a) compares the cumulative investments under three different prices $P= 0.5$, 1, and 2 (unit is dollar for running the VM for an hour). Unsurprisingly, the comparison result shows that a larger investment is needed to produce the same approximate result ($ar_1$ or $ar_2$) at a higher resource price. However, Figure 4.2(b) shows that when the resource price varies, the algorithm has invariant investment elasticity with respect to different prices.

Table 4.1. An example of eight approximate results

| Approximate result | Investment | Quality |
|:---:|:---:|:---:|
| $ar_1$ | 1 | 0.38 |
| $ar_2$ | 2 | 0.52 |
| $ar_3$ | 3 | 0.64 |
| $ar_4$ | 4 | 0.72 |
| $ar_5$ | 5 | 0.80 |
| $ar_6$ | 6 | 0.88 |
| $ar_7$ | 7 | 0.92 |
| $ar_8$ | 8 | 1.00 |

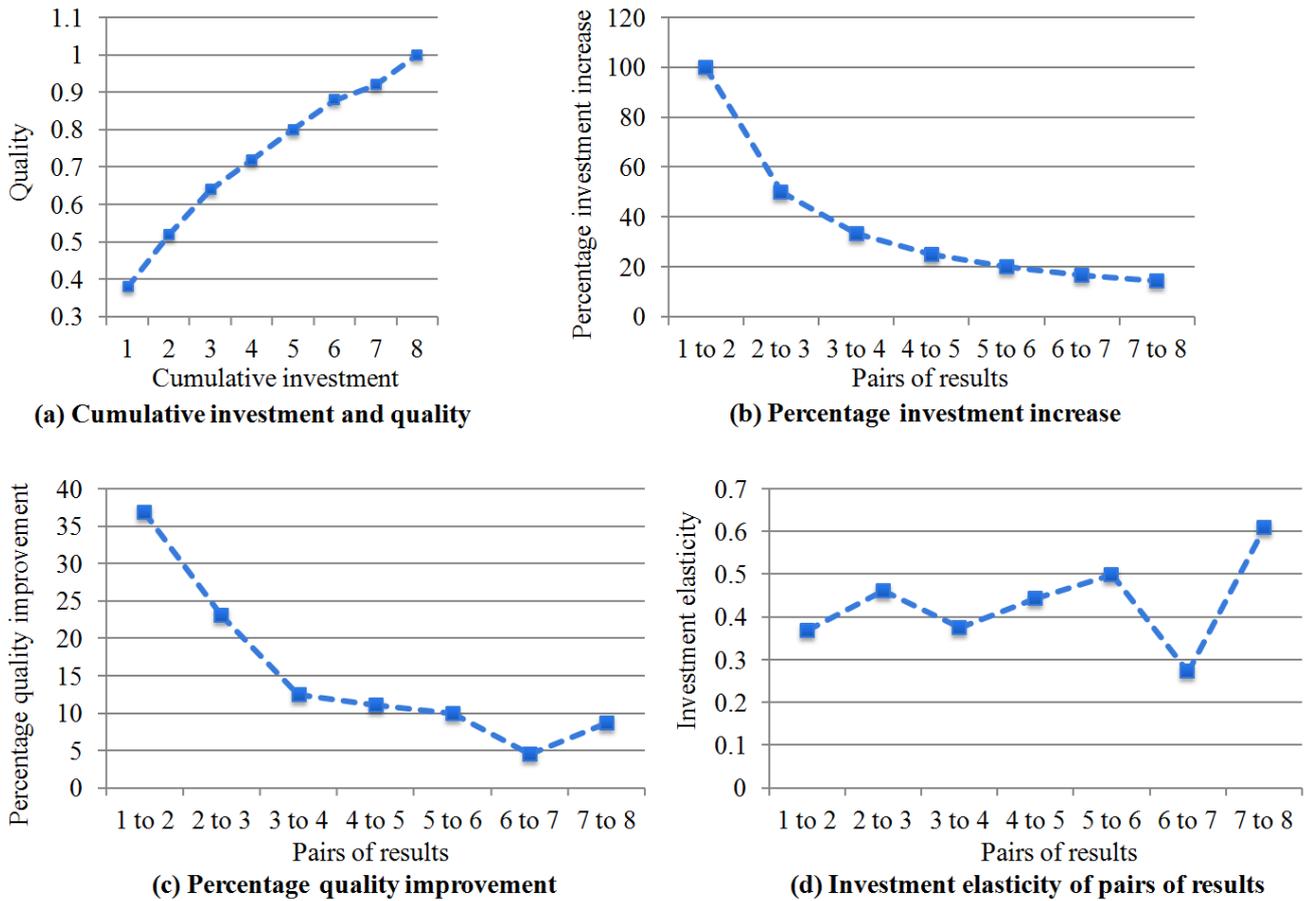

Figure 4.1: Comparison of investment, quality and investment elasticity.

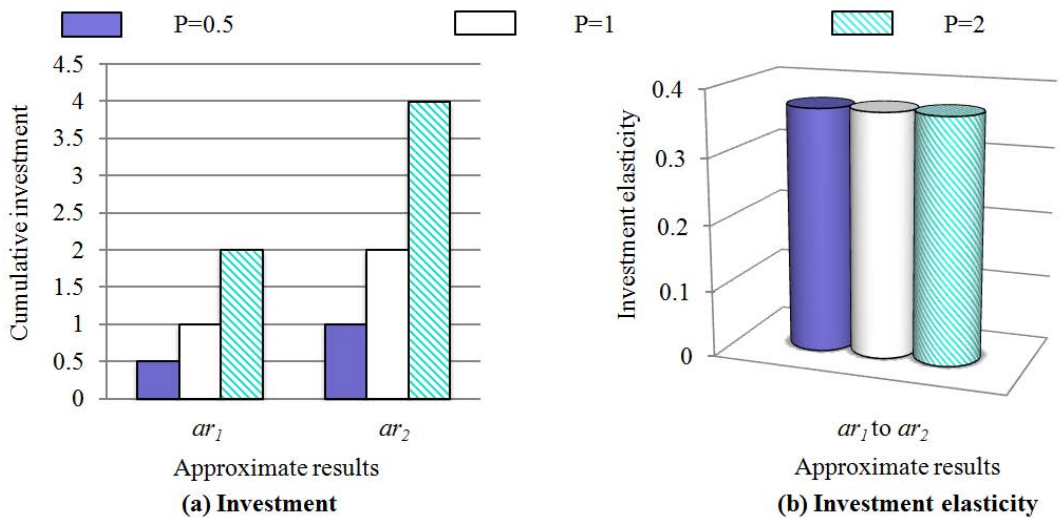

Figure 4.2: An example investment and investment elasticity using three prices.

## 4.6   Key Challenges in Algorithm Level Elasticity

At present, a key practical problem facing the processing of large-scale data in many applications is that computations are required to be conducted under varying resource and time constraints. To deal with this problem, application owners are usually willing to accept a useful approximate result from their computation that can be produced under the available time or resource budget. Typically, such results can be produced either by restricting the size of the input data fed to exact algorithms, or by using approximating algorithms over full datasets. Within this context, the concept of an *elastic algorithm* is proposed to study the problem of *algorithm level elasticity*. This elasticity organises the computational process to support a pay-as-you-go model, in which application owners are guaranteed not only a useful approximate result on a tight budget but also results of better quality if the budget is increased.

At the level of algorithmic elasticity, the key challenge we now face is how to develop software programs or algorithms that make use of such elasticity properties; that is, how to make our algorithms themselves elastic. Developing an elastic algorithm is not trivial since, in practice, many algorithms do not have a natural structure encapsulating the inherent elasticity of the algorithm. Two challenges need to be addressed when developing elastic algorithms:

- The first is being able to reason effectively about how the allocated budget affects the quality of results. Enabling such a trade-off between quality and computational cost is essential so that application owners can quantify their return on investment in cloud computations. This contrasts with the anytime algorithm methodology, in which application owners do not pay for resources and, even after an acceptable result is produced, may keep the algorithm running until their deadline expires.

- The second and more challenging issue is to design the elastic algorithm itself so that the quality measure improves as more resources are used. This monotonic improvement should indeed bring consistent increases in observable quality—such as prediction accuracy in classification or recommendation problems—as more investment is used. The importance of this second issue— not wasting compute resources—is not simply a matter of efficiency: it is becoming more important when considering the pay-as-you-go computing framework in cloud computing, in which consumed resources are measured in monetary terms.

Various paradigms do exist for developing time-adaptive algorithms that operate under time, budget, and other particular constraints, such as the inaccessibility of data or insufficiency of resources in incremental learning algorithms [82-92, 94], and the strict requirement to return a valid result whenever interrupted in the case of anytime algorithms [95, 96, 98, 105-109]. Little work has been done in such methodologies towards guaranteeing quality monotonicity with respect to investment. We address this issue in the next chapter by proposing a generic approach to develop elastic algorithms, which offers a promising methodology that provides a systematic means of guaranteeing quality monotonicity.

# Chapter 5

# A Framework for Developing Elastic Algorithms

## 5.1 Introduction

In this chapter, we investigate the development of elastic algorithms in the context of data mining, which aims to discover patterns and useful information in datasets to help people increase profits or reduce costs. Generally, data mining is an interdisciplinary activity that applies techniques from different areas including statistics, artificial intelligence, machine learning, and databases. In Section 5.2, we propose a generic approach to develop elastic data mining algorithms that guarantees quality monotonicity with respect to the allocated investment budget (that is, the maximal investment amount). In Section 5.3, we demonstrate the validity and practicality of our approach by designing an elastic version of the kNN classification algorithm. In Section 5.4, we perform extensive experimental evaluations on a number of real datasets to test the effectiveness of the elastic kNN algorithm. Note that from Section 5.2 to Section 5.4, we assume that resource prices are fixed during the execution of the elastic algorithm. Hence, we use *execution time* to represent *investment*, and *time budget* to denote *investment budget*. Finally, in Section 5.5, we demonstrate how to apply the elastic kNN algorithm in analysing large-scale datasets in clouds under two different resource pricing schemes. In Section 5.6, we discuss the constraints that a data mining algorithm needs to meet in order to be applicable to our proposed approach.

## 5.2 A Framework for Developing Elastic Data Mining Algorithms

In this section, we first present an overview of the framework by describing the two components of an elastic data mining algorithm: the coding component and the mining component. We then define the property that must be met by the coding component of any data mining algorithm in order to guarantee quality improvement with increased computations.

## 5.2.1   Two Components of an Elastic Data Mining Algorithm

Data mining can be viewed as a two-stage process. In the first stage, an observation dataset is coded using an assumed base (representation) for the mining that will follow. *Coding components* include transformation (such as wavelet-based transformation) or other feature-based representation (extraction) techniques. The coded data is then input into a *mining component* to solve a specific data mining problem or task, such as classification or making recommendations. The mining component outputs meaningful results, such as predicted class labels or rating values of test points. In an elastic mining scenario, a large-scale dataset needs to be processed within a limited time (resource) budget to produce a useful approximate result. The key to meeting this challenge is to represent the observation dataset in a proper hierarchical base using the coding component so that the mining can be elastic with respect to the time budget, as shown in Figure 5.1.

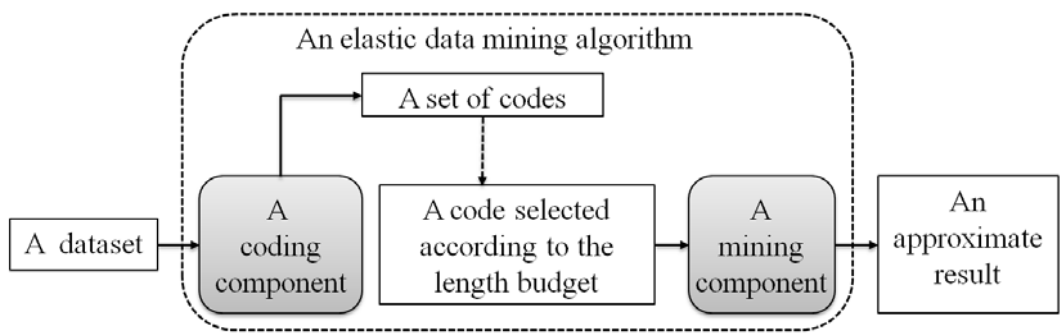

Figure 5.1: The two components of an elastic algorithm.

A **coding component** is a map from the dataset space to a code space, and is lossy if the corresponding map is not injective. The basic idea of lossy coding is to preserve the important information in a dataset while removing the unimportant parts. The remaining information is then organised using a suitable data structure. A **code** is the output of a coding component, and is a function of the dataset to be coded and the **length budget**, which denotes the *maximal length* of the code that can be processed by the mining component within the given time budget. Given a value $s$ of dataset, the coding component is only applied once to produce a set of codes of various lengths to be used in the elastic algorithm. Fixing the time budget $b$ and the mining component, the length budget $l$ can be estimated. Thus, a *code* is selected such that it has the *maximum length value* smaller than the *length budget $l$*.

Typically, there are two ways to compress a particular dataset: instance reduction and feature/dimensionality reduction. In instance reduction, the number of data points in the dataset is reduced. A typical component indexes these points in a hierarchy using tree data structures such as quadtree [119], cone tree [120], and R-tree [121]. At different levels of the tree, the information contained in the data points can be statistically summarised. In feature reduction, the feature number of data points is decreased. Many feature reduction techniques such as wavelet, principal component analysis (PCA), and hashing functions [122] can be applied in this type of compression.

A **mining component** is designed to solve a specific data mining problem, and needs to take a code as input and output an approximate result. Given a mining component and a code, we assume that the running time of the mining component can be primarily determined by the length of the code, and this running time should be less than or equal to the given time budget. Hence, the bound of an elastic data mining algorithm's running time can be controlled by tailoring the length of the input code.

## 5.2.2   A Property of the Coding Component

A quality-monotonic elastic data mining algorithm can generate a range of approximate results whose qualities are proportional to their consumptions of computational time. To this end, the coding component of the algorithm is designed to produce a set of codes of different lengths. More importantly, the coding component should guarantee that the resolutions (accuracies) of codes are proportional to their lengths (i.e. their computational complexities). Thus, application owners can still be guaranteed a usable result even on limited time budget by using a code of small length. Furthermore, application owners with larger time budgets can achieve better quality results by using codes of greater lengths; that is, by using more accurate codes.

We first define the resolution measure of a code based on Shannon entropy [123, 124]. Assume that a dataset $S$ is a variable that is drawn randomly from a prior distribution: $S \sim prior(s)$. Since the prior distribution $prior(s)$ of the dataset $S$ is known, the uncertainty of $S$ can be described as the Shannon entropy $H(S)$ of $S$; that is, $H(S)$ is the minimal number of bits of information needed to recover the value of $S$ exactly. Given a random variable $X$ with $n$ outcomes $\{x_1, ..., x_n\}$, where the probability of $X = x_i$ is $p(x_i)$ and $\sum_{i=1}^{n} p(x_i) = 1$, the entropy of $X$, denoted by $H(X)$, is calculated as $H(X) = -\sum_{i=1}^{n}(p(x_i) \times \log_b p(x_i))$. In other words, we have to obtain at least $H(S)$ bits of information to recover the value of $S$ exactly.

Given a length budget $l$, a code $C(S, l)$ is a random variable depending on variables $S$ and $l$. Given a coding component and a code value $c$ (a specific code) produced by this coding component, the conditional entropy $H(S|C = c)$ represents the number of bits required to recover the value of dataset $S$ exactly given an observed code value $c$. Thus, we can use the difference between $H(S)$ and $H(S|C = c)$, known as the information gain [123, 124], to measure the resolution of the code value $c$. This resolution can be seen as the benefit of the value $c$; that is, the decrease in the number of bits required to exactly recover the value of $S$ from a prior state to a state that takes a given value $c$.

**Definition 5.1 (Resolution measure for a code value c)**. *Given a value $c$ of code $C$, its resolution is defined as follows:*

$$R(c) = H(S) - H(S|C = c) \tag{5.1}$$

We can illustrate the meaning of the resolution measure using a simple example of integer guessing: guessing an integer between 1 and 100. In the prior state (no code value is known), the dataset $S$ has 100 possible outcomes and the probability of each outcome is 0.01. Hence, the entropy of $S$ can be calculated: $H(S) = -\sum_{i=1}^{100}(0.01 \times \log 0.01) = 2.00$. Given a code value $c$ that the integer is between 1 and 50, the dataset $S$ only has 50 possible outcomes and the probability of each outcome is 0.02. Hence, the entropy of $S$ given $c$ can be calculated: $H(S|c) = -\sum_{i=1}^{50}(0.02 \times \log 0.02) = 1.70$. According to Equation (5.1), the resolution of the code value $c$ (its information gain) is $Q(c) = H(S) - H(S|C = c) = 0.30$.

To support the quality-monotonic data mining algorithm, we need a coding component that produces codes whose resolutions increase with their length budgets. This *entropy monotonicity property* can be explained as follows: if a greater length budget $l$ is given, a coding component will output a more accurate code value; that is, a code value $c$ of higher resolution.

**Definition 5.2 (Entropy-monotonicity of a coding component)**. *Fixing a coding component and a value $s$ of dataset $S$, let there be two codes $c = C(S = s, l)$ and $c' = C(S = s, l')$, and let the two length budgets satisfy $l \leq l'$. A coding component is entropy-monotonic if:*

$$R(c) \leq R(c') \tag{5.2}$$

Thus, given a greater time budget, and thus a greater length budget $l$, the mining component operates on a code value of higher resolution and produces an approximate result that is closer to the exact result produced using the entire dataset, which usually results in quality improvement in the approximate result. Note that in the following passage, Section 5.3.5 proves the above property of entropy monotonicity, and we describe the uncertainty of a random variable $X$ and use a *value* of $X$ to represent a *specific outcome* of $X$—say, a value $s$ of dataset $S$ to represent a specific dataset and a code value $c$ to represent a specific code. Since no random variable is discussed in other sections, we omit "value" for simplicity.

We summarise the notation we have used in Table 5.1.

Table 5.1. Notations and definitions

| Notation | Description and Definition |
|:---:|:---:|
| $S$ | Dataset: a random variable representing possible variations of datasets |
| $s$ | A value of dataset: a specific dataset |
| $b$ | Time budget: total time allocated for a mining task |
| $l$ | Length budget: the maximal length of a code value that can be processed by the elastic algorithm within the time budget $b$ |
| $C(S, l)$ | Code: a random variable depending on dataset $S$ and length budget $l$ |
| $c$ | A value of code (or a code value): a specific code |
| $R(c)$ | Resolution of a code value $c$ |

## 5.3   An Example Elastic kNN Classification Algorithm

Based on the proposed framework, we develop an example elastic kNN algorithm. The core of the algorithm uses standard naïve kNN classification [16] over an R-tree coding component [18]. Given a training set, the codes produced by the R-tree are the nodes at different depths that successively approximate the training set at different levels of granularity. We first introduce some basic concepts and related work regarding kNN technique in Section 5.3.1. We then explain the two components of the algorithm in Sections 5.3.2 and 5.3.3, and explain how an approximate result is produced using the elastic kNN algorithm in Section 5.3.4. Finally, we discuss the properties of quality monotonicity and accumulative computation in the proposed elastic kNN algorithm in Sections 5.3.5 and 5.3.6.

## 5.3.1 Background and Related Work

Although conceptually simple, the kNN method is a classic approach [16] that provides a core function of many algorithms in fields such as statistical classification, pattern recognition, and recommender systems. This method also has many features that are common to a wide class of data mining algorithms. In the context of classification problems, the naïve kNN algorithm [122] classifies a test point $q$ by linearly scanning all data points in a training set with known class labels and setting the $k$ ones whose distances are closest to $q$ as its $k$ nearest neighbours. The algorithm then decides $q$'s class label according to the majority vote of its $k$ nearest neighbours; that is, it assigns $q$ to the same class as that of the majority of its nearest neighbours. In this work, we discuss a binary kNN classifier with a positive class $c_P$ and a negative class $c_N$, and our result can be extended to multiple classes.

In kNN classification, it is a major challenge when applying a training set with a large number of points to classify a set of test points to produce classification results. We now review existing work on dealing with this challenge.

**Linear/sublinear time kNN search techniques.**

At present, many kNN search techniques have been developed to improve the performance of searching for nearest neighbours. These techniques either restrict the size of the input dataset while searching for *exact* nearest neighbours, or use a threshold while searching the whole dataset in order to find some *approximate* nearest neighbours.

Many algorithms apply tree-based data structures such as kd-trees [125, 126], quad-trees [127], and R-trees [121, 128] to hierarchically index data points and accelerate the process of searching for a test point's *exact* nearest neighbours by using tree-pruning techniques. These techniques work well for low-dimensional data but their pruning power is significantly reduced when dealing with high-dimensional data. In addition, some kNN search techniques employ the triangle inequality in a metric space to remove data points that cannot be a test point's nearest neighbours, thus eliminating unnecessary distance computations. Example algorithms are walking [129], spatial approximation tree [130], and the linear approximating and eliminating search algorithm (LAESA) [131]. The running time of *exact* nearest neighbour search algorithms grows *linearly* with data size in the worst case.

Some other algorithms apply dimensionality/feature reduction techniques such as PCA [90], wavelet, and locality-sensitive hashing (LSH) [132, 133] to search for *approximate* nearest neighbours, especially in high-dimensional spaces. Such approximate search algorithms usually guarantee finding points within a specified distance to the test point, while improving search efficiency such that a *sublinear* growth of search time can be ensured in the worst case [132, 134, 135].

In conclusion, traditional kNN searching techniques run in linear/sublinear time [136]; that is, the time taken in searching for a test point's nearest neighbours increases linearly/sublinearly with the size of the dataset. These techniques organise computation as a non-interruptible process that only produces an all-or-nothing result and are not designed to adapt to a varying time budget.

**Anytime kNN classification algorithm.**

Existing anytime algorithms usually produce an acceptable approximate result under time constraints by restricting data size or number of computations. The worst-case time complexity of such algorithms is independent of input data size. In [107], Ueno et al. introduce an anytime kNN classification algorithm by ranking all training points according to their importance. The algorithm first uses more important points to produce an initial result and gradually adds less important points to update the result.

## 5.3.2   The R-tree Coding Component

The basic idea of the R-tree coding component [18] is to index the data points of a training set in a hierarchical way. In a $d$-dimensional R-tree, a node consists of a set of $m$ entries $\{e_1,\dots,e_m\}$ where $m \geq 1$. In a leaf node, each entry refers to a $d$-dimensional training point. In a non-leaf node $N$, each entry refers to one of node $N$'s child nodes. Each R-tree node has a *Minimal Bounding Rectangle* (*MBR*), which is the minimal $d$-dimensional rectangle bounding its enclosed training points. Let a $d$-dimensional $MBR=\{(low_1, upp_1),\dots,(low_d, upp_d)\}$, where $low_i$ represents the smallest value of the points enclosed by this $MBR$ at dimension $i$, and $upp_i$ represents these points' largest value at dimension $i$.

The R-tree index structure has three appealing features when it is employed as the coding component.

First, in the construction of an R-tree, training points close in space are allocated to the same leaf node. Leaf and non-leaf nodes are recursively grouped together following the same principle to preserve data similarity. In an R-tree, there is only one root node at the lowest depth (depth 0) and multiple leaf nodes at the deepest depth. Given a set of training points indexed by an R-tree, the root node corresponds to the entire training set itself and any other node corresponds to a set of training points.

Second, an R-tree is a depth-balanced tree. Each leaf node has the same distance to the root node; that is, all leaf nodes have the same depth. Thus, the nodes at the same depth of the R-tree represent training points at the *same level of granularity*. A node's level of granularity represents the possible number of training points enclosed by this node's *MBR*. It is easy to see that the larger the volume (covering area) of the *MBR*, the larger the possible number of points, and so the coarser the granularity. For example, Figure 5.2(a) shows 21 training points $\{z_1,\dots, z_{21}\}$ in 2-dimensional space and Figure 5.2(b) shows an R-tree used to index these points with three depths. It can be seen that points close in space are grouped to the same node. At depth 0 of the R-tree, the root node represents all 21 points at the coarsest level of granularity. At depth 2, each leaf node corresponds to three points, representing these points at the finest level of granularity.

Third, the R-tree is a dynamic index structure that allows runtime insertion and deletion of leaf nodes with small overheads, thus supporting the dynamic updating of the training set.

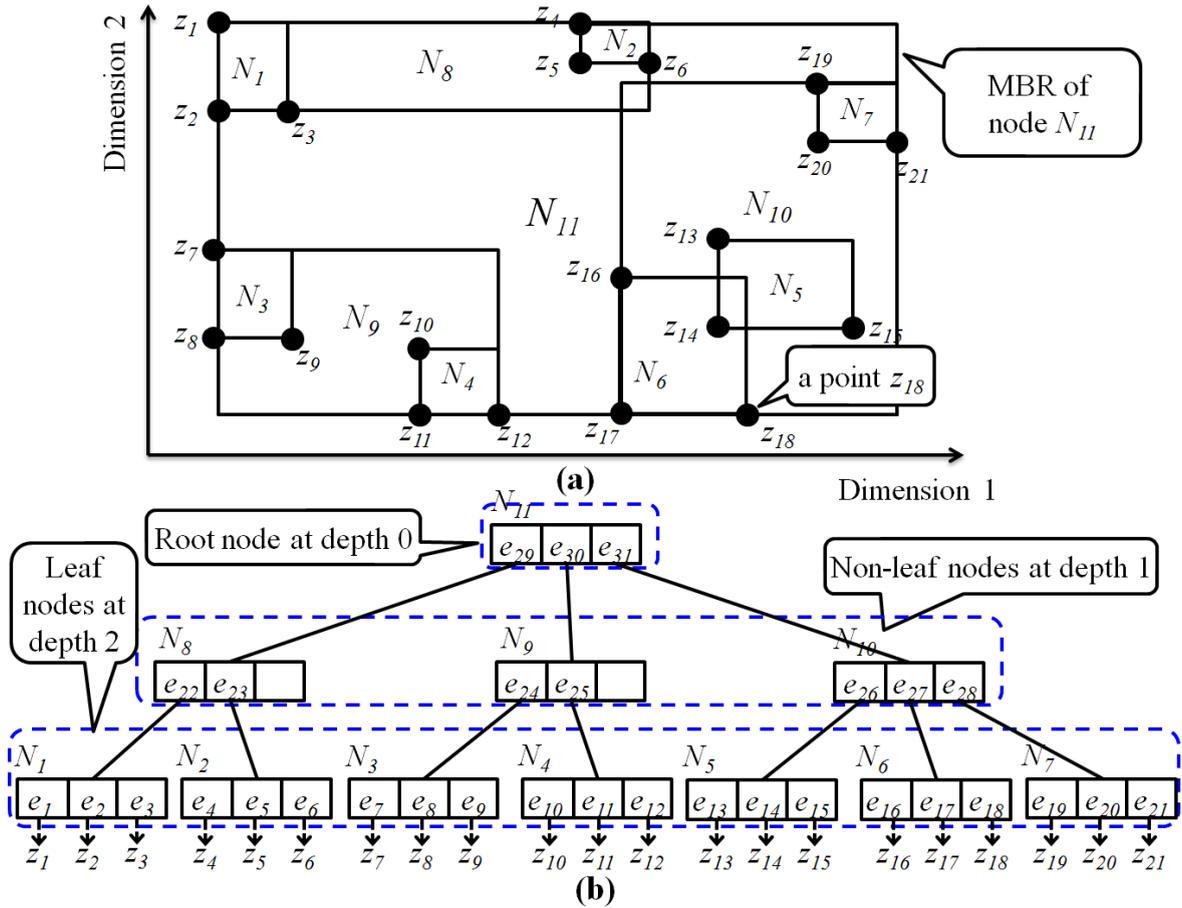

Figure 5.2: An example R-tree for indexing training points. We can see (a) a set of 21 2-dimensional training points and (b) the R-tree constructed to index all 21 points.

In the elastic kNN algorithm, given a training set, the R-tree coding component indexes all training points hierarchically using **two** R-trees: a positive R-tree indexes training points from the positive class, and another negative R-tree indexes training points from the negative class. Thus, one R-tree node corresponds to a set of training points from the same class, and this node has the same class label as these points. The output of this R-tree coding method is a set of *codes*, where each *code* corresponds to all the nodes at a particular depth of the two R-trees. For example, Figure 5.3 shows a positive R-tree and a negative R-tree generated by this R-tree coding method. There are three codes $c_0$={$N_{11}, N_{23}$}, $c_1$={$N_8$, $N_9$, $N_{10}$, $N_{20}$, $N_{21}$, $N_{22}$}, and $c_2$={$N_1$, ..., $N_7$, $N_{12}$, ..., $N_{19}$} indexing nodes at depth 0, 1, and 2 respectively of the two R-trees.

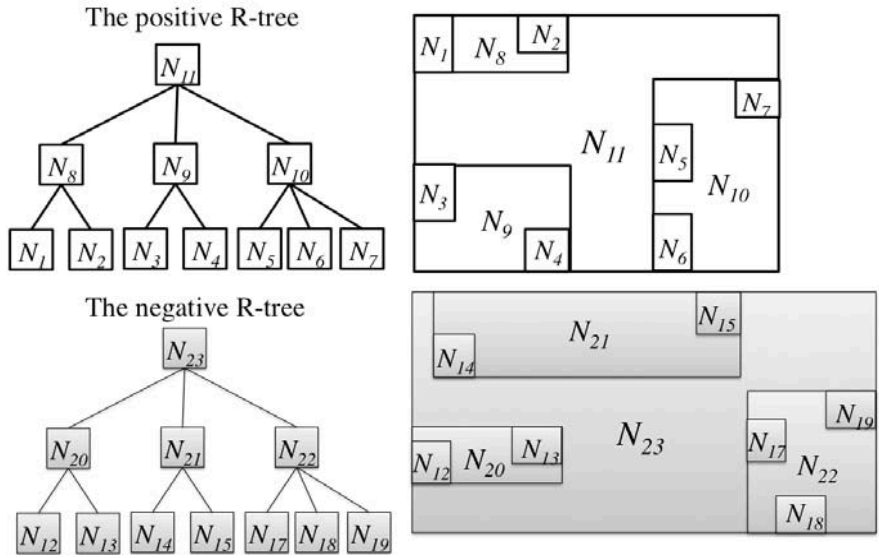

Figure 5.3: Two R-trees for indexing positive and negative training points.

Note that the R-tree coding component compresses the training set by reducing the number of training point instances, but keeps the dimensionality of data points unchanged. Hence, this coding component is applicable to datasets with different numbers of features. In the generated codes, each R-tree node represents a set of training points using the rectangle information from the node's *MBR*. Similar to traditional tree-based kNN techniques, the prediction accuracy and pruning power of the elastic kNN algorithm are influenced by the dimensionality of data. We will discuss this in detail in Section 5.4's experimental evaluation.

### 5.3.3   The Naïve kNN Classification Component

The naïve kNN classification component takes three inputs: a code $c$ consisting of a set of R-tree nodes; a test point $q$; and a starting state. This component outputs an **approximate result** $ar$: the $k$ R-tree nodes in $c$ that have the smallest distances to $q$. The *distance* between an R-tree node $N$ and a point $q$, denoted as $dist(q, N)$, is calculated as the *maximal Euclidean distance* from $q$ to any point in $N$'s *MBR*. Using these $k$ nodes with known class labels, $q$'s class label can be predicted. In addition, the starting state can either be the *initial state* or a *state* corresponding to the information retained in some data structure representing an *acquired result*. The elastic kNN algorithm, therefore, can start from the state of any previously obtained result and produce a better classification result if more budget is committed.

### 5.3.4 Calculation of an Approximate Result

Before delivering classification services, the elastic kNN algorithm first applies the R-tree coding component to a training set and generates a set of codes. Once a time budget is given, the algorithm estimates the code *length budget*; that is the *maximal number* of nodes that can be linearly scanned by the naïve kNN classification component within the time budget. A code $c$ is then selected that has the *maximum* length value smaller than the length budget. Using $c$ and a test point $q$, an approximate result $ar$ can be produced, either starting from an initial state or a state of some previously obtained result. Finally, the algorithm uses a data structure $s_{ar}$ to maintain some of the nodes in $c$ as $ar$'s state, which can be used as *the starting state* in future calculations.

The process of maintaining $ar$ is given below. Initially, the algorithm sets state $s_{ar} = c$ (line 2) and sets the *pruning threshold $dist_{max\_kNN}$* as the largest distance from $q$ to its $k$ closest nodes in code $c$ (i.e. $q$'s $k$ nearest neighbours) (line 3). Next, any node $N$ in $s_{ar}$ is removed if the minimal distance $dist_{min}(q, N)$ between $q$ and node $N$'s MBR is larger than $dist_{max\_kNN}$ (line 4 to 6). This is because such a node $N$ and its child nodes cannot be $q$'s nearest neighbours as proved in Proposition 5.1. Thus, all these nodes are pruned from the R-tree.

**Maintaining an approximate result $ar$ using state $s_{ar}$**

**Input:** A code $c$, a test point $q$, an approximate result $ar$.

**Output:** A state $s_{ar}$.

1. **Begin**

2.   Set state $s_{ar}=c$;

3.   Set $dist_{max\_kNN}=\max_{1 \leq i \leq k}\{dist(q, N_i)\}$ where $N_i \in ar$;   // $dist_{max\_kNN}$ is the pruning threshold

4.   **for** each node $N$ in set $s_{ar}$

5.     **if** $(dist_{min}(q, N) > dist_{max\_kNN})$, **then**

        //all $N$'s child nodes cannot be point $q$'s nearest neighbours;

6.       Set $s_{ar} = s_{ar} \backslash \{N\}$;   //remove node $N$ from set $s_{ar}$

7.     Return $s_{ar}$.

8. **End**

**Proposition 5.1: Nodes to be pruned from R-trees.** *Given an approximate result $ar$ produced using a code $c$ and a test point $q$, let $N$ be a node in $c$ and let the minimal Euclidean distance between $N$'s MBR and $q$ be $dist_{min}(q, N)$. Let the pruning threshold be $dist_{max\_kNN}=\max_{1 \le j \le k}\{dist(q, N_j)\}$ where $N_j \in ar$. Node $N$ is pruned from the R-tree if $dist_{min}(q, N) > dist_{max\_kNN}$.*

**Proof**. For any node $N_j \in ar$ $(1 \le j \le k)$, $N_j$'s distance to $q$ is *less than* or *equal to* $dist_{max\_kNN}$. Since $dist_{min}(q, N) > dist_{max\_kNN}$, $dist_{min}(q, N)$ is greater than node $N_j$'s distance to $q$:

$$dist_{min}(q, N) > dist(q, N_j). \tag{5.3}$$

Since a non-leaf node's $MBR$ covers all its child nodes' $MBRs$, any child node $N_j^{CHILD}$ of any node $N_j$ has distance $dist(q, N_j^{CHILD})$ to $q$ *less than* or *equal to* the distance $dist(q, N_j)$ between $N_j$ and $q$:

$$dist(q, N_j) \ge dist(q, N_j^{CHILD}). \tag{5.4}$$

For any child node $N^{CHILD}$ of node $N$, its minimal distance $dist_{min}(q, N^{CHILD})$ to $q$ is *greater* than or *equal* to the minimal distance between $q$ and $N$:

$$dist_{min}(q, N^{CHILD}) \ge dist_{min}(q, N). \tag{5.5}$$

According to Equations (5.3), (5.4), and (5.5), we have:

$$dist_{min}(q, N^{CHILD}) > dist(q, N_j^{CHILD})$$

Let the distance between $q$ and the node $N^{CHILD}$ be $dist(q, N^{CHILD})$. Since $dist(q, N^{CHILD}) \ge dist_{min}(q, N^{CHILD})$,

$$dist(q, N_j^{CHILD}) < dist(q, N^{CHILD}) \tag{5.6}$$

For any code $c'$ consisting of nodes from a deeper depth of the two R-trees, suppose node $N_j$ has $n_j$ child nodes $(n_j \ge 1)$ in $c'$. According to Equation (5.6), for any node $N_j^{CHILD}$ among these $n_j$ child nodes, its distance $dist(q, N_j^{CHILD})$ to $q$ is *less than* the distance $dist(q, N^{CHILD})$ between $q$ and any child node $N^{CHILD}$ of node $N$ in the same code value $c'$. Considering $q$'s $k$ closest nodes in $c$, there are $\sum_{i=1}^{k} n_i = \frac{k \times (k+1)}{2} > k$ nodes in $c'$ whose distances to $q$ are *less* than the distance between $q$ and any child node of $N$ in $c'$. Hence none of $N$'s child nodes can be $q$'s $k$ closest nodes in $c'$, and $N$ and its child nodes should be pruned from the R-tree.  ■

If extra budget is allocated, a code $c'$ can be selected. It is required that the nodes in $c'$ are from a deeper depth of the R-trees than the nodes in state $s_{ar}$, thus representing the training set at a finer level of granularity. Using $c'$, the elastic kNN algorithm can produce a refined result $ar'$ by starting from the state $s_{ar}$ of the obtained result $ar$. The process of producing $ar'$ starting from $s_{ar}$ is given below. The algorithm first removes any node $N \in c'$ whose parent node $N^{PARENT} \notin s_{ar}$, so $N^{PARENT}$ and its child nodes are pruned (line 2 to 4). The updated code $c'$ is then linearly scanned to find the test point $q$'s $k$ closest nodes (line 5) and these nodes are returned as result $ar'$.

**Producing a refined approximate result $ar'$ starting from the state $s_{ar}$**

**Input:** A code $c'$, a test point $q$, a state $s_{ar}$.

**Output:** An approximate result $ar'$.

1. **Begin**

2.     **for** each node $N$ in code $c'$

3.         **if** ($N$'s parent node $N^{PARENT} \notin s_{ar}$), **then**

            //all $N^{PARENT}$ and its child nodes are pruned from the R-tree

4.             Set $c' = c' \backslash \{N\}$;  //remove node $N$ from code $c'$

5.     Set $ar' = Na\ddot{\imath}ve\_kNN(c', q)$;

        //the function $Na\ddot{\imath}ve\_kNN(c', q)$ linearly searches code $c'$ to find $q$'s $k$ closest nodes $\{N'_1, \ldots, N'_k\}$ in $c'$

6.     Return $ar'$.

7. **End**

Figure 5.4 shows an example elastic 3NN algorithm ($k$=3). In Figure 5.4(a), an approximate result $ar$ is produced using a code $c = \{N_8, N_9, N_{10}, N_{20}, N_{21}, N_{22}\}$, which consists of six nodes at depth 1 of the two R-trees. Taking a test point $q$, nodes $\{N_9, N_{10}, N_{21}\}$ are selected as its three closest nodes (nearest neighbours), and $dist_{max\_kNN} = dist(q, N_{21})$. The *minimal* distances between $q$ and nodes $N_{20}$ and $N_{22}$ are greater than $dist_{max\_kNN}$, so these two nodes are removed from $s_{ar}$, and so $s_{ar} = \{N_8, N_9, N_{20}, N_{22}\}$. Figure 5.4(b) shows that, starting from $s_{ar}$, five nodes $N_{12}$, $N_{13}$, $N_{17}$, $N_{18}$ and $N_{19}$ in code value $c' = \{N_1, \ldots, N_7, N_{12}, \ldots, N_{19}\}$ are not used to produce result $ar'$ because their parent nodes do not belong to $s_{ar}$.

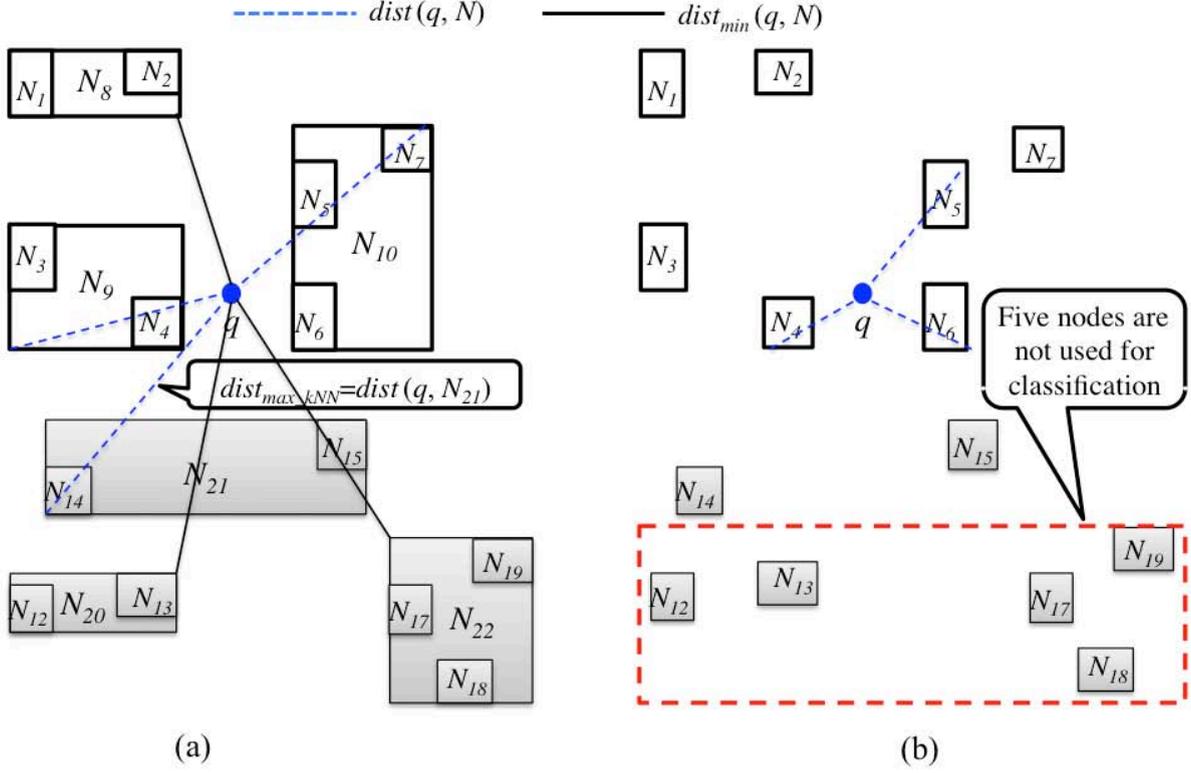

(a)                                              (b)

Figure 5.4: Producing a refined approximate result $ar'$ by staring from the state $s_{ar}$. We can see (a) six nodes in code $c$ used to produce $ar$, and (b) the nine nodes in code $c'$ used to produce $ar'$.

We now analyse the time complexity of the R-tree coding component and the naïve kNN classification component in the following.

**Proposition 5.2: Time complexity of constructing two R-trees.** *The time complexity of using the R-tree coding component to build an R-tree with $n_P$ positive training points and an R-tree with $n_N$ negative training points is $O(n_P \times logn_P + n_N \times logn_N)$.*

**Proof**: Using the standard R-tree construction algorithm [18], the construction time of the positive R-tree indexing $n_P$ training points is bounded by $O(n_P \times logn_P)$. Similarly, the construction time of the negative R-tree is bounded by $O(n_N \times logn_N)$. Hence, the total construction time is $(n_P \times logn_P + n_N \times logn_N)$. ∎

**Proposition 5.3. Time complexity of classifying a test point.** *Let $m$ be the maximum number of entries in one R-tree node. Using the naïve kNN classification method to classify a code $c$ consisting of nodes at depth $j$ of the two R-trees, the time complexity of classifying a test point $q$ is $O(m^j)$.*

**Proof**. At depth $j$ of one R-tree, there are at most $m^j$ nodes. Considering both R-trees, there are at most $2m^j$ nodes in $c$. Note that starting either from the initial state or from an obtained result, the upper bound of $c$'s length (number of nodes) is $2m^j$. The time complexity of using the naïve kNN classification method to find point $q$'s $k$ closest nodes in $c$ is $O(2m^j)$ [136], or namely $O(m^j)$. ∎

## 5.3.5 Discussion of the Property of Quality Monotonicity

In this section, we first provide theoretical proofs that the R-tree coding component has the property of entropy monotonicity, following by a discussion of how this property supports the quality monotonicity of the elastic kNN algorithm. We begin the proof with an observation: when a node is descended in an R-tree, its *MBR* is divided into the several smaller *MBRs* of its child nodes at a deeper depth of the tree.

**Observation 5.1:** *Let $c$ and $c'$ be two code values. If the length of $c$ is less than the length of $c'$, then we have: (i) the nodes in $c'$ are from a deeper depth of the R-tree; (ii) the MBRs of $c$'s nodes enclose the MBRs of $c'$'s nodes; (iii) the total volume of the MBRs of $c$'s nodes is greater than or equal to the total volume of the MBRs of $c'$'s nodes.*

For example, in Figure 5.5's two R-trees, let the code value $c$ contain six nodes ($N_8$ to $N_{10}$ and $N_{20}$ to $N_{22}$) at depth 1 and let the code value $c'$ contain 14 nodes ($N_1$ to $N_7$ and $N_{12}$ to $N_{19}$) at depth 2. We can see that if a node at depth 1 is descended, its *MBR* is divided into the multiple *MBRs* of its child nodes at depth 2. At depth 1, the six nodes' *MBRs* enclose the *MBRs* of the 14 nodes at depth 2.

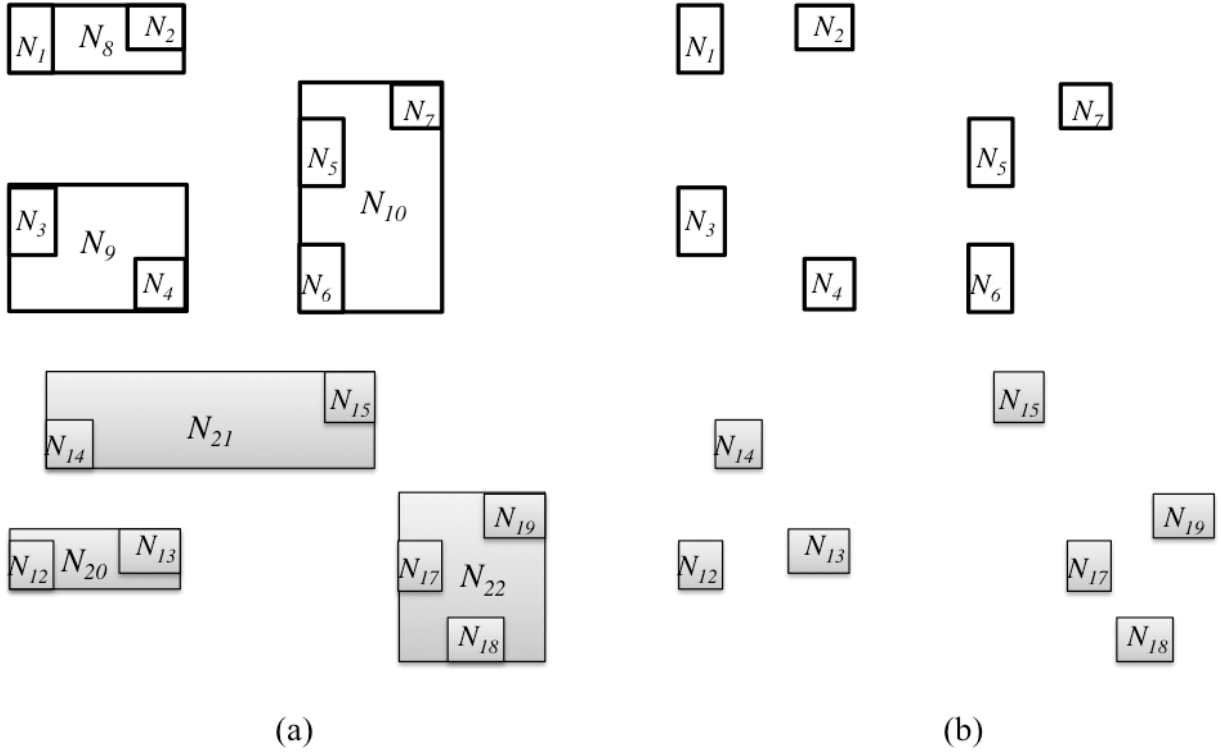

(a)                                                                    (b)

Figure 5.5: Two example code values $c$ and $c'$. We can see (a) the code value $c$ consisting of six nodes is selected according to a time budget $b$ and (b) the code value $c'$ consisting of fourteen nodes is selected according to a larger time budget $b'$.

**Theorem 5.1.** *The R-tree coding method satisfies entropy monotonicity.*

**Proof.** Let $l$ and $l'$ be two length budgets. Given a value $s$ of training set, the two selected code values are $c = C(S = s, l)$ and $c' = C(S = s, l')$, which correspond to two sets of nodes from two depths of the R-trees. If $l = l'$, then both $c$ and $c'$ contain the nodes from the same depth. Thus, we have $H(S|C = c) = H(S|C = c')$; equivalently, $R(c) = R(c')$.

If $l < l'$, then the total *MBR* volume of $c$'s nodes is *greater* than the total *MBR* volume of $c'$'s nodes according to Observation 5.1. Suppose training set $s$ has $m$ training points ($m \geq 1$) and the number of possible points in a node's *MBR* is proportional to its volume. Let the number of possible points in the *MBRs* of nodes in $c$ and $c'$ be $n$ and $n'$, respectively. We then have $n > n'$. For any possible value $s$ of training set $S$, $s$'s $m$ data points are enclosed by the *MBRs* of nodes in both $c$ and $c'$. The number of possible values of training set $S$ in $c$ and $c'$ are $\binom{n}{m}$ and $\binom{n'}{m}$ respectively, where $\binom{n}{m}$ represents a binomial coefficient.

We assume every possible point enclosed in a node's *MBR* has an equal probability of being selected. Thus, the probabilities of each value of the training set in $c$ and $c'$ are $1/\binom{n}{m}$ and $1/\binom{n'}{m}$, respectively. We can then calculate the conditional entropy of set $S$ given a code value $c$:

$$H(S|C=c) = \binom{n}{m} \times \frac{1}{\binom{n}{m}} \times log\binom{n}{m} = log\binom{n}{m}$$

Since $n > n'$,

$$H(S|C=c) - H(S|C=c') = log\binom{n}{m} - log\binom{n'}{m} > 0$$

Equivalently,

$$H(S|C=c) > H(S|C=c'),$$

$$H(S) - H(S|C=c) < H(S) - H(S|C=c'),$$

$$R(c) < R(c') \qquad\qquad\qquad \blacksquare$$

Furthermore, Proposition 5.4 indicates that if the length of a code value $c'$ is greater than that of another code value $c$—that is, $c'$ consists of more R-tree nodes—then the maximal distance between a test point $q$ and its $k$ closest nodes selected from $c'$ is *smaller*.

As illustrated in Figure 5.6, let $w$ denote the area of intersection of the *MBRs of nodes* in $c$ and the *sphere* whose centre is point $q$ and whose radius is $dist_{max\_kNN}$. Let $w'$ denote the area of intersection of the *MBRs of nodes* in $c'$ and the *sphere* whose centre is point $q$ and whose radius is $dist_{max\_kNN}'$. Since $c$'s *MBRs* enclose $c'$'s *MBRs* according to Observation 5.1 and $dist_{max\_kNN} \geq dist_{max\_kNN}'$ according to Proposition 5.4, it must be the case that the area $w$ encloses the area $w'$. In addition, both areas $w$ and $w'$ cover $q$'s $k$ exact nearest neighbours in the training set. Thus, in using the code value $c'$ with a greater length (i.e. a higher resolution), the naïve kNN classification component can produce an approximate result $ar'$ (i.e. nodes $N_4$, $N_5$, and $N_6$ in Figure 5.6(b)) that represents the exact result $er$ ($q$'s $k$ exact nearest neighbours selected from the entire training set) at a finer level of granularity. Compressed codes usually result in the prediction accuracy of the approximate result being lower than that of the exact result produced using the entire training set. The smaller discrepancy between results $ar'$ and $er$ indicates the better accuracy of result $ar'$.

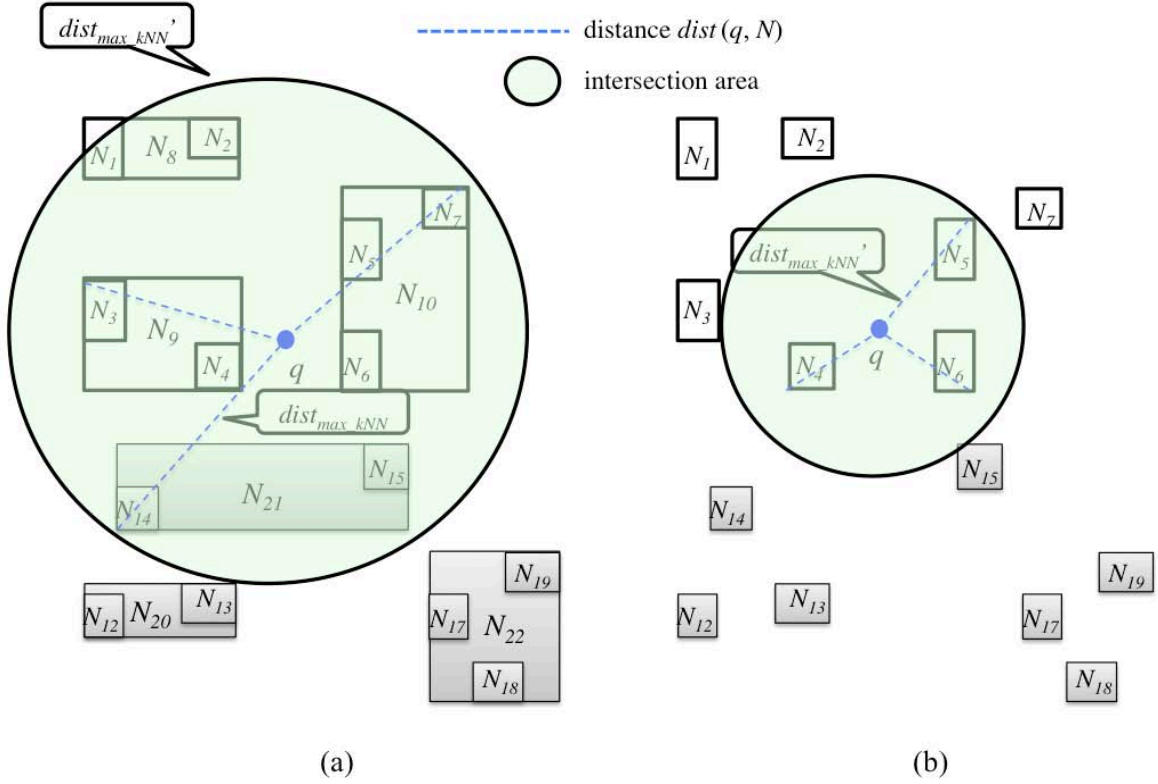

(a)                                        (b)

Figure 5.6: An example elastic 3NN classification algorithm ($k$=3). We can see (a) the area of intersection $w$ of the *MBRs* of nodes in $c$ and the sphere whose centre is point $q$ and whose radius is $dist_{max\_kNN}$; (b) the area of intersection $w'$ of the *MBRs* of nodes in $c'$ and the sphere whose centre is point $q$ and whose radius is $dist_{max\_kNN}'$. In this example, $c$={$N_8$, $N_9$, $N_{10}$, $N_{20}$, $N_{21}$, $N_{22}$} and $c'$={$N_1$, ..., $N_7$, $N_{12}$, ..., $N_{19}$}, and $dist_{max\_kNN} > dist_{max\_kNN}'$. We can see that the area $w$ encloses the area $w'$.

**Proposition 5.4**. *Let $c$ and $c'$ be two code values, and let their lengths be $len$ and $len'$ where $len \leq len'$. Given a test point $q$, let the maximal distance between $q$ and its $k$ closest R-tree nodes in $c$ and $c'$ be $dist_{max\_kNN}$ and $dist_{max\_kNN}'$, respectively. Then $dist_{max\_kNN}' < dist_{max\_kNN}$.*

**Proof**: If $len = len'$, then both code values $c$ and $c'$ contain the nodes selected from the same depth of the tree. Thus, $dist_{max\_kNN} = dist_{max\_kNN}'$.

If $len < len'$, the nodes in $c$ are from a lower depth of the R-trees, and these nodes' *MBRs* enclose the *MBRs* of $c'$'s nodes according to Observation 5.1. Now, for any node $N$ among $q$'s $k$ closest nodes in $c$, its distance $dist(q, N)$ to $q$ is less than or equal to $dist_{max\_kNN}$: $dist(q, N) \leq dist_{max\_kNN}$. Node $N$ has at least one child node $N^{CHILD}$ in $c'$ whose distance to $q$ is less than $dist(q, N)$: $dist(q, N^{CHILD}) < dist(q, N)$. Therefore, $dist(q, N^{CHILD}) < dist_{max\_kNN}$. Considering $q$'s $k$ closest nodes in $c$, there are at least $k$ nodes in $c'$ whose distances to $q$ are *less* than $dist_{max\_kNN}$. Hence, we have $dist_{max\_kNN}' < dist_{max\_kNN}$. ∎

## 5.3.6   Discussion of the Property of Accumulative Computation

Let $c$ and $c'$ be two codes of lengths $len$ and $len'$. Let $ar$ and $ar'$ be two approximate results produced using codes $c$ and $c'$, respectively. If $len' > len$, Theorem 5.1 proves that code $c'$'s resolution is greater: $R(c') > R(c)$ (the R-tree coding method satisfies entropy monotonicity). Hence, we can expect the result $ar'$ to have a higher prediction accuracy—that is, better quality—than the result $ar$. Let $s_{ar}$ and $s'_{ar}$ be the states for maintaining $ar$ and $ar'$, respectively. The property of accumulative computation indicates that starting from state $s'_{ar}$, a smaller computation budget is needed to produce a refined result $ar''$. We prove this in Proposition 5.5.

**Proposition 5.5**. Let $c''$ be the code used to produce the result $ar''$. Let the nodes in codes $c$, $c'$, and $c''$ be from depths $j$, $j'$, and $j''$ respectively of the two R-trees, where $j'' > j$ and $j'' > j'$. Let $b$ and $b'$ be the budgets used to produce result $ar''$ by starting from states $s_{ar}$ and $s'_{ar}$, respectively. We have $b' \geq b$ if $len' > len$.

**Proof**: Since the nodes in codes $c$ and $c'$ are selected from the same R-trees, and since $len' > len$, we have $j' > j$; that is, $c'$ consists of nodes at a deeper depth of the R-trees. Given a test point $q$ and a node $N$ at a lower depth $j$, if $N \in c$ but $N \notin s_{ar}$, all $N$'s child nodes at a deeper depth $j'$ cannot be included in set $s'_{ar}$. This is because all $N$'s child nodes cannot be $q$'s $k$ closest nodes in $c'$ according to Proposition 5.1. In contrast, given a node $N'$ at a deeper depth $j'$, if $N' \in c'$ but $N' \notin s'_{ar}$, $N'$'s parent node $N^{PARENT}$ at a lower depth $j$ may belong to set $s_{ar}$, because the node $N^{PARENT}$ may have other child nodes that are among $q$'s $k$ closest nodes. If $N^{PARENT} \in s_{ar}$, $N^{PARENT}$'s child nodes in code $c''$ can be used for classification. Hence, starting from $s_{ar}$, there may be more nodes in code $c''$ used for classification; that is, an equal or larger investment $\Delta I$ is needed to produce the same result $ar''$: $b' \geq b$ . ∎

## 5.4 Experimental Evaluation and Comparison of the Elastic kNN Algorithm

In this section, we experimentally evaluate the elastic kNN algorithm with three objectives.

First, Section 5.4.2 evaluates the property of quality monotonicity. We design experiments to produce a list of approximate results using the elastic kNN algorithm and demonstrate how their qualities, namely prediction accuracy and Area under the Receiver Operating Characteristic (ROC) Curve (AUC) [137, 138], gradually improve when more computations are conducted. We also discuss the connection between a code's resolution (*information gain* in Definition 5.1) and the quality (*accuracy* and *AUC*) of an approximate result produced using this code. The connection between an approximate result's quality and the quality of the exact result is also discussed.

Second, experiments in Section 5.4.3 demonstrate the property of accumulative computation. Since accumulative computation is achieved by pruning some R-tree nodes using information from an obtained result, we also present a detailed discussion of how this pruning is influenced by the data dimensionality.

Finally, Section 5.4.4 compares the elastic kNN algorithm against existing anytime kNN classification algorithms. The comparison results illustrate the qualities of approximate results produced by different anytime kNN algorithms under the same allocated time budgets.

### 5.4.1 Experimental Setup and Tested Datasets

**Experimental platform**. The elastic and the compared kNN algorithms are implemented using Java and compiled using NetBeans IDE 6.9.1. We run all the experiments on a Linux-based machine with four 2 GHz CPU cores and 4 GB memory.

**Tested datasets**. We experiment on five public datasets with a diverse range of sizes (from 351 to 245057) and features (from 2 to 128). These datasets all have two classes and are selected from the LIBSVM dataset repository [139] and the UCI machine learning repository [140]. Details of these datasets are given in Table 5.2.

Table 5.2. Five tested datasets

| Dataset | Size | Features | Source | Brief introduction |
|---------|------|----------|--------|--------------------|
| *fourclass*[1] | 862 | 2 | LIBSVM | Data points were created in [141] with irregular distributions that are not linearly separable. |
| *skin*[2] | 245,057 | 3 | UCI | Each point describe the various age groups, race grounds and genders of face images. |
| *credit*[3] | 690 | 14 | UCI | Each point describes information of credit card application using a mixture of six numerical attributes and eight categorical attributes. |
| *ionosphere*[4] | 305 | 34 | UCI | Each point has 34 continuous attributes used for judging whether a radar is "good" (showing evidence of structures in the ionosphere) or "bad". |
| *gas*[5] | 13,910 | 128 | UCI | Each point employs 128 continuous attributes to describe gases at various levels of concentrations. The values of attributes range from -16757.5986 to 670687.3477. |

**Quality measures of kNN classification**. Two widely applied assessment metrics for classification problems, namely prediction accuracy and AUC [137, 138], are used as quality measures in experiments. Briefly, in a binary kNN classifier, the *prediction accuracy* denotes the proportion of test points that are correctly classified. The *AUC* represents the probability that a randomly chosen positive test point (whose actual class label is positive) will have a greater chance of being predicted as positive than a randomly chosen negative test point.

**Definition of prediction accuracy.**

Let $c_P$ and $c_N$ be the positive and negative class labels, let $n$ be the number of test points, and let $q$ be a test point whose actual class label is $y \in \{c_P, c_N\}$. In kNN classification, let $k_p$ and $k_N$ be the number of $q$'s *positive* and *negative* nearest neighbours respectively, with $k = k_p + k_N$. The predicted class label is $y' = c_P$ if $k_p > k_N$; otherwise, it is $y' = c_N$. The accuracy of a binary kNN classifier, denoted by $Accuracy$, is the proportion of predictions it makes that are correct:

$$Accuracy \ = \ 1 - \frac{1}{n}\sum_{i=1}^{n} \ell(q_i, y'_i, y_i). \tag{5.7}$$

---

[1] http://www.csie.ntu.edu.tw/~cjlin/libsvmtools/datasets/binary.html
[2] http://archive.ics.uci.edu/ml/datasets/Skin+Segmentation
[3] http://archive.ics.uci.edu/ml/datasets/Statlog+(Australian+Credit+Approval)
[4] http://archive.ics.uci.edu/ml/machine-learning-databases/ionosphere/ionosphere.names
[5] http://archive.ics.uci.edu/ml/datasets/Gas+Sensor+Array+Drift+Dataset

Here, $y_i$ is test point $q_i$'s actual class label and $\ell(q_i, y'_i, y_i) = \begin{cases} 0, if \ y'_i = y_i \\ 1, if \ y'_i \neq y_i \end{cases}$.

**Definition of AUC.**

Let $n_P$ and $n_N$ be the number of positive and negative test points respectively, with $n = n_P + n_N$. We rank all test points in increasing order according their *estimated probabilities* of belonging to the positive class. In other words, the point with the greatest estimated probability has the highest ranking order. Let $q$ be a positive test point, and suppose its ranking order is $i$ among the $n_P$ positive points ($1 \leq i \leq n_P$) and $rank_i$ among all $n$ points ($1 \leq rank_i \leq n$). There are ($rank_i$-$i$) negative points whose ranks are lower than $q$. For example, Table 5.3 lists a test set of four positive points $\{q_{P,1}, q_{P,2}, q_{P,3}, q_{P,4}\}$ and four negative points $\{q_{N,1}, q_{N,2}, q_{N,3}, q_{N,4}\}$, with 8 being the highest ranking order. We can see positive point $q_{P,1}$'s ranking order is 1 among the four positive points and 2 among all eight points. This means $q_{P,1}$'s ranking order is greater than that of one negative point, $q_{N,1}$.

Table 5.3. An example test set with eight test points

| Test points | $q_{N,1}$ | $q_{P,1}$ | $q_{N,2}$ | $q_{P,2}$ | $q_{P,3}$ | $q_{N,3}$ | $q_{N,4}$ | $q_{P,4}$ |
|---|---|---|---|---|---|---|---|---|
| Ranking order $i$ among positive points | | 1 | | 2 | 3 | | | 4 |
| Ranking order $rank_i$ among all points | 1 | 2 | 3 | 4 | 5 | 6 | 7 | 8 |

When considering all $n_P$ positive points, we can summarise that positive points in general have higher ranks—that is, greater estimated probabilities of belonging to the positive class—than negative points:

$$\sum_{i=1}^{n_P}(rank_i - i) = \sum_{i=1}^{n_P} rank_i - \sum_{i=1}^{n_P} i = \sum_{i=1}^{n_P} rank_i - n_P(n_P + 1)/2.$$

In the ideal situation, each positive point has a higher rank than all $n_N$ negative points. Thus, $n_P \times n_N$ represents this situation, in which the ranking orders of all $n_P$ positive points are higher than those of all $n_N$ negative points. Thus, we can calculate the AUC quality as [142]:

$$AUC = \frac{\sum_{i=1}^{n_P} rank_i - n_P(n_P+1)/2}{n_P \times n_N}. \tag{5.8}$$

In the kNN classification problem, given a test point $q$ and its number $k$ of nearest neighbours, $q$'s estimated probability of belonging to the positive class, denoted by $p(c_P|q)$, is decided by its number $k_P$ of positive nearest neighbours [143]:

$$p(c_P|q) = \frac{k_P}{k}.$$

Intuitively, this means that the greater the number $k_P$ of positive nearest neighbours, the greater $q$'s estimated probability of belonging to the positive class. Thus, point $q$'s ranking order $rank$ is an increasing function of $k_P$:

$$rank = Rank\big(p(c_P|q)\big) = Rank\left(\frac{k_P}{k}\right). \tag{5.9}$$

For example, a test point $q$'s ranking order $rank$ is greater than another test point $q'$'s ranking order $rank'$ if $q$ has more positive nearest neighbours: $rank = Rank\left(\frac{k_P}{k}\right) > rank' = Rank\left(\frac{k'_P}{k}\right)$ if $k_P > k'_P$.

By substituting Equation (5.9) into Equation (5.8), we can calculate the AUC metric in the kNN classification problem:

$$AUC = \frac{\sum_{i=1}^{n_P} Rank(\frac{k_P^i}{k}) - n_P \times (n_P+1)/2}{n_P \times n_N} \tag{5.10}$$

where $q_i$ is a positive test point, $k_P^i$ is its number of positive nearest neighbours, and $k$ is its number of nearest neighbours.

**Setting of test sets**. In kNN classification, measures of both accuracy and AUC can be calculated using the numbers of positive and negative nearest neighbours of all test points according to Equations (5.7) and (5.10). Both quality measures are used to determine whether quality monotonicity can be guaranteed in a series of approximate results. Since using a small number of test points (e.g. 10) to calculate these quality measures results in greater variation in quality than using a large number of test points (e.g. 1000), to make our comparisons fair, we randomly selected 100 points from each dataset to form its test set and used the remaining points to form the training set, thus providing a common benchmark to assess the quality monotonicity of different kNN algorithms.

**Computational costs**. In the naïve kNN algorithm, the running time of classifying one test point using a training set (or a code) is determined by the number of points in the training set (or the number of R-tree nodes in the code). For each dataset, we use this number to represent the *computational cost* of producing exact and approximate results.

## 5.4.2   The Property of Quality Monotonicity

**Evaluation settings.** In the experiments that follow, we construct two R-trees for each dataset that provide five codes corresponding to nodes at depth 1 to 5 of the R-trees. The R-tree construction time is 0.73, 46.65, 0.71, 0.80, and 119.86 seconds for the *fourclass*, *skin*, *credit*, *ionosphere*, and *gas* datasets, respectively. In all evaluations, we set the number of nearest neighbours $k$=5. The two root nodes at depth 0 are not used because they are smaller than the value of $k$ and insufficient to produce a result. Using codes $c_1$ to $c_5$, five approximate results $ar_1$ to $ar_5$ are produced.

**Exact results and computational costs.** For each dataset, we first use the naïve kNN algorithm to linearly scan all the training points to produce the *exact result er*. Table 5.4 lists the accuracy and AUC for each exact result and the computational cost for producing the result.

Table 5.4. Qualities and computational costs of exact results for the five datasets

| Dataset | *fourclass* | *skin* | *credit* | *ionosphere* | *gas* |
|---|---|---|---|---|---|
| Accuracy | 1.00 | 1.00 | 0.75 | 0.80 | 0.92 |
| AUC | 1.00 | 1.00 | 0.78 | 0.91 | 0.96 |
| Computational cost | 763.00 | 204,957.00 | 590.00 | 205.00 | 13,810.00 |

The elastic kNN algorithm is designed to produce a series of approximate results $ar_1$ to $ar_5$ that gradually approach the exact result $er$. The codes $c_1$ to $c_5$ that are used to produce results $ar_1$ to $ar_5$ also gradually approach the training set that is used to produce the exact result $er$. Hence, it is reasonable to say that the quality of an approximate result not only depends on the resolution of the code used to produce it, but is also subject to the quality of the exact result to be approximated. We now discuss these two factors.

**Discussion 1: The connection between a code's resolution (information gain) and an approximate**

**result's qualities (accuracy and AUC).**

In this experiment, the elastic kNN algorithm is set to start from the *initial state* and produce all five results $ar_1$ to $ar_5$. As established in Theorem 5.1, the R-tree coding component has the property of entropy monotonicity: the five codes $c_1$ to $c_5$ have increasing resolution (information gain); that is, increasing levels of approximation to the training set. Hence, the *qualities* of results $ar_1$ to $ar_5$, which are produced using codes $c_1$ to $c_5$, should gradually approach the qualities of the exact result.

The experimental results shown in Figure 5.7 support the above claim. Here, the *x* axis lists represents the number of R-nodes in codes $c_1$ to $c_5$ as computational costs, and the *y* axis represents the accuracy and AUC of $ar_1$ to $ar_5$ as quality measures. The results show that the elastic kNN algorithm guarantees the monotonicity of quality in all cases: if application owners invest more computational cost, they can achieve a result with higher values of accuracy and AUC. By comparing Table 5.4 and Figure 5.7, we can see that these values stepwise approach the accuracy and AUC of the exact result. We can also see that the exact result and the approximate result $ar_5$ have very similar qualities (accuracy and AUC), but the computational cost of $ar_5$ is significantly smaller. This is because the R-tree nodes in $c_5$ have a high level of approximation to the original training points.

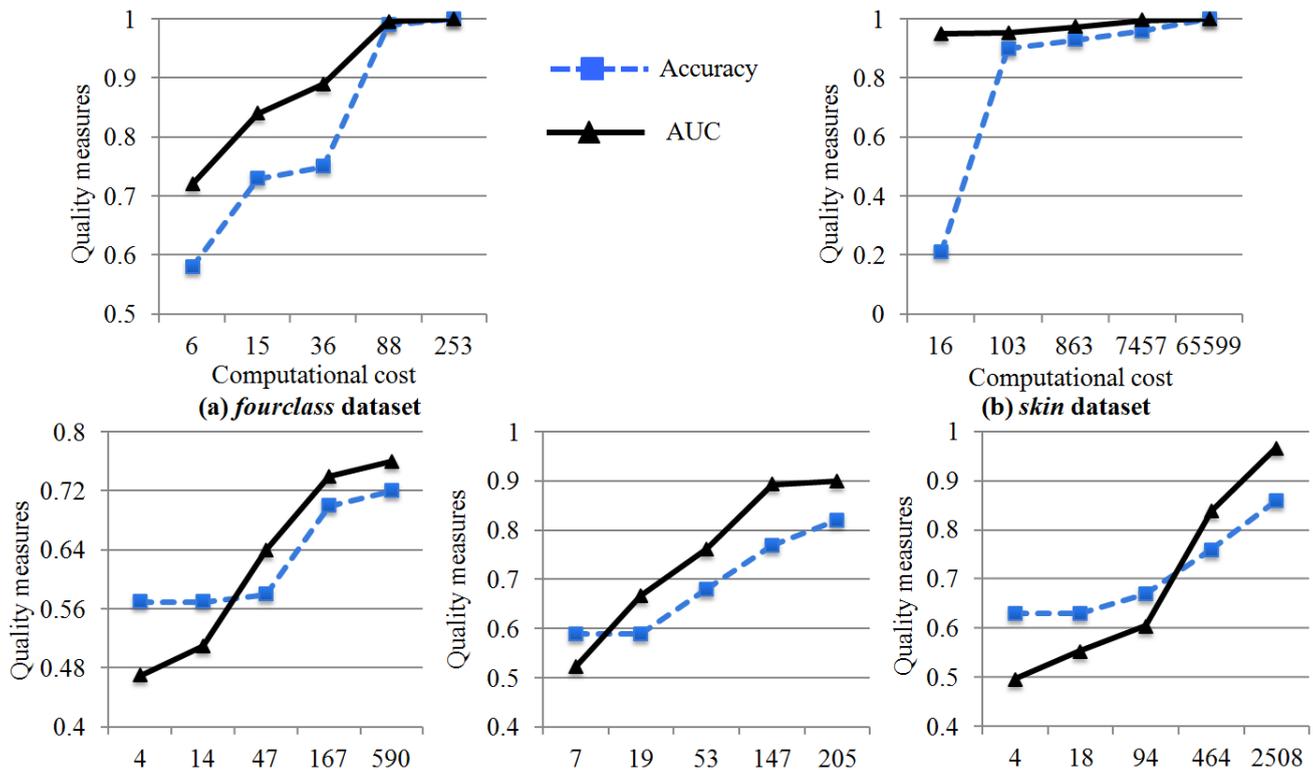

Figure 5.7: Five approximate results for the five datasets using the elastic kNN algorithm.

**Discussion 2: The connection between an approximate result's quality and the exact result's quality.**

We take the *ionosphere* dataset as an example and design two experiments to demonstrate how variations in the exact result's quality can influence the quality of approximate results.

**Design of two experiments**. Table 5.4 shows that even in *exact result*s produced using entire training sets, there are still some prediction errors. In kNN classification, these prediction errors can be divided into two types. First, the *intrinsic* or *irreducible error* comes from the noise in training and test points themselves. Given a fixed pair of training and test sets, this error is the lower bound of error in prediction. Second, the *bias error* comes from the kNN prediction model, such as the choice of the number $k$ of nearest neighbours. For example, suppose one positive training point and one negative training point have the same distance to a test point $q$. Under some value of $k$, if either training point can be selected as $q$'s nearest neighbour, then an incorrect selection may result in an incorrect prediction. Based on the above observation, we design two experiments to show how variations in training and test points (Experiment 1) and differing numbers $k$ of nearest neighbours (Experiment 2) influence the result quality.

**Experiment 1.** We create three pairs of training and test sets: each test set consists of 100 points randomly selected from the 305 points in the *ionosphere* dataset, and the remaining 205 points form the accompanying training set. The naïve kNN algorithm is first applied to each of the three pairs of datasets, producing three exact results ($er_1$, $er_2$, and $er_3$) whose qualities are listed in Table 5.5. Next, the elastic kNN algorithm is applied to each training/test set pair using codes $c_1$ to $c_5$, producing five approximate results $ar_1$ to $ar_5$. Figure 5.8(a) shows that each code $c_i$ ($1 \leq i \leq 5$) used in classification contains a very similar number of R-tree nodes in each of the three training/test set pairs. Thus, using the same code $c_i$, the results $ar_i$ produced from different training/test set pairs have very similar levels of approximation to the exact result. However, since the exact results $er_1$, $er_2$, and $er_3$ have different qualities, the same approximate results $ar_i$ from different training/test set pairs also have different qualities. Figures 5.8(b) and (c) display the accuracies and AUCs of these approximate results; that is, their qualities. We can see that the five approximate results produced from training/test set pair 3 have the best quality in most cases, because the exact result $er_3$ has the best quality.

Table 5.5. Accuracy and AUC of exact results vs. training/test sets for the *ionosphere* dataset

| Training/test set pairs | Exact result | Accuracy | AUC |
|---|---|---|---|
| 1 | $er_1$ | 0.78 | 0.86 |
| 2 | $er_2$ | 0.80 | 0.91 |
| 3 | $er_3$ | 0.84 | 0.97 |

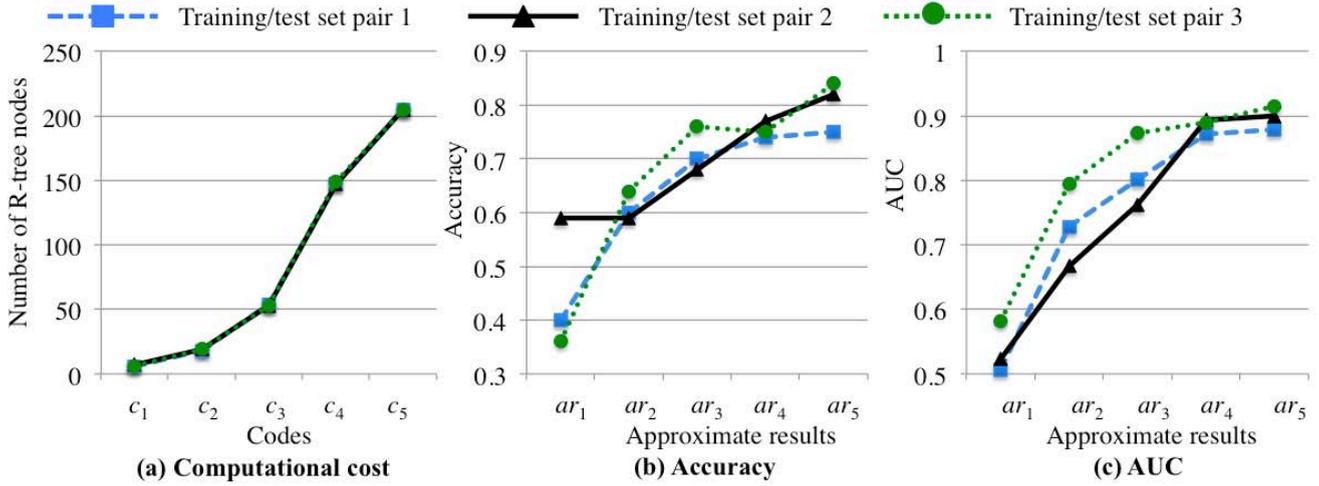

Figure 5.8: Evaluating quality vs. training/test set pairs for the ionosphere dataset.

**Experiment 2**. We vary the number $k$ of nearest neighbours to generate three kNN classifiers, and use them to test the same training/test set pairs from the *ionosphere* dataset. Three values (65, 25, and 5) of $k$ are tested. Table 5.6 lists the qualities of the three exact results $er_1$ to $er_3$ produced by these kNN classifiers. Figure 5.9 displays the qualities of five approximate results $ar_1$ to $ar_5$ for each kNN classifier. We can see that, similarly to *Experiment 1*, an approximate result $ar_i$ has better quality (i.e. higher values of accuracy and AUC) if its corresponding exact result has better quality. This is because if one kNN classifier (e.g. the 5NN classifier) can produce a better exact result using a training set than another kNN classifier (e.g. the 65NN classifier), it can also produce a better approximate result $ar_i$ using the same code $c_i$.

Table 5.6. Accuracy and AUC of exact results vs. value $k$ of nearest neighbours for the *ionosphere* dataset

| The number of nearest neighbours $k$ | Exact result | Accuracy | AUC |
|:---:|:---:|:---:|:---:|
| 65 | $er_1$ | 0.61 | 0.85 |
| 25 | $er_2$ | 0.78 | 0.88 |
| 5 | $er_3$ | 0.80 | 0.91 |

Moreover, we can see that if one kNN classifier can produce earlier approximate results (e.g. result $ar_1$) with better quality, this classifier has a good chance of producing later approximate results (e.g. result $ar_5$) with better quality. This is because the nodes in code $c_5$ are the child nodes of the nodes in code $c_1$; thus, the result $ar_5$ preserves the predictions made by the result $ar_1$. Since the computational cost of producing $ar_1$ is significantly lower, it is possible to introduce heuristic methods to determine the optimal value of $k$ in the elastic kNN algorithm. For example, we can test different values of $k$ in the calculation of the result $ar_1$, choose a value $k^*$ that maximises $ar_1$'s quality, and use $k^*$ to produce later approximate results.

***Results of Section 5.4.2****. The experimental results show that the elastic kNN algorithm does indeed exhibit quality monotonicity in practice under both accuracy and AUC quality measures. Experiment results also show that the quality of an approximate result can increase when either the resolution of the code used to produce the result increases (i.e. the code's information gain increases), or the quality of the exact result improves.*

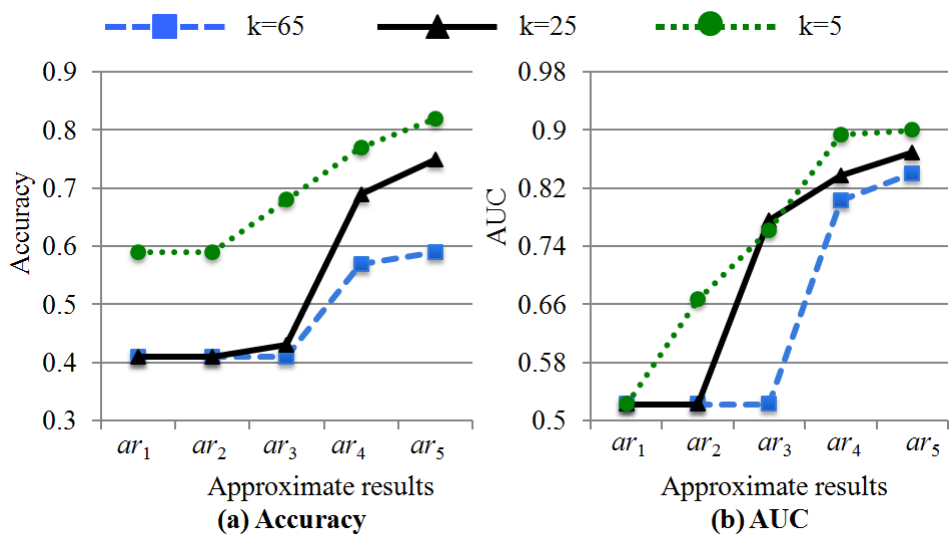

Figure 5.9: Evaluating quality vs. value $k$ of nearest neighbours for the *ionosphere* dataset

### 5.4.3   The Property of Accumulative Computation

**Evaluation settings.** We repeat Section 5.4.2's experiment, which produces five approximate results $ar_1$ to $ar_5$ for each of the five datasets. Rather than just starting from the *initial state*, different starting states are now tested. Using the *fourclass* dataset as an example, Table 5.7 lists the computational costs (the numbers of R-tree nodes) of producing five approximate results $ar_1$ to $ar_5$ when starting from different states. We can see that for each result, the computational cost is greatest when starting from the *initial* state, and this cost decreases when starting from a result with better quality. For example, the cost of producing result $ar_5$ is least when the starting state is $s_{ar_4}$.

Table 5.7. Computational costs of producing approximate results from different starting states for the *fourclass* dataset

|  |  | Produced approximate result | | | | |
|---|---|---|---|---|---|---|
|  |  | $ar_1$ | $ar_2$ | $ar_3$ | $ar_4$ | $ar_5$ |
| Starting state | The initial state | 6 | 15 | 36 | 88 | 253 |
|  | $s_{ar_1}$ |  | 15 | 36 | 88 | 253 |
|  | $s_{ar_2}$ |  |  | 28 | 69 | 201 |
|  | $s_{ar_3}$ |  |  |  | 31 | 92 |
|  | $s_{ar_4}$ |  |  |  |  | 36 |

Figure 5.10 shows the experimental results from all five datasets: here, the x axis lists the five starting states and the y axis represents the computational cost of producing result $ar_5$. The results show that the elastic kNN algorithm satisfies the property of *accumulative computation* in all datasets: if the algorithm starts from the state of an obtained result with better quality, it can produce result $ar_5$ with less computational cost. The above experimental results also verify the proof of Proposition 5.5: that starting from a result produced using a code with a longer length (i.e. a higher resolution) leads to less computation being needed.

**Discussion of the influence of data dimensionality.**

Figure 5.10's results show that for the two low-dimensional datasets (*fourclass* and *skin*), the computational costs appear to decrease significantly when starting from the state of an obtained result with better quality. In contrast, the costs decrease by only a small amount in the other three datasets whose data dimensions are greater than 10. This is because the elastic kNN algorithm applies an R-tree as the coding component, in which each R-tree node uses the rectangle information (upper bounds and lower bounds of nodes' *MBRs*) to represent the training points enclosed by the node. Therefore, similar to other spatial access techniques, the *pruning power* of the algorithm is reduced due to the curse of dimensionality. Specifically, as the dimensionality of data increases, the volume of space increases so that data points become sparser and the *MBRs* of R-tree nodes become highly overlapped. This overlap is problematic when the algorithm tries to prune nodes from the R-tree: a high degree of overlap means few nodes can be pruned, and so little can be saved in terms of computational cost. Using the low-dimensional *fourclass* dataset and the high-dimensional *gas* dataset as examples, we demonstrate how the curse of dimensionality influences pruning power.

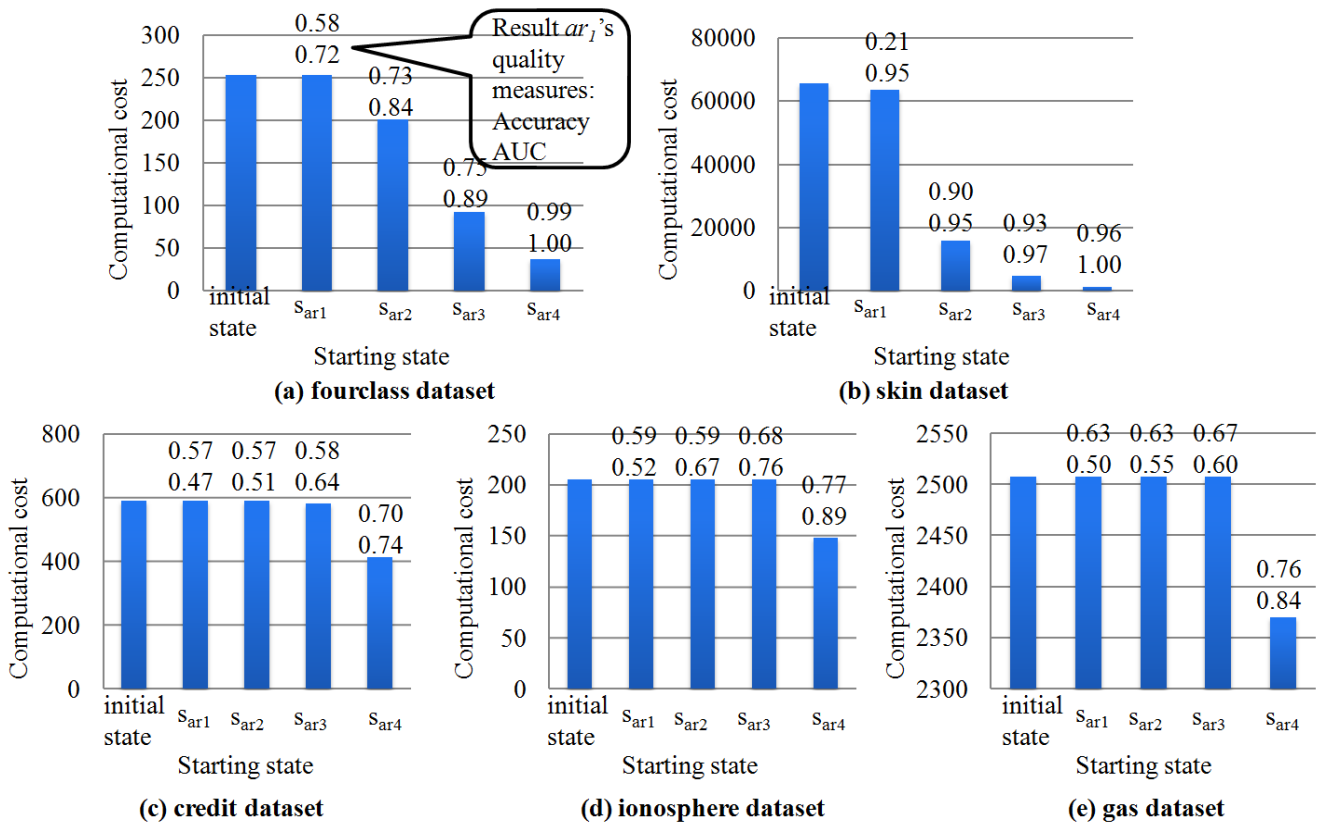

Figure 5.10: The computational costs of producing result $ar_5$ starting from five states: the initial state and the states $s_{ar_1}$ to $s_{ar_4}$. We have experimental results for: (a) the *fourclass* dataset; (b) the *skin* dataset; (c) the *credit* dataset; (d) the *ionosphere* dataset; and (e) the *gas* dataset.

Using the *fourclass* dataset and a test point $q$ as an example, Figure 5.11 shows the nodes in code $c_i$ that are kept in $s_{ar_i}$ and the nodes that are pruned from the R-trees when the elastic kNN algorithm starts from states $s_{ar_i}$ of results $ar_i$ ($1 \leq i \leq 4$). The child nodes of the pruned nodes are also pruned from the R-trees, and are not used in further classifications. In Figure 5.11, the x axis lists each node $N$ in code $c_i$ denoted by a number (e.g. "1" means the first node in $c_i$) and the y axis represents node $N$'s minimal distance $dist_{min}(q, N)$ to the point $q$. The pruning threshold $dist_{\max\_kNN}$ (the maximal distance between $q$ and its $k$ closest nodes in $c_i$) is plotted as a dashed line. If the minimal distance $dist_{min}(q, N)$ between a node $N$ and $q$ is *greater* than $dist_{\max\_kNN}$, $N$ should be pruned from the R-tree as proved in Proposition 5.1. For example, in Figure 5.11(a), the minimal distances between the six nodes in $c_1$ to the point $q$ are 0.00, 81.00, 13,689.00, 1,521.00, 1,570.00, and 169.00, and the pruning threshold $dist_{\max\_kNN}$ is 25,210.00. Thus, no node is pruned. In contrast, Figure 5.12(b) shows that if the starting state is $s_{ar_2}$, three nodes are removed. The value of the pruning threshold $dist_{\max\_kNN}$ decreases from result $s_{ar_1}$ to $s_{ar_4}$, resulting in more and more nodes being pruned from the R-trees.

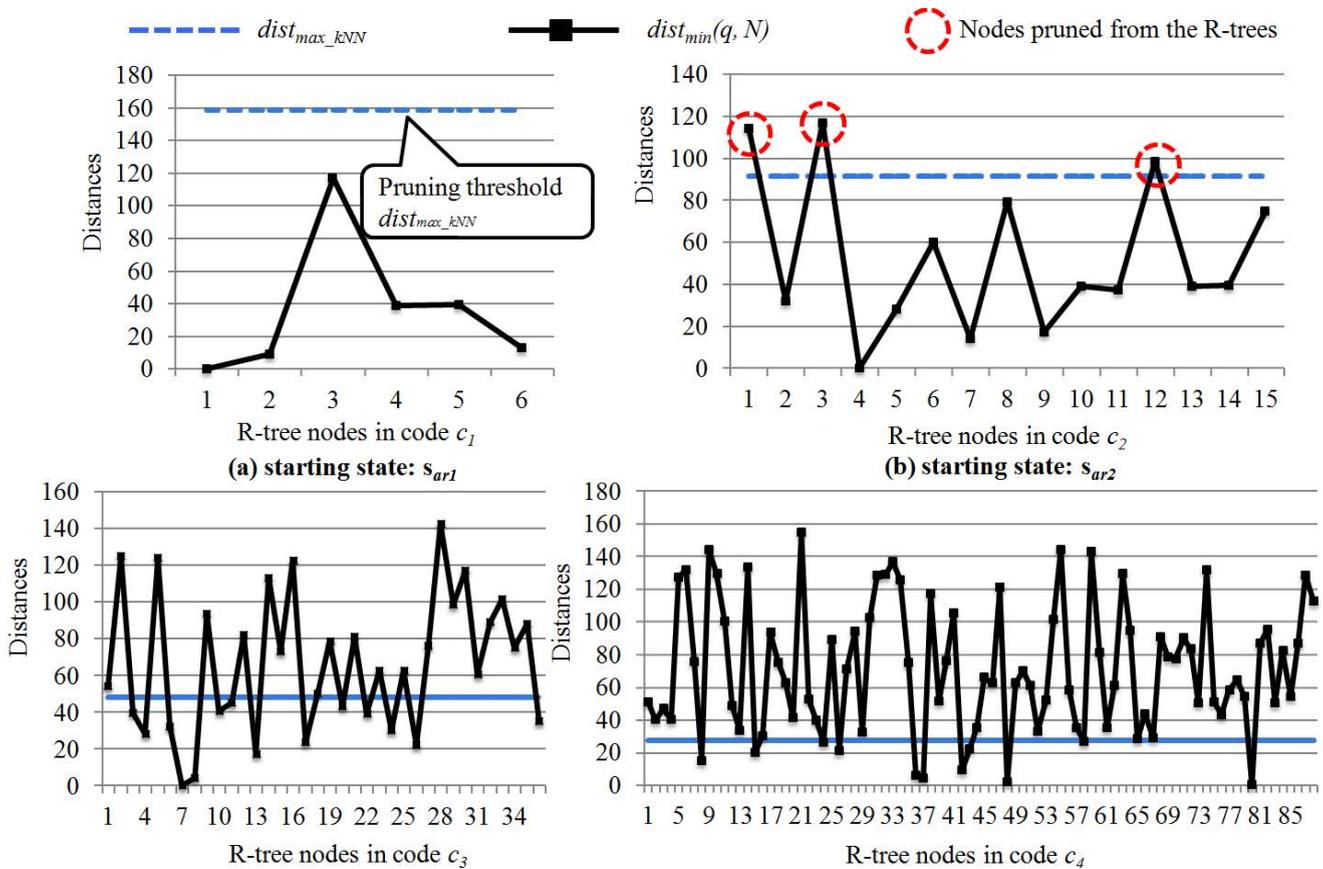

Figure 5.11: The pruning of R-tree nodes when starting from different states for the *fourclass* dataset.

We repeat the above experiment on the 128-dimensional *gas* dataset and Figure 5.12 shows the results. Given such high-dimensional training points, the volume of the R-tree nodes' $MBRs$ becomes very large and this has a twofold impact. First, the maximal distance between a test point $q$ and a node $N$'s $MBR$ increases; that is, the pruning threshold $dist_{\max\_kNN}$ increases. Secondly, the minimal distance $dist_{min}(q, N)$ between $q$ and $N$'s $MBR$ decreases. Therefore, the pruning condition that $dist_{min}(q, N)$ is greater than $dist_{\max\_kNN}$ becomes more difficult to meet. Figure 5.12 shows that no node is removed when the starting state is $s_{ar_1}$, $s_{ar_2}$, or $s_{ar_3}$, and only 13 nodes (out of 464 nodes in code $c_4$) are pruned from the R-trees when starting from state $s_{ar_4}$. Hence, computational cost can be saved only when starting from the state $s_{ar_4}$.

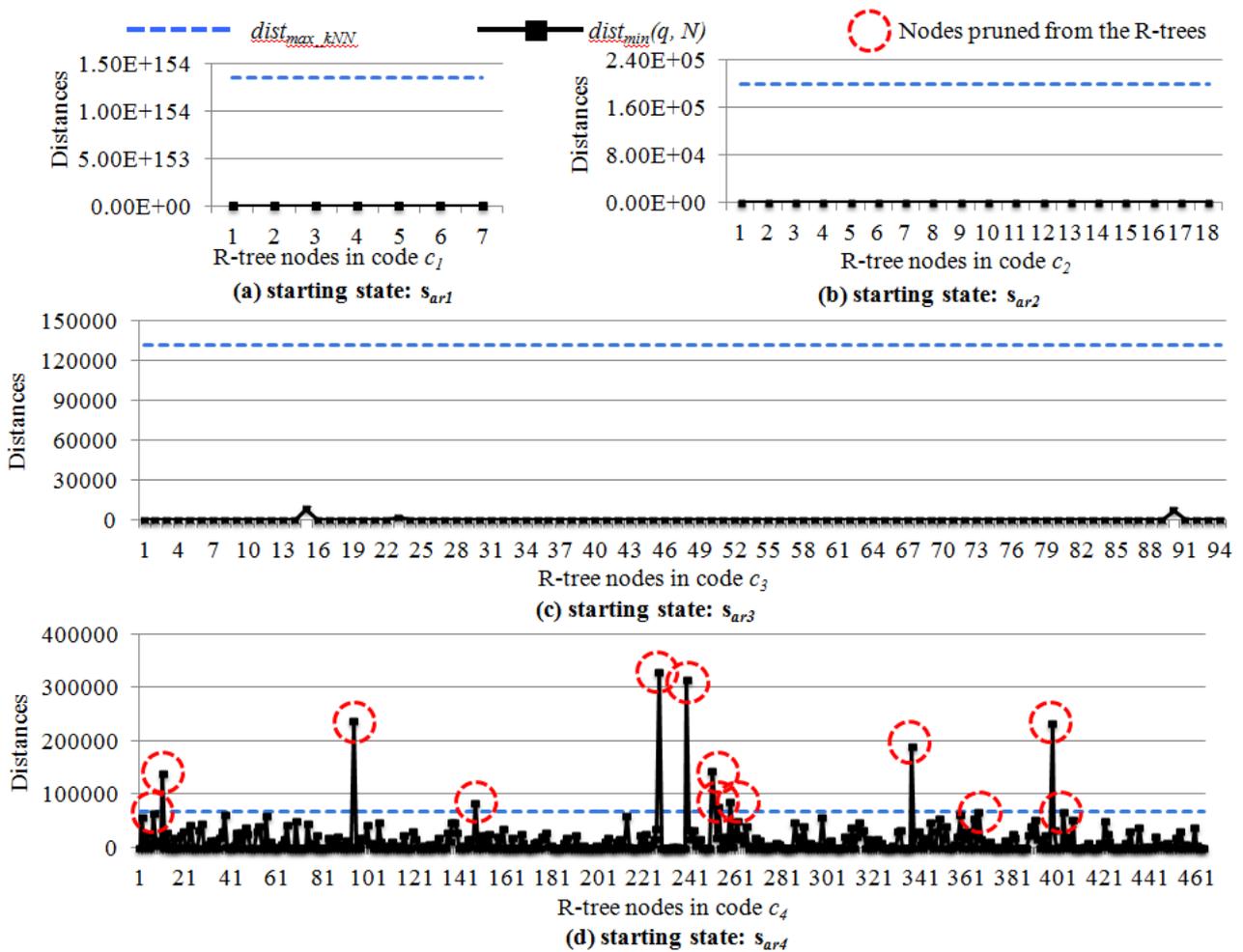

Figure 5.12: The pruning of R-tree nodes when starting from different states for the *gas* dataset.

**Results of Section 5.4.3.** *The experimental results show that the elastic kNN algorithm has the property of accumulative computation. The algorithm works well for low-dimensional data but high dimensionality* *reduces its pruning power. A possible solution is to apply dimension reduction techniques to preprocess the data before they are fed to the algorithm [144]. Many techniques such as PCA and clustering can be applied to reduce the data dimensionality while minimising information loss, thus minimising the decrease in prediction accuracy.*

### 5.4.4 Comparison to Existing Anytime kNN Classification Algorithms

**Comparison settings.** Following the experimental settings of previous sections, five approximate results $ar_1$ to $ar_5$ are sequentially produced for each dataset. In this section's experiment, once a result $ar_i$ is obtained, the next result $ar_{i+1}$ is produced by starting from $ar_i$'s state $s_{ar_i}$ so as to minimise the computational cost ($1 \leq i \leq 4$). The time required to produce each result $ar_i$ using the elastic kNN algorithm, denoted as $b_i$, is allocated as the time budget $b_i$ for producing the same result $ar_i$ using the compared anytime kNN algorithms. Thus, the results produced by all algorithms are compared under the same allocation of time budgets, as listed in Table 5.8.

Table 5.8. Time budgets for producing five approximate results

| Time budget (seconds) | *fourclass* | *skin* | *credit* | *ionosphere* | *gas* |
|:---:|:---:|:---:|:---:|:---:|:---:|
| $b_1$ | 0.03 | 0.06 | 0.04 | 0.07 | 0.51 |
| $b_2$ | 0.11 | 0.69 | 0.14 | 0.25 | 1.72 |
| $b_3$ | 0.15 | 2.06 | 0.52 | 0.75 | 6.02 |
| $b_4$ | 0.18 | 11.96 | 0.35 | 3.07 | 96.35 |
| $b_5$ | 0.21 | 65.00 | 14.68 | 3.17 | 2518.44 |

**The two compared anytime kNN algorithms.**

*The ranking-based anytime kNN algorithm* [107]. Before classification, this anytime algorithm employs a ranking method to sort all the training points in ascending order according to their distances to the closest training point of the same class: the smaller the distance, the higher the rank. The higher ranked points are viewed as more important points, and lower ranked points as less important points. In classification, this anytime kNN algorithm first uses more important points to produce an initial result and gradually adds less important points to produce refined results. Since the number $k$ of nearest neighbours is 5, the first classification result can be produced after five training points are used. In the experiment, this anytime kNN algorithm continuously adds more training points in classification and is interrupted to produce an approximate result $ar_i$ once the time budget $b_i$ is exhausted.

*The R-tree-based anytime kNN algorithm.* In one anytime Bayes classification algorithm, an R-tree is also applied to index training points [108], using three strategies to descend R-tree nodes: breadth-first search (BFS); depth-first search (DFS); and optimal-first search (OFS). The BFS strategy descends R-tree nodes at the same depth in turn before proceeding to the next depth. The DFS strategy descends nodes as far as possible along one R-tree branch. In the OFS strategy, the R-tree node with the smallest distance to a test point is descended first. Using these three descending strategies, we implement three versions of an R-tree-based anytime kNN algorithm: BFS, DFS, and OFS. This anytime kNN algorithm also applies one R-tree to index positive training points and one R-tree to index negative training points. An initial result can be produced using the nodes at depth 1 of the two R-trees as training points. At any new anytime iteration, one R-tree node is descended in both the positive and negative R-trees and the two descended nodes are replaced by their child nodes as training points. The algorithm is interrupted to produce an approximate result $ar_i$ once the time budget $b_i$ is exhausted.

**Experiment results**. Figures 5.13 to 5.17 show the comparison results from the five datasets, in which the accuracies and AUCs of the results are compared. We can see that for most of the datasets, the anytime kNN algorithms cannot guarantee producing a result with better quality (higher values of accuracy and AUC) when additional time budget is allocated. For example, consider the three versions of the R-tree-based anytime algorithm: the BFS strategy only guarantees quality monotonicity for the *ionosphere* dataset (Figure 5.16); the DFS strategy only guarantees quality monotonicity for the *fourclass* dataset (Figure 5.13); and the OFS strategy only guarantees quality monotonicity for the *fourclass* and *gas* datasets (Figures 5.13 and 5.17).

**Discussion of the influence of data dimensionality on prediction results**. We can see that for the low-dimensional datasets *fourclass* and *skin* (Figures 5.13 and 5.14), a majority of the results produced by the elastic kNN algorithm have better qualities than the results produced by the anytime kNN algorithms under the same time budgets. In contrast, for the datasets *credit*, *ionosphere*, and *gas*, whose data dimensionalities are greater than 10 (Figures 5.15 to 5.17), some early results produced by the anytime algorithms (e.g. results $ar_1$ and $ar_2$) have higher values of accuracy or AUC than the results produced by the elastic kNN algorithm. This is because as the data dimensionality increases, the R-tree nodes from low depths of the R-trees (e.g. nodes in code $c_1$ and $c_2$) become highly overlapped. This overlap is problematic when the kNN algorithm tries to use rectangle information from R-tree nodes to select a test point's nearest neighbours. Incorrect selection of nearest neighbours may cause incorrect predictions, thus decreasing prediction accuracy. However, the anytime kNN algorithms cannot guarantee monotonic improvement in their result quality. Hence, the later results $ar_4$ and $ar_5$ produced by the elastic kNN algorithm always have better qualities for each dataset. The experimental results show that even in high-dimensional space, the R-tree nodes from deep depths of the R-trees (e.g. nodes in codes $c_4$ and $c_5$) still provide sufficiently high resolution in their rectangle information to allow the algorithm to select the correct nearest neighbours.

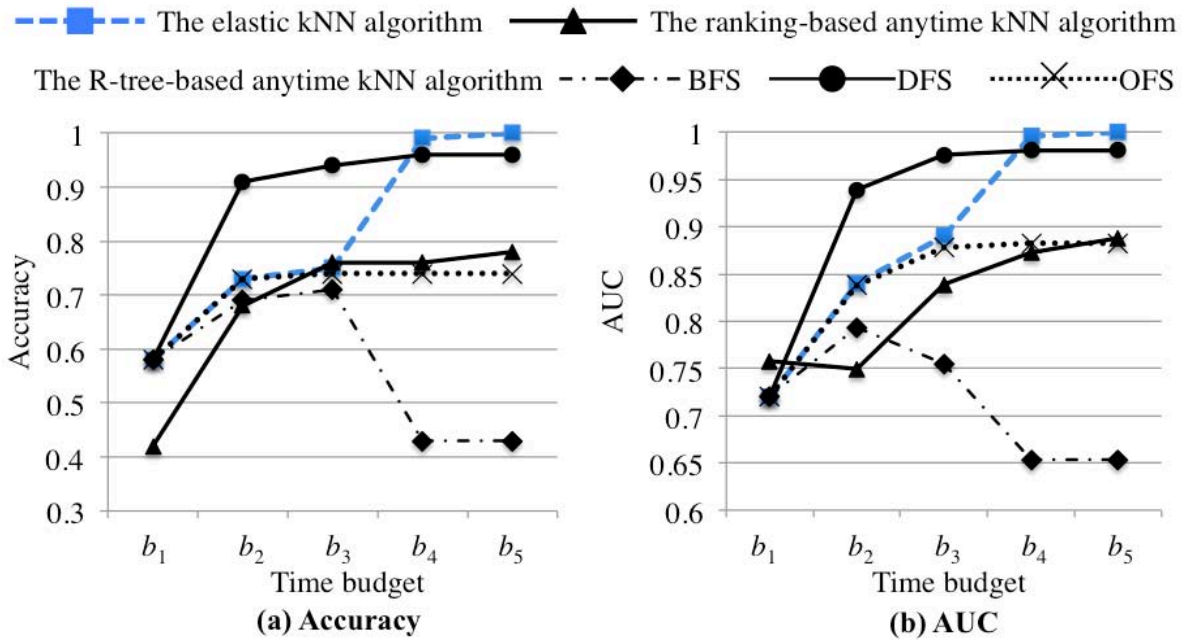

Figure 5.13: Comparison of the elastic and anytime kNN algorithms for the *fourclass* dataset.

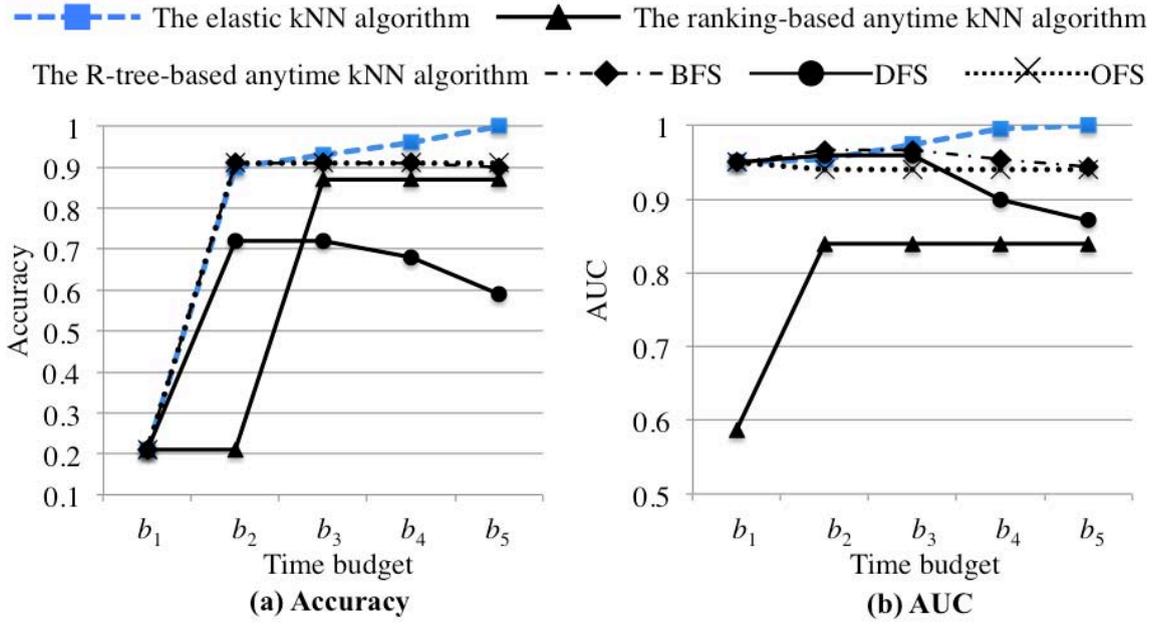

Figure 5.14: Comparison of the elastic and anytime kNN algorithms for the *skin* dataset.

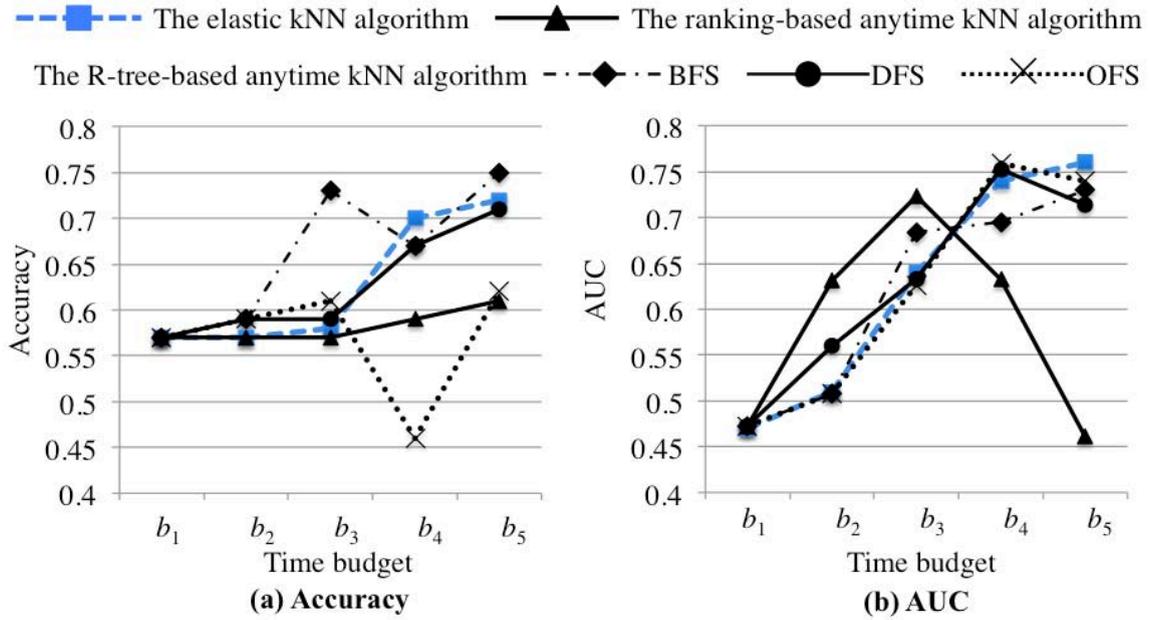

Figure 5.15: Comparison of the elastic and anytime kNN algorithms for the *credit* dataset.

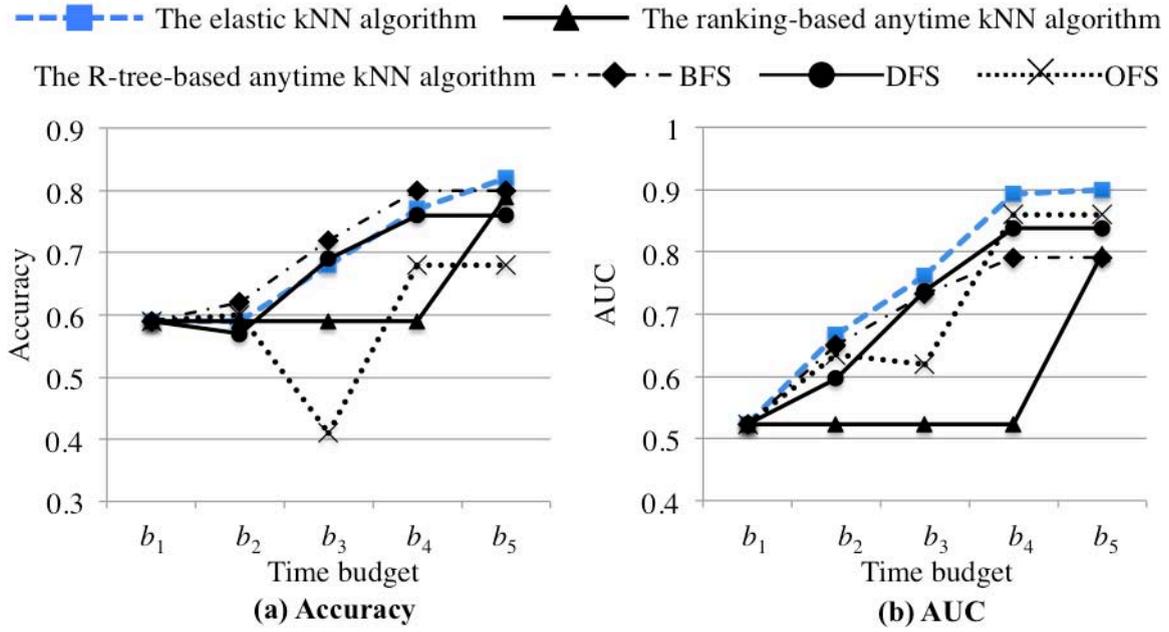

Figure 5.16: Comparison of the elastic and anytime kNN algorithms for the *ionosphere* dataset.

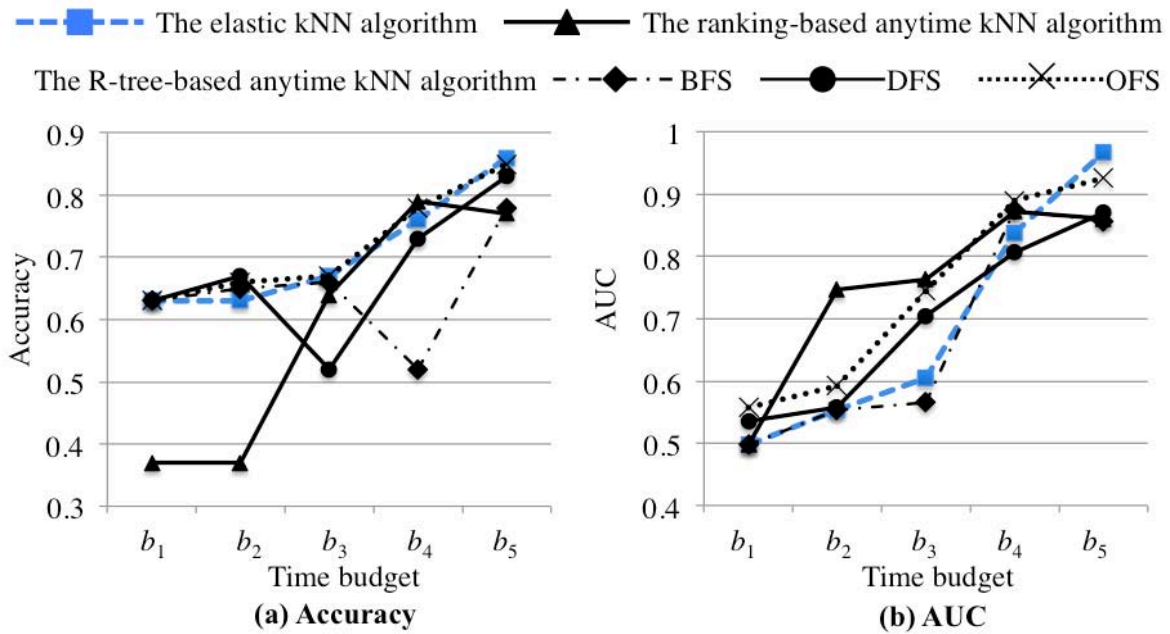

Figure 5.17: Comparison of the elastic and anytime kNN algorithms for the *gas* dataset.

**Discussion of the refinement strategies of different kNN algorithms**. Anytime algorithms need to return a useful result whenever they are interrupted. Thus, the *ranking-based anytime kNN algorithm* is designed to produce useful results at early stages: the initial result can be produced after the $k$ training points are tested. If extra time budget is available, the anytime algorithm tries to refine the existing result by taking more training points and updating a test point $q$'s nearest neighbours using closer training points.

However, in the kNN classification model, these closer nearest neighbours sometimes improve prediction accuracy, but sometimes worsen it. Similarly, at each anytime iteration, the R-tree-based anytime algorithm replaces one non-leaf R-tree node with its child nodes. This replacement can be viewed as refining some of the training points, because the child nodes represent the training points at a finer level of granularity. Such partial refinement of the training set causes great variance in result quality so the monotonicity of quality cannot be guaranteed. In contrast, the elastic algorithm employs a code—namely, the nodes at one depth of the R-trees—to represent the aggregate statistical information from all training points. Each new result is produced using nodes at a deeper depth of the R-trees and these nodes represent a complete refinement of the entire training set at a finer level of granularity, thus achieving steady quality improvement.

***Results of Section 5.4.4***. *The comparison results show that the elastic kNN algorithm is a better technique for guaranteeing quality monotonicity. Furthermore, the results indicate that when allocated the same budget, the elastic kNN algorithm outperforms the existing anytime kNN algorithms by producing later approximate results with better qualities.*

## 5.5    Applying Elastic Algorithms Under Different Cloud Pricing Schemes

Cloud computing offers a cost-effective approach for analysing large-scale scientific datasets. It provides on-demand metered access to `compute` resources under different pricing models. These include *fixed* pricing schemes that guarantee access to resources at an agreed price and *dynamic* pricing schemes in which resource prices change dynamically. In this section, we demonstrate how to apply the elastic mining algorithm under different cloud pricing schemes. The key idea of an elastic algorithm is to structure the computation appropriately, thus offering application owners a list of optional approximate results and guaranteeing the production of results with better quality if more computations are used. Thus, application owners can flexibly utilise the cloud pricing scheme to obtain a result meeting their minimum quality requirement while incurring small overheads. Subsequently, application owners can make trade-offs between computational cost, result delivery time and, more importantly, the quality of the analysis result. We illustrate this by employing the elastic kNN algorithm, and demonstrate how to apply this algorithm to control application owners' investment and quality of results in kNN data classification tasks under different cloud pricing schemes and user requirements.

## 5.5.1    Context and Motivation

In a cloud environment, the consumption of computational resources is effectively transformed into an application owner's investment from their own budget. For many scientific applications, using the pay-as-you-go paradigm means application owners should be able to selectively invest more of their budget in improving some analysis results but not others. In addition, by taking into account available pricing schemes, application owners should be able to make trade-offs easily between computational cost and result delivery time.

Taking the kNN classification algorithm as a motivating example, we discuss the problem of massive data analysis in a cloud environment. Suppose an application owner has 200,000 data points to be classified using cloud resources under a fixed pricing scheme for resource usage: 0.5 dollars per hour for access to a VM. Based on the available budget, the owner may decide to use a less computationally intensive algorithm (e.g. 3 hours of computation costing 1.5 dollars) to obtain classification results with low but acceptable quality (e.g. predictive accuracy of 70%). Alternatively, they may decide to invest more budget and therefore computation in a more expensive algorithm (e.g. 40 hours costing 20 dollars) to obtain a higher quality result (e.g. predictive accuracy of 90%). With many algorithms, the algorithm itself may also offer a wide range of options regarding investment vs. quality of result. In order to make an informed choice between the different options, the application owner needs to be able to reason, a priori, about how the quality of result varies with the investment.

This problem is compounded when considering different cloud pricing schemes. Most cloud IaaS providers offer a **fixed pricing scheme** that guarantees access to resources at a pre-specified price. Providers such as AWS also offer a **spot pricing scheme** which allows application owners to use the provider's available excess resources at a significantly lower price, usually 30% of the declared fixed price. This spot price is dynamically updated according to the current supply of excess resources and the demand for resources from cloud consumers (application owners). An application owner can bid for a spot

VM and get this instance once their bid price is *greater than*, or *equal to*, the offer spot price. However, under this scheme, the IaaS providers reserve the right to instantly terminate the application owner's access to resources without notice once the offer price is greater than the bid price. To make effective use of the spot price model, the application owner's algorithm needs to be able to save its intermediate results at regular checkpoints and also to be able to resume its computation from pause points.

## 5.5.2   Applying the Elastic kNN Algorithm in Analysing Massive Datasets

We investigate the problem of executing an elastic kNN data classification task in the cloud under both *fixed* and *spot* pricing schemes, and demonstrate how application owners can reason about the relationship between the quality of results vs. investment used in executing their classification algorithm to answer various questions, including the ones below.

- **Fixed pricing scheme**

  Application owners need to be able to address practical questions such as: "Given a certain budget, what kind of result can we obtain from my data classification algorithm, and does this result satisfy the minimal quality requirement?" They should also be able to address questions such as: "If more budget is available and we want to improve the quality of the existing result, what is the best quality we can achieve within the budget?"

- **Spot pricing scheme**

  Although application owners can make a low bid to use cheap resources under a spot pricing scheme, they run the risk of failing to meet their task completion deadline. This is because the low bid may result in many interruptions in algorithm execution and thus seriously delay the computation process. Thus, under a spot pricing scheme, it is more desirable for application owners to address questions such as: "Given a deadline constraint, what is the minimal investment that will meet the quality requirement within the deadline?" or "What is the best quality we can achieve while meeting both the budget and deadline constraints?"

- **Both fixed and spot pricing schemes**

  Even when application owners have a large budget, they still need to control how they spend this budget by assessing the benefit of paying for more computation. Thus, the question application owners want to address is: "Is the quality improvement we will achieve by increasing our investment worth that extra investment?" or alternatively, under a spot pricing scheme, "If we know that we will only gain an X% improvement on the existing result, how much should my bid price for spot resources be?"

**Experimental environment.**

We test a publicly available large-scale dataset, called cod-rna[6], collected from bio-medical applications to evaluate the behavior of the elastic kNN algorithm. The training set has 59,535 8-dimensional points used to build the R-trees. The test dataset has 200,000 test points and we produce four approximate results for illustration purposes. Our experiments utilise a VM with four 2.40 GHz CPU cores and 4G memory, and running the Ubuntu Linux operating system in the IC Cloud workstation. The fixed price for the VM is 0.5 dollars/hour. We also simulate spot price fluctuations over the course of 24 hours between 0.10 and 0.30 dollars/hour.

In order to predict the behavior of the algorithm at runtime—that is, the AUC quality of approximate results and the execution time—we conduct profiling runs to predict the expected quality improvement and the proportion of execution time required for each result. We use a subset consisting of 5% of the test set (10,000 test points) which only costs 0.3 dollars (0.6 hours) to classify. We note that classifying the entire test set costs 20 dollars (40 hours). Figure 5.18(a) indicates that the discrepancies between the predicted qualities based on the profiling runs and the actual qualities of results produced using the whole test set are less than 1%. Figure 5.18(b) shows that our profiling also provides accurate predictions of execution times.

---

[4] http://www.csie.ntu.edu.tw/~cjlin/libsvmtools/datasets/binary.html#cod-rna

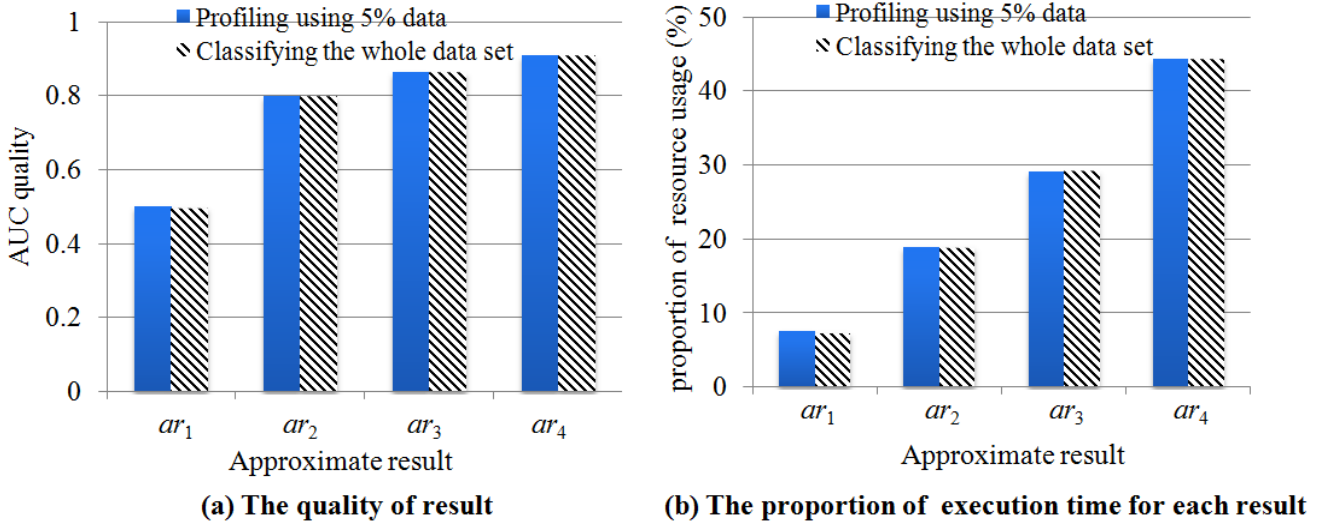

**(a) The quality of result**        **(b) The proportion of execution time for each result**

Figure 5.18: Comparison of profiling and actual analysis results for different approximate results.

**Controlling elastic data analysis under the fixed pricing scheme.**

Under the fixed pricing scheme, an application owner gets guaranteed access to VMs by paying a fixed resource price and can run analysis tasks without interruption. Supposing that $ET$ is the execution time and $PR_{FIX}$ is the fixed price, we can calculate the application owner's investment:

$$I_{FIX} = ET \times PR_{FIX}$$

Note that in the following scenarios, the four approximate results $ar_1$ to $ar_4$ are produced sequentially: result $ar_{i+1}$ is produced by starting from the state $s_{ar_i}$ of result $ar_i$ ( $1 \leq i \leq 3$ ). In addition, the *investments* listed in experiment results are the *cumulative investments*. For example, the *cumulative investment* of result $ar_2$ is 5.3 dollars, meaning that the application owner must spend 5.3 dollars to produce results $ar_1$ and $ar_2$.

**Scenario 1**. *Under the fixed pricing scheme, given a required quality $Q_{REQ}=0.8$ and a budget $B=20$ dollars (i.e. the maximum investment the application owner can make), can the application owner obtain a result whose quality is not worse than 0.8?*

As shown in Figures 5.19(a) and (b), the application owner can spend 5.3 dollars to obtain results $ar_1$ and $ar_2$, and the quality of the latter result is 0.8. Thus, the application owner can satisfy the required quality within their budget.

**Scenario 2**. *If the application owner meets the required quality $Q_{REQ}$=0.8 and still has a remaining budget of $B_{REM}$=14.7 dollars, what is the best quality the application owner can achieve within this budget?*

The elastic kNN algorithm enables application owners to continuously refine the quality of results by making more investments. Figures 5.19(a) and (b) indicate that by spending the remaining 14.7 dollars, the application owner can obtain results $ar_3$ and $ar_4$, thus achieving a result with quality 0.91.

**Scenario 3**. *Suppose that the application owner has a budget of $B$=20 dollars, which is enough to obtain all the four approximate results. The application owner wishes to spend their money in a cost-efficient manner by requiring an investment elasticity of $E_I^Q \geq 10\%$. What is the best quality they can achieve in this case?*

According to Equation (4.1) in Definition 4.2, the requirement of 10% investment elasticity means that if the application owner increases the current cumulative investment by 100% to obtain the next result, the quality of this result should be 10% greater than that of the previous result. Figure 5.19(c) lists the three investment elasticities between the three pairs of results. We can see that the application owner can only obtain the first two results $ar_1$ and $ar_2$ within the required investment elasticity; thus, the maximum quality they can achieve is 0.8.

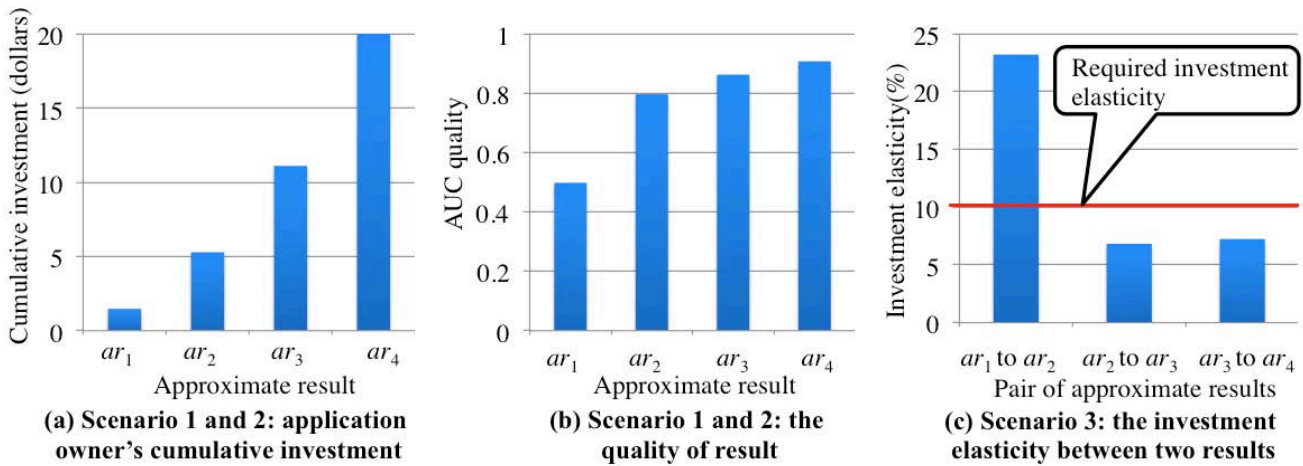

(a) Scenario 1 and 2: application owner's cumulative investment

(b) Scenario 1 and 2: the quality of result

(c) Scenario 3: the investment elasticity between two results

Figure 5.19: The experiment results under the fixed pricing scheme.

**Controlling elastic data analysis under the spot pricing scheme.**

Mainstream cloud infrastructure providers such as AWS usually provide line charts showing their spot price history over the past several days, weeks, or months, thus assisting application owners in making bids to use spot VMs. Table 5.9 lists an example distribution of spot prices over a 24-hour period. We can see that the spot price is highest (0.3 dollars/hour) at 12:00 noon when the greatest number of application owners are requesting resources, and it is lowest (0.1 dollars/hour) at 00:00 midnight when the greatest amount of excess resources is available. Using this historical information, once an application owner makes a bid, they can estimate when their algorithm can execute (i.e. the periods when their bid is *greater than* or *equal to* the spot price) and when their algorithm will be suspended (i.e. the periods when their bid is less than the spot price). For example, if an application owner makes a bid of 0.16 dollars/hour, they can estimate that, from Table 5.9's history information, they will be able to use the spot VM for 10 hours each day: from 00:00 to 04:00 and from 19:00 to 23:00.

Table 5.9. An example distribution of spot prices over a 24-hour period

| Hour in the day | Spot price (dollars/hour) | Hour in the day | Spot price (dollars/hour) |
|---|---|---|---|
| 0 | 0.10 | 12 | 0.30 |
| 1 | 0.11 | 13 | 0.28 |
| 2 | 0.12 | 14 | 0.26 |
| 3 | 0.14 | 15 | 0.24 |
| 4 | 0.16 | 16 | 0.22 |
| 5 | 0.18 | 17 | 0.20 |
| 6 | 0.20 | 18 | 0.18 |
| 7 | 0.22 | 19 | 0.16 |
| 8 | 0.24 | 20 | 0.14 |
| 9 | 0.26 | 21 | 0.12 |
| 10 | 0.28 | 22 | 0.11 |
| 11 | 0.30 | 23 | 0.10 |

Under the spot pricing scheme, an application owner gains access to spot instances and can run their elastic algorithm when their bid exceeds or equals the current spot price; otherwise, their elastic algorithm is suspended. During the running of the algorithm, the application owner must pay their *bid price* for resource usage. Note that each time an elastic algorithm is suspended and resumed, it incurs overheads. These overheads are usually much cheaper than the data analysis cost. For example, one suspension and resumption of the elastic kNN algorithm costs less than 0.006 dollars, while the total analysis cost is 40 dollars. Thus, we do not include these overheads when calculating an application owner's investment.

In practice, application owners usually have a deadline by which they must complete the analysis task. Without loss of generality, we assume that an analysis task's deadline $T_{DL}$ is longer than the task's execution time: an example of this would be a task which takes 40 hours to complete, and the deadline for which is in 48 hours (2 days). Supposing that $ET$ is the execution time (EA's running time), $ST$ is the suspended time, and $PR_{BID}$ is the application owner's bid, we can calculate the application owner's investment:

$$I_{BID} = ET \times PR_{BID} \text{ where the total elapsed time } ET + ST \leq T_{DL}.$$

**Scenario 4**. *Under the spot pricing scheme, given a budget B=20 dollars and a deadline constraint $T_{DL}$= 2 days, what is the minimum investment the application owner can make while meeting the required quality $Q_{REQ}$=0.8?*

Figures 5.20(a) and (b) show that the application owner needs 10.6 hours to obtain result $ar_2$ with the required quality 0.8. This means the application owner must execute the algorithm for 5.3 hours per day to meet the 2-day deadline. Thus, the minimum bid the application owner can make is 0.12 dollars/hour according to the distribution of spot prices in Table 5.9. This bid guarantees an execution time of 6 hours per day, resulting in a total execution time of 12 hours for the elastic algorithm. Using this bid, the application owner spends only 1.44 dollars to achieve the required quality, whereas the required investment is 5.3 dollars under the fixed pricing scheme in Scenario 1.

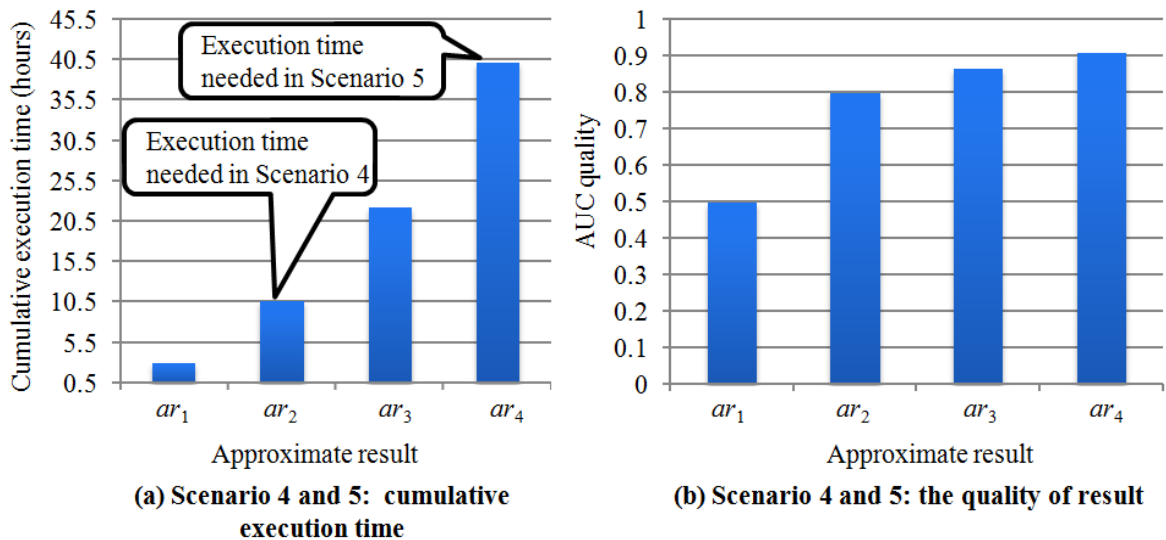

**(a) Scenario 4 and 5:  cumulative execution time**

**(b) Scenario 4 and 5: the quality of result**

Figure 5.20: The experiment results under the spot pricing scheme.

**Scenario 5**. *If the required quality $Q_{REQ}$ is met within the budget B=20 dollars and the application owner*

*still has budget remaining, can the application owner obtain the result $ar_4$ with quality 0.91 while meeting the deadline constraint of $T_{DL}$=2 days under the spot pricing scheme*?

Figures 5.20(a) and (b) show that the application owner needs approximately 40 hours to obtain the result $ar_4$ with quality 0.91. To meet the two-day deadline, the application owner should run the algorithm for 20 hours per day. Thus, the minimum bid they can make is 0.26 dollars/hour. The application owner must spend 10.4 dollars in making this bid, an investment that is just 52% of their budget (20 dollars). In contrast, the application owner spends their entire budget of 20 dollars to obtain the result $ar_4$ under the fixed pricing scheme in scenario 2 (where the fixed price is 0.5 dollars/hour).

**Controlling elastic data analysis under both the fixed and spot pricing schemes.**

**Scenario 6**. *Using both the fixed and spot pricing schemes, and given a required investment elasticity $E_I^Q \geq 10\%$, what is the best quality $Q_{max}$ an application owner can achieve within the budget $B$=20 dollars, and can the application owner still meet the deadline constraint of $T_{DL}$=2 days?*

Figure 5.21(a) shows that under the fixed pricing scheme, the application owner can only obtain the first two results if the required investment elasticity is 10%. Hence, the best quality they can achieve under this constraint is 0.8. In contrast, under the spot pricing scheme, the application owner can obtain all the four results and thus obtain a result with quality 0.91. This is because the application owner can make lower bids when producing results $ar_3$ and $ar_4$, thus making smaller investments, and so achieving higher investment elasticities and hence meeting the investment elasticity requirement.

Figure 5.21(b) indicates that the percentage quality improvement %$\Delta Q$ varies between different pairs of approximate results. According to the definition of investment elasticity ($E_I^Q = \frac{\%\Delta Q}{\%\Delta I}$), application owners need to make different bids to achieve different investment increases %$\Delta I$ in order to meet the required investment elasticity $E_I^Q$. For example, the percentage quality improvement from result $ar_2$ to result $ar_3$ is %$\Delta Q$ =7.5%. Thus, the percentage investment increase should be %$\Delta I$ =75% under the required investment elasticity $E_I^Q$=10%. Given the formula %$\Delta I = (I_3 - I_2)/I_2$ and given that the cumulative investment required to produce result $ar_2$ is $I_2$=3.18 dollars, we have result $ar_3$'s investment $\Delta I_3$=($I_3$ − $I_2$)=2.38 dollars. The execution time $ET_3$ for result $ar_3$ is 11.64 hours; thus, the application owner's bid for result $ar_3$ is calculated to be $PR_{BID} = \Delta I_3/ET_3$=0.2 dollars/hour. From Table 5.9, we know that a bid of 0.20 dollars/hour allows the algorithm to run for 14 hours per day, so the algorithm can complete

before the deadline.

Figures 5.21(c) and (d) show the application owner's bids and cumulative investments for each approximate result under both pricing schemes. The spot pricing scheme offers the application owner cheaper resource usage and allows them to make smaller investments compared to the fixed pricing scheme. However, as shown in Figure 5.21(e), the spot pricing scheme also requires the application owner to tolerate a longer elapsed time in completing their algorithm execution. This means that if the application owner requires an investment elasticity of 10% and wants to obtain a result with quality 0.91, it will take them 81.49 hours to obtain result $ar_4$, and they therefore cannot meet the two-day deadline.

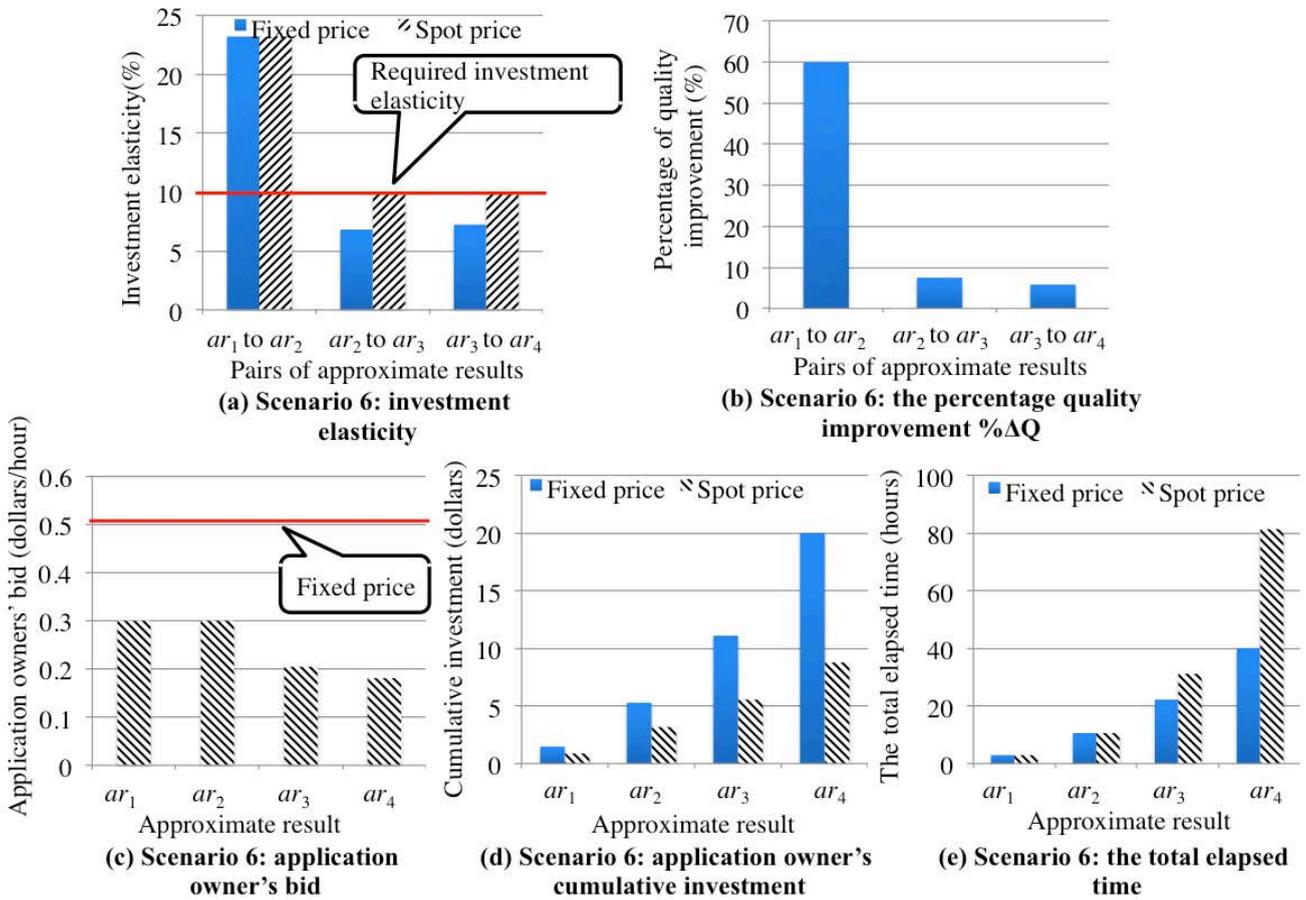

Figure 5.21: The experiment results under both fixed and spot pricing schemes.

Finally, we summarise the results of the above six scenarios. Generally, the elastic kNN algorithm allows application owners to make trade-offs between investment and quality of results in large-scale data analysis. Under the fixed pricing scheme, scenarios 1, 2, and 3 show that if an application owner makes more investments, they can expect to obtain better quality results. Under the spot pricing scheme, scenarios 4, 5, and 6 show that an application owner can make further trade-offs between the total elapsed time of the elastic kNN algorithm and their investment. In other words, an application owner can spend less money to get the same quality result if they can sacrifice more elapsed time.

## 5.6 Discussion of the Applicability of the Framework to Data Mining Algorithms

In this section, we discuss two requirements in applying the proposed elastic algorithm framework to develop an elastic version of a data mining algorithm. Each requirement is illustrated by an example data mining algorithm.

### 5.6.1 The Requirement of the Input Data

The first **requirement** is that the whole input data of a data mining algorithm is available and the coding of this data, either at an instance space or a feature space, can help *reduce* the computational complexity of this algorithm. We illustrate this requirement using the Approximate Bayesian computation (ABC) Rejection algorithm [122]. We first introduce some background information of this algorithm including the Markov Chain (MC) method and the ABC MC method. The ABC Rejection algorithm is then explained using an animal migration process as an example. Finally, we discuss the applicability of the framework to this algorithm.

**MC method**

MC methods [122] represent a class of algorithms that reply on iteratively sampling from a give distribution $f(x)$ to obtain a target numerical result, which cannot be calculated analytically or the calculation is very computationally expensive. Formally, given a dataset $D$ with known distribution $f(x)$ where $x \in D$. Suppose $h(x)$ is the target function and its expectation $E(h(x))$ on dataset $D$ is the value that people are interested to know:

$$E(h(x)|D) = \int h(x)f(x|D)dx = \frac{\int h(x)f(D|x)p(x)dx}{\int f(D|x)p(x)dx},$$

where $p(x)$ is the probability of $x$. For example, $E(h(x))$ can be the expected benefit for a stock between 2012 March to April or the expected reaction time of a Chemistry experiment.

However, in many cases, the calculation of $E(h(x))$ is intractable due to two reasons. First, the integral of function $f(x)$ or $h(x)$ cannot be evaluated analytically. Secondly, the calculation of expectation is infeasible due to the high dimensionality or large size of data set $D$. To this end, MC method [122] is proposed to approximate the posterior expectations of $h(x)$ using simulated samples:

$$E(h(x)|D, f(x)) \approx \frac{1}{N} \sum_{i=1}^{N} h(x_i),$$

where $N$ is the number of simulated samples drown from distribution $f(x)$.

**ABC MC method**

In standard MC methods, the acceptance or rejection of candidate samples are decided by the likelihood function that is calculated based $f(x)$. However, the distribution $f(x)$ of dataset $D$ is not known in many cases. For such an issue, ABC MC methods [145] are proposed to accept or reject samples according to the result of comparing the distance between a simulated dataset $D'$ and the observed dataset $D$ and a required threshold $\varepsilon$. Specifically, given a candidate sample (that is, a judgment is needed to decide whether this sample should be accepted or not), the simulated dataset $D'$ is generated by some simulation tools [146] by taking this candidate sample as input. If the distance between datasets $D'$ and $D$ is smaller than the threshold $\varepsilon$, the sample is accepted; otherwise it is rejected.

**The ABC Rejection algorithm**

The ABC Rejection algorithm is a representative ABC MC method, which has been applied in areas including computer graphics and vision; speech and audio processing; and decision theory. The pseudocode of this algorithm is presented below. At each iteration, the ABC Rejection algorithm draws a candidate sample from the proposed (prior) distribution $g(x)$ (line 4) and generates a simulated dataset $D'$ using this sample (line 5). The sample is accepted if the distance between $D'$ and the observed dataset $D$ is smaller than the threshold $\varepsilon$ (line 6). The algorithm keeps running until $n$ samples are accepted. These samples can be used to approximate the actual (posterior) distribution $f(x)$. In practice, a good approximation of distribution $f(x)$ requires a small value of threshold $\varepsilon$, which indicates the accepted samples are very close to the actual data. However, a small threshold might incur a large proportion of candidate samples being rejected. This means the ABC Rejection algorithm needs a large number of iterations (e.g. millions of iterations) to accept the required number of samples; that is, the algorithm is very time consuming.

**The ABC Rejection Algorithm**

**Input**: a observed dataset $D$, a proposed (prior) distribution $g(x)$, a threshold $\varepsilon$, and a number $n$ of required samples

**Output**: the accepted $n$ samples.

1. **Begin**

2. Set $i$=0;  // $i$ represents the number of accepted samples

3. while ($i < n$)

4.     $x \sim g(x)$;  //draw a candidate sample $x$ from distribution $g(x)$

5.     Generate the simulated dataset $D^{'}$ using the simulation tool by inputting sample $x$;

6.     **if** $d(D, D^{'}) < \varepsilon$, **then**

7.         Accept sample $x$;

8.         Set $i = i + 1$.

9. **End**

**Illustration of the ABC Rejection algorithm using an animal migration process**

We demonstrate the algorithm using an example of animal migration. This migration simulates the arrival process of Microtus to Europe after the ice age. The feature that people are interested to know in the migration process is the *population size* of the Microtus. There are three target parameters in the population: N_ANCESTRAL (the population size in the past), T_SHRINK (the change of population size generations ago), and T_NOW (the current population size). Since we only have some statistics of the Microtus such as the mean number of alleles in the observed dataset $D$, the distribution of these parameters are unknown so that their expectations cannot be calculated. Hence, we use the ABC Rejection algorithm to collect sufficient samples to approximate the probability distribution of these parameters so as to estimate their expectations.

We use parameter "T_NOW" as an example to illustrate the algorithm. The inputs of the algorithm are a observed dataset $D$ (listed in Table 5.10), a prior distribution $g(x)$ (uniform distribution between 2 and 4), a threshold $\varepsilon$=0.5, and a required number of sample $n$=1,000. Suppose a sample N_NOW=2.20038 is drawn from the distribution $g(x)$. The simulated dataset $D^{'}$ (listed in Table 5.11) is generated using the simulation tools in the ABCtoolbox [146]. We can calculate the distance between datasets $D$ and $D^{'}$ according to the Euclidean distance function:

$$d(D, D^{'}) = \sqrt{\sum_{i=1}^{8} (Statistics_i - Statistics_i^{'})} = 0.5685$$

Since $d(D, D^{'}) > \varepsilon$, the sample is rejected.

We run the ABC Rejection algorithm to obtain 1,000 accepted samples in a VM with two 2.40 GHz CPU cores and 4 GB memory. The whole process takes 2,876,583 iterations and the execution time is 192,416 seconds$\approx$53.45 hours.

Table 5.10. The observed dataset $D$

| $Statistics_1$ | $Statistics_2$ | $Statistics_3$ | $Statistics_4$ | $Statistics_5$ | $Statistics_6$ | $Statistics_7$ | $Statistics_8$ |
|---|---|---|---|---|---|---|---|
| 1.2 | 0.4103 | 0.05592 | 0.142394 | 0.91667 | 1.25 | 0.5 | 0.0714286 |

Table 5.11. The simulated dataset $D^{'}$

| $Statistics_1^{'}$ | $Statistics_2^{'}$ | $Statistics_3^{'}$ | $Statistics_4^{'}$ | $Statistics_5^{'}$ | $Statistics_6^{'}$ | $Statistics_7^{'}$ | $Statistics_8^{'}$ |
|---|---|---|---|---|---|---|---|
| 1.2 | 0.410391 | 0.0586939 | 0.135303 | 1 | 1 | 0 | 0.0468085 |

**Discussion of the applicability of the framework to the ABC algorithm**

As illustrated in the above example, the ABC Rejection algorithm is time-consuming and it usually needs millions of iterations to obtain the required number of samples. Hence, it is necessary to develop an elastic version of this algorithm. However, the input data (i.e. dataset $D$ in Table 5.10) only has one instance with eight features and the coding of dataset $D$ does not reduce the computational complexity of the algorithm. This is because the computational complexity, namely the number of iterations to complete the sampling process, mainly depends on another input parameter: the proposed distribution $g(x)$ used to draw samples. If distribution $g(x)$ and the actual distribution $f(x)$ are very similar, the drawn sample $x$ is close to the actual data and the distance $d(D, D^{'})$ between the simulated dataset $D^{'}$ and the observed dataset $D$ is small. This means the candidate sample $x$ has a large chance of being accepted; that is, less iterations are needed to obtain the required number of samples. Since distribution $g(x)$ cannot be coded either at an instance space or a feature space, the elastic algorithm framework is inapplicable to the ABC Rejection algorithm.

## 5.6.2 The Requirement of the Produced Exact Result

The second **requirement** is that fixing the input data, a data mining algorithm should output a deterministic exact result. We employ the SVM training algorithm [122] as a counter example to explain this requirement.

**SVM classifier**

SVM classifiers have a wide range of applications in practice. Initially proposed by Vapnik and Chervonenkis [147], SVM is a margin-based classifier that applies a discriminant function to construct hyper-planes in the feature space, thus classifying data points into different categories. As shown in Figure 5.22, a binary SVM classifier uses the discriminant function $f(x) = w^T x + b$ to map a 2-dimensional input pattern $x$ into the predicted class label $y \in \{c_P, c_N\}$: $y = c_P$ if $f(x) \geq 1$ and $y = c_N$ if $f(x) \leq -1$. Intuitively and empirically, the larger the margin (width) between the plus-plane ($f(x)=+1$) and the minus-plane ($f(x)=-1$), i.e. the larger the safe zone, the higher the prediction accuracy of the SVM classifier. Hence, given a set $D$ of training points, the SVM training is an optimal process that seeks the largest-margin classifier to separate the training points. The majority of computations in SVM training are spent in solving the quadratic programming problem in order to separate the support vectors from the remaining training points. Once the support vectors are obtained, they can be used to construct the discriminant function $f(x)$ used for classification.

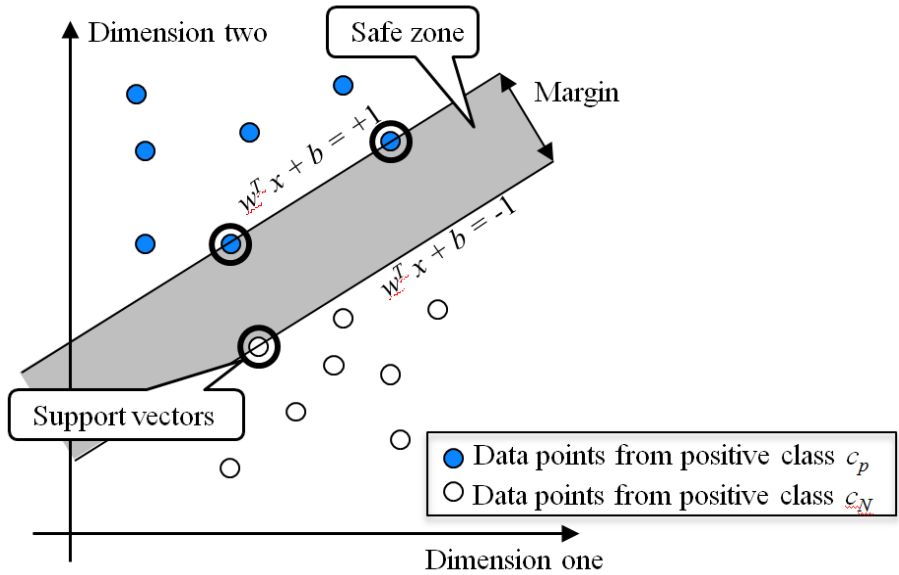

Figure 5.22: An example SVM classifier in the 2-dimensional space.

The SVM classification technique is later developed by adding kernel trick [148] and introducing soft margins to tolerate an error $\xi$ in its training [149]. The SVM classifier discussed here is the $C$-SVM [149], where $C$ is the cost parameter to control over-fitting by penalizing the error $\xi$. It is important to note that fixing a training set, conceptually there only exists one parameter combination $(C, \gamma)$ that can generate the best SVM classifier in the training. A practical method used in SVM training to find this optimal parameter combination is to perform grid-search on parameters $C$ and $\gamma$. The k-fold cross-validation is used in grid search to find the best SVM classifier with the highest cross-validation accuracy.

**The SVM training algorithm**

The pseudocode of the SVM training algorithm is presented below. The *input* of this algorithm includes three parts: (1) a training set $D$; (2) a kernel function such as linear, polynomial or Gaussian (Radial-Basis Function(RBF)) kernel. For example, the RBF kernel $K(\boldsymbol{x_i}, \boldsymbol{x_j})$=exp $(-\gamma \parallel \boldsymbol{x_i} - \boldsymbol{x_j} \parallel^2)$; (3) two sets of parameters $\{C_1, C_2, ..., C_m\}$ and $\{\gamma_1, \gamma_2, ..., \gamma_n\}$ to be searched. For example, $\{C_1, C_2, ..., C_m\}$=$\{2^{-5}, 2^{-3}, ..., 2^{13}\}$, $\{\gamma_1, \gamma_2, ..., \gamma_n\}$=$\{2^{-15}, 2^{-13}, ..., 2^3\}$, and a $10 \times 10$ grid search is performed for these parameters. The *output* of the algorithm is the SVM classifier with the highest cross-validation accuracy (i.e. the best SVM classifier $f^*(x)$). For example, $(C, \gamma)$=$(2^{13}, 2^{-10})$ is the optimal parameter combination that yields the best SVM classifier with the highest 10-fold cross-validation accuracy 88.7%.

**The SVM training Algorithm**

**Input**: a training set $D$, a kernel function, two sets of parameters $\{C_1, C_2, ..., C_m\}$ and $\{\gamma_1, \gamma_2, ..., \gamma_n\}$ to be searched.

**Output**: the best SVM classifier $f^*(x)$.

1.  **Begin**

2.  **for** $i$=1 to $m$

3.          Set $C=C_i$;

4.      **for** $j$=1 to $n$

5.              Set $\gamma=\gamma_j$;

6.          Construct the dual form of Lagrangian formulation according to the existing parameter combination $(C, \gamma)$;

7.           Solve the quadratic programming problem to derive all support vectors;

8.          Calculate $w$ and $b$ from the support vectors and construct the discriminant function $f(x)$;

            // $f(x)$ is the generated SVM classifier

9.          **if** ($f(x)$ is the first generated discriminant function), **then**

10.                 Maintain $f(x)$ as the best SVM classifier $f^*(x)$;

11.         **else**

12.                 Compare classifier $f(x)$ and the current best SVM classifier $f^*(x)$ by conducing the $k$-fold cross-validation. Maintain the classifier with a higher cross-validation accuracy.

13. **End**

**Illustration of the SVM training algorithm using a real dataset**

We demonstrate the algorithm by applying it in training a real dataset, call *ijcnn1*, obtained from the LIBSVM data repository [139] . This dataset has 99,701 data points with two classes (positive and negative) and each data point has 22 features. We randomly select 8,000 points to form the training set. The training is executed in a VM with eight 2.40 GHz CPU cores and 8 GB memory and its operating system is Linux Ubuntu.

The whole training process consists of four iterations, and different granularities of grid search on parameters $C$ and $\gamma$ are conducted. Specifically, at each iteration, a grid search is performed on five values of parameters $C$ and $\gamma$; that is, 25 different parameter combinations are tested. In grid search, the 10-fold cross-validation is used to select the optimal parameter combination $(C^*, \gamma^*)$ that generates the SVM classifier with the highest cross-validation accuracy. At iteration 1, the algorithm starts the training process with a coarse granularity of grid search: five values of parameter $C$ ($2^{-5}, 2^{-0.5}, 2^4, 2^{8.5}, 2^{13}$) and five values of parameter $\gamma$ ($2^{-15}, 2^{-10.5}, 2^{-6}, 2^{1.5}, 2^3$) are tested; that is, the search scope is $2^{-5}$ to $2^{13}$ for parameter $C$, $2^{-15}$ to $2^3$ for parameter $\gamma$; and the search granularity is $2^{4.5}$ for both parameters. In the following iterations, the search scope and granularity of both parameters are changed to a finer granularity; that is, 50% of the values of the previous iteration. The search scope is also set in the neighbourhood of the optimal parameter combination $(C^*, \gamma^*)$ of the previous iteration. The cross-validation accuracy of the SVM classifier at each iteration is listed in Table 5.12.

Table 5.12. Four iterations of SVM training for the *ijcnn1* dataset

| Iteration | 1 | 2 | 3 | 4 |
|---|---|---|---|---|
| 10-fold cross-validation accuracy | 97.53% | 97.80% | 97.80% | 97.93% |

**Discussion of the applicability of the framework to the SVM training algorithm**

Given a training set $D$, the SVM training algorithm cannot output a deterministic result, i.e. a SVM classifier with the highest k-fold cross-validation accuracy. This is because the SVM training algorithm applies a grid search to find the optimal parameter combination that generates the SVM classifier with the highest cross-validation accuracy. However, only limited number of parameter combinations can be searched, and the algorithm can always finds a SVM classifier with a higher cross-validation accuracy by refining the search granularity; that is, the global optimal combination of parameters $C$ and $\gamma$ cannot be found. For example, Table 5.12 shows that at each new iteration, when the algorithm searches parameters at a finer level of granularity, the generated SVM always has a higher cross-validation accuracy. Hence, the SVM training algorithm does not satisfy the second requirement and the framework is inapplicable to this algorithm.

In contrast, we note that many data mining algorithms can output a deterministic result if the input data is fixed. For example, in classification algorithms such as decision tree classifiers or Bayes classifiers, a deterministic discriminant function can be generated based on the training set. The elastic versions of these classification algorithms, therefore, can be developed using the elastic algorithm framework.

# Chapter 6

# Elastic Collaborative Filtering Recommendation in E-commerce Sites

## 6.1    Introduction

In this chapter, we first introduce motivations for offering elastic recommendation services in e-commerce sites by developing an elastic neighbourhood-based CF algorithm (Section 6.2). In Section 6.3, we review some basic concepts of neighbourhood-based CF and discuss related work. In Section 6.4, we introduce an elastic CF algorithm that applies R-tree data structures as the coding component and basic neighbourhood-based CF as the mining component. Finally, Section 6.5 introduces our experimental settings; Section 6.6 presents the evaluation and comparison results; and Section 6.7 present some discussion of the elastic CF algorithm.

## 6.2    Context and Motivation

### 6.2.1    Recommender Systems in E-commerce Sites

Today, e-commerce is a major industry in which the sale and purchase of products are conducted over the Internet. A key purpose of current e-commerce sites is to enable businesses to provide mass customisation of their products or services conveniently and cost-effectively. Thus, millions of products can be sold on e-commerce sites to meet a variety of customer needs.

In this context, recommender systems are widely applied in many e-commerce sites, helping customers to choose between millions of items for sale, such as books, songs, and movies [150]. By recording and analysing large amounts of information regarding the preferences of many users, recommender systems can be used effectively to personalise the online experiences of individual customers. Thus, recommender systems aim to enhance the mass customisation of e-commerce sites to benefit both product/service providers and customers [150]:

- Product/service providers (i.e. application owners): Recommender systems help these providers to sell their products or services more efficiently and enhance their revenues. By collecting data on and learning from customer behaviors, recommender systems can recommend products that are most likely to meet customers' needs, thus increasing sales.

- Customers (i.e. end users): Recommender systems provide customers with information that guides them towards particular items among millions of products or services, such as books and movies, on e-commerce sites, thus giving them useful options and satisfying their heterogeneous needs.

## 6.2.2   Neighbourhood-based CF Recommendation Techniques

Collaborative filtering is a traditional and popular data-driven recommendation technique that predicts an *active user*'s rating (preference score) for a *target item* based on existing ratings from similar users [17, 151]. The key idea behind CF-based recommendation is that the personal tastes of similar-minded users are correlated; thus, an active user is more likely to give a high rating to an item if users with similar profiles also preferred the item. In recommender systems, users' ratings of different items are collected and stored in a data structure called a *user-item rating matrix*. By profiling the ratings in the matrix, CF methods can predict an active user's rating for an item based on the existing ratings from similar users [151, 152]. For example, Table 6.1 shows an example user-item rating matrix with five users and four items, in which the rating for an item is either "Like" or "Dislike". In this example, the active user $u_5$'s rating for the *Movie* item is predicted using CF methods, and this prediction is built upon the existing ratings in the matrix. Users $u_2$ and $u_3$ have similar rating behaviours to user $u_5$ and their ratings for the *Movie* item are both "Dislike"; hence, the predicated rating is "Dislike".

Table 6.1. An example of predicting user $u_5$'s rating of the *Movie* item using CF methods

| User | Item | | | |
|---|---|---|---|---|
| | Picture | Book | Movie | Game |
| $u_1$ | Like | Dislike | Like | Like |
| $u_2$ | | Like | Dislike | Dislike |
| $u_3$ | Like | Like | Dislike | |
| $u_4$ | Dislike | | Like | |
| $\boldsymbol{u_5}$ | Like | Like | ? | Dislike |

CF techniques typically fall into two categories: memory-based CF methods that utilise the entire user-item rating matrix to generate a prediction; and model-based methods that first build a model using the rating data and then make a prediction using the model [17]. The neighbourhood-based CF method is a popular type of memory-based CF technique which has been used in many recommender systems and shows good prediction accuracy in practice [17]. Formally, given a user-item rating matrix with $m$ users and $n$ items, the entry $r_{u,i}$ denotes the user $u$'s rating of item $i$ ($1 \leq u \leq m$ and $1 \leq i \leq n$), and the neighbourhood-based CF algorithm scans the entire user-item matrix to make rating predictions [17]. This user-oriented method has two major phases: the neighbourhood selection phase and the recommendation phase. Given an *active user* $u$ and a *target item* $i$, this method predicts $u$'s rating of $i$. At the neighbourhood selection phase, this method calculates the *similarity* or *weight* $w(u, v)$ between user $u$ and any other user $v$ who has rated the same item $i$ in the matrix. One widely applied similarity measure in the CF research community for calculating this weight is Pearson's correlation coefficient,

$$w(u, v) = \frac{\sum_{j \in I}(r_{u,j} - \overline{r_u}) \times (r_{v,j} - \overline{r_v})}{\sqrt{\sum_{j \in I}(r_{u,j} - \overline{r_u})^2} \times \sqrt{\sum_{j \in I}(r_{v,j} - \overline{r_v})^2}}$$

(6.1)

where $I$ is the set of items that both users $u$ and $v$ have rated, $r_{u,j}$ denotes user $u$'s rating of item $j$, and $\overline{r_u}$ is the average rating of all items rated by user $u$.

After calculating the weights, at the second *recommendation* phase, the algorithm generates the prediction $p(u, i)$ of user $u$'s rating of item $i$ by taking a weighted average of all ratings of item $i$ from user $u$'s

neighbourhood users:

$$p(u, i) = \bar{r_u} + \frac{\sum_{v \in I} w(u,v) \times (r_{v,i} - \bar{r_v})}{\sum_{v \in I} |w(u,v)|} \tag{6.2}$$

where $I$ is the set of all users that have rated item $i$, and $|w(u, v)|$ denotes the absolute value of the weight $w(u, v)$.

### 6.2.3   Problem Analysis

At present, many approaches to implementing CF methods have been developed [17, 153], both in research projects (e.g. [154, 155]) and in commercial recommender systems (e.g. Amazon.com [156]). Neighbourhood-based CF is a prominent type of CF algorithm which shows good prediction accuracy in practice [17]. However, it also suffers from being computationally expensive since it has to process a large-scale user-rating matrix. Today's recommender systems deployed in large-scale e-commerce sites usually need to handle hundreds of thousands of requests per second, and each request needs to receive a satisfactory recommendation within a short real-time period. However, the exact predictions produced by a neighbourhood-based CF algorithm are based on processing the entire matrix and this cannot be performed under a tight time budget if the matrix is large. Hence, when on a limited time budget, a recommender system may settle for producing approximations of the exact result by reducing the amount of computation performed.

In this context, the key challenge is to develop effective time-adaptive CF approaches that can produce approximate predictions with high accuracies when on limited resource budgets. Moreover, such approaches need to guarantee the improvement of the prediction accuracy of results when given more time budget so as to avoid both wasting computational resources and result deterioration.

From the perspective of a recommender system, a high-quality CF prediction result needs to have two properties with regard to the characteristics of a user-item rating matrix:

- **High prediction accuracy under limited budgets**. The rating matrix is often very large and sparse. For example, the Netflix dataset [157] for movie recommendations is a 48,019×17,700 rating matrix in which over 98% of the rating values are missing. It is necessary to develop efficient CF algorithms that produce approximate results with high prediction accuracies even with large user-item matrices and on limited time budgets.

- **Low variation in prediction accuracies for different recommender system users**. The numbers of items rated by different users in the rating matrix are highly unbalanced. For example, the number of rated items per user ranges from 1 to 3,420 in the Netflix dataset [157]. Nonetheless, the recommender system is still expected to give each user fair treatment and offer the best possible recommendations for all, or most, users within the limited time budget. Since different users' test sets of items have different numerical scales of ratings and the prediction results may have different numerical scales of *prediction accuracies*, the *quality* of an approximate result in this case can be measured by the **relative error (RE)** [158], which has been widely used to compare approximations of values of different sizes. Let the *absolute error* be the discrepancy between the accuracies of *approximate* and *exact* predictions, where the *exact* result is produced by full computation over the entire rating matrix. The *RE* is the *absolute error* divided by the *exact* prediction accuracy. Hence, a lower value of *RE* denotes a higher prediction accuracy of the approximate result. Moreover, when considering the prediction accuracies of multiple approximate results for a range of users, a low variation in quality can be represented by a small mean and variance of the REs for these users. For example, suppose there is one approximate result for each user and 48,019 approximate results in the whole Netflix dataset. If the mean of these results' REs is $\mu$=0.05 and the variance $\sigma^2$ of these REs is 0.0001 (standard deviation $\sigma$=0.01), the *discrepancy* between the approximate and exact prediction accuracies is between 0.02 (i.e. $\mu$-3$\sigma$) and 0.08 (i.e. $\mu$+3$\sigma$) for the majority of users. If the values of REs follow the normal distribution, over 99% of RE values will lie in the interval [$\mu$-3$\sigma$, $\mu$+3$\sigma$]. Thus, most users of recommender systems can expect to get an approximate prediction that is very close to the exact prediction.

## 6.2.4   The Elastic Approach Proposed to Address the Problems

A number of CF algorithms that produce approximate results have been proposed in the literature [17, 155, 159-163]. These algorithms mainly operate by restricting the size of the input dataset fed to the algorithm; for example, by using sampling or clustering techniques. However, none of them guarantees improvements in prediction accuracies when allotted greater time budgets.

To address this issue, we propose an elastic neighbourhood-based CF algorithm designed to produce approximate predictions whose accuracy increases monotonically with the used computational time. As the entire user-item rating matrix potentially contributes to the prediction of an item, the algorithm employs an aggregation model that summarises the statistics of the user-item rating matrix in a hierarchical fashion. The model indexes similar users in the rating matrix while aggregating their rating information. Thus, this approach can adapt efficiently to variations in the available time budget while ensuring the production of high accuracy predictions using larger budgets.

## 6.3 Related Work to Handle Large-Scale Recommendation Problems

### 6.3.1 Computation Reduction Techniques in CF Methods

Since the basic neighbourhood-based CF method suffers from being computationally expensive when processing large-scale rating matrix, many dimensionality reduction techniques have been proposed to reduce the amount of computations by reducing the size of rating matrix [17]. They either remove insignificant or unrepresentative users or items, or map data into a space with a smaller dimensionality (e.g. those based on matrix factorization [164-166], PCA [155], or subgroup techniques [167, 168]). Although these techniques improve prediction accuracy or reduce computational complexity, their time of generating a prediction still increases with the size of the user-item matrix. Hence, they are not designed to generate a prediction adapting to varying time budgets.

### 6.3.2 Time-adaptive CF Algorithms

The work most related to this work includes approximate CF algorithms that work by allocating each active user to a group of users and use only the ratings for these users in the computation rather than using the entire rating matrix to produce predictions. By varying the size of such group the algorithms can control the amount of computation conducted to fit within a limited time budget. We call such algorithms

"time-adaptive" CF algorithms.

One simple approach for building such time-adaptive CF algorithms is to select a subset of users from the entire matrix using sampling techniques [17, 159, 160] and to adjust the sample size based on the available time budget. Other time-adaptive CF approaches include those based on clustering techniques, such as k-means clustering [159, 161, 162] and hierarchical clustering [163]. These approaches partition the whole set of users into multiple clusters and associate each active user with a group of like-minded users based on their rating preferences. The expensive computation of cluster construction is pre-processed offline and the online recommendation can then adapt to the available time budget by controlling the number of users in the cluster, i.e. the number of users used for generating predictions.

For example, k-means [169] has been applied as the clustering technique in existing time-adaptive CF algorithms [159, 161, 162]. This clustering method can allocate a user (a data point) to a cluster whose centroid (the mean of all the data points in this cluster) is the closest to the user. Formally, let $m$ be the number of users in the rating matrix and $k$ be the number of clusters ($k \leq m$). The k-means clustering method aims to assign $m$ users $\{u_1, ..., u_m\}$ into $k$ clusters $\{\omega_1, ..., \omega_k\}$ in order to minimise the within-cluster sum of squares (WCSS):

$$WCSS = \sum_{i=1}^{k} \sum_{u \in \omega_i} d(u, \bar{u}_i)$$

where $\bar{u}_i$ is the centroids in cluster $\omega_i$, and $d(u, \bar{u}_i)$ is the distance between user $u$ and centroid $\bar{u}_i$ (e.g. Euclidean distance $\| u - \bar{u}_i \|^2$). Hence using k-means, the number of users in a cluster can be controlled by adjusting the number $k$ of clusters to be generated.

Although these time-adaptive CF algorithms can adapt their computation to time budget constraint, they only use part of the rating information in prediction and ignore most of the data. For these approaches, given only a small number of users (i.e. a small portion of rating information) being used for prediction, their accuracy may become much lower than that of the exact prediction using the entire rating matrix. Furthermore, when the budget is increased, these algorithms apply only a simple data accumulation strategy by adding more users to refine their prediction results. Obviously, adding a batch of poorly correlated users may not necessarily improve the prediction accuracy and, in the worst case, can result in decreasing it drastically. In general, this means reasoning about, and guaranteeing, quality monotonicity of results with respect to resources used by these algorithms is difficult.

In addition, these time-adaptive CF algorithms usually requires long preprocessing times, especially when dealing with large rating matrices. This time-consuming preprocessing is a major stumbling block for the applicability of such algorithms to real-world recommender systems. For example, in the time-adaptive algorithms based on k-means clustering [159, 161, 162], the clustering construction is typically implemented as an iterative refinement process; that is, a number of iterations are used to reduce WCSS. At each iteration, the k-means method reassigns each user to the cluster whose centroid has the smallest distance to this user. The clustering process completes when no user changes the cluster between two consecutive iterations or a specified number of iterations are completed. Hence, using k-means to construct clusters, most of the time is spent in calculating the distances between users and cluster centroids. Suppose one operation of such calculation costs $O(1)$ (constant time). At each iteration, there are $m \times k$ calculations of distances to allocate $m$ uses into $k$ clusters. Hence, the time complexity of one iteration is $O(m \times k)$. Suppose the specified number of iteration is $i$, the time required to construct the clusters is $O(i \times m \times k)$. For example, given the user-item rating matrix in Netflix dataset [170] with 48,019 users, 17,770 items, and 9,566,400 ratings, suppose the clustering is executed in a VM with four 2.40 GHz CPU cores and 8 GB memory. It takes the k-means clustering method 87.75 hours to cluster 48,019 users into 8,000 clusters using 10 iterations.

## 6.4   The Elastic Neighbour-based CF Algorithm

### 6.4.1   Overview of the Elastic CF Algorithm

As opposed to traditional time-adaptive CF algorithms that simply use part of the available rating information in making predictions, we propose an elastic CF algorithm that groups similar users and stores their aggregated statistical rating information in a **code**. The proposed algorithm can generate a range of approximate results based on the aggregation of the entire rating matrix in a hierarchical fashion using the R-tree data structure [21].  In the construction of the R-tree, each R-tree node includes a set of original users from the rating matrix, and all R-tree nodes at each level of the model form a **code**. Thus, a code at a lower level of the model only contains a small number of aggregated users but still represents the statistical information from the entire rating matrix. Using this code, an accurate prediction can be produced using small space and time consumptions. As the tree is descended, codes at deeper levels consist of more R-tree nodes and represent the rating matrix at finer levels of granularity. Although this requires longer processing times, it results in more accurate predictions.

Following the elastic algorithm framework proposed in Section 5.2, we design an elastic CF algorithm with two stages, as shown in Figure 6.1. At the *preprocessing* stage, the algorithm applies an R-tree coding component to transform a rating matrix into a list of codes. The *length* of a code is its number of R-tree nodes. Note that this preprocessing stage is only applied once to produce a set of codes before delivering the elastic recommendation service. Next, once a time budget is given, a *length budget*, which denotes the maximal length of the code that can be processed by the neighbourhood-based CF component within the budget, can be estimated. A code is then selected that has the maximum length value smaller than the length budget.

At the *online recommendation* stage, the neighbourhood-based CF component takes the selected code, a target item from an active user, and a starting state as inputs, and outputs an *approximate prediction result*. Fixing the last two inputs, we assume that the running time of the CF component can be primarily determined by the length of the code, and this running time should be smaller than or equal to the given time budget. Hence, the bound of the elastic CF algorithm's running time can be controlled by tailoring the length of the input code. Note that in order to enable the incremental refinement of results if more processing budget is made available, the CF component also takes the starting state as an input. This starting state can be either the *initial state* or a *state* representing some obtained approximate prediction from a previous R-tree level. In practice, this state corresponds to the information retained in some data structure representing an acquired result.

Hence, the proposed CF approach is elastic with respect to a given time budget. Given a specific budget, the recommender system can decide which level of the model to start its computation from, determining the granularity of the computation as well as the accuracy of the resultant prediction. Users are guaranteed good prediction results even with limited resource consumption. If more resources are put towards the computation, this approach produces better results.

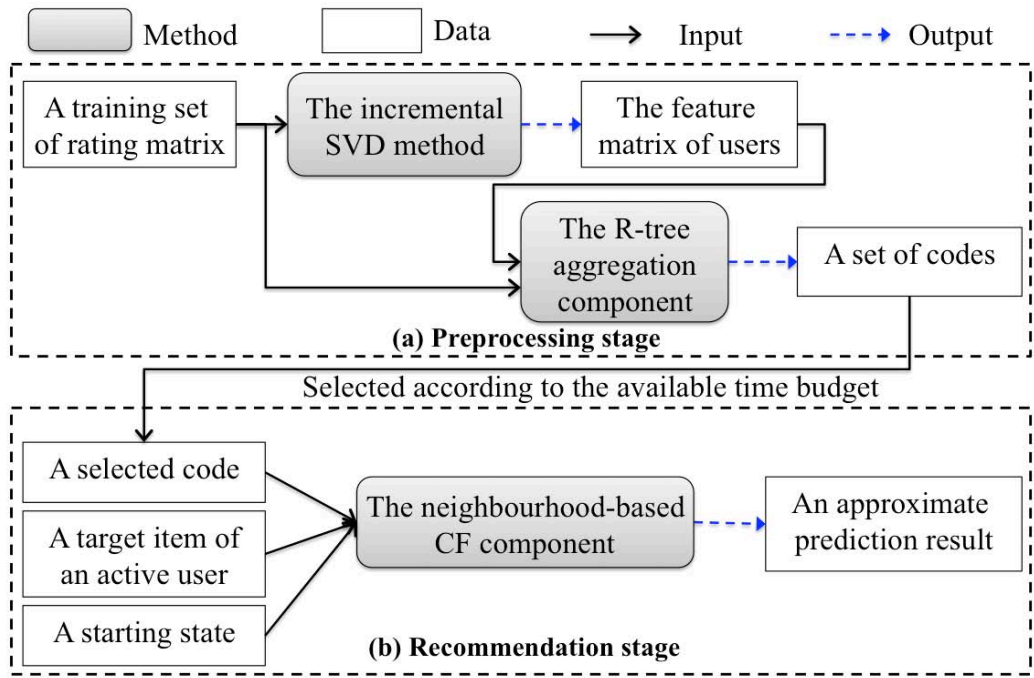

Figure 6.1: The two stages of the elastic CF algorithm.

Note that the proposed algorithm groups similar users in the rating matrix. This is because in most practical scenarios, the calculation of user-user weights is the most computationally complex task in the neighbourhood-based CF method, while the number of users is often greater than the number of items in the rating matrix—for example, in the Netflix dataset [157], there are 48,019 users and 17,700 items. Our approach is also applicable to item-based CF algorithms [17], which group similar items in the matrix and aggregate their ratings.

In the following two sections (6.4.2 and 6.4.3), we introduce the R-tree coding component. Since the R-tree index model works effectively in low-dimensional spaces, the coding process consists of two steps. First (Section 6.4.2), the incremental SVD method [171, 172] is employed to transform the sparse rating matrix into a dense feature matrix of users, in which each user is represented by a low-dimensional user vector (i.e. a row in the feature matrix). Next (Section 6.4.3), an R-tree is constructed by grouping similar user vectors in the *reduced* feature space while the aggregated ratings are calculated using the *original* rating matrix. We then explain how the neighbourhood-based CF method can use the aggregated user ratings to generate predictions in Section 6.4.4.

## 6.4.2   The Incremental SVD Method

SVD is a widely applied feature reduction technique in recommender systems, and is based on the matrix factorisation technique [165]. The SVD method decomposes the rating matrix into two feature matrices of users and items [173, 174]. Formally, given a rating matrix $R \in \mathbb{R}^{m \times n}$ with $m$ users and $n$ items, the SVD method generates three reduced matrices: a $m \times d$ feature matrix of users $U_d$, a $d \times d$ rectangular diagonal matrix $S_d$, and a $d \times n$ feature matrix of items $I_d^T$ (the conjugate transpose of $I_d$). In the dimensionality reduction process, the SVD method guarantees that the reconstructed rating matrix $R_d = U_d \times S_d \times I_d^T$ is the closet rank $d$ matrix to the original rating matrix $R$; that is, $R_d$ has the smallest Frobenius norm $\parallel R - R_d \parallel_F$ among all rank $d$ matrices.

The matrix factorisation process of SVD is traditionally time-consuming with time complexity $O(m^3)$ for a $m \times n$ rating matrix [173, 174]. We therefore use the incremental SVD method developed in [171, 172] to generate the reduced feature matrices $U_d$ and $I_d$. This method treats matrix factorisation as a gradient descent optimisation problem with the aim of minimising the objective function; that is, minimising the Frobenius norm $\parallel R - R_d \parallel_F$. The running time of the incremental SVD method is *independent* of the size of the rating matrix. Hence, the matrix factorisation can be completed quickly even when dealing with a large-scale matrix.

Given a target dimensionality $d$ and an $m \times n$ rating matrix ($d < n$), the *output* of the incremental SVD method is an $m \times d$ dense user feature matrix $U_d$, in which each user is represented by a $d$-dimensional vector—a row in matrix $U_d$. The incremental SVD method guarantees that two user vectors will have similar feature values if these users have similar rating preferences. For example, Table 6.2(a) shows a 12×5 user-item rating matrix. The incremental SVD method can decompose this rating matrix into a 12×2 feature matrix of users, as shown in Table 6.2(b). We can see that users $u_1$, $u_2$, and $u_3$ have assigned item $i_1$ similar rating values. Thus, these users have similar feature values in their two-dimensional user vectors. Note that the incremental SVD method requires presetting the number of features. In our elastic CF algorithm this is decided by application owners, but all our experimental results presented in the following sections are based on using three features.

Table 6.2. An example rating matrix and feature matrix

(a) A 12×5 rating matrix  (b) A 12×2 feature matrix

| User | Item | | | | | | User | Feature | |
|------|------|------|------|------|------|---|------|---------|------|
| | $i_1$ | $i_2$ | $i_3$ | $i_4$ | $i_5$ | | | $f_1$ | $f_2$ |
| $u_1$ | 5.00 | | | | | | $u_1$ | 1.47 | 2.60 |
| $u_2$ | 4.00 | | | | | | $u_2$ | 1.47 | 2.20 |
| $u_3$ | 5.00 | | 3.00 | | | | $u_3$ | 2.07 | 2.20 |
| $u_4$ | 3.00 | 2.00 | | | | | $u_4$ | 0.76 | 2.60 |
| $u_5$ | 2.00 | | 3.00 | | | | $u_5$ | 0.76 | 1.80 |
| $u_6$ | 3.00 | | | | | | $u_6$ | 0.85 | 1.80 |
| $u_7$ | | | 3.00 | | 2.00 | | $u_7$ | 0.88 | 0.70 |
| $u_8$ | | | 2.00 | | 3.00 | | $u_8$ | 0.88 | 0.20 |
| $u_9$ | | | | 3.00 | 3.00 | | $u_9$ | 1.45 | 0.20 |
| $u_{10}$ | | | 2.00 | | 1.00 | | $u_{10}$ | 0.15 | 1.18 |
| $u_{11}$ | | | | | 2.00 | | $u_{11}$ | 0.15 | 0.78 |
| $u_{12}$ | | | 2.00 | | 1.00 | | $u_{12}$ | 0.44 | 0.78 |

## 6.4.3   The R-tree Coding Component

The R-tree [18] is a bottom-up index model that assigns similar user vectors in the feature matrix to the same leaf node. Leaf and non-leaf nodes are then recursively grouped together following the same principle to preserve data similarity. In a leaf R-tree node, each entry refers to a $d$-dimensional user vector. In a non-leaf node $N$, each entry refers to one of $N$'s child nodes. For example, Figure 6.2(a) shows Table 6.2(b)'s 12 two-dimensional user vectors $\{u_1,\ldots, u_{12}\}$ and Figure 6.2(b) shows an R-tree with three depths constructed to index these vectors. Thus, three codes $c_0=\{N_7\}$, $c_1=\{N_5, N_6\}$, and $c_2=\{N_1, N_2, N_3, N_4\}$ for indexing nodes at depths 0, 1, and 2 of the R-tree respectively are generated.

In addition to a set of entries, an R-tree node $N$ also comprises an identifier $I_N$ used to point to a record that aggregates users' rating data. Let node $N$'s $MBR$ enclose a set $U$ of users. For some item $i$, suppose a subset $U_i \subseteq U$ of users have rated $i$. We say that node $N$ has rated item $i$ if $U_i \neq \phi$. Node $N$'s rating $r_{N,i}$ of $i$ and the average rating $\overline{r_{N,i}}$ depending on $i$ can be calculated as:

$$r_{N,i} = \sum_{u=1}^{|U_i|} r_{u,i} / |U_i| \tag{6.3}$$

and

$$\overline{r_{N,i}} = \sum_{u=1}^{|U_i|} \overline{r_u} / |U_i| \tag{6.4}$$

where user $u \in U_i$, $r_{u,i}$ is $u$'s rating of item $i$, $\overline{r_u}$ is $u$'s average rating over all items, and $|U_i|$ is the number of users in the set $U_i$. Note that node $N$ has a distinct average rating $\overline{r_{N,i}}$ for each item $i$ depending on the set $U_i$ of users in $U$ that have rated $i$.

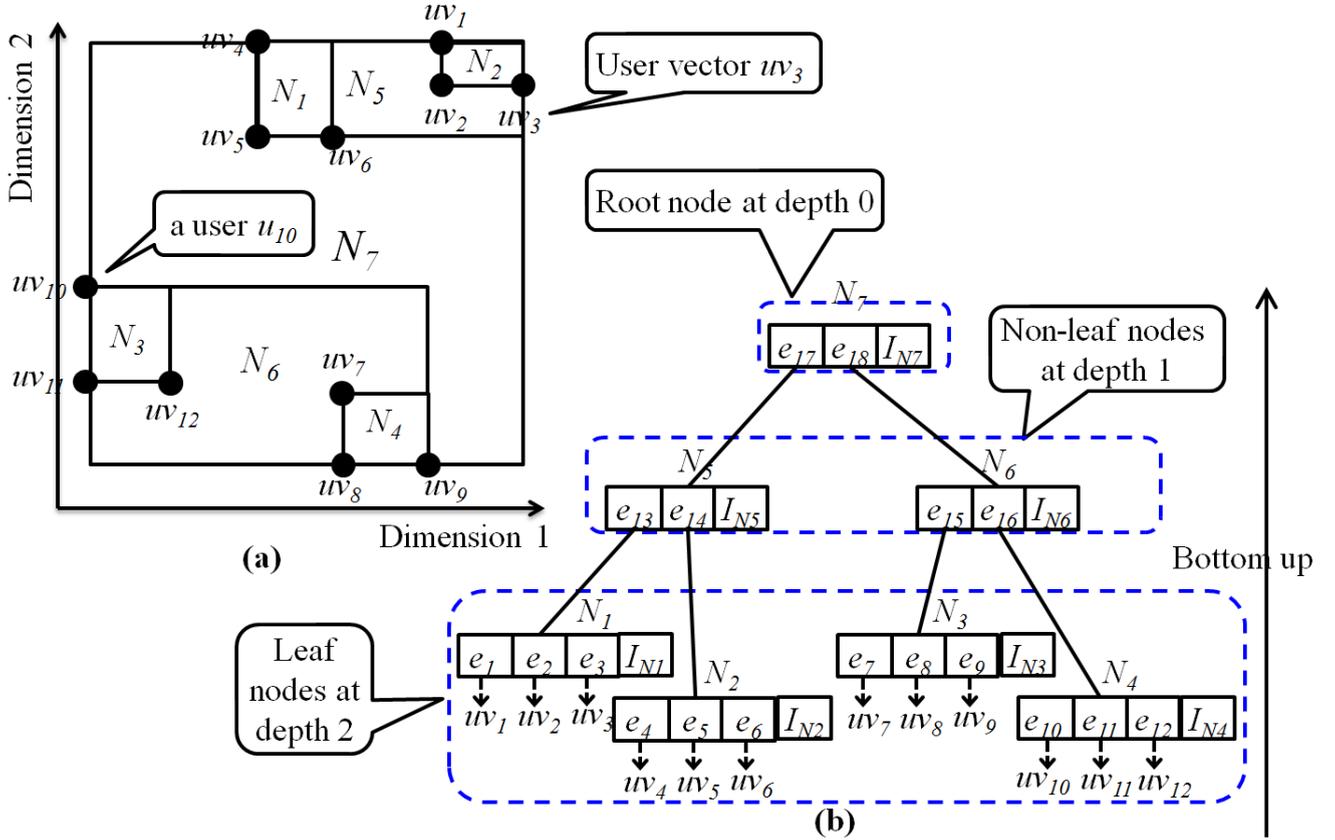

Figure 6.2: An example R-tree for indexing user vectors. We have (a) a set of 12 two-dimensional user vectors and (b) a constructed R-tree.

Using Table 6.2(a)'s rating matrix and Figure 6.2's R-tree as an example, Figure 6.3 shows three types of R-tree nodes and their records storing aggregated rating information. Consider the leaf node $N_1$ (Figure 6.3(a)): its $MBR$ encloses users $u_1$, $u_2$, and $u_3$. None of these users have rated item $i_2$; thus, $N_1$ also has no rating for $i_2$. Only user $u_3$ has rated item $i_3$; thus, $N_1$'s rating of $i_3$ is 3.00 (i.e. $u_3$'s rating of $i_3$) and its average rating depending on $i_3$ is 4.00 (i.e. $u_3$'s average rating) according to Equations (6.3) and (6.4). Similarly, all three users have rated item $i_1$; thus, $N_1$'s rating of item $i_1$ is 4.67 and its average rating depending on $i_1$ is 4.33. Figure 6.3(b) shows the non-leaf node $N_5$'s rating and average rating for each item, which are calculated using the rating information from the six users $u_1$ to $u_6$ enclosed by $N_5$'s $MBR$. Finally, Figure 6.3(c) shows the root node $N_7$, whose aggregated ratings are calculated using all 12 users' rating information from the entire rating matrix, and which has ratings for all five items.

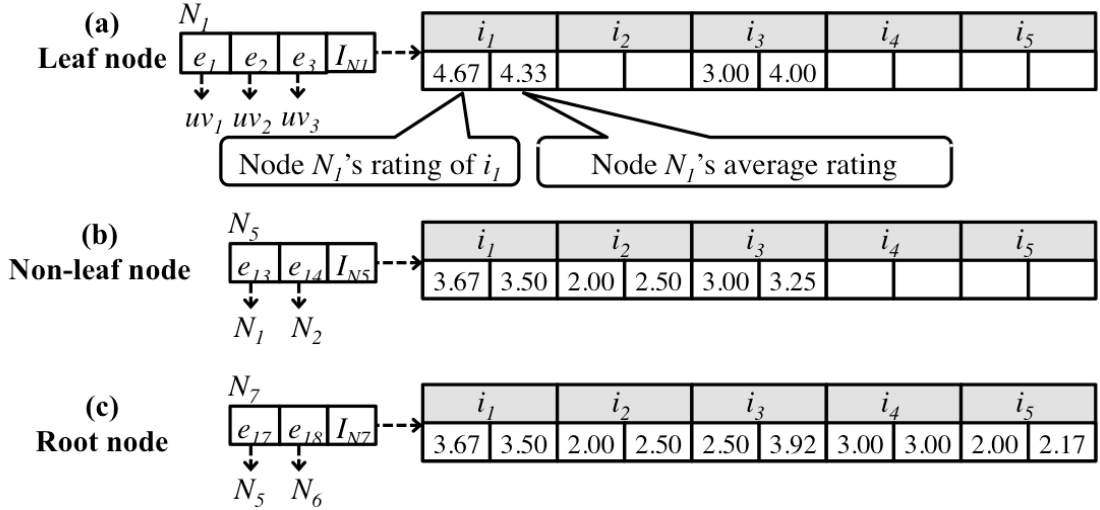

Figure 6.3: Records for storing aggregated rating and average rating information. We have (a) a leaf node $N_1$, (b) a non-leaf node $N_5$, and (c) the root node $N_7$.

### 6.4.4 The Neighbourhood-based CF Component

Given a target item $i$ of an active user $u$, the neighbourhood-based CF method uses all the R-tree nodes in a code $c$ to generate an **approximate prediction result** $ar$: all the nodes in $c$ that have rated item $i$. Using these nodes, a prediction of the rating of $i$ can be generated. Specifically, given a node $N \in ar$ that has rated item $i$, the weight between node $N$ and user $u$ is first calculated. Since node $N$ has a distinct average rating $\overline{r_{N,j}}$ for each item $j$, the weight calculation in Equation (6.1) is redefined as:

$$w(u,N) = \frac{\sum_{j\in I}(r_{u,j}-\overline{r_u})\times(r_{N,j}-\overline{r_{N,j}})}{\sqrt{\sum_{j\in I}(r_{u,j}-\overline{r_u})^2}\times\sqrt{\sum_{j\in I}(r_{N,j}-\overline{r_{N,j}})^2}} \qquad (6.5)$$

where $r_{N,j}$ is node $N$'s rating of item $j$ and $\overline{r_{N,j}}$ is node $N$'s average rating depending on $j$.

Given a set $U$ of users to be aggregated, the ratings $r_{N,j}$ and $\overline{r_{N,j}}$ can be calculated according to Equations (6.3) and (6.4). By substituting the weight $w(u, N)$ and ratings $r_i$, $r_{N,i}$, and $\overline{r_{N,i}}$ into Equation (6.2), a prediction of user $u$'s rating of item $i$ can be generated,

$$p(u, i) = \overline{r_u} + \frac{\sum_{N \in I} w(u,N) \times (r_{N,i} - \overline{r_{N,i}})}{\sum_{N \in I} |w(u,N)|} \qquad (6.6)$$

where $I \subseteq c$ is the set of R-tree nodes in code $c$ that have rated item $i$.

After obtaining a result $ar$, the elastic CF algorithm uses a data structure $s_{ar}$ to maintain all the nodes in $ar$ as its state. Thus, $s_{ar}$ stores the nodes in code $c$ that have rated item $i$. If extra budget is allocated, a code $c'$ consisting of nodes at a deeper depth of the R-tree can be used to produce a refined result $ar'$ by starting from the state $s_{ar}$. The process of producing $ar'$ starting from $s_{ar}$ is given below. The algorithm first removes any node $N \in c'$ whose parent node $N^{PARENT} \notin s_{ar}$; that is, $N^{PARENT}$ and its child nodes are pruned (lines 2 to 4). This is because neither node $N^{PARENT}$ nor any of its child nodes $N$ have rated item $i$ and so they will not be used for prediction. The updated code $c'$ is then used to produce a result $ar'$ (line 5). Result $ar'$ can then be used to generate a refined prediction result for item $i$.

**Producing an approximate result $ar'$ starting from the state $s_{ar}$**

**Input:** A code $c'$, a target item $i$, a state $s_{ar}$

**Output:** An approximate result $ar'$

1. **Begin**

2.     **for** each node $N$ in code $c'$

3.         **if** ($N$'s parent node $N^{PARENT} \notin s_{ar}$), **then**

           // $N^{PARENT}$ and its child nodes are pruned from the R-tree

4.         Set $c' = c' \backslash \{N\}$;   //remove node $N$ from code $c'$

5.     Set $ar' = CF(c', i)$;

        //the function $CF(c', i)$ returns all nodes in code $c'$ that have rated item $i$

6.     Return $ar'$.

7. **End**

Let $ar$ and $ar'$ be two approximate results produced using codes $c$ and $c'$. Let $s_{ar}$ and $s'_{ar}$ be the states for maintaining $ar$ and $ar'$. Let $b$ and $b'$ be the time budgets required to produce a refined result $ar''$ using code $c''$ by starting from states $s_{ar}$ and $s'_{ar}$, respectively. Using the above algorithm, we have $b' \leq b$ if the length $len'$ of code $c'$ is greater than the length $len$ of code $c$, as proved in Proposition 6.1.

**Proposition 6.1**. *Let the nodes in codes $c$, $c'$, and $c''$ be from depths $j$, $j'$, and $j''$ of the R-tree, with $j'' > j$ and $j'' > j'$. We have $b' \leq b$ if $len' > len$.*

**Proof**. Since the nodes in codes $c$ and $c'$ are selected from the same R-tree and since $len' > len$, we have $j' > j$; that is, $c'$ consists of nodes from a deeper depth of the R-tree. Given a target item $i$, for *any* node $N$ at a lower depth $j$, if $N \in c$ but $N \notin s_{ar}$, none of $N$'s child nodes at a deeper depth $j'$ can be included in the set $s'_{ar}$. This is because none of $N$'s child nodes have rated item $i$. In contrast, given a node $N'$ at a deeper depth $j'$, if $N' \in c'$ but $N' \notin s'_{ar}$, $N'$'s parent node $N^{PARENT}$ at depth $j$ may belong to set $s_{ar}$, because the node $N^{PARENT}$ may have other child nodes that have rated item $i$. If $N^{PARENT} \in s_{ar}$, $N^{PARENT}$'s child nodes in code $c''$ can be used for prediction. Hence, starting from $s_{ar}$ may result in there being more nodes in code $c''$ used for prediction; therefore, an equal or greater time budget $b$ is needed to produce the same result $ar''$: $b \geq b'$. ∎

## 6.4.5  Space and Time Complexities

Given a rating matrix with $m$ users, $n$ items, and $r$ ratings, the basic neighbourhood-based CF method keeps the whole set of ratings in memory, so its space (memory) complexity is $O(r)$. This method needs to scan the entire matrix for each prediction, so its time complexity is $O(m \times n)$. In the following propositions and lemmas, $h$ is the number of codes in an R-tree, and $c$ is a code consisting of $v$ R-tree nodes $\{N_1, \ldots, N_v\}$ ($v \leq m$).

**The preprocessing stage.**

**Proposition 6.2**. *The space complexity of the R-tree coding model is $O(m + r)$.*

**Proof**. $O(m)$ space is required to construct an R-tree to index $m$ user vectors, and $O(r)$ space to store the rating matrix. Thus, the total space consumption is $O(m + r)$. ∎

**Lemma 6.1**. *The time required to aggregate rating information for $h$ aggregation sets is $O(h \times m \times n)$.*

**Proof**. Let an R-tree node $N_j$ ($1 \leq j \leq v$) in the code $c$ contain $|N_j|$ users. The time required to calculate $N_j$'s aggregated rating and average rating for one item is $O(|N_j|)$ according to Equations (6.3) and (6.4), and so the aggregation time for all $n$ items is $O(|N_j| \times n)$. Thus, the total aggregation time for the code $c$ is $\sum_{j=1}^{v} O(|N_j| \times n)$. Since $\sum_{j=1}^{v} |N_j| = m$, we have $\sum_{j=1}^{v} O(|N_j| \times n) = O(m \times n)$. Hence, the total time required to aggregate $h$ codes is $O(h \times m \times n)$. ■

**Lemma 6.2**. *The construction time of an R-tree with $h$ depths is $O(m \times logm + h \times m \times n)$.*

**Proof**. We proved in Lemma 6.1 that the rating aggregation time for $h$ codes is $O(h \times m \times n)$. In addition, using the standard R-tree construction algorithm [18], the tree construction time is bounded by $O(m \times logm)$. Hence, the total R-tree construction time is $O(m \times logm + h \times m \times n)$. ■

**Proposition 6.3**. *The coding process can be completed in $O(d \times i \times \bar{n} + m \times logm + h \times m \times n)$ time using an R-tree.*

**Proof**. The $m \times d$ feature matrix of users can be produced in $O(d \times i \times \bar{n})$ time using the incremental SVD method [175], where $i$ is the number of iterations (epochs) per feature. As established in Lemmas 6.1 and 6.2, the R-tree construction time is $O(m \times logm + h \times m \times n)$. Hence, the total execution time of the offline preprocessing stage is $O(d \times i \times \bar{n} + m \times logm + h \times m \times n)$. ■

**The recommendation stage.**

**Proposition 6.4**. *Using the code $c$ for prediction, the space complexity of the neighbourhood-based CF component is $O(v \times n)$.*

**Proof**. The code $c$ can be represented by a $v \times n$ rating matrix, and so it occupies $O(v \times n)$ space.

Since the R-tree groups similar users into one R-tree node, and since these users' rated items are largely overlapped, the code $c$ is usually sparse. This means $c$'s actual space consumption is usually smaller than $O(r)$. ■

**Proposition 6.5**. *Using code $c$, an approximate result can be produced in $O(v \times n)$ time using the neighbourhood-based CF component.*

**Proof**. The neighbourhood-based CF method scans all $v$ users in code $c$ to generate a prediction for an item, which takes $O(v \times n)$ time. ■

## 6.5   Experimental Settings

We conduct a set of experiments to evaluate the effectiveness of the proposed elastic CF algorithm. The elastic and the compared time-adaptive CF algorithms are implemented in Java by extending the recommender systems library [176] and compiled using NetBeans IDE 6.9.1. We run all the experiments on a Linux-based VM with four Intel(R) Xeon(R) 2.40 GHz CPUs and 8 GB memory. In this section, we introduce the experimental settings in detail.

### 6.5.1   Description of Datasets

Our experiments are performed on two real datasets: MovieLens [170] and Netflix [157]. For each dataset, we randomly select 20% of users as the active users. For each active user, we further randomly select 20% of items to form the test set, while the remaining 80% of items form the training set. The *MovieLens* dataset has 1,208 active users and the number of rated items per user (the ratings kept for training) ranges from 16 to 1,480. The *Netflix* dataset has 9,603 active users and the number of rated items per user ranges from 1 to 3,420. Table 6.3 summarises the global statistics of the training and test sets for the two datasets.

**Two types of test set**. We conduct experiments on two types of test sets. In the **first type**, all active users are merged into one test set; that is, there is one test set for each of the MovieLens and Netflix datasets. These test sets are designed to evaluate prediction accuracy for all active users.

In the **second type**, each test set only contains target items from one active user; thus, there are 1,208 test sets for the *MovieLens* dataset and 9,603 test sets for the *Netflix* dataset. In the experiments, we evaluate the prediction accuracies of these test sets separately, thus demonstrating the effectiveness of the CF approach on active users with different numbers of rated items. Moreover, for each dataset, we divide the test sets into **three groups** according to the active users' numbers of rated items in the rating matrix (training set). For the MovieLens dataset, these numbers range from 16 to 40 for the first group, 41 to 100 for the second group, and 101 to 1480 for the third group. These three groups are denoted as *MovieLens_16To40* (comprising 405 active users), *MovieLens_41To100* (comprising 349 active users), and *MovieLens_101To1480* (comprising 454 active users), respectively. For the Netflix dataset, these numbers range from 1 to 40 for the first group, 41 to 140 for the second group, and 141 to 3420 for the third group. These three groups are denoted as *Netflix_1To40* (comprising 3,364 active users), *Netflix_41To140* (comprising 3,000 active users), and *Netflix_141To3420* (comprising 3,239 active users), respectively. These groups represent users that have rated small, medium, and large numbers of items, respectively, in real-world recommender systems.

Table 6.3. Statistics of the MovieLens and Netflix datasets

|  | MovieLens | Netflix |
|---|---|---|
| **Training set** |  |  |
| Number of users | 6,040 | 48,019 |
| Number of items | 3,953 | 17,770 |
| Number of ratings | 962,319 | 9,566,400 |
| Values of ratings | 1~5 | 1~5 |
| Density of rating matrix | 4.03% | 1.12% |
| **Test set** |  |  |
| Number of active users | 1,208 | 9,603 |
| Number of testing ratings | 37,890 | 392,088 |

## 6.5.2   Evaluation Metrics

**Quality measure**. We use the *root-mean-square error* (RMSE) [17], a weighted average error, to measure the prediction accuracy for all the target items in a test set $T$:

$$RMSE = \sqrt{\frac{\sum_{i \in T}(p(u,i) - r_{u,i})^2}{n_T}}$$

where $n_T$ represents the number of items in set $T$, $p(u,i)$ is the item $i$'s predicted rating and $r_{u,i}$ is its actual rating. RMSE provides a simple value measurement of successful prediction that measures the *errors* between the predicted and actual values of ratings.

Multiple test sets of the **second type** are used to evaluate the prediction performance for different active users. Since different test sets may have different numerical scales of $RMSE$ values, we employ the RE to measure the quality of approximate results. Let $ar$ be an *approximate prediction result* and let $er$ be the *exact prediction result* produced using the entire rating matrix. Let $ar$'s prediction accuracy be $RMSE_{ar}$ and let $er$'s prediction accuracy be $RMSE_{er}$. The result $ar$'s RE $RE$ is the difference between $RMSE_{ar}$ and $RMSE_{er}$ divided by $RMSE_{er}$ [158]:

$$RE = \frac{RMSE_{ar} - RMSE_{er}}{RMSE_{er}}$$

Intuitively, $RE$ shows how much an approximate result $ar$ deviates from the exact result $er$. Since $RMSE_{er}$ is fixed for each test set for a given rating matrix, a lower value of $RE$ denotes a higher prediction accuracy for the result $ar$. Using RE, the prediction accuracies of different users' test sets can be compared and analysed. For example, given a user $u_1$'s test set, if the RMSE of the *exact* prediction result is 0.90 and the RMSE of the *approximate* prediction result is 0.99, then the relative error is 0.10. Given another user $u_2$'s test set, if the RMSE of the *exact* result is 4.00 and the RMSE of the *approximate* result is 4.20, then the *relative error* is 0.05. The *approximate* result from $u_2$'s test set is better because its prediction accuracy is closer to that of the *exact result*. Hence, when comparing different users' test sets, a lower value of *relative error* denotes a higher prediction accuracy—that is, better quality—of the approximate result.

**Computational cost**. As established in Proposition 6.5, the time complexity (upper bound of running time) of generating a prediction is $O(v \times n)$, where $v$ is the number of R-tree nodes in the code and $n$ is the number of items in the rating matrix. Fixing a rating matrix (i.e. $n$ is fixed), this time complexity is decided by the *length $v$* of the code used. Hence, we use this length to represent computational cost.

### 6.5.3   Traditional Time-adaptive CF Algorithms

We now describe the three compared CF algorithms.

*The CF algorithm using random sampling*. This algorithm adapts to a given time budget to produce an approximate prediction using a randomly selected subset of users [17, 159, 160]. If extra budget is made available, a refined approximate result can be produced by adding more users to the subset.

*The CF algorithm using k-means clustering*. This algorithm first applies a clustering method to partition the entire set of users into multiple clusters. Given a target item of an active user, the algorithm then assigns the user to their most closely related cluster and only employs the users in this cluster to generate a prediction [159, 161, 162].

*The CF algorithm using RectTree*. This algorithm applies a clustering method to construct a hierarchy of clusters, called *RectTree* [163]. Given a test item of an active user, the algorithm assigns the user to their most closely related cluster from the bottom level of the RectTree and only uses users in this cluster for prediction. In the RectTree, the number of users in a cluster can be controlled by adjusting the number of levels (layers) in the tree.

Given a time budget, the two clustering-based algorithms create clusters of suitable sizes to produce an approximate prediction result. If a larger budget is allocated, they need to reconstruct the clusters to increase the cluster sizes, thus involving more users in producing refined results. For simplicity, we will call the three compared algorithms *Sampling*, *Clustering*, and *RectTree* respectively, and our proposed algorithm will be called *elastic algorithm*. In the experiments, *k-means clustering* [169] is applied to construct clusters for the *Clustering* and *RectTree* CF algorithms.

## 6.6   Experimental Results

In this section, we experimentally evaluate the elastic CF algorithm with three objectives.

First, Section 6.6.1 examines the behaviours of the elastic CF algorithm. A list of five approximate prediction results are produced to evaluate the properties of quality monotonicity and accumulative computation in the elastic CF algorithm.

Then, Section 6.6.2 compares the elastic CF algorithm against current time-adaptive CF algorithms. The comparison results illustrate the qualities of approximate results produced by different algorithms under similar computational cost limits.

Finally, Section 6.6.3 further compares different CF algorithms under different settings: different numbers of nearest neighbours and different densities of rating matrices are used.

### 6.6.1   Behaviour of the Elastic CF Algorithm

**The preprocessing stage**. We first apply the incremental SVD method to decompose the training set into a feature matrix of users, in which each user is represented by a *three-dimensional* user vector. In the generation of these user vectors, the learning rate is set to 0.001 and the number of iterations per feature (dimensionality) is 120 in the incremental SVD method. Next, for both datasets, five codes $c_1$ to $c_5$ are generated. The total coding time (including SVD transformation, R-tree construction, and aggregation of user ratings) is 136.35 seconds for MovieLens and 4276.62 seconds for Netflix.

**Space consumptions and computational costs at the recommendation stage**. Table 6.4 shows the space consumptions (in megabytes) of the five codes and the user-item rating matrix used as the training set for each of the two datasets. We can see that, for both datasets, all codes have space consumptions less than that of the training set. Table 6.5 lists the computational costs (number of R-tree nodes) required to produce approximate prediction results $ar_1$ to $ar_5$ using codes $c_1$ to $c_5$.

Table 6.4. Space consumptions in megabytes of the five codes for the two datasets

| | Code and training set | | | | | |
|---|---|---|---|---|---|---|
| | $c_1$ | $c_2$ | $c_3$ | $c_4$ | $c_5$ | Training set |
| MovieLens | 0.14 | 0.38 | 0.88 | 1.73 | 3.05 | 12.43 |
| Netflix | 1.08 | 4.66 | 16.68 | 48.44 | 115.45 | 138.38 |

Table 6.5. Computational costs of the five codes for the two datasets

|  | Code and training set | | | | | |
|---|---|---|---|---|---|---|
|  | $c_1$ | $c_2$ | $c_3$ | $c_4$ | $c_5$ | Training set |
| MovieLens | 2 | 6 | 14 | 27 | 56 | 6,040 |
| Netflix | 2 | 10 | 55 | 301 | 1564 | 48,019 |

**Evaluation of quality monotonicity**. We first experiment on test sets of the *first type*. For each test set, five approximate results $ar_1$ to $ar_5$ are produced using codes $c_1$ to $c_5$, and the exact result $er$ is produced using the entire training set. In this experiment, the elastic algorithm is set to start from the initial state. The prediction error (measured by RMSE) of the exact result $er$ is 0.91 for the MovieLens dataset and 0.90 for the Netflix dataset. Figure 6.4 shows the RMSE (y axis) of each approximate result. We can see that as more computations are conducted, the RMSEs of the five results $ar_1$ to $ar_5$ gradually decrease; that is, the prediction accuracies increase, thus exhibiting the property of *quality monotonicity*. This verifies the proof of Theorem 5.1: codes $c_1$ to $c_5$ have increasing resolution (information gain); thus, the qualities (prediction accuracies) of the results $ar_1$ to $ar_5$ produced using codes $c_1$ to $c_5$ gradually approach the quality of the exact result.

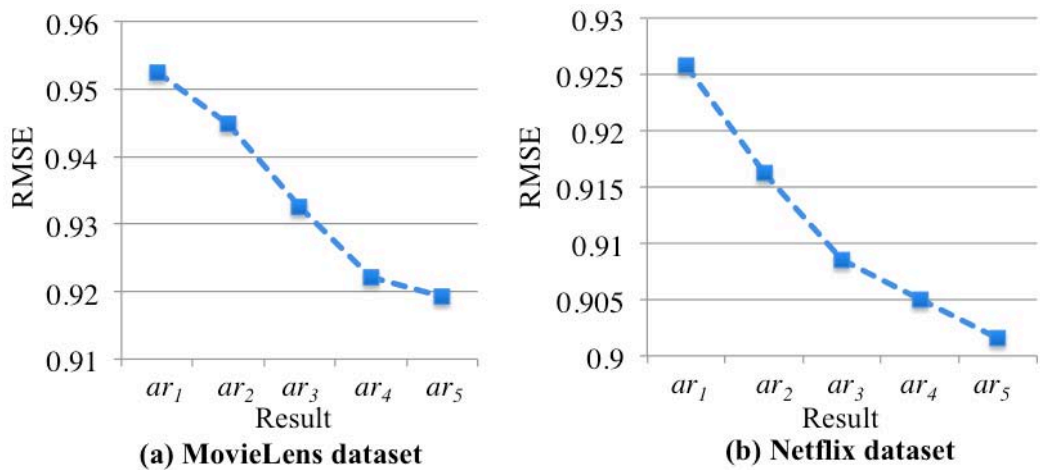

**(a) MovieLens dataset**　　　**(b) Netflix dataset**

Figure 6.4: Qualities of the five approximate results for the two datasets.

Under the same experimental settings as above, we experiment on test sets of the *second type*: the 1,208 test sets in the MovieLens dataset and the 9,603 test sets in the Netflix dataset. For each test set, five approximate results $ar_1$ to $ar_5$ are produced and the REs of these results are used to compare their prediction accuracies to those of the other test sets. Figure 6.5 displays the distribution of REs using their mean and variance. The experiment result shows that the mean of REs is less than 0.06 and the variance of REs is less than 0.14 for both datasets. This indicates that the discrepancies between most of the approximate prediction results and the exact prediction results are small. We can also see that as more computations are conducted, the REs of the five approximate results $ar_1$ to $ar_5$ gradually reduce to 0, thus exhibiting the property of *quality monotonicity*: the values of RE represent the discrepancies between approximate and exact results, and they decrease to 0. This means for most of the users, the qualities of the approximate results approach the quality of the exact result when more computations are conducted.

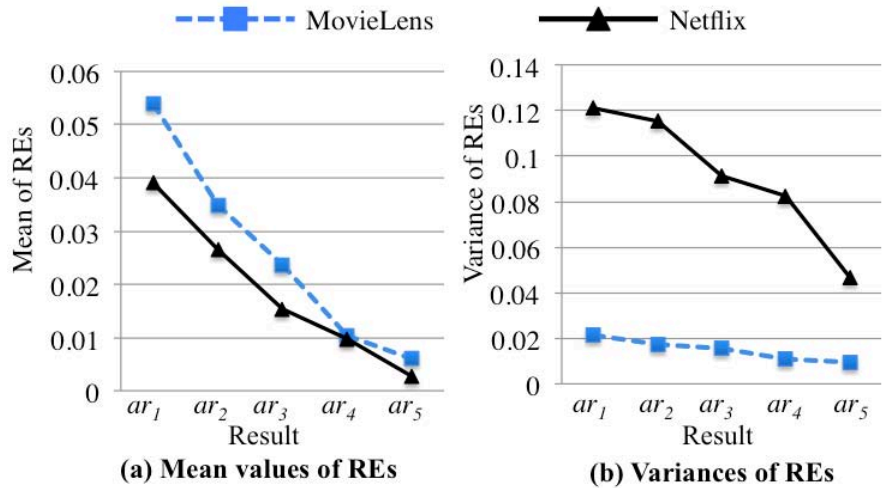

(a) **Mean values of REs**

(b) **Variances of REs**

Figure 6.5: Mean and variance values of REs of the five approximate results for the two datasets.

**Evaluation of accumulative computation**. In generating the five approximate results $ar_1$ to $ar_5$, the algorithm can start either from the *initial state* ($S_{ar} = \phi$) or from the *state* of a previously obtained result. Table 6.6 lists the computational costs of producing the five results when starting from different states. If starting from the initial state, all the nodes in the codes are used for prediction. If starting from a previously obtained result's state, different test items have different R-tree nodes that can be pruned. Considering all test items, the *average number* of remaining (unpruned) R-tree nodes in a code $c_j$ is used to denote the computational cost of producing a result $ar_j$. The results in Table 6.6 indicate that if the algorithm starts from the state of a prediction result produced from nodes at a deeper depth of the R-tree, the algorithm can produce the same refined result with less computational cost. Hence, the algorithm satisfies the property of *accumulative computation*.

***Results of Section 6.6.1.*** *The elastic CF algorithm can produce good prediction results using small overheads (in terms of both space and time consumptions), and further increases prediction accuracy with* *more computations. It also organises the computation in a cost-efficient manner that allows users to start from a previously obtained result to save on computations.*

Table 6.6. Computational costs of producing five approximate results from different starting states

| Dataset | Starting state | Approximate results | | | | |
|---------|----------------|--------|--------|--------|--------|--------|
| | | $ar_1$ | $ar_2$ | $ar_3$ | $ar_4$ | $ar_5$ |
| MovieLens | $\phi$ | 2.00 | 6.00 | 14.00 | 27.00 | 56.00 |
| | $s_{ar_1}$ | | 5.65 | 13.65 | 26.65 | 35.65 |
| | $s_{ar_2}$ | | | 13.18 | 26.18 | 35.18 |
| | $s_{ar_3}$ | | | | 24.89 | 33.33 |
| | $s_{ar_4}$ | | | | | 32.18 |
| Netflix | $\phi$ | 2.00 | 10.00 | 55.00 | 301.00 | 1,564.00 |
| | $s_{ar_1}$ | | 10.00 | 54.99 | 300.94 | 1,563.69 |
| | $s_{ar_2}$ | | | 54.82 | 300.00 | 1,558.80 |
| | $s_{ar_3}$ | | | | 293.39 | 1,526.11 |
| | $s_{ar_4}$ | | | | | 1,411.78 |

## 6.6.2   Comparison to Time-adaptive CF Algorithms

In this section, we repeat Section 6.6.1's experiments and compare the elastic CF algorithm to the three time-adaptive CF algorithms: *Sampling*, *Clustering*, and *RectTree*. Five approximate results $ar_1$ to $ar_5$ are produced using the elastic algorithm by starting from the initial state, with computational costs of 2.00, 6.00, 14.00, 27.00, and 56.00, respectively.

**Space consumptions and computational costs of the three time-adaptive algorithms**.

To make our comparisons fair, the same or slightly greater computational costs are permitted in generating the five approximate results with the three compared algorithms. This is achieved by adjusting the numbers of users used to generate predictions. The *Sampling* algorithm has no preprocessing time (the sampling process is very fast and its consumed time is not considered). Given a time budget, the two clustering-based algorithms (*Clustering* and *RectTree*) create clusters of suitable sizes to produce an approximate result. If a larger budget is allocated, they need to *reconstruct* these clusters to increase the numbers of users in them, thus including more users to produce refined results. This means they need to construct clusters five times to produce the five results $ar_1$ to $ar_5$. In clustering construction, the value of the iteration number $i$ is set to 10. This is because over 10 iterations, the k-means can reach either complete convergence or a state that is very close to convergence, in which further iterations have a small effect on the clustering result. This setting also follows the default settings of k-means clustering in popular open source tools for machine learning and data mining such as R (http://www.r-project.org/) and RapidMiner (http://sourceforge.net/projects/rapidminer/). For the *Clustering* algorithm, the clustering construction takes 54,026.69 seconds for the MovieLens dataset and 1,300,870.90 seconds (approximately 15 days) for the Netflix dataset. For the *RectTree* algorithm, the total clustering process takes 6,075.41 seconds for the MovieLens dataset and 241,523.52 seconds for the Netflix dataset. Both the *Clustering* and *RectTree* algorithms have much longer preprocessing times than the elastic algorithm.

At the recommendation stage, the *Sampling* algorithm operates on the entire training set and thus needs to load the entire training set into memory. The *Clustering* and *RectTree* algorithms also need to keep the whole training set and all constructed clusters in memory. Since the training set occupies more space than any of the five aggregation sets, the three time-adaptive algorithms incur greater space consumptions than the elastic algorithm.

**Comparison using the RMSE metric.**

We first experiment on test sets of the first type. Figure 6.6 shows five approximate results (x axis) produced by different CF algorithms and their RMSEs (y axis). We can see that all the results produced by the elastic algorithm have higher prediction accuracies (i.e. lower RMSEs) than the results produced by the three time-adaptive algorithms. Moreover, as more computations are performed, the prediction accuracies of the results produced by the elastic algorithm *monotonically* increase (i.e. the RMSE decreases). In contrast, the three time-adaptive algorithms cannot guarantee this monotonic increase in accuracy with computations performed. This is because in the time-adaptive algorithms, the newly added users only represent a small proportion of the whole user base, and the partial refinement of results using these users may actually decrease prediction accuracy. In the elastic algorithm, a code represents the aggregated statistical information of the entire rating matrix. Hence, a code of greater length represents the rating matrix at a finer level of granularity, which guarantees a higher prediction accuracy. The results in Figure 6.6 also show that although the *Sampling* algorithm has no preprocessing, it also has the lowest prediction accuracy among all four algorithms.

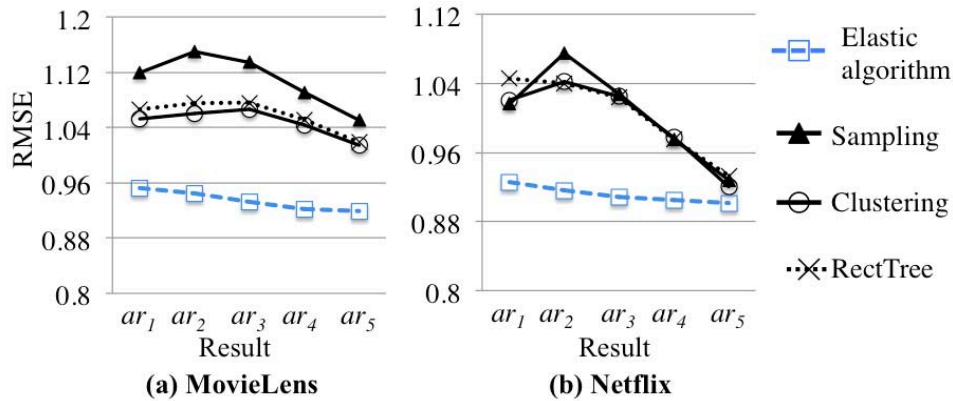

Figure 6.6: Comparison of RMSEs for the two datasets

**Comparison using the RE metric.**

We evaluate test sets of the second type using the RE metric. The 1,208 test sets in MovieLens and 9,603 test sets in Netflix are divided into the *three groups* described in Section 6.5.1. Since each group of test sets has multiple RE values, Figure 6.7 displays the distribution of REs using their mean and variance. The experimental results show that the REs of results generated by the elastic algorithm have the least mean and variance in all cases. We can also see that for the three time-adaptive algorithms, results that are produced using more computations can have REs with greater mean or variance. Since low RE values denote high prediction accuracies, the above observations indicate that the time-adaptive algorithms can produce results with lower predication accuracies when performing more computations. More importantly, the elastic algorithm can produce results with higher prediction accuracies for a majority of users' test sets.

In support of the above claim, Figure 6.8 shows the percentages of active users' test sets whose results produced by the elastic algorithm have higher prediction accuracies than those produced by the time-adaptive algorithms. It can be seen that in the groups MovieLens_16To40 (Figure 6.8(a1)) and Netflix_1To40 (Figure 6.8(b1)), active users have rated only a small number of items: these users can be regarded as new users in recommender systems. Although it is difficult to make accurate predictions for these users due to the lack of substantial numbers of rated items, the elastic algorithm still produces results with higher prediction accuracies for approximately 70% of test sets. Moreover, for active users with medium numbers of rated items (MovieLens_41To100 and Netflix_41To140), approximately 80% of test sets in results $ar_1$ to $ar_3$ have higher prediction accuracies as shown in Figures 6.8(a2) and (b2). This percentage further increases to 90% for active users with larger numbers of rated items (MovieLens_101To1480 and Netflix_141To3420 in Figures 6.8(a3) and (b3)). We can also see that as more computational time is used, the percentages of better test sets gradually reduces to 60-80%. This is because the prediction results produced by different algorithms gradually approach the exact prediction result.

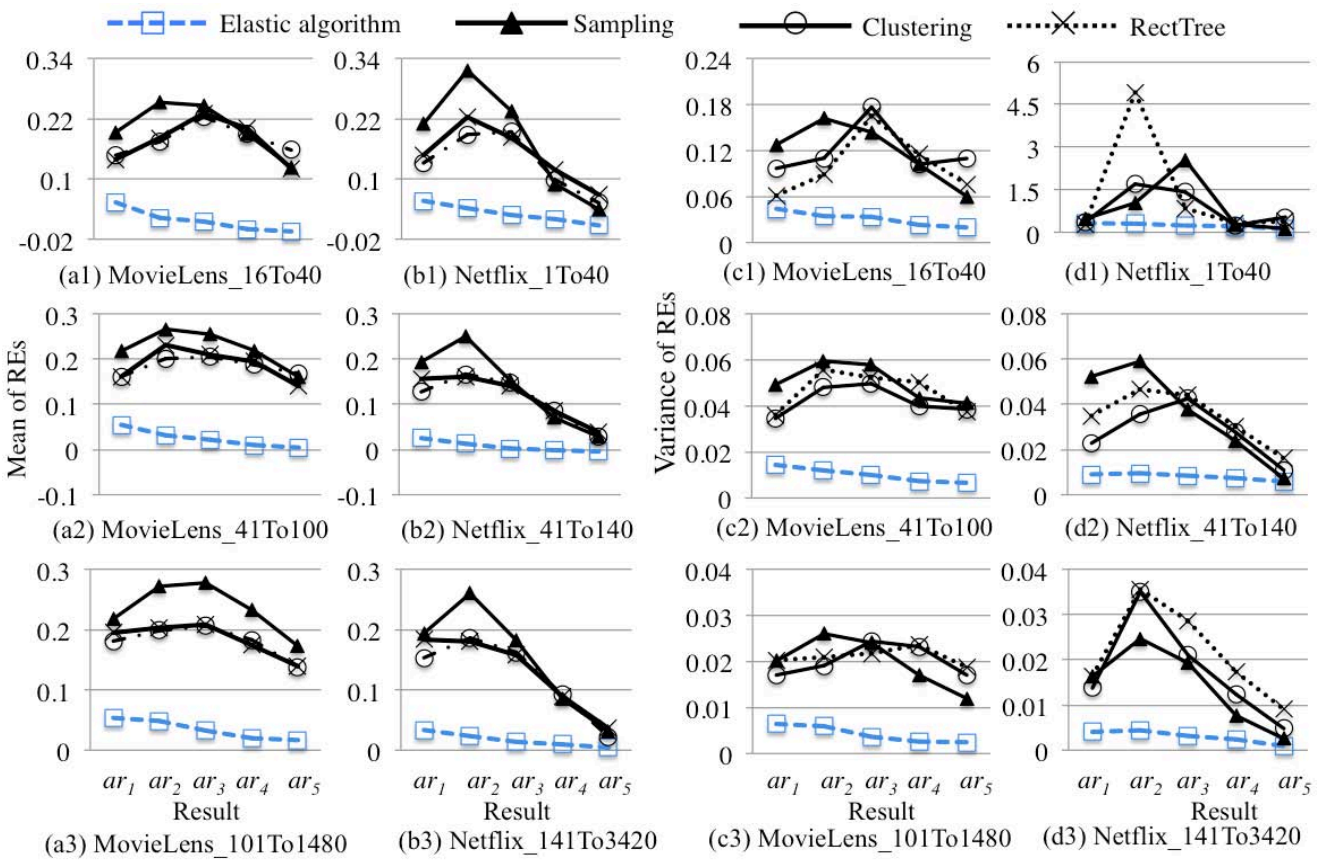

Figure 6.7: Comparison of mean and variance of REs for the two datasets.

**Results of Section 6.6.2**. *When consuming the same computational costs and less memory space, the elastic CF algorithm displays obvious superiority over existing time-adaptive algorithms: it can produce approximate results with higher prediction accuracies, and guarantees better predictions for a majority of users.*

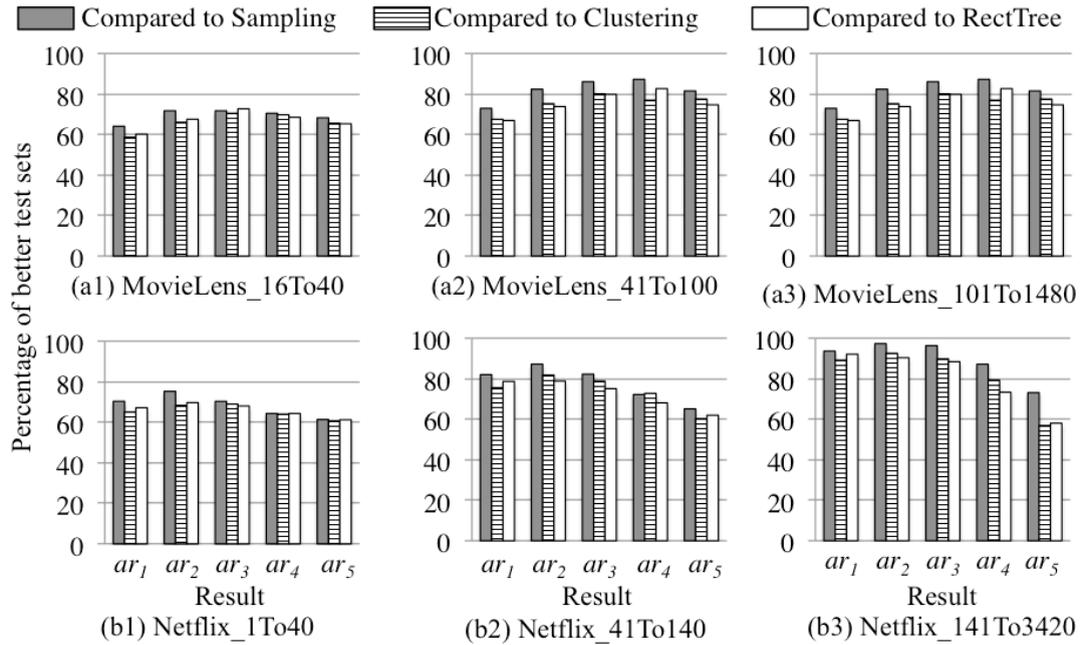

Figure 6.8: Percentages of test sets with higher prediction accuracies for the two datasets.

## 6.6.3 Comparison Using Different Experimental Settings

To compare algorithms thoroughly, we conduct two experiments under varying experimental settings.

**Experiment 1: Comparison under different numbers of nearest neighbours**.

One typical prediction mechanism in the neighbourhood-based CF method [17] is to employ an active user $u$'s $k$ nearest neighbours (i.e. the $k$ users with the greatest weights/similarities with respect to $u$) in making a prediction [17]. Following the experimental settings of Section 6.6.2, we experiment on test sets of the *first type* from MovieLens under four different values of $k$: 10, 20, 40, and 100. Thus, only the most similar 10, 20, 40, and 100 users are used for prediction. The experimental result, shown in Figure 6.9, shows that under different values of $k$, the results produced by the elastic algorithm always have greater prediction accuracies than those of the other algorithms. This indicates that although only the $k$ most similar users are used for prediction, the aggregated rating information used in the elastic algorithm still performs best.

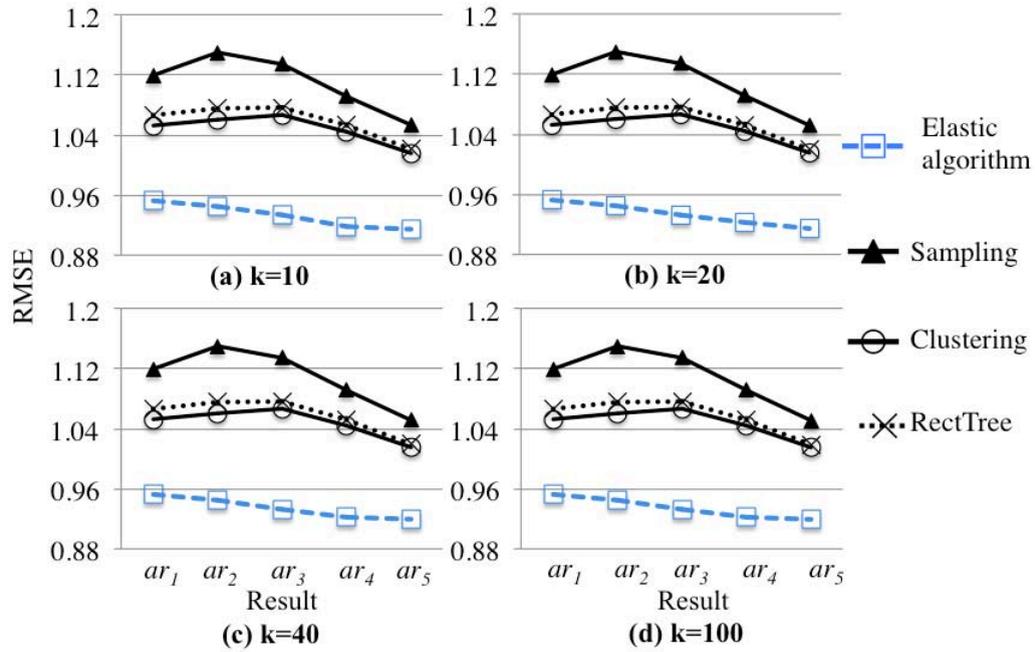

Figure 6.9: Comparison of RMSEs using different numbers of nearest neighbours.

**Experiment 2: Comparison under different densities of rating matrix**.

In this experiment, we randomly select 100%, 80%, 50%, and 20% of the training data from the MovieLens dataset for prediction. These four training sets represent four rating matrices with different degrees of density: 4.03%, 3.21%, 2.00%, and 0.80%. Table 6.7 presents the statistics of the four training sets. For simplicity, we denote these four training sets as 100%, 80%, 50%, and 20%, respectively.

Table 6.7. Statistics of the four training sets representing different densities of rating matrix

|  | **Training set** | | | |
|---|---|---|---|---|
|  | 100% of data | 80% of data | 50% of data | 20% of data |
| **Number of ratings** | 962,319 | 767,446 | 479,735 | 190,082 |
| **Density of rating matrix** | 4.03% | 3.21% | 2.00% | 0.80% |

**Space consumptions and computational costs**.

Table 6.8 lists the space consumptions of the five codes generated using the four training sets of rating matrices. We can see that all these codes occupy spaces smaller than those of the original training sets. Table 6.9 lists the preprocessing time for each algorithm. In all cases, the preprocessing time of the elastic algorithm is much smaller than the clustering construction time of the *Clustering* and *RectTree* algorithms.

Table 6.10 lists the computational costs of producing five approximate results using the five codes. We can see that the *Sampling* algorithm has the same computational cost as the elastic algorithm, and the *Clustering* and *RectTree* algorithms have slightly greater computational costs.

Table 6.8. Space consumptions in megabytes of the five codes for the four training sets

| | Code and training set | | | | | |
|---|---|---|---|---|---|---|
| | $c_1$ | $c_2$ | $c_3$ | $c_4$ | $c_5$ | Training set |
| 100% | 0.14 | 0.38 | 0.88 | 1.73 | 3.05 | 12.43 |
| 80% | 0.22 | 0.52 | 1.76 | 3.18 | 5.16 | 9.91 |
| 50% | 0.27 | 0.75 | 1.95 | 3.18 | 5.64 | 6.20 |
| 20% | 0.19 | 0.43 | 0.81 | 1.38 | 1.97 | 2.46 |

Table 6.9. Preprocessing times of three CF algorithms for the four training sets

| Algorithm | | Training set | | | |
|---|---|---|---|---|---|
| | | 100% | 80% | 50% | 20% |
| **Preprocessing time (seconds)** | Elastic algorithm | 136.35 | 119.28 | 104.68 | 81.69 |
| | Clustering | 54,026.69 | 27,076.92 | 15,681.62 | 12,586.93 |
| | RectTree | 6,075.41 | 3,391.89 | 2,832.68 | 2,595.00 |

Table 6.10. Computational costs of the five codes generated using the four training sets

| | Code | | | | |
|---|---|---|---|---|---|
| | $c_1$ | $c_2$ | $c_3$ | $c_4$ | $c_5$ |
| 100% | 2 | 6 | 14 | 27 | 56 |
| 80% | 2 | 6 | 36 | 71 | 127 |
| 50% | 5 | 13 | 41 | 81 | 188 |
| 20% | 2 | 6 | 15 | 34 | 66 |

**Comparison using the RMSE metric.**

We first conduct experiments on test sets of the **first type** from the MovieLens dataset. Figure 6.8 compares the RMSEs of the four algorithms using training sets with different densities. We can see that, similarly to Experiment 1, the RMSEs of the five results produced by the elastic algorithm have the highest prediction accuracies in every case. This comparison result indicates that the elastic algorithm still performs best when dealing with rating matrices of different densities. In addition, the result shows that as the rating matrix becomes sparser (from Figure 6.10(a) to (d)), all algorithms generally produce approximate results with lower prediction accuracies (i.e. RMSE increases). This is because the accuracy of exact results decreases when less rating information is used: the RMSEs of the exact results are 0.91, 0.91, 0.93, and 1.00 when using 100%, 80%, 50%, and 20% of the rating data, respectively. We can also see that the amount of accuracy deterioration is the least with the elastic algorithm.

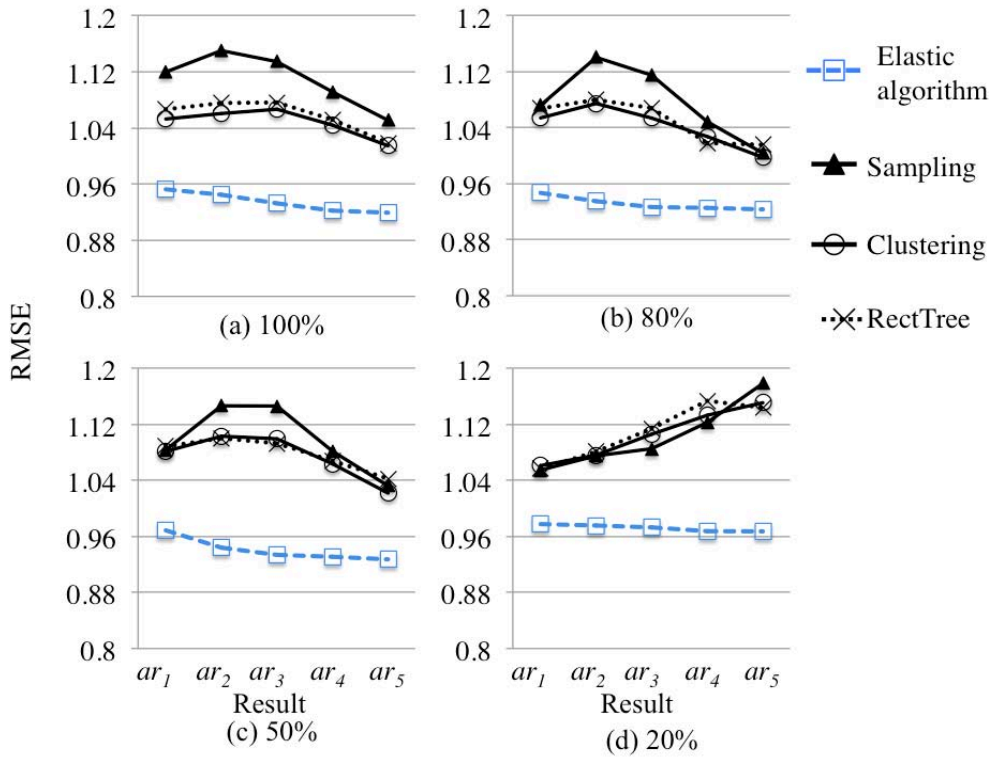

Figure 6.10: Comparison of RMSEs for different densities of rating matrix.

We further evaluate our algorithms using test sets of the **second type**: the 1,208 test sets from the MovieLens dataset. In Section 6.6.2's experiment, there is only one training set and each test set has only one exact prediction result. In this experiment, there are four training sets and each training set results in one exact prediction result for each test set. Each training set also results in four approximate results for each test set. Figure 6.11 shows the mean and variance of the 1,208 RMSEs for exact and approximate results (the average RMSE of five approximate results is reported). Unsurprisingly, as the rating matrix becomes sparser, the mean and variance of RMSEs increase (prediction accuracy decreases) for both exact and approximate results. We can see that the elastic algorithm has increases in mean and variance similar to those of the exact result. In contrast, the three time-adaptive algorithms have smaller increases in mean and variance as the rating matrix becomes sparser, because they already had large mean and variance of RMSEs when using the matrix with 100% of the rating data.

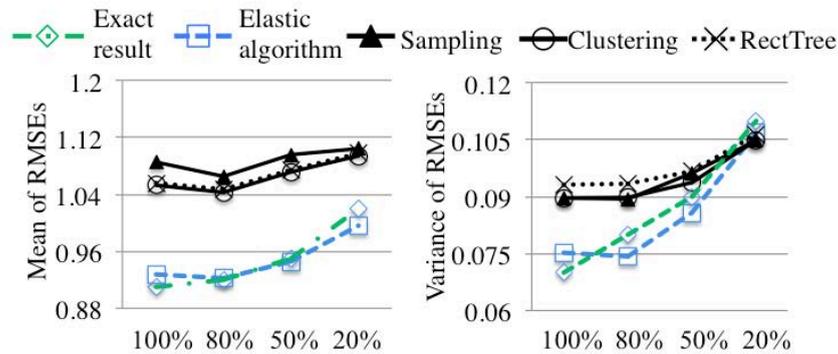

Figure 6.11: Mean and variance of 1,208 RMSEs for different densities of rating matrix.

Based on the above observation, we further use the RE metric to evaluate how the approximate results approach the exact results produced under different densities of rating matrix. Figure 6.12 compares the mean and variance of the five approximate results' 1,208 REs for each of the four algorithms. We can see that, similarly to Experiment 1, the REs of results generated by the elastic algorithm have the smallest mean and variance. As seen in Figures 6.12(a1) to (a4), the mean of REs decreases for all algorithms as the rating matrix becomes sparser, which indicates that the discrepancies between the approximate and exact prediction accuracies become smaller. This is because as the rating matrix becomes sparser, the exact result has a greater increase in mean than the approximate results, as illustrated in Figure 6.11(a).

Moreover, Figures 6.12(b1) to (b4) show that as the density of matrix decreases, all algorithms have greater fluctuations in the variances of their REs. In particular, Figure 6.12(b4) shows that results $ar_1$ and $ar_2$ produced by the elastic algorithm have greater variances than those of the other algorithms. This is because when only 20% of the rating data is used, the matrix is very sparse and 99.20% of the rating information is missing. Hence, the codes used to produce $ar_1$ and $ar_2$, which represent the statistical information from the remaining 0.80% of ratings, are not very accurate. In contrast, the three time-adaptive algorithms only use a smaller number of users to produce results $ar_1$ and $ar_2$. These users have little influence on prediction accuracy, which indicates that the 1,208 results produced by these users are similar to each other. Hence, in Figures 6.12(a4) and (b4), the results $ar_1$ and $ar_2$ produced by the three time-adaptive algorithms have larger means, but smaller variances, of REs. Furthermore, Figures 6.12(a4) and (b4) show that the results $ar_4$ and $ar_5$ produced by the elastic algorithm have smaller means and variances of REs than those produced by the other algorithms. In the three time-adaptive algorithms, the mean and variance of REs increase as more users are included, because the partial refinement of results using the newly added users decreases the prediction accuracy. In conclusion, the comparison results indicate that, generally, the sparseness of the rating matrix has the least impact on our elastic algorithm, and our algorithm still performs better than the other algorithms when dealing with rating matrices of different densities.

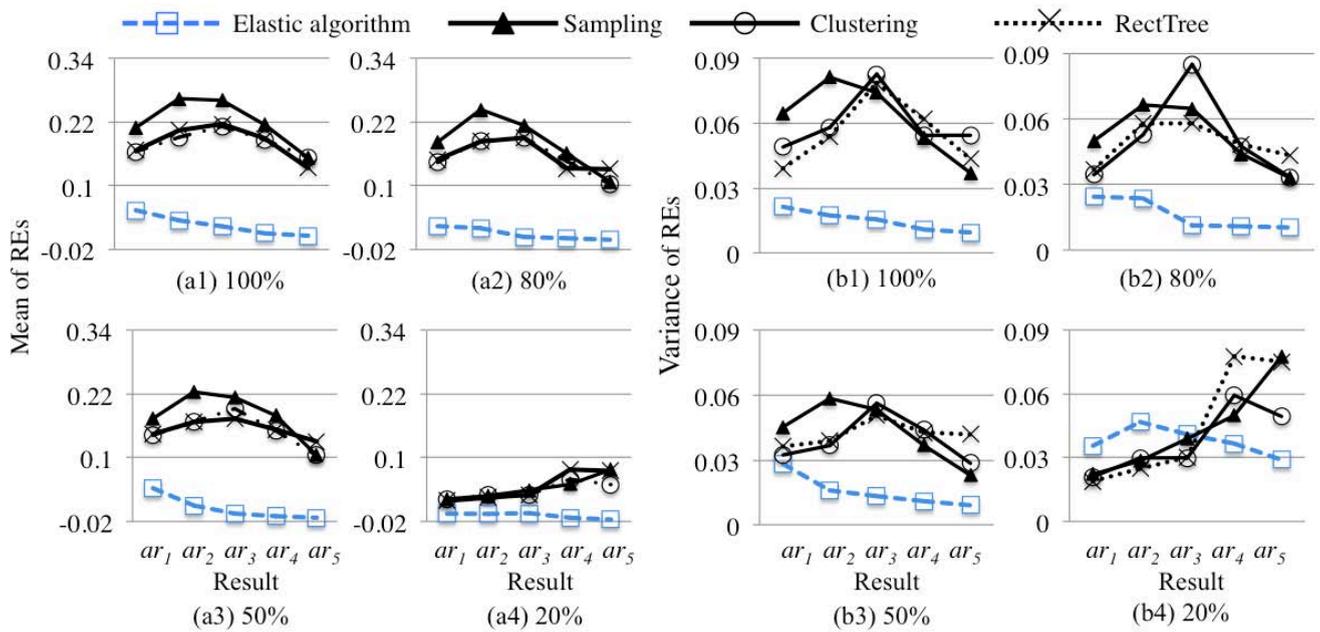

Figure 6.12: Comparison of REs for different densities of rating matrix.

**Percentages of users with better prediction quality**. To verify the above claim, Figure 6.13 presents the percentages of users with better prediction quality when comparing the elastic algorithm against the other three time-adaptive algorithms. The comparison results show that under four densities of rating matrix, the elastic algorithm produces better prediction results for approximately 80% of 1,208 test sets.

*Results of Section 6.6.3. The elastic CF algorithm is the best technique overall for producing approximate prediction results on limited time budgets. In particular, this algorithm can consistently produce results with higher prediction accuracies under different neighbourhood sizes and different densities of rating matrix.*

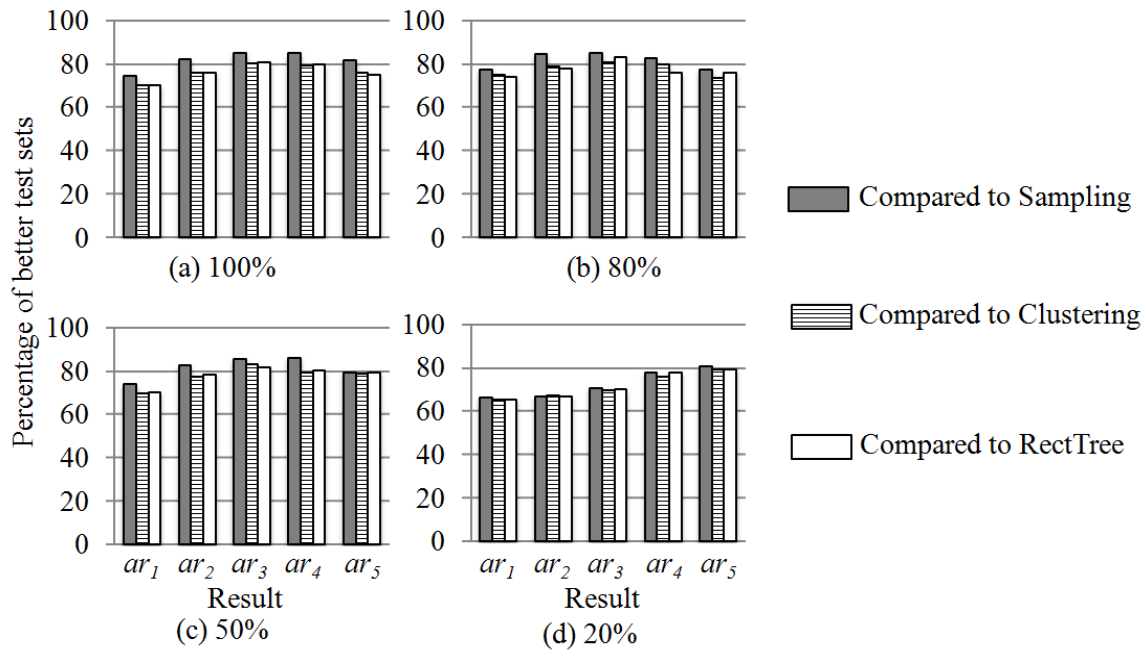

Figure 6.13: Percentages of test sets with higher prediction accuracies for different densities of rating matrix.

## 6.7 Discussion of the Elastic CF Algorithm

### 6.7.1 Discussion of Other Coding Techniques

The elastic CF algorithm can be used for large-scale data analysis problem that is required to produce predictions in real-time under different time budget allocations. Its effective use is based on first structuring the large data offline using appropriate coding components that summarise its input information at multiple levels of granularity, and then choosing the appropriate granularity for run-time. The R-tree coding component we propose in this chapter is an agglomerative model [13] that groups similar user vectors in a bottom-up fashion. The elastic CF algorithm is generic and other coding techniques can also be applied to summarise user rating information. To illustrate this, we present another coding component that represents the divisive (top-down) grouping of similar users in the rating matrix.

This divisive coding component is implemented as a hierarchical clustering model, which starts from the whole set of users and splits them recursively as moving down the hierarchy. Each split is implemented as a k-means clustering, which partitions a set of users to a specified number of clusters. In the construction of the hierarchical clustering model, the number of clusters at a certain level of the model can be controlled by adjusting the value of $k$ in k-means clustering. For example, Figure 6.14 displays a hierarchical clustering model used to group the 12 users in Table 6.2(a)'s rating matrix using 2-means clustering ($k$=2). Initially, there is only one cluster $C_1$ at level 1 of the model. After one split, there are two clusters at level 2. We can observe that users with similar rating preferences are assigned the same cluster: users $u_1$ to $u_6$ belong to cluster $C_2$ and users $u_7$ to $u_{12}$ belong to cluster $C_3$. These 12 users are further divided into four clusters at level 2 of the model.

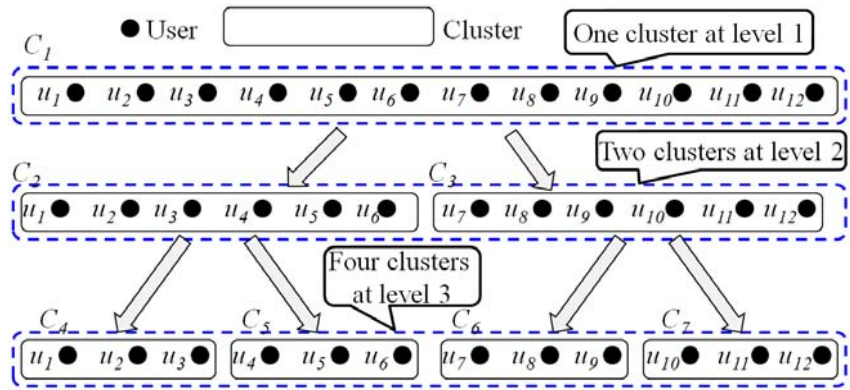

Figure 6.14: An example hierarchical clustering model for grouping 12 users in the rating matrix.

In the elastic CF algorithm, each cluster consists of a set $U$ of users and stores their aggregated statistical information: the aggregated rating and average rating of these users can be calculated using Equations (6.3) and (6.4). All the clusters at a particular level of the model form a code.

For convenience, we call the two elastic algorithms, which are developed based on the coding components of hierarchical clustering and R-tree, *Elastic K-means* and *Elastic R-tree* respectively. We evaluate both algorithms using the two types of test sets in the MovieLens and Netflix datasets described in Section 6.5.1. Figure 6.15 shows the results of test sets of the first type and Figure 6.16 shows the mean and variance of REs in evaluating test sets of the second type (the 1,208 test sets in the MovieLens dataset and the 9,603 test sets in the Netflix dataset). We can observe that as more computational is performed, the prediction accuracy of the results produced by both algorithms *monotonically* increases (i.e. the RMSE and RE decrease). In addition, although the *Elastic K-means* algorithm needs longer preprocessing time (1,035.02 seconds for the MovieLens dataset and 36,640.08 seconds for the Netflix dataset), this algorithm also produces more results with higher accuracies (lower RMSEs and REs) than those of the *Elastic R-tree* algorithm. To verify this claim, Figure 6.17 presents the percentages of 1,208 test sets in the MovieLens dataset with higher prediction accuracies when comparing the two elastic algorithms against the three time-adaptive algorithms (Sampling, Clustering, and RectTree described in Section 6.5.3). The comparison results show that both elastic algorithms produce better prediction results for 75% to 85% of the 1,208 test sets. In addition, the *Elastic K-means* algorithm always achieves slightly higher percentages of better test sets compared to the *Elastic R-tree* algorithm.

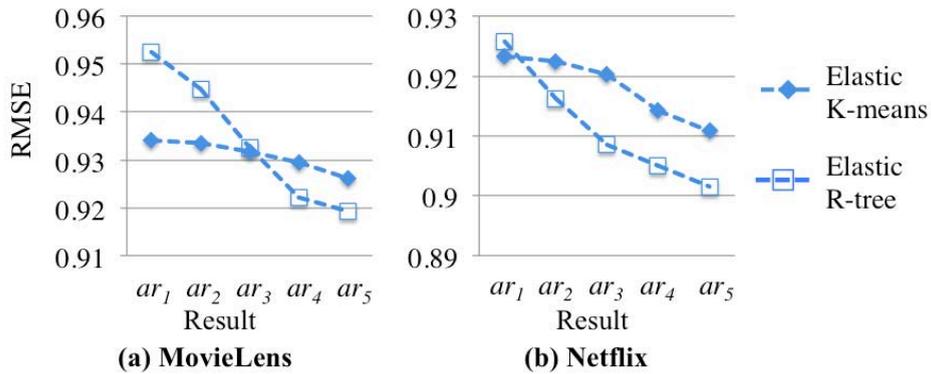

**(a) MovieLens**  **(b) Netflix**

Figure 6.15: Comparison of RMSEs using the *Elastic K-means* and *Elastic R-tree* algorithms.

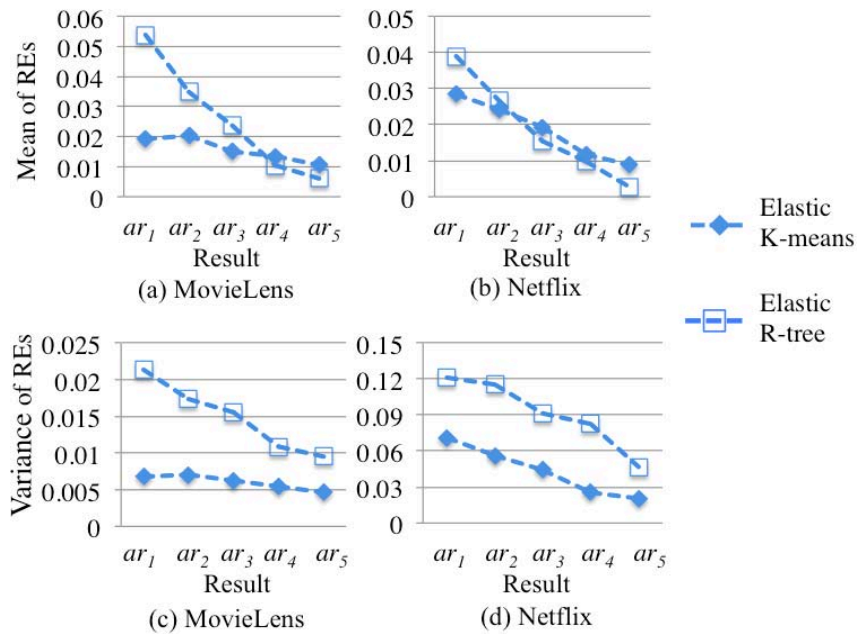

(a) MovieLens  (b) Netflix

(c) MovieLens  (d) Netflix

Figure 6.16: Comparison of mean and variance of REs using the *Elastic K-means* and *Elastic R-tree* algorithms.

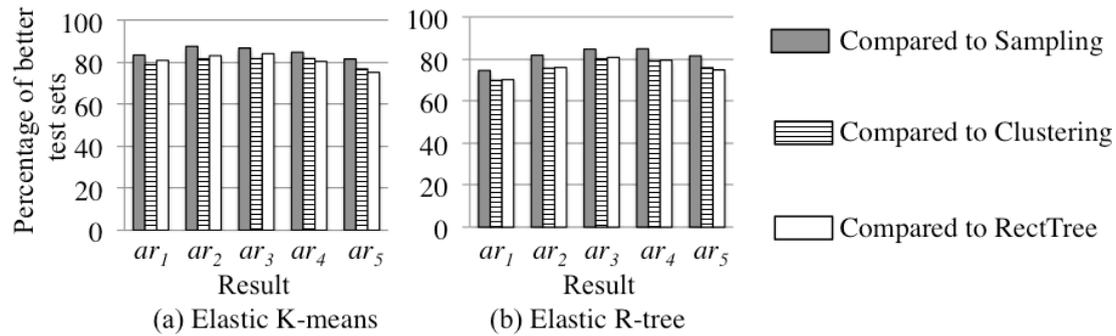

Figure 6.17: Percentages of test sets with higher prediction accuracies using the *Elastic K-means* and *Elastic R-tree* algorithms.

In the future, we plan to explore the use of other coding techniques including wavelets and hierarchical locality sensitive hashing that perform compression by reducing data dimensionality.

## 6.7.2 Discussion of the Size and Density of Rating Matrix

In e-commerce recommender systems, the user-item rating matrix used for prediction is usually extremely large and sparse, which is major challenge faced by existing CF techniques. The elastic and time-adaptive algorithms (*Sampling*, *Clustering* and *RectTree*) discussed in this chapter also suffer from the large size and low density of the rating matrix. Using a huge rating matrix with over 1 million users and 43 million ratings, we now demonstrate how the performance and prediction accuracy of these algorithms are influenced by the size and density of the matrix.

**The Million Song dataset**

The Million Song dataset is a huge collection of 1,018,563 users' ratings on 382,473 songs (items). There are 43,399,187 ratings in the matrix and the data density of this matrix is 0.01%. The value of ratings ranges from 1 to 5. In experiment, we randomly select 1,000 users as the active users. For each active user, we further randomly select 20% of the items to form the testing set, while the remaining 80% of the items form the training set. Thus, the training set includes 43,363,151 observed ratings and the test set has 36,036 target items to be predicted. Similar to the experiments in Section 6.6, we generate two types of test sets. In the *first type*, all active users are merged into one test set; that is, there is one test set for the Million Song dataset. In the *second type*, each test set only contains target items from one active user; thus, there are 1,000 test sets for the Million Song dataset.

**Experimental settings**

We conduct experiments in a VM with eight 2.40 GHz CPU cores and 16 GB memory. Five codes $c_1$ to $c_5$ are generated at the preprocessing stage. For each test set, five approximate results $ar_1$ to $ar_5$ are produced using codes $c_1$ to $c_5$. The numbers of R-tree nodes in codes $c_1$ to $c_5$ are 8, 45, 250, 1249, and 6,442, respectively.

**Influence of the rating matrix size on preprocessing time**

Since in elastic and time-adaptive algorithms, the recommendation time is independent of the size of the rating matrix, only their preprocessing time replies on the matrix size. Following the experiment settings of Section 6.6 that in the elastic algorithm, the 3-dimensional user vectors are generated by the incremental SVD method by setting the learning rate to 0.001 and the iterations per feature to 120. The five codes $c_1$ to $c_5$ are generated in 146.28 hours. In contrast, the estimated preprocessing time (i.e. the cluster construction time) takes seven years for the *Clustering* algorithm and one year for the *RectTree* algorithm. These two time-adaptive algorithms, therefore, are not applicable to deal with the huge rating matrix of the Million Song dataset.

**Influence of the data sparsity on prediction accuracy**

We compare our elastic algorithm to the *Sampling* algorithm using the two types of test sets in the Million Song dataset under the same computational cost. The comparison results are presented in Figure 6.18 (the test set of the first type) and Figure 6.19 (the test sets of the second type). As shown in Figure 6.18(a), Figure 6.19(a), and Figure 6.19(c), the prediction accuracy of results $ar_1$ to $ar_5$ produced by the elastic algorithm gradually increases (that is, RMSE and RE decrease). This indicates the elastic algorithm still guarantees quality monotonicity to the used computations even when dealing with very sparse rating matrix (99.99% of rating information is missing in the Million Song dataset). This data sparsity also means the small difference between two neighbouring codes $c_i$ and $c_{i+1}$ ($1 \le i \le 4$), thus resulting in the small discrepancy of the prediction accuracies of two neighbouring results $ar_i$ and $ar_{i+1}$. For example, Figure 6.18(a) shows that the RMSE of results $ar_1$ is 1.1055 and the RMSE of results $ar_2$ is 1.1049. In contrast, the experimental results for the *Sampling* algorithm in Figure 6.18 (b), Figure 6.19(b), and Figure 6.19(d) show that a result $ar_{i+1}$ produced using more computations always has *lower* prediction accuracy than that of another result $ar_i$ produced using less computations. This is because the rating matrix is very sparse, the newly added users used to produce result $ar_{i+1}$ are poorly correlated with the existing users used to produce result $ar_i$, thus deteriorating the prediction accuracy of result $ar_{i+1}$.

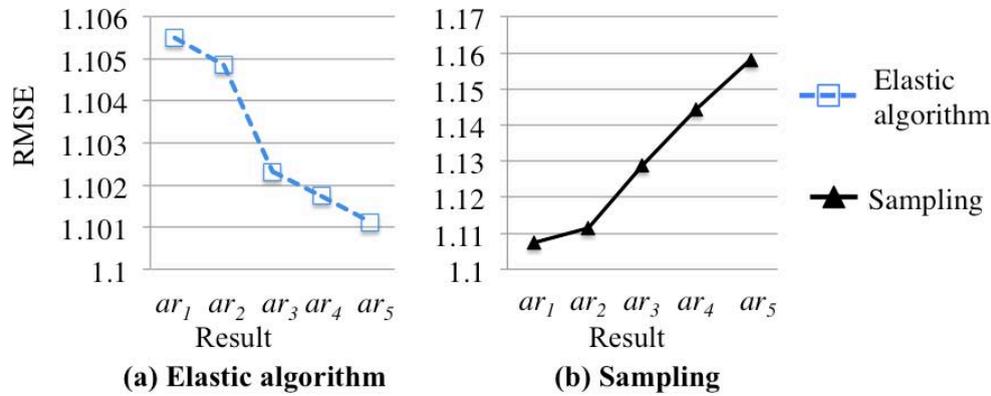

**(a) Elastic algorithm**          **(b) Sampling**

Figure 6.18: Comparison of RMSEs using the elastic and *Sampling* algorithms.

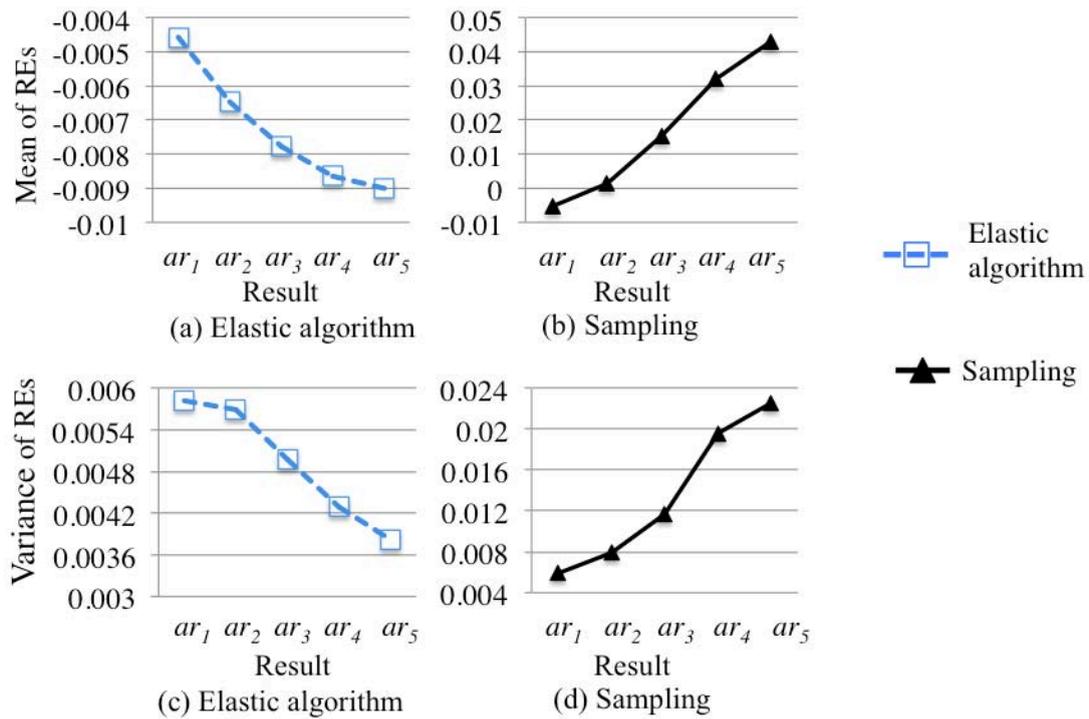

(a) Elastic algorithm          (b) Sampling

(c) Elastic algorithm          (d) Sampling

Figure 6.19: Comparison of mean and variance of REs using elastic and *Sampling* algorithms.

# Chapter 7

# Conclusions

## 7.1  Contributions

Cloud computing has emerged as a cost-effective means of delivering metered compute resources. It supports a pay-as-you-go model of computation in which application owners (cloud consumers) pay only for the resources used. Within this context, managing resource elasticity has become an active topic of investigation in the research community. A major focus of this research has been investigating various approaches to support the dynamic provision of resources so as to match computational demands. In this thesis, we explored cloud elasticity management and studied techniques to the need for elasticity of cloud services at both application and algorithm levels. The concrete contributions of this thesis are as follows:

**Application level elasticity.**

- A prototype of an intelligent scaling platform was implemented as a service on the IC Cloud workstation to support the automatic scaling of cloud applications. In this platform, we presented extensions to the TOSCA framework to enable platform-independent specification of cloud applications, and provided a list of service components to support the dynamic scaling of cloud applications.

- We proposed an elastic scaling approach that makes use of cost-aware criteria to detect and analyse the bottlenecks within multi-tier cloud-based applications. We presented an adaptive scaling algorithm that reduces the costs incurred by consumers of cloud infrastructure services by allowing them to scale their applications only at bottleneck tiers. We also studied the approach that operates fine-grained scaling at the resource level itself (CPUs, memory, I/O, etc.) in addition to VM-level scaling.

**Algorithm level elasticity**

- We introduced the concept of elastic algorithms for cloud computing. We described a class of such algorithms that work by generating successive approximate results over large datasets and discussed their desirable properties. We also provided a formal definition of algorithmic elasticity.

- We proposed a generic approach to guaranteeing quality monotonicity for data mining in a pay-as-you-go computing model. This quality monotonicity enables an elastic mining algorithm to produce approximate results whose quality, based on some metric, improves as the allocated time budget increases. We presented the two basic components of such algorithms: the coding component and the mining component. Furthermore, we defined the entropy monotonicity property that the coding component of any elastic algorithm must satisfy in order to support the quality monotonicity of the algorithm.

- We presented two detailed case studies in which elastic kNN classification and CF algorithms were designed, and theoretical proofs that their coding components satisfy the property of entropy monotonicity were provided. Moreover, extensive experiments on and comparisons of both algorithms were performed. The experimental results showed that the two elastic algorithms indeed guarantee the steady increase of quality metrics as time budget increases. The experimental comparison results further indicate that these algorithms outperform the existing time-adaptive algorithms by producing results with better qualities in most cases.

## 7.2   Future Work

Our work in this thesis focused on investigating the foundations of enabling cost-effective elasticity management in cloud computing. We plan to investigate the development of a more general framework for supporting such elasticity across a wider class of applications. Developing such a framework requires the investigation of the different factors that affect elasticity, including: the type of computation conducted; its required performance/quality properties; the budget constraints; the cost models used; and the general strategy for allocating resources. Making full use of such a framework, however, requires the addressing of a number of key challenges that we summarise below.

**Application level elasticity**

- *Cost-aware criteria*: We plan to study the cost-aware criteria and make the cost function more comprehensive. In the current work, the cost model assumed that all VMs had fixed costs per unit of time. The problems with this resource cost model are compounded when considering different cloud pricing schemes. Most cloud providers offer a fixed pricing scheme that guarantees access to resources at a fixed price. Providers such as Amazon EC2 [2] also offer a spot pricing scheme which allows application owners to use the IaaS provider's available excess resources at a significantly lower price, usually 30% of the declared fixed price. This spot price is dynamically updated according to the current supply of excess resources and the demand for resources from application owners. An application owner can bid for a spot machine instance and get this instance once the offer price is equal to, or less than, their bid price. However, under this scheme, infrastructure providers reserve the right to instantly terminate the application owner's access without notice once the offer price is greater than the bid price. Suitable scaling approaches therefore need to be developed to make effective use of the spot price model.

- *Power-aware cost function*: Another good extension of this work would be to investigate the power-aware cost function by considering servers' resource utilisation and power consumption, thus enabling cost-aware scaling in the context of power-aware resource management.

- *Multiple classes of requests*: We plan to extend our CAS algorithm to support multiple classes of requests, thus enabling the scaling up and down of cloud applications to meet the time requirements of different classes of users. For example, an e-commerce site could divide its incoming requests into three classes: requests making financial transactions, forming the *high priority* class; requests making product inquiries, forming the *medium priority* class; and browsing requests, forming the *low priority* class. By simply extending the FIFO queueing discipline applied in the M/M/1 queueing systems to a priority queueing discipline, the CAS algorithm can preferentially process requests from higher priority classes. Specifically, the idea of priority queueing is to divide requests into multiple FIFO queues in a queueing system. Each queue maintains requests from one class and has the same priority as these requests. Within each queue, requests are still managed using the FIFO queueing discipline. In this queueing system, the server always tries to process requests from the highest-priority and non-empty queue, thus guaranteeing shorter response times for higher-priority requests.

- *Other types of applications*: We have only considered transaction-based applications that execute at a single cloud provider. In these applications, computations are based on conducting a fixed set of transactions serving an end user browsing an e-commerce site. The desired performance requirement was to maintain the QoS of the application (measured as response time for each transaction) when serving multiple users. In future work, we will consider more complex applications such as high-performance scientific applications that can be scheduled, or coordinated, across different cloud IaaS providers: addressing their requirements would be another interesting avenue for extending this research.

- *Scaling multiple applications*: We have described scaling techniques in the context of a single application. This single-application requirement could be relaxed, and scaling in the context of multiple applications belonging to the same application owner could be investigated. This is appealing for two reasons. First, coordinating resources within one application owner's applications does not affect other application owners' applications. Second, the scaling up or down of resources can be completed within milliseconds and this approach could enable real-time scheduling to allow multiple applications to meet their QoS requirements simultaneously. However, further investigation is needed to consider how resources can be scheduled between applications with different QoS requirements in this situation.

**Algorithm level elasticity**

- *Scheduling elastic algorithms under budget and deadline constraints*: In this thesis we did not investigate the problems associated with scheduling multiple elastic algorithms to minimise their total execution costs while meeting user deadlines. We also did not investigate how to deal with resource price fluctuations in dynamic pricing schemes. We note here that traditional resource scheduling algorithms are designed to produce the shortest possible execution schedule for a program with the resources available, assume that idle resources are free and that the price of resources does not change with time, and have no notion of result quality. Thus, in a pay-as-you-go paradigm, new scheduling algorithms need to be designed to take into consideration the properties of both elastic algorithms and cloud environments. Our work provides a foundation for studying these problems and for investigating different strategies and algorithms. For example, one user may be more inclined to settle for an early result with adequate quality to save their available budget rather than waste it on diminishing quality returns. In a spot price model, another user may be willing to bid for resources at a higher price during earlier iterations to ensure that the computation produces an acceptable result before a deadline, and then to bid for resources at a much lower price for extra improvements. Investigating such different scenarios and strategies is an active focus of our current research, and we believe it could open a whole new area of research into scheduling computations in cloud environments.

- *Developing a reusable elastic algorithm development environment*: In this thesis, we demonstrated how to use hierarchical data structures (R-trees) with varying granularities to support the development of an elastic algorithm. In practice, developing and using such data structures from scratch may be beyond the capabilities of the average programmer. Going mainstream with this approach requires the development of a reusable algorithm development environment that provides programmers with a variety of data structures, programming libraries, and associated tools that simplify the development of elastic algorithms. It would also require developing offline modelling tools and real-time performance monitoring that would support application owners in profiling the behaviour of their programs, making investment decisions in real time, and evaluating the practicality of the proposed approach.

- *Applicability to wider problems*: The approach presented in this thesis is, so far, a theoretical framework, and is well suited to applications whose owners are willing to settle for approximate answers if the cost of generating the full results exceeds their available computation budget. As discussed in the thesis, this approach naturally applies to a wide range of domains, including numerical, scientific, and engineering computations, statistical estimation and prediction in data mining applications, heuristic search applications, database query processing applications, and multimedia applications. However, many other applications may not lend themselves as readily to such a paradigm. For example, a traditional payroll application that calculates and transfers employees' pay may not fit the framework at first glance: paying half the employees only, or approximating the salary of employees, could have detrimental consequences on the business itself. Even for such a payroll application, one can see that the first result of the program must perform a mandatory task with a minimum acceptable quality. One can also see that further refinement iterations can produce associated management reports at different levels of granularity and quality. It would be interesting to investigate how the elastic algorithm approach can be applied effectively in such applications.